# PPT : New Low Complexity Deterministic Primality Tests Leveraging Explicit and Implicit Non-Residues

## A Set of Three Companion Manuscripts

**PART/Article 1 :** **Introducing Three Main New Primality Conjectures:**
**Phatak's Baseline Primality (PBP) Conjecture** ,
and its extensions to
**Phatak's Generalized Primality Conjecture (PGPC)** ,
and
**Furthermost Generalized Primality Conjecture (FGPC)** ,
**and New Fast Primailty Testing Algorithms Based on the Conjectures and other results.**

**PART/Article 2 :** **Substantial Experimental Data and Evidence**[1]

**PART/Article 3 :** **Analytic Proofs of Baseline Primality Conjecture for Special Cases**

**Dhananjay Phatak** **(phatak@umbc.edu)**

and

Alan T. Sherman[2] and Steven D. Houston and Andrew Henry
(CSEE Dept. UMBC, 1000 Hilltop Circle, Baltimore, MD 21250, U.S.A.)

---

[1] **No counter example has been found**

[2] Phatak and Sherman are affiliated with the UMBC Cyber Defense Laboratory (CDL) directed by Prof. Alan T. Sherman

**First identification of the Baseline Primality Conjecture @ $\approx 15^{\text{th}}$ March 2018**

**First identification of the Generalized Primality Conjecture @ $\approx 10^{\text{th}}$ June 2019**

**Last document revision date  (time-stamp) = August 19, 2019**

---

**Color convention used in (the PDF version) of this document :**

All clickable hyper-links to external web-sites are brown : For example :
**G. E. Pinch's excellent data-site that lists of all Carmichael numbers** $< 10^{(18)}$ .

clickable hyper-links to references cited appear in magenta. Ex : G.E. Pinch's web-site mentioned above is also accessible via reference [1]

All other jumps within the document appear in darkish-green color. These include Links to :

Equations by the number : For example, the expression for `BCC` is specified in Equation (11)   ;

Links to Tables, Figures, and Sections or other arbitrary hyper-targets.
Examples: Table 2 ; Figure 4 ; and **Phatak's Baseline Primality Conjecture (PBPC)** .

Following the usual web document conventions, most of the hyperlinks (except those to the references cited) are underlined.

**Notation and symbols used**

For the most part the symbols are the same as those used in the literature. For the sake of clarity, we summarize those below.

| | | |
|---:|:---:|:---|
| **NUT** | $\equiv$ | Number Under Test (for primality) = input $N$ |
| s.t. | $\equiv$ | such that |
| w.r.t. | $\equiv$ | with respect to |
| l.h.s.  or  (LHS) | $\equiv$ | left hand side |
| r.h.s.  or  (LHS) | $\equiv$ | right hand side |
| gcd | $\equiv$ | greatest common divisor |
| lcm | $\equiv$ | least common multiple |
| lof($z$) | $\equiv$ | the Largest Odd Factor of integer $z$ |
| $\therefore$ $\longleftrightarrow$ | | therefore |
| $\because$ $\longleftrightarrow$ | | because (or since) |
| $x \Rightarrow y$ | $\equiv$ | One-way implication : if $x$ is true then $y$ is true |
| $x \Longleftrightarrow y$ | $\equiv$ | Both-way implication $\equiv$ **iff**, i.e., $x \Rightarrow y$ AND $y \Rightarrow x$ |
| $\lfloor x \rfloor$ | $\equiv$ | floor($x$) = largest integer $\leq x$ $\equiv$ round-toward $-\infty$ |
| $\lceil x \rceil$ | $\equiv$ | ceil($x$) = smallest integer $\geq x$ $\equiv$ round-toward $+\infty$ |
| $\stackrel{\Delta}{=}$ | $\equiv$ | Definition of a function of variable |
| $\square$ | $\equiv$ | end of a proof; or a part thereof; or an analytic argument/derivation. |

---

| | | |
|---:|:---:|:---|
| **QNR** | $\equiv$ | Quadratic Non Residue |
| **QR** | $\equiv$ | Quadratic Residue |
| **NR** | $\equiv$ | Non Residue of any order, i.e., a value that does not exist as an integer modulo-$N$.  for example, root of a polynomial that has no integer solutions modulo-$N$ |
| **Jacobi_Symbol**$(m,n)$ or **JS**$(m,n)$ | $\equiv$ | the **Jacobi_Symbol** of integer $m$, w.r.t. integer $n = \left(\frac{m}{n}\right)$ <br><br> Note that **JS** $\in \{-1, 0, 1\}$; the **Jacobi_Symbol** is evaluated using the quadratic-reciprocity-results and extended gcd-like computations; without factoring either $m$ or $n$. |
| $\left({}^N C_k\right)$ | $\equiv$ | Binomial Coefficient "N_choose_k" = $\frac{N \times (N-1) \times \cdots \times (N-k+1)}{1 \times 2 \times \cdots \times k}$ |
| Non trivial binomial coeff. | $\equiv$ | any binomial coefficient whose value is $> 1$ $\equiv$ $\left({}^N C_k\right)$ where $k \neq 0$ and $k \neq N$ |

$\texttt{ECC}$   $\equiv$   a function of two variables $q$ and $N$ that indicates whether
     $q$ and $N$ satisfy the "Euler-Criterion"; (hence the name "Euler-Criterion Check"
     and the abbreviation $\texttt{ECC}$ ).
     A formulae/expression for $\texttt{ECC}$ is specified below, in Eqn. (5) .

$\texttt{BCC}$   $\equiv$   Analogous function for the "Binomial-Congruence Check" (see Eqn. (11) )

$\texttt{modexp}(x, e, N)$   $\equiv$   $x^e \bmod N$   ; wherein ; the modular exponentiation is done
     via repeated square–and–reduce\_modulo\_$N$ operations.

$\lg N$   $\triangleq$   $\log_2 N$

$O(\cdot)$   $\triangleq$   "order of" metric for complexity of algorithms.

$\text{polylog}(x)$   $\triangleq$   A polynomial of the argument = $(\log x)$
     Note that throughout this set of articles the argument of
     $\text{polylog}(\cdot)$ is always $(\log N)$, which is
     therefore tantamount to a polynomial of the argument = $(\log(\log N))$

# OVERALL ABSTRACT FOR THE SET OF 3 ARTICLES

In this set of three companion manuscripts/articles, we unveil our new results on primality testing and reveal new primality testing algorithms enabled by those results.

The results have been classified (and referred to) as lemmas/corollaries/claims whenever we have complete analytic proof(s); otherwise the results are introduced as conjectures.

In **Part/Article 1** , we start with the **Baseline Primality Conjecture (PBPC)** which enables deterministic primality detection with a low complexity = $O\left((\log N)^2 \left(\text{polylog}(\log N)\right)\right)$ ; when an explicit value of a **Quadratic NON Residue (QNR)** modulo-$N$ is available (which happens to be the case for an overwhelming majority = $\left(\frac{11}{12}\right)$ = **91.67 %** of all odd integers).

We then demonstrate **Primality Lemma 1** , which reveals close connections between the state-of-the-art Miller-Rabin method and the renowned Euler-Criterion. This Lemma, together with the Baseline Primality Conjecture enables a synergistic fusion of Miller-Rabin iterations and our method(s), resulting in hybrid algorithms that are substantially better than their components.

Next, we illustrate how the requirement of an explicit value of a **QNR** can be circumvented by using relations of the form: **Polynomial**$(x)$ mod $N \equiv 0$ ;
whose solutions implicitly specify **NON−RESIDUEs** modulo-$N$.
We then develop a method to derive low-degree canonical polynomials that together guarantee implicit **NON−RESIDUEs** modulo-$N$ ; which along with the Generalized Primality Conjectures enable algorithms that achieve a
worst case deterministic polynomial complexity = $O\left((\log N)^3 \left(\text{polylog}(\log N)\right)\right)$ ; unconditionally, for any/all values of $N$.

In **Part/Article 2** , we present substantial experimental data that corroborate all the conjectures[3].
For a striking/interesting example, wherein a combination of well-known state-of-the-art probabilistic primality tests applied together fail to detect a composite (which is easily/instantly detected correctly to be a composite number by the Baseline Primality Conjecture), see **Section 23 in Part/Article 2.**

The data show that probabilistic primality tests in software platforms such as Maple are now extremely robust. I have not yet found a single composite number $N$ that fools Maple's Probabilistic primality tests; but gets correctly identified as a composite by our deterministic tests. Construction of (or a search for) such integers (assuming that those exist) is an interesting open problem which we plan to investigate in the future.

Finally in **Part/Article 3** , we present analytic proof(s) of the Baseline Primality Conjecture that we have been able to complete for some special cases (i.e. for particular types of inputs $N$).

We are optimistic that full analytic proofs of all conjectures introduced herein will be generated in the near future. That will be a big step toward moving the problem of primality detection into the category of problems that have been "solved and retired".

---

[3] **As repeatedly mentioned throughout the entire set of articles, No counter example has been found.**

# Contents

## Part/Article 1

# Part/Article 2

# Part/Article 3

# List of Figures

# List of Tables

# PPT : New Low Complexity Deterministic Primality Tests Leveraging Explicit and Implicit Non-Residues

## A Set of Three Companion Manuscripts

**PART/Article 1 :** **Introducing Three Main New Primality Conjectures:**
**Phatak's Baseline Primality (PBP) Conjecture** ,
**and its extensions to**
**Phatak's Generalized Primality Conjecture (PGPC)** ,
**and**
**Furthermost Generalized Primality  Conjecture (FGPC)** ,
**and  New Fast Primailty Testing Algorithms Based on the Conjectures and other results.**

**Dhananjay Phatak     (phatak@umbc.edu)**

and

Alan T. Sherman   and   Steven D. Houston   and   Andrew Henry
(CSEE  Dept. UMBC, 1000 Hilltop Circle, Baltimore, MD 21250, U.S.A.)

First identification of the Baseline Primality Conjecture @ $\approx 15^{th}$ March 2018
First identification of the Generalized Primality Conjecture @ $\approx 10^{th}$ June 2019

Last document revision date  (time-stamp) = August 19, 2019

## § Section 1 : Introduction[4]

The renowned "Euler Criterion" [3, 4, 5, 6] stipulates that

If $N$ is a prime number; then $\forall$ integers $a$ that are co-prime w.r.t. $N$;

$$
\begin{aligned}
\texttt{Jacobi\_Symbol}(a,N) &\triangleq \texttt{Jacobi-Symbol of} \quad a \quad \texttt{w.r.t.} \quad N \\
&= \texttt{the Legendre-Symbol of} \quad a \quad \texttt{w.r.t.} \quad N \\
&= a^{(\frac{N-1}{2})} \mod N
\end{aligned}
\tag{1}
$$

The well known Jacobi–Symbol [7] is a generalization of the Legendre–Symbol and can be computed extremely fast via extended GCD like computations, coupled with Quadratic Reciprocity results [7, 8]. For example, Maple's [9] built-in function "$\texttt{Jacobi}(x,y)$" from the "$\texttt{numtheory}$" tool-box can evaluate the the Jacobi–Symbol of integer $x$ w.r.t. integer $y$ extremely fast (in a time interval on the order of human reaction time $\approx$ 10s of milli-seconds; even when the operands are 100,000 decimal digits long).

In general

$$
\texttt{Jacobi\_Symbol}(x,y) = \begin{cases} 0 & \text{if } \gcd(x,y) > 1 \\ \pm 1 & \text{otherwise} \end{cases}
\tag{2}
$$

If the Jacobi–Symbol of any integer say $q$ w.r.t. $N$ evaluates to $0$; then $q$ and $N$ share a common factor $> 1$ and therefore $N$ is not a prime number.
**In this case, we refer to "$q$" as the "Jacobi-Witness" to the composite nature of $N$**

This test: ($\texttt{Jacobi\_Symbol}(q,N) \overset{?}{=} 0$) is an important step in our new primality testing algorithms
(the " **PP–Testing Algorithm using explicit $\texttt{QNR}$ (PPTA\_EQNR)** " ; as well as the
" **PP–Testing Algorithm using implicit $\texttt{NON-RESIDUE}$ (PPTA\_INR)** " ;

and ends up resolving (i.e., correctly detecting) a non trivial number of composite input values.

To analyze the remaining cases; we therefore assume that

$$
\texttt{Jacobi\_Symbol}(q,N) \neq 0 \quad \Rightarrow \texttt{Jacobi\_Symbol}(q,N) = \pm 1
\tag{3}
$$

The contra-positive of the Euler Criterion is :

**If** $\left[ \exists \text{ int } z \text{ s.t. } \texttt{Jacobi\_Symbol}(z,N) \neq \left( z^{(\frac{N-1}{2})} \mod N \right) \right]$ **then** $N$ is not a prime (4)

and that case, $z$ is said to be an "Euler-Witness (EW)" to the composite-ness of the NUT $N$.

---

[4] We provide minimal background material to facilitate the illustration of the main results as soon as possible. Accordingly, this section is a short prelude to the main results ; it is not a comprehensive or exhaustive survey of the literature.

For brevity and clarity, we define a function of the two arguments/variables of interest; $q$ and $N$ as:

$$\texttt{ECC}(q,N) \quad \triangleq \quad \left[ q^{\left(\frac{N-1}{2}\right)} \mod N \right] - \texttt{Jacobi\_Symbol}(q,N) \tag{5}$$

so that

$$\texttt{ECC}(q,N) \begin{cases} = 0 & \text{if the Euler Criterion is satisfied;} \quad \therefore \ N \text{ can be prime or composite} \\ \neq 0 & \text{if the Euler Criterion is NOT satisfied;} \quad \therefore \ N \text{ cannot be prime} \end{cases} \tag{6}$$

If any integer $q$ satisfies the Euler Criterion modulo $N$; then it must also satisfy
Fermat's Little Theorem (FLT) modulo-$N$. To verify this fact,
square Eqn. ([1]) , i.e., square both sides of that equation ; to obtain

$$\left[ q^{\left(\frac{N-1}{2}\right)} \right]^2 = \left[ \texttt{Jacobi\_Symbol}(q,N) \right]^2 = \left[ \pm 1 \right]^2 \quad \Rightarrow \quad q^{(n-1)} = 1 \mod N \tag{7}$$

However, the converse is not true : in other words
even if an integer $k$ does satisfy FLT modulo $N$; it may or may not satisfy the Euler Criterion.

**For example let** $q = 2$ **and** $N = 341$
**then it can be verified that**

$2^{(340)} \mod 341 = 1 \ \Rightarrow$ **Fermat's Little Theorem (FLT) is satisfied**
    **however ;**

$\left. \begin{array}{l} 2^{(170)} \mod 341 = +1 \ \neq \ \texttt{Jacobi\_Symbol}(2,341) = -1 \\ \textbf{or equivalently ;} \qquad \texttt{ECC}(q,N) \neq 0 \\ \quad \Rightarrow \ \textbf{the Euler Criterion is not satisfied} \\ \quad \Rightarrow \ N = 341 \ \textbf{must be a composite} \end{array} \right\} \tag{8}$

**Thus, the Euler Criterion is more discriminating ; it can detect many composites that satisfy/fool FLT.**
**All of our experimental data show that out of all paths/checks/tests within the PPT algorithms; it is the**
**"Euler Criterion Check" that detects the majority of composites.**

Likewise, it is a well known fact that if $N$ is a prime number, then the nontrivial binomial coefficients satisfy the condition :

$$\left(^{N}\boldsymbol{C}_{k}\right) \mod N = 0 \quad \text{for} \quad 1 \leq k \leq N-1 \tag{9}$$

The preceding constraints lead to an extremely important and useful identity :

$$\textbf{If} \quad N \text{ is a prime number} \quad \textbf{then} \quad (1+x)^{N} \mod N = 1 + \left((x)^{N} \mod N\right) \tag{10}$$

which refer to as the "Modular Binomial Expansion Congruence (`MBEC`)" [5]

Note that the `MBEC` holds for arbitrary entities $x$ ; as long as multiplication is commutative and associative in the underlying domain (which is in turn necessary for the binomial expansion to be valid).

For example, Relation ( 10 ) holds even when the indeterminate "$x$" is a scalar real or complex number ; or it is a polynomial with coefficients modulo-$N$ ; or even when $x$ is an arbitrary square matrix and "1" is substituted by the Identity matrix of the same dimension as $x$. Note that the product of the Identity matrix with any other matrix is commutative (as a matter of fact the first/early versions of the PP Conjecture as well as the primality testing algorithms evolved via Matrices, see **Section 15** for further details.)

Analogous to the function `ECC` ; we define a second function `BCC(`$x, N$`)` as follows :

$$\texttt{BCC}(x,N) \quad \overset{\Delta}{=} \quad \left[(1+x)^{N} - 1 - (x)^{N}\right] \mod N \tag{11}$$

Then, the contra-positive of Relation ( 10 ) (i.e., the `MBEC`) can be stated in terms of `BCC(`$x, N$`)` as the following fact :

$$\textbf{If} \quad \texttt{BCC}(x,N) \neq 0 \mod N \quad \textbf{then} \quad N \textbf{ must be a composite} \tag{12}$$

and in that case we refer to $x$ to be a "`Binomial-Witness`" to the compositeness of $N$.

Analogous to Relations ( 6 ), the corresponding relations for `BCC` are :

$$\texttt{BCC}(x,N) \begin{cases} = 0 & \text{if the Binomial Congruence Check is satisfied} \\ & \quad \Rightarrow \quad N \text{ can be either prime or composite} \\ \\ \neq 0 & \text{if the Binomial Congruence Check is NOT satisfied} \\ & \quad \Rightarrow \quad N \text{ cannot be a prime, it must be a composite} \\ & \text{In this case "}x\text{ " is a "Binomial Witness" to the compositeness of } N \end{cases} \tag{13}$$

With all the relevant functions defined, we dive into the first main result in the next section.

---

[5]This remarkable identity happens to be the starting point of the now famous AKS deterministic primality test [10].

**§   Section 2 :   New number theoretic result on which the ensuing algorithms are based**

<u>**Phatak's Baseline Primality Conjecture (PBPC)**</u> :    Given a positive integer   $N$ (to be tested for primality); suppose that we find an integer  $q$ ; that together with $N$ satisfies the following conditions :

**C–1 :**    $N > 1$  is an  odd  integer and is not a perfect square of any other integer.

**C–2 :**   The `Jacobi-Symbol` of  $q$  w.r.t  $N$  $\overset{\triangle}{=}$  `Jacobi_Symbol(`$q, N$`)` $= -1 \Rightarrow q$ is a `QNR`  modulo-$N$

Note that $q$ can be any  `QNR`  (Quadratic Non Residue) ; except one restriction/exclusion :

**C–3 :**   $q \neq -1 \bmod N$
In other words, only the largest value in  $Z_N^* = (N-1)$   is not acceptable as a  `QNR`
Any other integer in the closed interval   $[2, (N-2)]$   that is a  `QNR` modulo-$N$ works.

**C–4 :**   $q$  satisfies the  **Euler Criterion**  modulo-$N$

$$q^{\left(\frac{N-1}{2}\right)}  \bmod N = \texttt{Jacobi\_Symbol(}q, N\texttt{)} = -1  \bmod N \tag{14}$$

**C–5 :**   $\sqrt{q}$  also satisfies the Modular Binomial Expansion Congruence (`MBEC`) :

$$(1 + \sqrt{q})^N  \bmod  N  =  1 + \left[ (\sqrt{q})^N  \bmod  N \right] \tag{15}$$

**The claim is that if all of the above conditions are satisfied , then $N$ must be a prime number, it cannot be a composite.**

In other words :
<u>**NO COMPOSITE INTEGER CAN (SIMULTANEOUSLY) SATISFY ALL CONDITIONS C–1 thru C–5.**</u>

Numerical examples, i.e., specific numbers $N$, that demonstrate that one or more of the conditions   <u>C–3</u>   thru  <u>C–5</u>   are necessary to correctly detect whether $N$ is a prime or not,   are on   **article 1–page 26**   and   **§ Section 19 in Part/Article 2**

Note that the above claim is **independent** of
**1 :** whether or not $q$ is the smallest  `QNR` modulo-$N$
**2 :** whether or not $q$ is a prime number
**3 :** whether or not $q$ is square-free
**4 :** whether or not $q < \sqrt{N}$ or any other range restriction (except the one single exclusion $q \neq -1$).
**5 :** $q > N$ also works (this fact could be useful to quickly find a `QNR` ).

This independence allows us to cleanly separate the computational issue of
"how to most efficiently/quickly find a  `QNR`  $q$" from the rest of the issues.  Therefore, we can fully leverage the new result in the PPT Algorithm (using explicit  `QNR`  value $q$) specified in the next section.

It should be noted that if $N$ is a prime number; then it is trivial to prove [11] that all the conditions **C–1** thru **C–5** are satisfied by $N$ and any/every integer $q$ which is a `QNR` modulo-$N$.
(including $q = -1$ ; if $-1$ happens to be a `QNR` modulo-$N$; which in turn happens if the prime number $N$ is of the form $(4i+3)$   ).

In other words; the proofs in one direction (typically called as the "forward" direction, i.e.,
if $N$ is a prime number then ...)
turn out to be trivial/straightforward.

The main highlight of the new result is that it proposes conditions that are
sufficient in the   REVERSE   direction.

§ Section 3 : Specification of Phatak's Primality Testing Algorithm using Explicit QNR (PPTA_EQNR)

line 1⟩    **Algorithm   PPTA_EQNR** ($N :: \texttt{posint}$)

line 2⟩    `local s , q ;`

line 3⟩    **Begin_Algorithm**

line 4⟩    **if** $(N \bmod 2 == 0)$ **then**        **return   Composite ;**   **fi ;**

line 5⟩    **if** $\big[ (N \bmod 8 == 3)$ **or** $(N \bmod 8 == 5) \big]$ **then**        // $q = 2$ is a QNR ; no need to search line 6⟩        $q := 2$ ;

line 7⟩    **else**                        // 2 is NOT a QNR

line 8⟩        **if** $(N \bmod 4 == 3)$ **then**        // $(-1)$ is a QNR $\Rightarrow 2 \times (-1) = -2$ is a QNR ; no need to search line 9⟩            $q := -2$ ;                // equivalent to setting $q := (N-2)$ ;

line 10⟩       **else**                        // need to search for a QNR ∴ rule-out perfect-squares line 11⟩            $s := \big\lfloor \sqrt{N} \big\rfloor$        // exact maple command is $s := \texttt{floor(evalf[length(N)](sqrt(N)))}$ ;

line 12⟩            **if** $(s^2 == N)$ **or** $\big( (s+1)^2 == N \big)$ **then** **return   Composite ;** **fi ;**

line 13⟩            $(\texttt{detection\_done\_flag} , q) := \texttt{Find\_QNR}(N , s)$ ;

line 14⟩            **if** $(\texttt{detection\_done\_flag} > 0)$ **then**    **return   Composite ;**    **fi ;**

line 15⟩        **fi ;**

line 16⟩    **fi ;**

line 17⟩    // If the algorithm has not already exited by this point; then it has found a QNR $= q$

line 18⟩    // ∴ Euler Criterion Check followed by (if needed) one Binomial Congruence Check complete the detection

line 19⟩    $\texttt{ECC} := \Big[ \texttt{modexp}\Big( q , \big(\tfrac{N-1}{2}\big) , N \Big) + 1 \Big] \bmod N$   ;

line 20⟩    **if** $(\texttt{ECC} \neq 0)$ **then**            // Euler Criterion is violated $\Rightarrow N$ must be composite; $q$ is the Euler-witness

line 21⟩            **return   Composite ;**

line 22⟩    **else**        // Euler Criterion is satisfied line 23⟩            // ∴ Binomial Congruence Check is the last and " **iff** " **determiner of primality assuming PBPC holds**

line 24⟩        **if (** $\texttt{BCC}(\sqrt{q} , N) \neq 0$ **) then**

line 25⟩            **return   Composite ;**

line 26⟩        **else**

line 27⟩            **return   Prime ;**

line 28⟩        **fi ;**

line 29⟩    **fi ;**

line 30⟩    **End_Algorithm**

Next, we specify the pseudo-code for the `Find_QNR` procedure called from within the algorithm. For the sake of clarity; we present the simplest version of the `Find_QNR`($N$, $k$) procedure/function that appears in the `PPTA_EQNR` algorithm

line 1⟩    **Procedure    Find_QNR** ($N$::posint , Iter_Limit::posint)

line 2⟩        // If no errors occur, then the procedure returns 2 parameters :

line 3⟩        // 1 : an integer set to a value $> 0$ if Jacobi-Symbol = 0 $\Rightarrow$ $N$ is a composite ;    set to 0 otherwise line 4⟩        // 2 : an integer $p$ = the first argument of the Jacobi-Symbol evaluation function in the final iteration line 5⟩        local i, j, p ;

line 6⟩        $p := 2$ ;

line 7⟩        **for** $i$ **from** $1$  **while** ($i <$ Iter_limit)  **do**

line 8⟩            $p := \text{nextprime}(p)$ ;

line 9⟩            $j := \text{Jacobi}(p, N)$ ;

line 10⟩            **if**  ($j == 0$)  **then**

line 11⟩                **return    (1 , p) ;**

line 12⟩            **elif**  ($j == -1$)  **then**

line 13⟩                **return    (0 , p) ;**

line 14⟩            **fi ;**

line 15⟩        **od ;**

line 16⟩        **Error("Did not find a QNR");**

line 17⟩        **return    NULL ;**

line 18⟩    **End_Procedure**

This completes the specification of the Algorithm and all the procedures it invokes.

Enhanced versions of the PPTA are illustrated in  **Section 7** .

**<u>Remark 3.1</u> :**

Note that in the modular exponentiations as well as all other computations; "$\sqrt{q}$" is treated as a "symbol" and any coefficients of this symbol as well as all other integers are always reduced modulo-$N$. Therefore the output of the function `BCC(`$\sqrt{q}$, $N$`)` is always an expression of the form

$$\text{BCC(}\sqrt{q}, N\text{)} = A + B(\sqrt{q}) \qquad \text{wherein ;} \; A \text{ and } B \text{ are modulo-}N \text{ integers} \tag{16}$$

so that

$$\text{BCC(}\sqrt{q}, N\text{)} \begin{cases} = 0 & \text{If} \quad (A = 0 \mod N) \text{ AND } (B = 0 \mod N) \\ \neq 0 & \text{otherwise} \end{cases} \tag{17}$$

**Remark 3.2 :**

The Binomial Congruence Check can also be implemented in the following equivalent manner :

Define a Divisor Polynomial of variable/indeterminate $x$ as $\quad \mathcal{D}(x) = \mathcal{D}_q(x) = (x^2 - q)$ $\qquad$ (18)

and then evaluate the polynomial remainder

$$\mathcal{R}(x) = \Big[ \Big( (1+x)^N - 1 - (x)^N \Big) \mod \mathcal{D}(x) \Big] \mod N \qquad (19)$$

so that for the specific divisor polynomial $\mathcal{D}_q(x) = (x^2 - q)$ ; the remainder takes the form

$$\mathcal{R}(x) = \mathcal{A}_1 + \mathcal{B}_1(x) \quad \text{wherein ;} \quad \mathcal{A}_1 = A \mod N \quad \text{and} \quad \mathcal{B}_1 = B \mod N \qquad (20)$$

wherein ; $A$ and $B$ are defined in Equation ( 16 )

Note that the Euler Criterion also be re-expressed in terms of the same divisor polynomial $\mathcal{D}_q(x)$ that gets used in the **MBEC** check (specified in Eqn. (18) ) as follows :

If $N$ is a prime, then $\quad \Big[ x^{(N-1)} \mod (x^2 - q) \Big] \mod N = \text{\textbf{Jacobi\_Symbol}}(q, N) \qquad (21)$

§ **Section 4 :** **Proof of correctness of** PPTA_EQNR Algorithm **assuming that** PBP Conjecture **is true**

Note that the Algorithm checks whether the input $N$ is a square (which subsumes all even powers) of some integer (say $K$), only if $N \bmod 8 = 1$. This is correct as demonstrated by the following well known fact:

**Basic Number Theory Background Fact 1 :** **Let** $N > 1$ **be an odd positive integer . Then,** $N$ **can be a square of some other integer only if** $N \bmod 8 = 1$

**Proof :** Suppose $N = K^2$

Since $N$ is an odd number; then $K$ must also be an odd number. Therefore, the remainder $r$ of $K$ w.r.t. 8 must be one of the four odd values less than 8; i.e.;

$r = K \bmod 8 \Rightarrow r \in \{1, 3, 5, 7\} \Rightarrow K = (8I + r)$ for some integer $I$.

Then it is straightforward to verify that

$N \bmod 8 = K^2 \bmod 8 = (8I + r)^2 \bmod 8 = r^2 \bmod 8 = 1$ for $r \in \{1, 3, 5, 7\}$ □

Accordingly, the check to screen-out a square is performed only if $N \bmod 8 = 1$; (within the scope of the "**else**" clause on `line 10 )"` in the algorithm specification on article 1−page 18 .

From that point on (i.e., from `line 13 )"` in the algorithm specification), the method simply finds a `QNR` $q$ and then checks if `ECC(`$q, N$`)` and `BCC(`$q, N$`)` are satisfied. If both are satisfied; then the number must be prime as per the PBP Conjecture. If any of those conditions is violated, then the number cannot be a prime; and this is exactly what the algorithm specifies; thereby demonstrating that the algorithm implements the checking of all conditions C−1 thru C−5 in the PBP Conjecture in a systematic and efficient manner (only when necessary) and then decides primality exactly as stated in the Conjecture. It therefore follows that if the PBP Conjecture is true; then the PPTA_EQNR Algorithm works correctly.

**Note that in the rest of Parts/Articles 1 and 2 ; it is assumed that the** first main Result (PBPC) **is true. As repeatedly mentioned throughout the document, no counter example has been found.**

## § Section 5 : **Complexity of PPTA_EQNR**

We first clarify that the phrase "`complexity of a computation`", means
"`the number of arithmetic and logic operations in a fixed/constant-radix number representation; that are required to perform that computation`".
We have frequently used the shorter term "computational effort" (or simply "cost") to refer to "the number of constant radix arithmetic/logical operations".

There are multiple distinct ways in which the baseline algorithm **PPTA_EQNR** can be substantially improved and those are illustrated starting from the next section (i.e. **Section 6** onward). To begin with, in this section, we first analyze the complexity of the simplest version of the PPTA specified on pages **article 1 – page 18** and **article 1 – page 19** .

Also, we deliberately separate the computational effort required
**AFTER a `QNR` is available as a modulo-$N$ integer** ;   from the worst case complexity of generating a `QNR` .

The main reason is that in later sections we build toward and unveil a generalized form of the primality conjecture that does not need explicit `QNR` (or higher order Non-Residue) value, and enables algorithms that achieve low complexity for any/all possible inputs $N$.

However, during a search of another canonical parameter (which is required by that generalized primality conjecture); if a `QNR` is encountered then we simply invoke the baseline conjecture as the preferred path of least complexity to quickly and efficiently complete the primality detection process.

It is therefore essential to have an accurate estimation of the complexity once a `QNR` is available.

## § Section 5.1 : **Complexity after an explicit `QNR` is available**

From the PPTA specification, it is clear that after a `QNR` $q \neq -1$ is available; the algorithm needs no more than 2 scalar modular exponentiations :
(1) To check whether the Euler Criterion is satisfied (i.e., to check whether `ECC(`$q,N$`) == 0`) ;   and if the Euler Criterion is satisfied ;
        then
(2) the second modular exponentiation with symbolic computation (using $\sqrt{q}$ as a "symbol")
to check the whether the modular Binomial Congruence is satisfied,
(i.e., to check whether `BCC(`$\sqrt{q},N$`) == 0`).

In other words,
**after a `QNR` is found; the number of modular exponentiations needed is at most "2"; which is a constant ;   or $O(1)$ ; independent of the size (or bit-length) of $N$** .

A modular exponentiation where the exponent is a positive integer $N$ requires exactly $\lceil \lg N \rceil$ iterations of a loop; wherein the operations performed within each iteration of the loop are a square followed by a remainder calculation and possibly another multiplication followed by another remainder evaluation. Using Montgomery's method and the FFT Multiply paradigm ; a (long word-length) multiplication and remaindering can be done with $O((\lg N)\lg\lg N)$ computational effort.

Therefore, The total complexity of 2 modular exponentiations to evaluate `ECC(`$q,N$`)` and `BCC(`$\sqrt{q},N$`)` is

$2\times$ (Number_of_iterations_per_modular_exponentiation $\times$ computations_per_ iteration) $=$

$$2\times(\lceil \lg N\rceil)\times O((\lg N)\lg\lg N) \quad \approx \quad O\left((\log N)^2\cdot\log\log N\right) \tag{22}$$

## § Section 5.2 : Complexity of generating explicit Quadratic (or higher order) Non Residue

A huge amount of literature exists on the topic of the construction of, or a search for a `QNR` (or more generally, an irreducible polynomial in arbitrary finite fields ).
See [12, 13, 14, 15, 16, 17, 18, 19, 20, 21, 22, 23, 24] and references therein for a sampling.

The literature demonstrates that if the Extended Riemann Hypothesis (ERH) is assumed to be true, then low complexity (non-trivially lower complexity than AKS) deterministic polynomial time algorithms exist to find/construct a Non Residue for a large number of input types/forms of the NUT-$N$.

However, the literature also indicates that there is no known deterministic polynomial cost method to construct an irreducible polynomial or equivalently, find a non-residue ; for any arbitrary composite integer $N$.

## § Section 5.3 : Worst Case Overall Complexity of Baseline Method

It is clear that the overall complexity is

Maximum $\left\{O\left((\lg N)^2\lg\lg N\right) \text{, complexity of generating a quadratic non-residue}\right\}$ (23)

As per the literature, the explicit determination of the smallest `QNR` for arbitrary composite $N$ is still an unsolved open problem and no deterministic polynomial cost method is known yet.

Nonetheless, for a comparison, in the AKS method [10] and its latest and fastest/most efficient derivatives/variants [6, 25, 26, 27, 28] ;  the number of modular exponentiations that need to be performed are on the order of
at least a 4-th degree polynomial of the bit-length of $N$; i.e.; $O\left([\lg N]^{(4+\epsilon)}\right)$ [6].

Each modular exponentiation in AKS has the same exponent $N$. Therefore the number of steps/iterations of the square-and-reduce loop (used to perform modular exponentiation) is $\lceil \lg N\rceil$.

Each iteration includes a square followed by polynomial remainder evaluation and potentially another polynomial multiplication followed by polynomial remainder evaluation.  The topic of polynomial multiplication, reciprocation and division over finite fields has been extensively analyzed in the literature (ex: see [29, 30, 31, 32, 33]). The results show that the square-and-remainder operations (on polynomials of degree $d$ with coefficients that are modulo-$N$ integers assuming that $d << N$) within each iteration of the loop can be performed with a

computational effort $\lesssim O\left((d \cdot \lg N)\,\text{polylog}(\log N)\right)$ $\qquad$ (24)

The divisor polynomial in the AKS algorithm is of the form
$\mathcal{D}(x) = (x^r - 1)$ where $r > (\lg N)^2$
Assuming that $r < 2(\lg N)^2$ (see [6, page 207] for details),
the total computational complexity of the best variants of the AKS method is

$$O((\lg N)^{4+\epsilon}) \times (\lg N) \times (r \lg N \,\text{polylog}(\log N)) \approx O\left((\log N)^{(8+\epsilon)} \cdot \text{polylog}(\log N)\right) \qquad (25)$$

The preceding comparison might appear unfair and premature because the AKS method is proven analytically and works in ALL cases. However, the main idea is just to get a quick assessment of the potentially how much improvement there could be, and how much slack is available (unless the methods turn out to be substantially faster than AKS derivatives, why bother ?).

Note that even if the computations required in the baseline conjecture are implemented as polynomial remainder evaluations, the computational effort is still the same,
because the DEGREE of our divisor polynomial $\mathcal{D}(x) = x^2 - q$ is a small fixed constant = 2 = $O(1)$ INDEPENDENT of the size of the number $N$. Therefore the overall complexity is

$$O(1) \times (\lg N) \times [O(1) \times O(\lg N \,\text{polylog}(\log N))] = O\left((\log N)^2 \cdot \text{polylog}(\log N)\right) \qquad (26)$$

In the following sub-section we give the reasons as to why the baseline conjecture is valuable.

§ **Section 5.4 : Why the Baseline Conjecture is extremely useful Nonetheless**

**Reason 1 :**     For a substantial majority of all odd integers, it is trivial to determine a `QNR` :

**5.1 :**   For 3 out of every 4 = 75% of all odd integers (all except those that satisfy $N \mod 8 = 1$); either $+2$ or $-2 \equiv (N-2)$ is a `QNR`.

**5.2 :**   The above condition can be trivially expanded to all odd integers except those satisfying

$$N \mod 24 = 1 \qquad\qquad (27)$$

In other words, out of the 25% of all odd integers not covered by $\pm 2$ being the `QNR`; 1/3rd are divisible by 3 and (additional/non-overlapping) 1/3rd have 3 as a `QNR`. Thus, the 75% coverage can be trivially extended to 11/12 ≈ 91.67% of all possible inputs being such that either 3 divides $N$ or a `QNR` is trivially available.

**5.3 :**   Moreover, the above condition can be further expanded[6] so that

$$\text{an explicit search for a } \texttt{QNR} \text{ is needed  only if }\quad N \mod 240 = 1 \qquad\qquad (28)$$

Thus, an overwhelming majority of all odd integers; i.e., $\frac{119}{120} \approx 99.16\%$ of all possible odd inputs do not require a search for a `QNR`; it is easy to generate a `QNR`.

In all these (strong majority) of cases, invoking the Baseline Primality Conjecture is the least complexity method of primality detection to the best of our knowledge.

**Reason 2 :**     Even when a search is needed, a substantial fraction of all integers in the interval $[2, N-2]$ are either `QNR`s or share a factor with $N$

**5.1 :**   The literature focuses mainly on the smallest `QNR` modulo arbitrary odd integer $N$. For our purpose, any `QNR` (except "$-1$") works irrespective of whether it is the smallest `QNR`, or whether it is a prime or square free etc.

**5.2 :**   In **Appendix 1** (In this manuscript, i.e., Part/Article 1), we have provided a simple proof to show that for any odd integer $N$ that is not a square of some other integer, the number of integers $a \in Z_N^*$ with Jacobi_Symbol$(a, N) = -1$
= number of integers with Jacobi_Symbol = +1
and therefore each of those values = $\frac{\phi(N)}{2}$ where $\phi(N)$ = the Euler Totient function of $N$. Therefore if a number is selected randomly in the interval $[2, N-2]$, the probability of hitting a `QNR` = $\frac{\phi(N)}{2(N-3)} > \frac{1}{2}\frac{\phi(N)}{N}$
which is substantial (especially in practice with the RSA crypto-system wherein $N$ is a product

---

[6]see **Section 8.1**  and [20] for details

of two large primes).

**Reason 3 :**    **Hidden `QNR`s :** In a large number of cases the overall `Jacobi_Symbol` of an integer $a \in [2, N-2]$ w.r.t. $N = +1$ ; even though multiple divisors of $a$ have their individual `Jacobi_Symbol`s $= -1$ (for example if an even power of a prime with `JS` $= -1$ divides $a$ ; and/or if, for every divisor of $a$ whose `JS` $= -1$ ; there is another divisor with `JS` $= -1$).

In such cases although `Jacobi_Symbol`$(a, N) = +1$ ; in reality $a$ is a `QNR` .

It turns out that in almost all such cases

**when the non-residue nature of $a$ is hidden from its Jacobi Symbol w.r.t. $N$ ; then**

$$\texttt{BCC(}\sqrt{a}, N\texttt{)} \neq 0 \mod N \tag{29}$$

In other words, even if `Jacobi_Symbol` is a +1; speculatively performing the `MBEC` check will produce a Binomial Witness in a substantial number of cases.

For example, consider the composite number

$$N = 2047 = 23 \times 89 \tag{30}$$

For this composite[7] it can be verified that

$$\texttt{Jacobi\_Symbol}(3, 2047) = -1 \quad \text{and} \quad \texttt{Jacobi\_Symbol}(5, 2047) = -1$$

and therefore

$$\texttt{Jacobi\_Symbol}(15, 2047) = +1$$

In other words, the **`Non-Residue`** nature of $a = 15$ is hidden from its `Jacobi_Symbol` w.r.t. $N = 2047$.

Accordingly, it turns out that

$$\texttt{BCC(}\sqrt{15}, 2047\texttt{)} = 1194 + 322\sqrt{15} \neq 0 \mod 2047 \tag{31}$$

as expected.

**Reason 4 :**    **Even more intriguing, it turns out that a sizable fraction of actual `Quadratic Residues` in $Z_N^*$ are also Binomial Witnesses for composite numbers $N$.**

For example, for the same composite number $N = 2047$ ; let $a = 2$

Then `Jacobi_Symbol`$(2, 23) = 1$ and `Jacobi_Symbol`$(2, 89) = 1$

so that $a = 2$ is a legitimate `QR` modulo-$N$,

it can be verified that $r = \{\pm 64, \pm 915\}$ are the square-roots of 2 modulo-2047.

Yet it is also the case that

$$\texttt{BCC(}\sqrt{2}, 2047\texttt{)} = 1196 + 1265\sqrt{2} \tag{32}$$

$\Rightarrow$  2 is a binomial witness to the composite nature of $N$ even though 2 is a `QR` modulo-$N$

---

In other words, note that the Baseline Conjecture says nothing about the case when a candidate integer $a \in Z_N^*$ is a `Quadratic Residue`; i.e., condition C–2 is not **necessary** for every input value $N$.[8]

**Reason 5 :** Second to last but not the least ; a substantial amount of experimental data demonstrate that a combination of deterministic heuristics by themselves or in combination with randomized searches has always yielded a `QNR` in a number of iterations that is $<< O((\lg N)^2)$ in every numerical example we have tried to date.[9]

These include

(i) every Carmichael number we could find in the literature; including the exhaustive list of every single Carmichael number $< 10^{18}$ from G.E. Pinch's excellent web-site [1, 34], and other larger special Carmichaels we could find in the literature that are of various lengths from 20 (decimal) digits to 397 decimal digits.

and

(ii) Cases from extremely large formula-generated numbers that are probable primes without a proof of primality. The PPTA_EQNR algorithm has generated the first known proofs for some of these numbers (assuming that the PBP Conjecture is true). See **Section 21** thru **Section 24** in **companion manuscript Part/Article 2** in this set for details.

**Reason 6 :** The second most important reason why the Baseline Conjecture is extremely useful is because using another result (that we prove in the next section), it can be combined with the state-of-the-art Miller Rabin Primality Testing method. The resulting hybrid method is substantially better than its components.

In other words, even if the development of the Generalized Primality Conjecture were to not happen, the enhancement of the Miller-Rabin method, by itself is a sufficiently strong reason to add the Baseline Primality Conjecture as a tool to the repertoire of primality-testing methods.

---

[8] which in no way contradicts or interferes with the claim of the conjecture; which is that the same condition, if satisfied together with the four other conditions; is **SUFFICIENT** to conclude that $N$ must be a prime number.

[9] the algorithms included a combination of the Euler Criterion Check, Miller-Rabin check (which includes the Fermat's Little Theorem check) and the extremely powerful Modular Binomial Expansion Congruence Test. Clearly the Miller-Rabin checks for Non-QNR values do help. We are not claiming that we can generate `QNR` in every case, rather we are reporting that in the relatively small number of cases that were not quickly resolved by the combination of other mechanisms and therefore had to go to the `MBEC` test as the last resort, we were always able to find a `QNR` value explicitly very quickly even by the brute-force method of marching thru all primes starting at a small value such as 5.

## § Section 6 :  New Result Enabling Fusion of Baseline Conjecture with the Miller-Rabin Method

In this section we prove an extremely useful result:

It turns out that when the argument $q$ is a **QNR** ; the Euler Criterion Check is at least as discriminating/powerful/effective as one iteration of the standard Miller-Rabin method with $q$ as the Miller-Rabin witness candidate (or in other words, the strong-pseudoprime test with $q$ as the base [35]).

Before we state and prove this fact, we first briefly describe what the Miller-Rabin method does [36, 37].

For any NUT $N$;  let

$$N - 1 \quad = \quad \Delta \times 2^T \quad \text{where,} \quad \Delta \text{ is odd} \tag{33}$$

Then, the Miller-Rabin method first evaluates the base value

$$B \quad = \quad (q)^\Delta \mod N \tag{34}$$

and  then  it examines the chain of squares (modulo-$N$) :

$$B^2, B^4, B^8 \cdots, B^{[2^{(T-1)}]} = q^{\left(\frac{N-1}{2}\right)}, \quad \text{and, finally;} \quad B^{[2^{(T)}]} = q^{(N-1)} \tag{35}$$

If the last element in the chain does not equal +1, then FLT is violated and $q$ is a Fermat-Witness to the composite-ness of $N$.

Moreover; if any occurrence of a +1 in the chain is not immediately preceded by a $\pm 1$ but say some other value say $B_k$ ;  Then $(B_k)^2 = 1 \mod N \Rightarrow (B_k)^2 - 1 = 0 \mod N$
$\Rightarrow (B_k + 1)(B_k - 1) = 0 \mod N$.

Additionally; if $B_k \neq \pm 1$ ; then $(B_k \pm 1) \neq 0 \mod N$ ; (then $B_k$ is called a non-trivial square-root of 1 modulo-$N$).  It is well known that a non-trivial square root of 1 modulo-$N$ exists only if $N$ is a composite number; and some non-trivial factor $f$ of $N$ (i.e., $1 < f < N$) divides $(B_k + 1)$; and the other non-trivial factor  $g = \left(\frac{N}{f}\right)$ divides $(B_k - 1)$.

In essence, the Miller-Rabin method checks for a non-trivial square-root of 1 modulo-$N$.

If such a root is found; it declares $N$ to be composite; otherwise it proceeds with the next base value (distinct from the current value $q$) to be tried in search of non-trivial roots of unity modulo-$N$.

With the preceding (brief) explanation of what the Miller-Rabin Method does, we state and prove the result next.

## Primality Lemma 1 (PL1) [10]

## If a QNR $q$ is a Miller-Rabin witness; then it must also be an Euler-Witness

**Proof :** If an integer $q$ is a Miller-Rabin witness to the composite-(ness/nature) of the NUT $N$ ; then one of the following conditions must hold:

<u>Condition 1</u> : The last element in the chain of squares $\neq 1$
In this case; the immediate predecessor element in the chain $\neq -1$
(because if the penultimate element $= -1 \mod N$ ; then
the last element $= (-1)^2 \mod N = 1 \mod N$ which contradicts the hypothesis in condition 1)

But the penultimate element $= q^{\left(\frac{N-1}{2}\right)} = -1 \mod N$ if $q$ is a QNR modulo-$N$ and satisfies the Euler Criterion.

Therefore if the penultimate element $\neq -1 \mod N$ ; then the Euler Criterion is violated; which completes the proof in case condition 1 holds. □

Or <u>Condition 2</u> : the last element is 1; but the penultimate element is $\neq -1$:
The hypothesis of this condition directly contradicts the requirement that
the penultimate element $= -1 \mod N$ if $q$ satisfies the Euler Criterion modulo-$N$.
Therefore if condition 2 is true; then the Euler Criterion must be violated. □

Or <u>Condition 3</u> : One of the prior elements

$$\left\{ B^2, B^4, B^8, B^{16}, \cdots, [B]^{[2]^{(T-2)}} \right\} = 1 \mod N \tag{36}$$

and its predecessor $\neq \pm 1 \mod N$

but this in turn implies that the penultimate element $=$

$[B]^{\{2\}^{(T-1)}}$ must be $= +1 \mod N$

Because, if some $k$-th element of the chain where $k \leq (T-2)$ is 1 then

all the subsequent elements in the chain-of-squares must also be $[1]^{(2)^m} = 1 \mod N$

$\Rightarrow$ the Euler Criterion is violated □ $\tag{37}$

**The practical implications of this result are substantial: it essentially says that if the base of the Miller-Rabin iteration is a QNR then the Euler Criterion Check obviates the Miller-Rabin checks.**

**This in turn implies that the Baseline Primality Conjecture (which includes the Euler Criterion Check (ECC)) obviates the Miller-Rabin Test if the candidate-base $q$ happens to be a QNR .**

**Indeed if we do encounter a QNR then the baseline conjecture is arguably the quickest way to complete the primality detection process; there is no need for Miller-Rabin (or any other type of) iterations.**

---

[10] I became aware of this result only via experimental observations

**In other words, a Miller-Rabin iteration brings additional capability only when the base-value is a `Quadratic-Residue` modulo-$N$.**

**Given that the computation of the Jacobi-Symbol requires very low computational effort, the preceding Lemma (PL1) makes it clear how to integrate the best aspects of the Miller-Rabin method with the Baseline Primality Conjecture.**

**Hybrid Algorithms that deploy a fusion of our results/methods with Miller-Rabin are unveiled in the next section.**

**<u>Remark 6.1</u> :**

The **converse** of **Primality Lemma 1** is the condition/statement :

**If a `QNR` $q$ is an Euler Witness then it is also a Miller-Rabin witness.**

It is not immediately clear how to prove (or disprove) the converse analytically.
However, experimentally, I have not found any integer that contradicts the converse statement.[11]

---

**§ Section 7 : Hybrid Algorithms**

We present two primality testing algorithms (PPTAs) :

1 : Enhances **PPTA_EQNR** by adding Miller-Rabin checks in the "**Find_QNR**" procedure
   if the prime turns out to be a `QR` ; i.e., `Jacobi_Symbol(`$p, N$`) = +1`

2 : The reverse process, to enhance the standard Miller-Rabin method by leveraging
   the **PBP Conjecture** and **Primality Lemma 1**

**§ Section 7.1 : `PPTA_EQNR variant that includes Miller-Rabin Checks`**

The only change in the algorithm specification on page **article 1 – page 18** is to invoke a modified version of the procedure that finds a `QNR` . Accordingly, replace

"  **line 13)**  **(detection_done_flag , $q$) := Find_QNR(**$N$ , $s$**);** "

with

"  line 13)  (detection_done_flag , $q$) := Find_QNR_with_miller_rabin_checks($N$ , $s$) ;  "

The specification of the modified procedure is as follows :

line 1) **Procedure**    **Find_QNR_with_miller_rabin_checks** ($N$::posint , Iter_Limit::posint)

line 2)    // If no errors occur, then the procedure returns 2 parameters :

line 3)    // parameter 1 : an integer flag set to a value $= \begin{cases} 1 & \text{if Jacobi-Symbol = 0 is encountered} \\ 2 & \text{if a nontrivial square root of 1 is found} \\ 3 & \text{if a Fermat-witness is found} \\ 0 & \text{otherwise} \end{cases}$ line 4)    // parameter 2 : an integer $= \begin{cases} \text{p} & = & \text{first argument of the Jacobi-Symbol if it evaluates to 0 or } -1 \text{ ,} \\ & & \text{or } p \text{ is a Fermat-Witness} \\ \text{B} & = & \text{the value of non-trivial square root if found} \end{cases}$

line 5)    `local i, j, k, p, B, S ;`

line 6)    $p := 2$ ;

line 7)    **for** $i$ **from** $1$ **while** ($i <$ **Iter_limit**)   **do**

line 8)      $p := \mathrm{nextprime}(p)$ ;

line 9)      $j := \mathtt{Jacobi}(p, N)$ ;

line 10)      **if** ($j == 0$)   **then**

line 11)        **return**    (1 , $p$) ;

line 12)      **elif** ($j == -1$)   **then**

line 13)        **return**    (0 , $p$) ;

line 14)      **else**                   // $p$ is a **QR** . therefore do a Miller-Rabin check line 15)        $B := \mathrm{modexp}(p, \Delta, N)$ ;        // where ;   $N - 1 = \Delta \times 2^{T}$ and $\Delta$ is odd

line 16)        **for** $k$ **from** $1$ **to** $T$ **do**

line 17)          $S := B^2 \bmod N$ ;

line 18)          **if** ($S == 1$) **and** ($B \neq \pm 1 \bmod N$) **then**

line 19)            **return**($2, B$) ;

line 20)          **fi** ;

line 21)          **if** $k < T$ **then**    $B := S$ ;    **fi** ;

line 22)        **od** ;

line 23)        **if**   $S \neq 1$    **then**    **return**($3, p$) ;    **fi** ;    // $p$ is a Fermat-Witness line 24)      **fi** ;

line 25)    **od** ;

line 26)    **Error("Did not find a QNR & did not find MR witness") ;**

line 27)    **return   NULL ;**

line 28) **End_Procedure**

line 1) **Algorithm  Enhanced_miller_rabin_with_PBPC(**$N$::**posint)**

line 2)    `local i, j, k, a, B, R, S, U, q::nonnegitiveint := 0 ;`

line 3)    $R := N \bmod 24 \; ;$

line 4)   **If** $R == 1$ **then**        // check for perfect squares (subsumes all even powers)

line 5)    $U := \left\lfloor \sqrt{N} \right\rfloor \;\; ;$

line 6)    **if** $(U^2 == N)$ **or** $\left((U+1)^2 == N\right)$ **then return Composite ; fi;**

line 7)    **for** $i$ **from** $1$ **while** $(i \leq U)$ **do**

line 8)    $a := \mathtt{random}(5..N-5) \; ;$

line 9)    $j := \mathtt{Jacobi}(a, N) \; ;$

line 10)   **if** $(j == 0)$ **then**

line 11)    **return Composite ;**

line 12)   **elif** $(j == -1)$ **then**

line 13)    **q := a ;**

line 14)    **break ;**    // execution jumps to "**line 35**"

line 15)   **else**                    // $a$ is a **QR** . therefore do the Miller-Rabin check line 16)    $B := \mathtt{modexp}(a, \Delta, N) \; ;$        // where ; $N - 1 = \Delta \times 2^T$ and $\Delta$ is odd

line 17)    **for** $k$ **from** $1$ **to** $T$ **do**

line 18)    $S := B^2 \bmod N \; ;$

line 19)    **if** $(S == 1)$ **and** $(B \neq \pm 1 \bmod N)$ **then**

line 20)    **return Composite ;**

line 21)    **fi ;**

line 22)    **if** $k < T$ **then** $B := S ;$    **fi ;**

line 23)    **od ;**

line 24)    **if** $S \neq 1$ **then** **return Composite ; fi;**    // $a$ is a Fermat-Witness

line 25)   **fi ;**

line 26)   **od ;** // end of the scope of the for loop with index $i$ that starts on "**line 7**"

line 27)  **Else**    // the else block of if on "**line 4**"

line 28)    // 2 or 3 is a factor of $N$ OR $q \in \{\pm 2, 3\}$ is a **QNR**

line 27⟩ **Else**    // line 27 near the bottom of the of previous page repeated for visual pseudo-code clarity

line 30⟩     **if**   $\gcd(R,6) > 1$ **then**   **return Composite ; fi ;**

line 31⟩     **if**   $(R \bmod 8 == 3)$ `OR` $(R \bmod 8 == 5)$   **then**    $q := 2$ ;

line 32⟩     **elif**   $R \bmod 8 == 7$   **then**    $q := N - 2$ ;

line 33⟩     **else**    $q := 3$ ;   **fi ;**

line 34⟩ **Fi ;**   // end of the scope of the else on line 27

line 35⟩   // if the execution reaches this point then a **QNR** has been found line 36⟩ **if**   $\big[\texttt{ECC}(q,N) \neq 0\big]$   `OR`   $\big[\texttt{BCC}(\sqrt{q},N) \neq 0\big]$    **then**

line 37⟩     **return Composite ;**

line 38⟩ **else**

line 39⟩     **return Prime ;**

line 40⟩ **fi ;**

line 41⟩ **End_Algorithm**

**Remark  7.1** :
Note that   if   **Jacobi_Symbol**$(a,N) == 0$ then   $a$ and $N$ must share a nontrivial common factor and therefore $N$ must be a composite. This simple check that is executed on "**lines 10 and 11**" in the preceding enhanced version is not part of the standard Miller-Rabin method and therefore represents an improvement.

**Remark  7.2** :
The main improvement is that if the randomly generated "base" = $a$ (at **line 8** ) in the standard Miller-Rabin method turns out to be a **QNR** then the algorithm breaks out of the loop that tries a new randomly generated base in each iteration; thereby potentially substantially reducing the number of iterations needed and also providing deterministic guarantees that are integral to the Baseline Primality Conjecture.

**Remark  7.3** :
It is extremely interesting and important to further investigate the hybrid methods and (at least experimentally) assess the improvements over their existing standard counterparts that are realized in practice. We intend to look at these issues in the near future.

However, up to this point, our main focus has been (as it should be) on completing an analytic proof of the Baseline Primality Conjecture; and Circa June of 2019 on developing and verifying the Generalized Primality Conjecture and the algorithms it enables.

## § Section 8: Circumventing computation of explicit integer non-residue value in underlying number field

All the good reasons and hybrid methods from the preceding sections still do not overcome the issue of worst-case complexity of generating a `QNR` value.

Fortunately it turns out that we can completely circumvent that issue because higher order non-residues that are implicitly specified as roots of polynomials can also work for primality detection. We develop these ideas in this and ensuing sections, starting with numerical examples and culminating in the generalized conjecture and the algorithm(s) it enables.

## § Section 8.1: Examples showing that Implicit Specification of non residues can also work

The baseline PPTA invokes a search for a `QNR` only when $N \bmod 8 \equiv 1$
i.e., when $N = 8I + 1$ for some integer $I$.

This condition can be improved and the search for a `QNR` can be avoided unless $N \bmod 24 \equiv 1$

The proof of the extension from **8** to **24** is trivial, since 3 either divides $N$ or is a `QNR` w.r.t. $N$ of the form $(8I + 1)$ unless the additional condition $N \bmod 3 \equiv 1$ is also satisfied.

Moreover, in the excellent article [20], Sze demonstrates that the condition can be substantially improved further by another factor of 10, so that

an explicit search for a `QNR` is needed only when $N \equiv 1 \bmod 240$ (38)

We focus on how Sze brings in the additional prime factor = 5 (which is needed to extend **24** to **240**). Without loss of generality, in order to illustrate Sze's argument, assume that $N$ is an odd prime $p$ of the form $(4J + 1)$, i.e., $N = p = 4J + 1$ for some integer $J$.

Next, for brevity and clarity we directly quote from paragraphs 3, 4 and 5 from the top of page number 2 in Sze's article [20]:

"Then (i.e., assuming $p \bmod 4 = 1$), $-1$ is a `QR`
and if $p \bmod 5 = 2$ or 3 ; then 5 is a `QNR`.
Suppose $p \equiv 4 \pmod 5$. In this case, 5 is a square mod $p$. Let

$$\zeta_5 = \frac{a + \sqrt{a^2 - 4}}{2} \quad \text{where} \quad a = \frac{-1 + \sqrt{5}}{2} \qquad (1.1)$$

Then, $\zeta_5$ is a primitive 5th root of unity. Note that $a \in \mathbf{F}_p$ but $\zeta_5 \notin \mathbf{F}_p$.
Therefore, $a^2 - 4$ must be a quadratic non-residue.
In conclusion, the problem of finding a quadratic non-residue in $\mathbf{F}_p$ is hard
only if $p \equiv 1 \pmod{16}$, $p \equiv 1 \pmod 3$ and $p \equiv 1 \pmod 5$,
which is $p \equiv 1 \pmod{240}$ "

The aspect of the preceding discussion from Sze's article that is relevant for our purpose is the fact that the non residue $q = (a^2 - 4)$ is not explicitly available as an integer modulo-$p$ (or modulo-$N$ in general).

In other words, as per the preceding analysis of Sze,
(assuming that $N \mod 4 = 1$ and $N \mod 5 = 4$)

$$q = (a^2 - 4) \quad \text{is a } \texttt{QNR} \text{ w.r.t. } N \text{ when} \qquad a = \frac{1 - \sqrt{5}}{2} \mod N \tag{39}$$

**Therefore, to explicitly evaluate the integer value of the `QNR` $q = (a^2 - 4)$;**
**the value of $(\sqrt{5} \mod N)$ must be computed first.**

We now demonstrate that such a computation of $\sqrt{5}$-modulo-$N$ in the underlying number field can be obviated by a straightforward modification/extension of how the `MBEC` test is implemented:
the `MBEC` check is extremely powerful and flexible and can be performed in at least three distinct but equivalent ways by directly leveraging any of the Equations
(in this case, any one of the 3 equivalent forms of the same equation(s) in (39))
`that implicitly specify the non residue, without the need to`
`explicitly compute the modulo-`$N$` roots in any of those Equations.`

**Circumvention method 1:** in Eqn. (19) plug in $x = \sqrt{q} = \sqrt{(a^2 - 4)} = \dfrac{\sqrt{-10 - 2\sqrt{5}}}{2}$ (40)

and perform the computation symbolically. [12]

Then, as per **Remark 3.2** an equivalent check is the following

**Equivalent circumvention method 2:** in Eqn (19) plug in

the divisor polynomial $\mathcal{D}(x) = (x^2 - q) = x^2 - (a^2 - 4) = x^2 + \dfrac{5 + \sqrt{5}}{2}$ (41)

Finally, to obtain one more equivalent method of avoiding explicit computation of a `QNR`
simplify Eqn. (40) to get rid of all the square roots (radicals in general) to obtain

$$x^4 + 5x^2 + 5 = 0 \tag{42}$$

and use the Left-Hand-Side of the preceding homogeneous Equation, i.e., the polynomial $x^4 + 5x^2 + 5$ as the divisor polynomial $\mathcal{D}(x)$ in (19):

**Equivalent Circumvention method 3:** in Eqn (19) plug in $\mathcal{D}(x) = x^4 + 5x^2 + 5$ (43)

---

[12] The software package "Maple" that we have used in all our experiments has a robust built-in support for such symbolic computations. This is one of the reasons we selected Maple for conducting our experiments. For further details, see **the Introduction to Part/Article 2** .

To complete the illustration of circumvention methods, we (arbitrarily) pick two values of $N$[13] :

a composite number $N_c = 589 = 19 \times 31$

and

a prime number $N_p = 569$

and illustrate each of the 3 circumvention methods applied to $N_c$ and $N_p$.

## § Section 8.1.1 : Illustration for a composite $N = N_c = 589$

**<1>** : First, we evaluate $(\sqrt{5} \mod N_c)$ explicitly by brute force; trying each number $r$ from $r = 2$ thru $r = \left( \frac{N_c - 1}{2} \right)$ in sequence, till the values that satisfy the congruence

$$r^2 \mod 589 = 5 \tag{44}$$

are found. [14]

It turns out that

$$\sqrt{5} \mod 589 \equiv \{ \pm 161 , \pm 180 \} \tag{45}$$

i.e.,

for $R \in \{ 161, 428, 180, 409 \}$ ; $R^2 \mod 589 = 5$

Accordingly, plugging in the preceding values of square roots of 5 into Relations (39) ; the **QNR** values implied by Sze's analysis are

$$q_c = \{ 506, 78, 202, 382 \} \tag{46}$$

respectively, corresponding to each of the $R$ values in Relation (45) .

Direct computation of the **Jacobi_Symbol** values confirms that

**Jacobi_Symbol**$(q_c, 589) = -1$ for $q_c \in \{ 506, 78, 202, 382 \}$

---

[13]The $N$ values must satisfy the conditions: $N \mod 4 = 1$ and $N \mod 5 = 4$

[14] It is well known that if non-trivial modulo-$N$ square-roots of
arbitrary integer $a < N$ (where $a$ is not a square of another integer)
can be evaluated for arbitrary composite number $N$; then $N$ can be trivially factored, once the square-roots are available. To the best of our knowledge, no polynomial cost method of extracting square root(s) modulo arbitrary composites is known; otherwise the RSA crypto-system would have not survived since it essentially depends on the hardness factoring product of two suitably chosen large primes.

Then it can be verified that for each of the four values of $q_c$ specified in Relation (46)

$\texttt{ECC}\,(q_c, N_c) = 0$   and therefore, as required by the Baseline Primality Conjecture,   $\texttt{BCC}\,(\sqrt{q_c}\,, N_c) \neq 0$

In particular, the non-zero  $\texttt{BCC}$  values can be verified to be

$$\texttt{BCC}\,(\sqrt{506}, 589) \;=\; 561 + 329\,\sqrt{506}$$
$$\texttt{BCC}\,(\sqrt{78}, 589) \;=\; 259 + 376\,\sqrt{78}$$
$$\texttt{BCC}\,(\sqrt{202}, 589) \;=\; 352 + 500\,\sqrt{202}$$
$$\texttt{BCC}\,(\sqrt{382}, 589) \;=\; 468 + 205\,\sqrt{382}$$

**<2>** :   Next we illustrate that each circumvention method also yields non-zero $\texttt{BCC}$ value; thereby proving that 589 is a composite.

The first circumvention method yields

$$\texttt{BCC}\left( \frac{\sqrt{-10 - 2\sqrt{5}}}{2}\,,\, 589 \right) \;=\; \left( 410 + 268\sqrt{5} \right) + \left( 29 + 299\sqrt{5} \right) \cdot \left( \sqrt{-10 - 2\sqrt{5}} \right)$$
$$\not\equiv\; 0 \mod 589 \tag{47}$$

Equivalently, the second circumvention method yields the non-zero polynomial remainder

$$\mathcal{R}\,(x) = \left[ \left( 410 + 268\sqrt{5} \right) + \left( 58 + 9\sqrt{5} \right) x \right] \mod 589 \not\equiv 0 \mod 589 \tag{48}$$

And the third circumvention method also yields the non-zero polynomial remainder

$$\mathcal{R}_{\text{alt}}\,(x) = \left[ 248 + 13x + 53x^2 + 571x^3 \right] \mod 589 \not\equiv 0 \mod 589 \tag{49}$$

This completes the demonstration for a composite.

**§ Section 8.1.2 : Illustration for a prime** $N = N_p = 569$

**<1>** : First by explicitly evaluating the **QNR** value(s) and then invoking the **PBP Conjecture** :
It can be verified that

$$\sqrt{5} \mod 569 \equiv \pm 104$$

the corresponding **QNR** values are: $q_p = \{233\,,\,337\}$

**Jacobi_Symbol(**$q_p$, 569**)** $= -1$

and

**ECC(**$q_p$, 569**)** $= 0$  AND  **BCC(**$q_p$, 569**)** $= 0$

**<2>** : Next, by each of the circumvention methods :
It can be verified that

$$\texttt{BCC}\left(\frac{\sqrt{-10-2\sqrt{5}}}{2}\,,\,569\right) = 0 \mod 569$$

$$\mathcal{R}\,(x) = 0 \mod 569$$

and

$$\mathcal{R}_{\text{alt}}\,(x) = \left[\left((1+x)^N - 1 - (x)^N\right) \mod (x^4 + 5x^2 + 5)\right] \mod 569$$

$$\equiv 0 \mod 569$$

That completes the illustration for a prime.

We close out this section with an important but obvious result for Primes.

## § Section 8.2 : `Modular Binomial Congruence Check works with ANY/ALL Non-zero divisor polynomials`

## Primality Lemma 2 (PL2) :

**Let** $\mathcal{R}(x)$ **be the finite-field remainder of the following division of polynomials (over the same finite field):**

$$\mathcal{R}(x) = \left\{ \underbrace{\left[ \left( (1+x)^N - 1 - (x)^N \right) \mod N \right]}_{\textbf{Dividend}} \mod \underbrace{\left[ \mathcal{D}(x) \mod N \right]}_{\texttt{Divisor}} \right\} \mod N$$

**If** $N$ **is a prime number, then**

$\mathcal{R}(x) \equiv 0 \mod N$ **For ANY/ALL non-zero divisor polynomials** $\left[ \mathcal{D}(x) \mod N \right]$ (50)

**Proof :**

Trivial, note that the dividend is the polynomial

$$\left[ (1+x)^N - 1 - (x)^N \right] \mod N \equiv 0 \mod N \qquad \text{when } N \text{ is a prime number} \tag{51}$$

and $0$ divided by anything non-zero must yield a $0$ remainder. □

The contra positive of the preceding result immediately suggests that one could use RANDOM polynomials and if the result is not zero then the divisor polynomial yielding a non-zero (Finite Field) remainder is a Binomial-Witness to the composite-ness of $N$.

The dangerous slippery slope (as it has been for eternity) is that the converse is not true, or in other words if the remainder is zero for some instance of divisor polynomial $\mathcal{D}(x)$ then that by itself does not necessarily imply that $N$ must be a prime; i.e.; there exist composites that will pass the polynomial congruence test for some instances of divisor polynomials. The question then becomes how many random polynomials should be tried and if all those produce a 0 remainder then the only outcome is "Probable Prime".

To resolve this issue, in the next section, we derive canonical divisor polynomials and then identify one more set of conditions (besides the Modular Binomial Congruence Check specified in **Primality Lemma 2** ) to be satisfied by the same divisor polynomials; that (when all satisfied together) guarantee that $N$ must be a prime. In other words no composite number can satisfy all the conditions specified in the Generalized Primality Conjecture unveiled in the ensuing sections.

## § Section 9 :     **Auxiliary Results that lead to Generalized form of PBP Conjecture**

The most critical point to note is that if the **QNR** value is explicitly known to be the integer $q$; then the "divisor polynomial" in the **MBEC** check can be trivially constructed to be $\mathcal{D}(x) = x^2 - q$.

The circumvention methods in the preceding section indicate the fact that any other polynomial say $\mathcal{H}(x)$ can also work effectively as the divisor polynomial in the **MBEC** check; as long as

$$\mathcal{H}(x) = 0 \mod N \qquad \text{does not have any integer solution/root.} \tag{52}$$

Therefore, in the next sub-section we derive low degree polynomials that do not have integer solutions modulo-$N$.

## § Section 9.1 :     **Derivation of Canonical Divisor Polynomials**

Let $N$ be an odd integer (to be tested for primality); and let
$m < N$ be an integer that satisfies satisfying the following conditions:

$$\gcd(m, N) = 1 \tag{53}$$

$$N \mod m \neq 1 \qquad \text{i.e., } m \text{ does not divide } (N-1) \tag{54}$$

and

$$m = (P_m)^k \quad \text{where the base } P_m \text{ is a prime number and the exponent is an integer } k \geq 1 \tag{55}$$
in other words, $m$ is either a prime number or a power of a single prime base
however,

$$\text{if } k > 1 \text{ then } (P_m)^{(k-1)} \text{ does divide } (N-1) \tag{56}$$

and finally

$$m \text{ is the smallest integer that satisfies the preceding four conditions (53)–(56)} \tag{57}$$

The final constraint that $m$ be the smallest integer (satisfying the stated conditions) is required in order to achieve the lowest possible complexity. It should be noted that the ensuing result(s) hold for any integer $m < N$, irrespective of whether or not it is the smallest one satisfying conditions (53)–(56).

Then, condition (54) implies that least one prime divisor; say $P_N$ of $N$ is such that[15]

$$P_N \mod m \neq 1 \tag{58}$$

The preceding relation in turn implies that $m$ does not divide $(P_N - 1)$ and therefore nontrivial primitive m-th order roots-of 1 (unity) modulo-$P_N$ do not exist, so that

$$x^m - 1 = 0 \mod P_N \quad \text{has no integer solution other than the trivial value } x = 1 \tag{59}$$

---

[15] If $N$ is a prime number then $P_N = N$

Since $x^m - 1 = (x - 1) \times \Phi_m(x)$     where, $\Phi_m(x)$ is the $m$-th cyclotomic polynomial [38]    (60)

Then, Relation (59) implies that

$\Phi_m(x) = 0 \mod P_N$    has no integer solution.                                 (61)

In other words,

$\Phi_m(x) \neq 0 \mod P_N$    for any integer $x$                                  (62)

**The preceding relation implies that**

$\left.\begin{array}{l} \Phi_m(x) \neq 0 \mod N \quad \textbf{for any integer } x \\[8pt] \qquad \textbf{or equivalently} \\[8pt] \Phi_m(x) = 0 \mod N \quad \textbf{has no integer solution.} \end{array}\right\}$    (63)

(otherwise, if there exists an integer $a$ such that

$\Phi_m(a) = 0 \mod N$

then the same equality holds modulo every divisor of $N$, so that

$\Phi_m(a) = 0 \mod P_N$    which contradicts Relation (62) )

It is well known that the Equation

$\Phi_m(x) = 0$                                                          (64)

can always be re-expressed as another monic-polynomial equation of the form [16]

$\Upsilon_m(t) = 0$    where                                        (65)

$t = \left(x + \dfrac{1}{x}\right)$    is a canonical transformation variable          (66)

The preceding variable-transformation Relation (66) can be re-arranged as the following quadratic equation in $x$:

$x^2 - t\,x + 1 = 0$                                          (67)

which has the roots

$x = \dfrac{t \pm \sqrt{t^2 - 4}}{2}$                                      (68)

---

[16] For example: for $m = 5$,    $\Phi_5(x) = 1 + x + x^2 + x^3 + x^4$

     $\Phi_5(x) = 0$   can be re-written as   $\left[x^2 + \dfrac{1}{x^2}\right] + \left(x + \dfrac{1}{x}\right) + 1 = 0$

     or as    $\left[\left(x + \dfrac{1}{x}\right)^2 - 2\right] + \left(x + \dfrac{1}{x}\right) + 1 = 0$

     which, in terms of $t$ becomes the equation   $t^2 + t - 1 = 0$

Let

$$u = \sqrt{t^2 - 4} \tag{69}$$

so that (68) can be written as

$$x = \frac{t \pm u}{2} \mod N \tag{70}$$

The preceding equation together with Relation (63) implies that

**at least one of the two variables $t$ and $u$ must not exist as an integer modulo-$N$** $\tag{71}$

**(Otherwise if both $t$ and $u$ exist as modulo-$N$ integers, then $x = (t \pm u)/2$ also exists as a modulo-$N$ integer; which contradicts (63) .)**

To obtain an explicit equation in terms of $u$ ; re-write (69) as

$$t = \sqrt{u^2 + 4} \tag{72}$$

and plug this expression for "$t$" in Eqn. (65) to obtain

$$\Upsilon_m \left( \sqrt{(u^2 + 4)} \right) = 0 \tag{73}$$

Expanding the polynomial on the LHS of the preceding equation and collecting even and odd powers, we obtain

$$\mathcal{C}_0 \left( m, u \right) + \mathcal{C}_1 \left( m, u \right) \cdot \left( \sqrt{(u^2 + 4)} \right) = 0 \tag{74}$$

where; $\mathcal{C}_0 \left( m, u \right)$ and $\mathcal{C}_1 \left( m, u \right)$ are themselves monic polynomials (without radicals) in the variable $u$.

If the polynomial $\Upsilon_m(\cdot)$ contains only even degree terms (which, for example, happens if $m = 2^k$ see the last entry in Table 1) then $\mathcal{C}_1 \left( m, u \right) = 0$ and no further simplification is needed. Otherwise, one squaring is required to get rid of the radical in Eqn. (74) .

The resulting polynomial equation in variable $u$ is

$$\Psi_m(u) = 0 \tag{75}$$

where

$$\Psi_m(u) = \begin{cases} \mathcal{C}_0 \left( m, u \right) & \text{if } \mathcal{C}_1 \left( m, u \right) = 0 \\ (u^2 + 4) \cdot [\mathcal{C}_1 \left( m, u \right)]^2 - [\mathcal{C}_0 \left( m, u \right)]^2 & \text{otherwise} \end{cases} \tag{76}$$

In summary, in this sub section (7.1), thus far, we have proved the following result :

## Primality Lemma 3 (PL3)

**Let $N$ be an odd integer being tested for primality. Then, given an $m < N$ satisfying Relations (53) thru (56);**
**there are two monic polynomials $\Upsilon_m(\cdot)$ and $\Psi_m(\cdot)$ ; both derived from the**
**cyclotomic polynomial $\Phi_m(\cdot)$ via the transformations specified in Relations (66) and (72) ;**
**such that roots of at least one of the two polynomials are non-residues**
**(i.e., do not exist as integers) modulo−$N$.**

Table 1 shows all the 3 polynomials corresponding to each value of $m$ for the first few (small) values of $m$.

| $m =$ $(P_m)^k$ | $P_m, k$ | Cyclotomic Polynomial $\Phi_m(x)$ | Polynomial $\Upsilon_m(t)$ | Polynomial $\Psi_m(u)$ |
|---|---|---|---|---|
| 3 | 3,1 | $x^2 + x + 1$ | $t + 1$ | $u^2 + 3$ |
| 5 | 5,1 | $x^4 + x^3 + x^2 + x + 1$ | $t^2 + t - 1$ | $u^4 + 5u^2 + 5$ |
| 7 | 7,1 | $x^6 + \cdots + x + 1$ | $t^3 + t^2 - 2t - 1$ | $u^6 + 7u^4 + 14u^2 + 7$ |
| 9 | 3,2 | $x^6 + x^3 + 1$ | $t^3 - 3t + 1$ | $u^6 + 6u^4 + 9u^2 + 3$ |
| 11 | 11,1 | $x^{10} + \cdots + 1$ | $t^5 + t^4 - 4t^3 - 3t^2 + 3t + 1$ | $u^{10} + 11u^8 + 44u^6 + 77u^4 + 55u^2 + 11$ |
| 13 | 13,1 | $x^{12} + \cdots + 1$ | $t^6 + t^5 - 5t^4 - 4t^3 + 6t^2 + 3t - 1$ | $u^{12} + 13u^{10} + 65u^8 + 156u^6 + 182u^4 + 91u^2 + 13$ |
| 16 | 2,4 | $x^8 + 1$ | $t^4 - 4t^2 + 2$ | $u^4 + 4u^2 + 2$ |

**Table 1.** Canonical divisor Polynomials $\Upsilon_m$ and $\Psi_m$ for the **MBEC** check, for first few values of $m$

This is table is only for the purpose of illustration. However, note that in principle an exhaustive table like this one can be pre-computed and be made available as a tool to further ease/assist/speed-up primality testing.

**This immediately suggests that using each of the two polynomials $\Upsilon_m$ and $\Psi_m$ as the divisor polynomial in the MBEC check should be an extremely strong discriminator test sniffing out composites from primes.**

**Indeed these checks are the first and integral part of the generalized form of the PBP Conjecture.**

**what is missing ?**

As the astute reader might have guessed, it is clear that a second independent condition similar to (or equivalent to) the Euler Criterion check is still to be identified.

The next sub-section addresses that issue.

## § Section 9.2 : `Analogues of Euler's Criterion for the Canonical Divisor Polynomials`

## Auxiliary Primality Conjecture 1 :

Given an odd integer $N$ ; let integer $m$ and the corresponding polynomials $\Upsilon_m$ and $\Psi_m$ be determined as specified in **Primality Lemma 3** .
Let

$$d \quad = \quad \textbf{degree of polynomial} \quad \Upsilon_m(.) \tag{77}$$

**If $N$ is a prime number, then the following identities hold:**

$$\left[ x^{\left( (N^d) - 1 \right)} \quad \mod \quad \Upsilon_m(x) \right] \quad \mod \ N \quad \equiv \quad \mathbf{1} \tag{78}$$

and

$$\left[ x^{\left( (N^d) - 1 \right)} \quad \mod \ \Psi_m(x) \right] \quad \mod \ N \equiv \begin{cases} 1 & \text{if } P_m = 2 \\ \texttt{Jacobi\_Symbol}(N, P_m) & \text{otherwise} \end{cases} \tag{79}$$

We do not yet (at the time of this writing) know how to prove in general (i.e., in all cases), the two identities in Eqns. (78) and (79). Therefore the results are introduced as a "conjecture".

However, in the special cases wherein the one or both of the polynomials $\Upsilon_m(x)$ and $\Psi_m(x)$ happen to be `irreducible` modulo-$N$; some parts of the proof are relatively straightforward. We therefore present a general result that holds for any divisor polynomial $\boldsymbol{D}(x)$ that is `irreducible`[17] modulo-$N$.

---

[17]Note that `irreducibility` is a stronger/stricter condition than having `no-integer-roots`:
If a polynomial $\boldsymbol{P}(x)$ is `irreducible` modulo-$N$ then $\boldsymbol{P}(x) \mod \ N = 0$ has no integer solution.

However, the converse is not true; i.e., a polynomial can have only non-integer roots but still be factorable into lower degree polynomials.
For example, it turns out that

$$\left. \begin{array}{l} \text{for } N = 11 \text{ and } m = 7, \text{ the cyclotomic polynomial } \Phi_7(x) \text{ has no integer roots modulo-}N, \\ \text{yet it is factorable modulo-}N : \\ \Phi_7(x) = x^6 + x^5 + x^4 + x^3 + x^2 + x + 1 = (x^3 + 5x^2 + 4x + 10)(x^3 + 7x^2 + 6x + 10) \mod \ 11 \end{array} \right\} \tag{80}$$

**Primality Lemma 4 (PL4) :**

Let $N$ be an odd prime number.

Let $D(x)$ be a polynomial (of the argument or indeterminate or variable $x$) of degree $\delta < N$, defined over the Finite Field $F_N$, i.e., the coefficient of each degree is a modulo-$N$ integer.

Suppose that $D(x)$ is **IRREDUCIBLE, i.e., it cannot be factored into product of lower degree polynomials.**

Let $\mathcal{S}$ be the set of all polynomials modulo–$\langle D(x), N \rangle$, (except "0" polynomial wherein each coefficient is 0).

Then, for any polynomial $P_\lambda(x) \in \mathcal{S}$

$$\left( P_{[\lambda]}(x) \right)^{(N^\delta - 1)} \mod \ \langle \boldsymbol{D}(x), N \rangle \ \equiv \mathbf{1} \tag{81}$$

**Proof :**

Note that the remainder of any polynomial w.r.t $\boldsymbol{D}(x)$ must be a polynomial of degree $\leq (\delta - 1)$ where each coefficient is a modulo-$N$ integer and can therefore assume any of $N$ distinct values from 0 thru $(N-1)$.

Therefore there are at most $N^\delta$ distinct remainder polynomials modulo-$\langle \boldsymbol{D}(x), N \rangle$.

Exclude the "0" polynomial which has all coefficients = 0.

That leaves $(N^\delta - 1)$ distinct remainder polynomials so that the set $\mathcal{S}$ can be enumerated :

$$\mathcal{S} = \Big\{ P_{[1]}(x), P_{[2]}(x), \cdots, P_{[N^\delta - 1]}(x) \Big\} \tag{82}$$

It turns out that this set of polynomials has 3 important properties (from which many identities can be derived).

<u>Property 1</u> : For any two polynomials (same or distinct)

$$\Big[ P_{[i]}(x) \times P_{[j]}(x) \mod \boldsymbol{D}(x) \Big] \mod \ N \not\equiv 0 \tag{83}$$

This property follows from the hypothesis that $\boldsymbol{D}(x)$ is irreducible modulo-$N$.

<u>Property 2</u> : Closure: the modulo-$\langle \boldsymbol{D}(x), N \rangle$ product of any two polynomials in the set $\mathcal{S}$ results in another polynomial in the same set $\mathcal{S}$, i.e.,

$$\Big[ P_{[i]}(x) \times P_{[j]}(x) \mod \boldsymbol{D}(x) \Big] \mod \ N \in \mathcal{S} \quad \forall \, i, j \in [1, N^\delta - 1] \tag{84}$$

This is true because the remainder w.r.t. the divisor polynomial must be a polynomial of degree $\leq (\delta - 1)$ ; wherein each coefficient can only assume values between 0 and $(N-1)$, and $\mathcal{S}$ includes all such polynomials (excluding "0", but the product cannot be 0 as per property 1).

<u>Property 3</u> :

$$\left[ P_{[i]}(x) \times P_{[j]}(x) \right] \quad \equiv \quad \left[ P_{[i]}(x) \times P_{[k]}(x) \right] \quad \text{iff} \quad j = k \tag{85}$$

(otherwise,

$$P_{[i]}(x) \times \left[ P_{[j]}(x) - P_{[k]}(x) \right] \quad \equiv \quad 0$$

which contradicts Property 1 and the irreducibility hypothesis.)

From here on we invoke the standard/usual (number-theory) argument:

Pick any one element say $P_{[\lambda]}(x)$ from the set $\mathcal{S}$ and note that
multiplying each element of the set $\mathcal{S}$ by $P_{[\lambda]}(x)$
results in a set $\mathcal{S}'$ which is merely a permutation of the set $\mathcal{S}$.

Therefore equating the product of all elements of $\mathcal{S}$ to the product of all elements of $\mathcal{S}'$ yields

$$\left\{ P_{[1]}(x) \times \cdots \times P_{[N^\delta - 1]}(x) \right\} \equiv \left\{ P_{[1]}(x) \times \cdots \times P_{[N^\delta - 1]}(x) \right\} \times \left( P_{[\lambda]}(x) \right)^{(N^\delta - 1)} \tag{86}$$

Canceling the common factor inside the curly-braces
(= the product of all polynomials in $\mathcal{S}$ )
from both sides yields the desired identity

$$\left( P_{[\lambda]}(x) \right)^{(N^\delta - 1)} \quad \text{mod} \quad \langle \boldsymbol{D}(x), N \rangle \quad \equiv \mathbf{1} \qquad \square$$

The reader can verify that additional identities analogous to standard ones in number theory also hold. For example

$$(\text{Product of all polynomials in } \mathcal{S}) \quad \text{mod} \quad \langle \boldsymbol{D}(x), N \rangle \quad \equiv \quad (N - 1) \quad \equiv \quad -1 \tag{87}$$

which is the analogue of Wilson's Theorem (which says that
$(P - 1)! \equiv -1 \mod P$ for all primes $P$.
Moreover, the set $\mathcal{S}$ turns out to be a cyclic group with non-trivial generator-polynomials, etc.

We now provide partial proofs for the two identities in the conjecture.

**Partial Proof of the First Identity in Eqn. (78) :**

If $\Upsilon_m(x)$ is `irreducible` modulo-$N$ then then identity (78) a special case of the main Identity, i.e.,
Eqn. (81) in **Primality Lemma 4** , wherein,

$$\left. \begin{array}{l} \boldsymbol{D}(x) = \Upsilon_m(x) \quad \text{and degree } \delta = d \\ \qquad \text{and} \\ P_{[\lambda]}(x) = x = \text{one specific polynomial in the set } \mathcal{S} \end{array} \right\} \qquad \square \tag{88}$$

Note that identity (78) holds for prime values of $N$ even if the polynomial $\Upsilon_m(x)$ does have non-trivial integer roots modulo-$N$. However, we have not yet been able to analytically prove it in that case.

**Remark 9.1** :

Experimental data indicates that $\Upsilon_m(x) \bmod N = 0$ has integer solutions
only when ($N \bmod m \equiv -1$)
(recall that $m$ is selected to ensure that $N \bmod m \neq 1$, so that case does not arise).
The fact that the $\Upsilon_m$ polynomial has integer roots for one value of remainder $r = N \bmod m$
does not contradict **Primality Lemma 3** , which only proves that one of the two polynomials $\Upsilon_m$ and
$\Psi_m$ is guaranteed to have no integer roots.

**Remark 9.2** :

If all that is needed is a polynomial that does not have integer roots, then the cyclotomic polynomial
itself is sufficient. Why bother deriving additional polynomials if some of the derived polynomials
might not have the same guarantee of NO INTEGER ROOTS modulo-$N$ that the cyclotomic
polynomial offers?

Three main reasons:

Reason 1 : The degree $d$ of polynomial $\Upsilon_m$ is 1/2 the degree of the cyclotomic polynomial. Later on,
in **Section 13** , the computational effort required is proved to be proportional to $d^2$. Therefore using
the lower degree polynomials drops the run-time by a factor of $2^2 = 4$.

Reason 2 : Unlike Identity (79), we do not yet (at the time of this writing) know of any relations that
connect remainders modulo-$\langle$`cyclotomic_polynomial`$_m(x)$, $N \rangle$ , to
the Jacobi-Symbols (also see remarks 3 and 4 at the end of this Section).

**Reason 3 : The most important reason is that the polynomial $\Psi_m(x)$ turns out to be essential.**

**In other words all combinations of tests that include only the cyclotomic polynomial $\Phi_m(x)$ and
the polynomial $\Upsilon_m(x)$ turn out to be foolable and therefore insufficient to guarantee primality.**

**See Section 13.1 and Section 13.2 for details.**

**Partial Proof of the Second Identity in the Conjecture, in Eqn. (79) :**

Experimental data indicates that $\Psi_m(x) \mod N = 0$ has integer solutions for some primes $N$, only when $m$ is a power of 2. However, we have not yet analytically proved this observation.

We split the arguments into two cases:

**case 1:** $P_m = 2$ :

In this case, from Relations (76), it can be seen that

$$\text{the degree of polynomial } \Psi_m(\cdot) = \text{the degree of polynomial } \Upsilon_m(\cdot) = d \tag{89}$$

(for a specific numerical/actual example, see the last row in Table 1, wherein both $\Upsilon$ and $\Psi$ polynomials have the same degree $d = 4$.)

Further, the right hand side of the identity to be proved is 1.

Therefore in all those cases, wherein, $\Psi_m(x)$ is `irreducible` modulo-$N$, the desired identity is a special case of the general Identity, i.e., Eqn. (81) in **Primality Lemma 4** , and can be deduced from it by substituting

$$\left.\begin{array}{c} \boldsymbol{D}(x) = \Psi_m(x) \quad \text{and degree } \delta = d \\ \text{and} \\ P_{[\lambda]}(x) = x \end{array}\right\} \quad \square \tag{90}$$

**Partial Proof for the remaining cases :**

The main point to note is that

$$\text{if } (P_m \neq 2) \text{ then} \quad \text{degree of polynomial } \Psi_m(\cdot) = 2 \times \text{degree of polynomial } \Upsilon_m(\cdot) = 2d \tag{91}$$

In this case we do not yet (as of the time of this writing) know how to prove Identity (79).

However, it is relatively straightforward to prove the identity which results by raising both sides of the Identity in Eqn. (79) to the $(N^d + 1)$–th power:

$$\left[ x^{\left( (N^d) - 1 \right)} \right]^{\left( 1 + N^d \right)} \quad \equiv \quad (\pm 1)^{\left( 1 + N^d \right)} \tag{92}$$

Since $N$ is odd, then $(N^d + 1)$ is an even number and therefore
the right hand side of the preceding equation $\equiv \boldsymbol{1}$.
Re-arranging the exponent on the left hand side of the preceding equation
gives us the identity we need to prove:

$$[x]^{\left[ \left( N^d - 1 \right) \times \left( N^d + 1 \right) \right]} \equiv 1 \tag{93}$$

or equivalently,

$$[x]^{\left( N^{2d} - 1 \right)} \equiv 1 \tag{94}$$

In all those cases wherein $\Psi_m(x)$ is `irreducible` modulo-$N$, the previous equation can be obtained from the general identity in Eqn. ([81](#))
with the following substitutions

$$\left.\begin{array}{c} \boldsymbol{D}(x) = \Psi_m(x) \quad \text{and degree } \delta = 2d \\ \text{and} \\ P_{[\lambda]}(x) = x \end{array}\right\} \quad \square \tag{95}$$

**Remark 9.3** :

The results presented in this sub-section were identified very recently (on or about the 10-th of June 2019); and have been experimentally tested pretty much round the clock, as thoroughly as possible since then. Unfortunately, in the limited amount of time since the identification; I have not been able to complete the preceding partial proofs.

Usually, in the forward direction (i.e., when $N$ is assumed to be a prime number, then) the results turn out to be relatively easier to prove analytically (such as the robust result that the `MBEC` check must always produce a 0 remainder if $N$ is a prime.)
Yet, it is not immediately clear as to how to complete the partial proofs. [18]

**Remark 9.4** :

It is somewhat counter-intuitive that the second identity in Auxiliary Primality Conjecture 1, (i.e. Eqn. ([79](#)) above) is independent of $k$; sniffing out the Jacobi-symbol of $N$ w.r.t. the base prime $P_m$ that divides $m$. This subsumes all sub-cases where $k$ is an even number thereby rendering $m$ into a perfect square of another integer; which in turn implies that Jacobi-symbol of $N$ w.r.t. $m$ is a +1, irrespective of the Jacobi-symbol of $N$, w.r.t. $P_m$.

The other counter-intuitive fact in that identity is that the order of variables in the Jacobi-Symbol seems reversed.
It is Jacobi symbol of $N$ w.r.t. $P_m$; not the other way around as might be expected; since the underlying finite field consists of modulo–$N$ integers.

---

[18] obviously, no counter-example has been found. See Manuscript Part 2, for experimental details.

With all the pieces in place we are ready to state the generalized primality conjecture.

## § Section 10: **`Phatak's Generalized Primality Conjecture(PGPC)`**

**Given an odd integer** $N$ **; let integer** $m = (P_m)^k$ **and the corresponding polynomials** $\Upsilon_m$ **and** $\Psi_m$ **be determined as specified in** <span style="color:green">**Primality Lemma 3**</span> **.**
**Let** $d$ **be the degree of the polynomial** $\Upsilon_m(\cdot)$.
**With the parameters as specified;**
**If the all of the following four conditions are satisfied, then** $N$ **is a prime.**

**PGPC-condition-1 :** $\left\{ \left[ (1+x)^N - 1 - (x)^N \right] \mod \Upsilon_m(x) \right\} \mod N \equiv 0$

**AND**

**PGPC-condition-2 :** $\left\{ \left[ (1+x)^N - 1 - (x)^N \right] \mod \Psi_m(x) \right\} \mod N \equiv 0$

**AND**

**PGPC-condition-3 :** $\left[ (x)^{(N^d-1)} \mod \Upsilon_m(x) \right] \mod N \equiv 1$

**AND**

**PGPC-condition-4 :** $\left[ (x)^{(N^d-1)} \mod \Psi_m(x) \right] \mod N \equiv \begin{cases} 1 & \textbf{if } P_m = 2 \\ \texttt{Jacobi\_Symbol}(N, P_m) & \textbf{otherwise} \end{cases}$

**In other words, no composite number can satisfy all the four conditions above.**

### Remark 10.1 :

Note that the generalized conjecture is independent of whether
$N$ is an exact square of another integer,    or
$N$ is a power $> 2$ of a single prime, etc.
It is robust and treats ALL composites uniformly; no special processing is needed for squares, or powers of single primes or Carmichael Numbers.

That does not mean that we should not perform a simple check to screen out exact squares of integers because such a simple square-root extraction check opens up the possibility of actually finding the explicit value of a **`QNR`** during the search for the canonical integer "$m$". If that happens, (i.e., an explicit value of a **`QNR`** is encountered during a search for $m$); then it is simpler (requires substantially lower computational effort) to invoke the Baseline Primality Conjecture to quickly complete the detection (see the algorithm presented next for details).

line 1)    **Algorithm    PPTA_INR** ($N$ :: **posint**)

line 2)    local s, q::int := 0, m::int := 0, detection_done_flag::int := 0 ;

line 3)    **Begin_Algorithm**

line 4)    **if**  ($N \bmod 24 \neq 1$)  **then**

line 5)        **if**  ($N \bmod 2 == 0$)  **or**  ($N \bmod 3 == 0$)  **then  return    Composite** ;  **fi ;**

line 6)        **if**  ($N \bmod 8 == 3$)  **or**  ($N \bmod 8 == 5$)  **then**    $q := 2$ ;

line 7)        **elif**  ($N \bmod 8 == 7$)  **then**        $q := N - 2$ ;

line 8)        **else**    $q := 3$ ;     **fi ;**

line 9)    **else**

line 10)        (detection_done_flag, ***q, m***)  := Find_qnr_or_PGPC_parameter_m ( $N$ ) ;

line 11)        **if**  (detection_done_flag $> 0$) **then**    **return  Composite** ;    **fi ;**

line 12)   **fi ;**

line 13)   // at this point the algorithm has found a   **QNR** $= q > 0$   or   the canonical integer "$m$"

line 14)   // accordingly, use the baseline conjecture or the generalized version line 15)   **if**  $q > 0$   **then**      // explicit  **QNR** $= q$  has been found $\therefore$ invoke the  **baseline primality conjecture**

line 16)        **if**  (**ECC**($q, N$) $\neq 0$ )  **or**  ( **BCC**($\sqrt{q}, N$) $\neq 0$ )  **then**

line 17)                **return  Composite** ;

line 18)        **else**

line 19)                        **return  Prime** ;

line 20)        **fi ;**

line 21)   **else**  // use  **Generalized Primality Conjecture** with the  "$m$"  value returned by procedure called on  **line 10**

line 22)        // polynomials $\Upsilon_m$ and $\Psi_m$ are assumed to be looked-up from a pre-computed table using that $m$ value

line 23)        **if** (**any of the four conditions in  PGP Conjecture don't hold** )  **then**

line 24)                **return  Composite** ;

line 25)        **else**

line 26)                        **return  Prime** ;

line 27)        **fi ;**

line 28)   **fi ;**

line 29)   **End_Algorithm**

Next, we specify the pseudo-code for the "Find_qnr_or_PGPC_parameter_m" procedure called from within the algorithm.

line 1)  **Procedure  Find_qnr_or_PGPC_parameter_m** ($N$::posint)

line 2)  // the procedure returns 3 integer parameters, only 1 of them is non-zero, depending on what is found line 3)  // par_1 : set to a value $> 0$ if $N$ is a square or a prime divides $N$;  set to 0 otherwise line 4)  // par_2 : set to 0 if (Par_1 > 0 or explicit **QNR** is NOT found) ; otherwise = value of **QNR** encountered line 5)  // par_3 : set to 0 if (Par_1 > 0 or Par_2 > 0) ; otherwise = value of the canonical PGPC parameter "$m$"

line 6)  `local s, r, i, j, m, k, deg, tdeg, myprod::int := 1, p::int := 2 ;`

line 7)  $s := \mathrm{round}\left(\sqrt{N}\right)$        // exact maple command is $s := \mathrm{round}(\texttt{evalf[length}(N)\texttt{](sqrt}(N)))$ ;

line 8)  **if**    $(s^2 == N)$    **then**    **return(s, 0, 0) ;**   **fi;**

line 9)  **for** $i$ **from** 1  **while** $(\texttt{myprod} \leq N-1)$    **do**

line 10)    $r := N \bmod p$ ;

line 11)    **if**    $(r == 0)$    **then**    **return(p, 0, 0) ;**

line 12)    **elif** $(r == 1)$    **then**        // $p$ divides $(N-1)$ $\therefore$ find the smallest $k$ s.t. $p^k$ does not divide $(N-1)$

line 13)       **for** $k$ **from** 2 **do**

line 14)          **if** $[(N-1) \bmod \left(p^k\right) \neq 0]$ **then**    break ;    **fi;**

line 15)       **od ;**

line 16)       **if**    $p == 2$    **then**                // at least 8 divides $(N-1)$ $\Rightarrow$ $k \geq 4$

line 17)          `deg := ` $2^{(k-2)}$ `;   m := ` $2^k$ `;`

line 18)       **else**

line 19)          `tdeg := ` $\left(\frac{p-1}{2}\right) * p^{(k-1)}$ `;`

line 20)          **if**  `tdeg` $<$ `deg`  **then**

line 21)             `deg := tdeg ;`      $m := p^k$ ;

line 22)          **fi ;**

line 23)       **fi ;**

line 24)       `myprod := myprod * ` $p^{(k-1)}$ `;`

line 25)       **if**     `myprod` $> N-1$    **then**    break;

line 26)       **else**   $p := \texttt{nextprime}(p)$ ;    **fi;**

line 27)    **else**    // found the smallest prime that does not divide $N-1$

line 28)       **if** `jacobi_symbol(r,p) == -1`    **then**    **return(0, p, 0) ;** **fi ;**

line 29)       **if**   `(p-1)/2` $<$ `deg`   **then**

line 30)          `deg := (p-1)/2 ;`     $m := p$ ;

line 31)       **fi ;**

line 32)       **break ;**

line 33)    **fi ;**

line 34)  **od ;**

line 35)  **return(0, 0, m) ;**

line 36)  **End_Procedure**

This completes the specification of the Algorithm and the procedure it invokes.

§   **Section 12 :**        **Proof of Correctness of PPTA_INR Algorithm assuming that**
                        **PGP Conjecture  is true**

We split the proof into two sub-sections:

The first sub-section proves that the flow control of the overall algorithm specification on  **article 1 – page 52**  is correct.

and

The next sub-section proves that the procedure specification on  **article 1 – page 53**  is correct and works as intended (which naturally leads to a proof of the overall complexity).

## §  Section 12.1 :  `Analysis of Flow Control in Overall Algorithm`

Please refer (preferably in a separate window) to the algorithm specification on  **article 1 – page 52** .

Start with the first "**if**" block whose scope begins on  **line 4**  and  ends on line  **line 8** .

Let    $\rho = N \mod 24$.
Then, if the execution reaches  **line 6**  then

$$(\rho \neq 1) \quad \textbf{and} \quad (\rho \ \text{is not even}) \quad \textbf{and} \quad (\rho \mod 3 \neq 0) \quad \Rightarrow \tag{96}$$

$$\rho \quad \in \quad \{5, 7, 11, 13, 17, 19, 23\} \quad \Rightarrow \tag{97}$$

$$N \mod 8 = \rho \mod 8 \quad \in \quad \{1, 3, 5, 7\}$$

Also note that only one value, viz., the value  "17" in the set (of possible values of remainder of $N$ w.r.t. 24, at line 6 in the code)  in Relation (97) is such that  $(N \mod 8 = 1)$.
Therefore all other cases except when  $(N \mod 24 = 17)$  are taken care of by the if-else-if condition statements on lines 6 and 7.
Thus, if the execution reaches the "else" on  **line 8** ,  then

$$N \mod 24 = 17 \quad \Rightarrow \tag{98}$$

$$N \mod 8 = 1 \quad \text{and} \tag{99}$$

$$N \mod 3 = 2 \quad \Rightarrow \quad \texttt{Jacobi\_Symbol(}N,3\texttt{)} = -1 \tag{100}$$

Finally, note that Eqns. (99) and (100) together with the well known quadratic-reciprocity result imply that

$$\texttt{Jacobi\_Symbol(}3,N\texttt{)} = \texttt{Jacobi\_Symbol(}N,3\texttt{)} \tag{101}$$

$$\Rightarrow \quad \texttt{Jacobi\_Symbol(}3,N\texttt{)} = -1 \quad \Rightarrow \quad q = 3 \quad \text{is a } \texttt{QNR}$$

and therefore the assignment on line 8 is correct.

In essence, we have proved that lines 4 thru 8 take care of the strong majority of cases
(91.67% of all odd integers = all possible inputs)
wherein, either 3 divides $N$ or $q = \{\pm 2, 3\}$ is a **QNR** modulo-$N$.                                    ☐

**<u>Remark 12.1</u>** :

Note that there is no need to be overtly aggressive and extend the coverage to 99.17%,
i.e.; all cases except those wherein $(N \bmod 240 = 1)$ ; because it may require the explicit
evaluation of square roots modulo-$N$.

Resuming the analysis of the algorithm, it is clear that
the (flow) control reaches the "else" block on **line 9** only when $N \bmod 24 = 1$ ,
which happens in 1 out of 12 cases (which is a small minority of all possible inputs).

Note that the only task that gets done in that
else $\cdots$ end_if block that includes lines numbered 9 thru 12,
is a to call the procedure "Find_qnr_or_PGPC_parameter_m($\cdot$)" which returns 3 parameters (see the
procedure specification on **article 1−page 53** ). If the procedure finds a divisor of $N$ then it sets the first
parameter returned to the value of that divisor.

Accordingly, **line 11** (which is the line immediately following the procedure call in the main/caller
program/module) simply checks if the procedure has found a (non trivial) divisor of $N$ and if that is
the case, the algorithm returns the result "Composite".                                                    ☐

That brings us to the last part of the code specification, lines 15 thru 29 on **article 1−page 52** .

This part is very simple: if the values returned by the procedure indicate that a **QNR** has been encoun-
tered (during a search for the canonical integer parameter $m$ which is required by the Generalized
Primality Conjecture), then the algorithm invokes the
Baseline Primality Conjecture.

Otherwise the procedure has found the canonical integer parameter $m$. Accordingly, the algorithm
invokes the Generalized Primality Conjecture to complete the detection.

This completes the proof that the flow-control of the overall algorithm as specified on **article 1−page 52**
is correct and therefore the algorithm will generate the correct result assuming that the conjectures are
true, as long as the procedure works as intended (which is proved in the next sub-section.)            ☐

## § Section 12.2 : **Procedure Find_qnr_or_PGPC_parameter_m() Analyzed**

This procedure constitutes the core of the new algorithm. Please refer (preferably in a separate window)
to the procedure code specified on **article 1−page 53** .

In order to facilitate the explanation, the procedure code is logically split into the following blocks of contiguous lines:

**Code–Block–1** : includes lines 1 thru 8 at the start,
                                 and
**Code–Block–2** : a second block of lines numbered 9 thru 11
                                 and
**Code–Block–3** : the third block of lines numbered 12 thru 26
                                 and
**Code–Block–4** : lines numbered 27 thru 33
                                 and finally
**Code–Block–5** : the last two lines numbered 35 and 36

Each of the above mentioned code-blocks is explained in a separate sub-sub-section.

To begin with, however, we explain the overall attributes of the procedure as a whole.

## § Section 12.2.1 : Return Parameters

As mentioned in the comments at the beginning of the code specification of that procedure, it returns 3 integer parameters wherein, only one of the 3 is non zero to indicate/signal to the caller (i.e., the overall algorithm) which one of the 3 conditions/cases has occurred for the input $N$ (the NUT for primality).

Case 1: The procedure finds a non-trivial divisor of $N$ via one of the two possibilities:
(i) $N$ is a square of another integer (in which case the square-root which is an integer divides $N$)
     or
(ii) During a search for $m$, one of the primes tested happens to be a divisor (factor) of $N$.

In this case the first parameter is set to the value of the divisor;
= the square-root (via the " `return(s, 0, 0) ;` " statement on **line 8** ,
     or
= the prime factor if one is encountered (via the " `return(p, 0, 0) ;` " statement on **line 11** ),
and the 2-nd and 3-rd parameters are set to 0.

This setting of parameters signals to the caller program that a non-trivial factor has been found and therefore the detection process is finished with the final result: input $N$ is a "Composite".

<u>Case 2</u>: The procedure encounters a `QNR` during the search for the canonical parameter $m$.
In this case the second parameter set to the value of the `QNR` and the other two return values are set to 0 , via the " **`return(0, p, 0) ;`** " statement on <span style="color:green">**line 28**</span>

Note that the Jacobi-Symbol test on line 28 is within the scope of the
`else ⋯ end_if` logic block spanning lines numbered 27 thru 33. That block is never executed for $p = 2$ because for odd inputs,
$r = N \mod 2 = 1.$
Thus, the Jacobi-Symbol test on line 28 is invoked only with odd values of both $N$ and $p$ and therefore

$$\textbf{Jacobi\_Symbol}(N, p) = \textbf{Jacobi\_Symbol}((N \mod p), p) = \textbf{Jacobi\_Symbol}(r, p) \qquad (102)$$

Moreover, since the procedure itself is called only when $N \mod 8 = 1$, then, by quadratic reciprocity,

$$\left.\begin{array}{l} \textbf{Jacobi\_Symbol}(N, p) = \textbf{Jacobi\_Symbol}(p, N) \\ \Rightarrow \quad p \text{ is a } \texttt{QNR} \text{ modulo-}N \text{ if } r \text{ is a } \texttt{QNR} \text{ modulo-}p \end{array}\right\} \qquad (103)$$

Therefore the parameters assigned by the return statement on line 28 are correct   □

<u>Case 3</u>: The procedure did not find a factor of $N$, neither did it come across a `QNR` during the search for the canonical parameter $m$ ; however ; it has found the canonical parameter $m$ that is required by the Generalized Primality Conjecture. In this case, the procedure returns the value of $m$ using the 3rd parameter (the first two parameters set to 0) via the " **`return(0, 0, m) ;`** " statement on <span style="color:green">**line 35**</span> at the very end of the procedure.

This setting of parameters signals to the caller program that none of the (relatively) lower-complexity paths will work for that input $N$ (= the input argument with which the procedure gets called) ; its time to invoke the generalized conjecture to complete the detection. This is the last-resort default exit out of the procedure, when every other opportunity to use a lower-complexity path turns out to be infeasible.

## § Section 12.2.2 : `Overall Structure`

Almost all of the procedure-code comprises of a conditional "**for**" loop that
starts on <span style="color:green">**line 9**</span> and ends on <span style="color:green">**line 34**</span> .

Note that 3 out the above mentioned Code-Blocks  are simply contiguous sets of lines within the loop.

Therefore the analysis of the loop is tantamount to the explanation of the code blocks and is done next, in the ensuing sub-sub sections.

Note that **line 6** in the first block declares all the variables used within the procedure. Maple allows the variables to be also type-cast and initialized at the time of declaration. This feature is used to initialize two key variables needed by the main loop:

`myprod` is type-cast as an int (integer) and is initialized to 1 (the identity of multiplication, analogous to initializing a sum to 0, which is the value inert to additions)

$p$ is initialized to the first prime number = 2

Besides the initialization, the only other task performed by Code-Block-1 is via the last two lines numbered 7 and 8 in that block.
It checks if $N$ is a square of another integer and if so, returns the value of the square-root as a non-trivial divisor of $N$.

There are at least 3 important optimizations achieved by performing the screening for a square inside the procedure and as the first check within the procedure:

**Optimization 1:** Note that performing the square-root screening any earlier in the main program would be tantamount to wasted computational effort: A strong majority of all possible inputs (91.67% of all odd numbers) do not even need this check because of the following fact:

**Basic Number Theory Background Fact 2: Let $N > 1$ be an odd positive integer that is not divisible by 3. Then, $N$ can be a square of some other integer only if $N \bmod 24 = 1$**

Equivalently, if $N \bmod 24 \neq 1$ and 3 does not divide $N$; then $N$ cannot be a square of another integer.

**Proof:** This is a trivial extension of **Basic number theory background fact 1** .
Let $N = K^2$ and note that if $K \bmod 3 \neq 0$ then $K \bmod 3 = \pm 1 \Rightarrow K^2 \bmod 3 = +1$
Combine the preceding condition with fact1 (that the square of every odd number $\equiv 1 \bmod 8$)
using the chinese remainder theorem to conclude that $K^2 \equiv 1 \bmod (8 \times 3 = 24)$    □

**Optimization 2:** While the generalized conjecture holds irrespective of whether $N$ is an exact square another integer; The preferred lowest-complexity path of invoking the Baseline Conjecture does require an explicit `QNR` value. Therefore, in order to keep-open the possibility of using the least-effort path; perfect squares must be ruled out.

Therefore, the check screening out square of another integer is exactly at the right place, yielding maximum returns for the computational effort spent (in extracting the square-root).

**Optimization 3 :** Last but not the least: the placement of the test to screen for squares at its chosen place substantially streamlines the actual code and improves clarity/readability without compromising computational efficiency.

Before proceeding with the line-by-line analysis of the code, we present the main idea behind what the loop is intended to achieve:

**Recall that we want to find the smallest integer $m$ of the form $m = (P_m)^k$ where $k \geq 1$ ;**
**that does not divide $(N-1)$. That is exactly what the loop does (with some minor additional checks).**

**To find the desired value we use the following obvious strategy:**

**Note that 2 (or a higher power of 2) must divide $(N-1)$.**
**Therefore starting with $p = 2$ ; step through successive primes $p = 2, 3, 5, 7, 11, 13, \cdots$**
**as long as each of those primes (or a power > 1 of each of those primes) divides $(N-1)$;**
**while also keeping track of the product of the divisors of $(N-1)$ seen up-to that point.**

**That is all (with some minor additional book-keeping to see if a divisor of $N$ can be found or if a `QNR` has been encountered) !!!**

We now resume the line-by-line analysis of the code.

**line 10 :** It evaluates the remainder of $N$ w.r.t. the current value of prime number $p$.

The following `if ··· else_if ··· else` logic blocks performs the obvious checks:

**line 11 :** if the remainder is zero, we have found a non-trivial divisor of $N$, so simply return the value.

That brings us to the next block.

First, note that this entire block corresponds to the case when the remainder
$r$ (computed on line 10) assumes the value 1, which implies that
the current value of the prime number denoted by the local variable $p$ divides $(N-1)$.

Therefore, as mentioned in the comment on line 12, the next 3 lines perform the first task:

**lines 13, 14, 15:** find that smallest power $k$ of $p$ such that $p^k$ does not divide $(N-1)$.

**lines 16 and 17:** if $p == 2$ then set the initial values of the variables "`deg`"[19] and $m$.

Note that this initialization is guaranteed to happen for odd values of $N$; which is in-turn guaranteed because the procedure gets called only when
$(N \bmod 24 = 1) \; \Rightarrow \; N \;$ is an odd number.

**lines 18 thru 23:** Otherwise, if the current values of $p$ and $k$ are such that they result in a lower degree of $\Upsilon_m(\cdot)$ then update the values of `deg` and $m$ by setting them equal to the lower values.

**line 24** Update the value of the variable "`myprod`" (which equals the product of divisors of $(N-1)$ that are all pair-wise co-prime w.r.t. each other because each divisor is some power of a distinct prime-base) to include the latest divisor found $= p^k$.

**lines 25 and 26** if `myprod` exceeds $(N-1)$ then exit the main loop because the product of relatively co-prime divisors of $(N-1)$ can never exceed $(N-1)$ itself.

Otherwise step to the next-prime (via the $p = \text{nextprime}(p)$ statement on line 26, that uses maple's built-in/native "nextprime()" function.[20]

## § Section 12.2.6 : Analysis of Code–Block–4 : Lines 27 thru 33 on article 1–page 53

This entire block corresponds to the case when the remainder $r$ (computed on line 10) satisfies the conditions $(r \neq 0)$ **AND** $(r \neq 1)$ which implies that the current value of the prime number $p$ does not divide $(N-1)$.

At this point only a couple of checks/updates remain to be performed before breaking out of the loop and those are done by the remaining lines in this block:

**line 28** since $r \neq 1$, there is a possibility that the prime $p$ is a **QNR** w.r.t. $N$.
That important check is performed on this line, and if it turns out that $r$ is a **QNR** w.r.t. $p$ then as explained under the heading **"Case 2" at the top of article 1–page 57** , because $N \bmod 8 = 1$, then quadratic reciprocity implies that if $r$ is a **QNR** w.r.t. $p$, then $p$ is a **QNR** w.r.t. $N$.
If an explicit value of **QNR** has been encountered in this manner, then it is best (lowest effort path) to invoke the Baseline Primality Conjecture to complete the detection process and that is what the procedure signals the caller program to do.

**lines 29 thru 31:** Check if the value of $p$ is such that it yields a lower degree for the canonical divisor polynomial $\Upsilon_p(\cdot)$; and if so, then update the values of `deg` and $m$ by setting them equal to the lower values.

**lines 32:** All the required computations are done so simply break-out-of/exit the loop

---

[19] `deg` = obvious short-form for the "degree" of the first canonical divisor polynomial $\Upsilon_m(\cdot)$

[20] Note that this apparently "circular" dependence could be replaced by using a pre-computed array holding successive prime numbers upto a value no bigger than $2(\lg N)$, see **Primality Lemma 5** .

That brings us to the last two lines. The next sub-sub section therefore considers those along with some concluding remarks.

§  **Section 12.2.7 : Analysis of final (trivial) `Code-Block-5`**

The second-last line (number 35) is the default return that corresponds to **"Case 3"** ; wherein; all the lower-complexity paths have been tested and found to be not applicable and therefore it is time to use the Generalized Primality Conjecture using the value of $m$, and that is what the procedure signals the caller program to do.

## Primality  Claim 1 (PCL-1) :

**Based on the analysis presented in this section, we conclude that Algorithm `PPTA_INR()`**
**as specified on   article 1 – page 52 ,   together with the procedure it calls, which is specified on**
**article 1 – page 53 ;**
**work as intended and correctly perform primality detection,**
**assuming that the Primality Conjectures unveiled herein are true.**

**Remark 12.2 :**
**Note that the preceding claim does not say anything about the  complexity  of the algorithm.**
**It claims only what has been demonstrated thus far; i.e.; correctness of the code-specification**
**assuming the underlying conjectures are true.**[21]

**Proving the worst-case bounds on the computational effort is the subject of the next section.**

---

[21] as repeatedly mentioned throughout this set of manuscripts, no counter example has been found.

## Primality  Lemma 5  (PL5) :  Logarithmic bound on $m$

**For (asymptotically/sufficiently) large $N$; the canonical parameter $m$ satisfies the bound**

$$m \quad < \quad 2\,(\lg N) \tag{104}$$

**Proof :**

To derive the worst case bound on $m$,

(i): ignore the fact that at least $2^3 = 8$ must divide $(N-1)$.

(ii): Likewise, also ignore the fact that in general, higher powers of one or more of the odd primes $\{3, 5, 7, 11, 13, \cdots\}$ can also divide $(N-1)$.

Then, suppose that

all primes from 2 thru say $P_h$ divide $(N-1)$. \hfill (105)

Then, since $m$ is defined as "the smallest prime or prime-power that does not divide $(N-1)$", it follows that

$$m \leq \ (\text{the next prime after/following } P_h) \tag{106}$$

From Relation (105), by definition of integer division, we obtain

$$2 \times 3 \times 5 \times 7 \times 11 \times \cdots \times P_h \leq (N-1) < N \tag{107}$$

Analogous to the definition of a **factorial**, which is typically denoted as "$Q\,!$";

A **primorial**, is another well known function [39] which typically denoted as "$[Q\#]$" and defined for positive integer argument $Q$ as

$$[Q\#] = \begin{cases} 1 & \text{if } Q = 1 \\ \text{Product of all primes } \leq Q & \text{for } Q \geq 2 \end{cases} \tag{108}$$

Using the preceding definition of the primorial function, Relation (107) can be stated as

$$[P_h \#] \leq (N-1) < N \tag{109}$$

The primorial function asymptotically grows as [39, 40, 41]

$$Q\# \quad \approx \quad e^Q \quad \text{for large } Q \tag{110}$$

and the primorial also satisfies well-known identities

$$2^Q < [Q\#] < 4^Q \tag{111}$$

for large $Q$.

More precisely, it turns out that

$$[Q\#] < 4^Q \quad \text{right from the get go for } Q \geq 1 \tag{112}$$

However,

$$2^Q < [Q\#] \quad \text{holds for all integers} \quad Q \geq 29 \tag{113}$$

Therefore we assume[22] that $N > 2^{29}$.

Then, from Relations (113) and (109), we obtain

$$2^{P_h} < [P_h \#] \leq (N-1) < N \tag{114}$$

The preceding relation yields

$$P_h < \lg N \tag{115}$$

Finally, by the Bertrand-Chebychev theorem [42],

$$(\text{next prime after } P_h) < 2P_h \tag{116}$$

and therefore

$$m \leq (\text{next prime after } P_h) < 2P_h < 2(\lg N) \tag{117}$$

which completes the proof that the parameter $m$ satisfies the logarithmic bound claimed in the Lemma.  □

---

[22] the conjectures as well as the algorithm has been exhaustively verified numerically for all 32-bit long odd integers
⇒   the results are true for all positive odd integers

**Corollary 1 : Bound on Number of iterations to find $m$**

**It takes no more than $O(\log N)$ steps to find $m$.**

**Proof :** The bound follows directly from the well-known "prime-counting-function" [43] which is defined as

$$\pi(x) = \text{number of primes} \leq x \tag{118}$$

and it satisfies the constraint

$$\pi(x) < 1.25506 \left(\frac{x}{\ln x}\right) \quad \text{for} \quad x > 1 \tag{119}$$

Plugging $x = m$ in the preceding relation yields

$$\text{number of iterations to find } \; m < 1.25506 \left(\frac{m}{\ln m}\right) < O(m) = O(\log N) \qquad \square \tag{120}$$

# Primality Lemma 6 (PL6) : $\; O\left((\log N)^4 \,(\text{polylog}(\log N))\right)$ Complexity Bound

**For (asymptotically/sufficiently) large $N$ ; the worst case computational effort required by algorithm `PPTA_INR` is $\;O\left((\log N)^4 \,(\text{polylog}(\log N))\right)$**

**Proof :**

Note that the power in the modular-exponentiation required to check **PGPC condition 3** as well as **PGPC condition 4** is

$$N^d - 1 \tag{121}$$

Therefore the number of steps/iterations of the standard square-and-reduce loop in the modular exponentiation is

$$\lg\left(N^d - 1\right) \quad < \quad \lg\left(N^d\right) \quad = \quad d\left(\lg N\right) \tag{122}$$

As mentioned earlier, the topic of polynomial multiplication, reciprocation and division over finite fields has been extensively analyzed in the literature (ex: see [29, 30, 31, 32, 33]). The results show that the square-and-remainder operations (on polynomials of degree $d$ with coefficients that are modulo-$N$ integers) within each iteration of the loop can be performed with a

$$\text{computational effort} \; \lessapprox O\left((d \cdot \lg N) \cdot \text{polylog}(\log N)\right) \tag{123}$$

Therefore the total complexity of checking PGPC-conditions 3 and 4 is

$$O(d \lg N) \times O\left[(d \lg N) \cdot \text{polylog}(\log N)\right] = O\left(d^2 \cdot (\lg N)^2 \cdot \text{polylog}(\log N)\right) \tag{124}$$

Finally using the constraints $d < m$ and
the logarithmic bound on $\; m < 2(\lg N)$ established in **Primality Lemma 5**
yields

$$\text{Overall Complexity} \; \lessapprox O\left((\log N)^4 \,(\text{polylog}(\log N))\right) \qquad \square \tag{125}$$

**Remark 13.1** :

In practice, substantial amount of experimental data indicate that merely checking one single condition, viz., **PGPC-condition 2** is sufficient to distinguish a composite from a prime.

In other words it appears that a shorter form of Generalized Conjecture with only the second condition in it also holds.[23]

Accordingly we state the most aggressive result as the final conjecture.

## § Section 13.1 : Furthermost Generalized Primality Conjecture (FGPC)

**Given an odd integer $N$ ; let integer $m = (P_m)^k$ and the corresponding polynomial $\Psi_m$ be determined as specified in Primality Lemma 3 .**
**With the parameters as specified;**
**If the following single condition is satisfied, then $N$ is a prime.**

$$\text{FGPC-condition :} \quad \left\{ \left[ (1+x)^N - 1 - (x)^N \right] \mod \Psi_m(x) \right\} \mod N \equiv 0 \qquad (126)$$

**Primality Lemma 7 (PL7) :** $O\left( (\log N)^3 \left( \text{polylog}(\log N) \right) \right)$ **Complexity Bound**

**Assuming that the Furthermost Generalized Primality Conjecture (FGPC) is true, for (asymptotically/sufficiently) large $N$ ; the worst case computational effort required by algorithm PPTA_INR is** $O\left( (\log N)^3 \left( \text{polylog}(\log N) \right) \right)$

**Proof :** Note that the power/exponent in the modular-exponentiations in the FGPC-condition is $N$; which is substantially smaller than the exponent $= (N^d - 1)$ that is required to check PGPC-conditions 3 and 4.

As a result, the complexity of checking the FGPC-condition is substantially smaller; and therefore lowers the worst case total computational effort required by the PPTA_INR algorithm to the following bound :

$$\lg N \times O((d \log N) \cdot \text{polylog}(\log N)) = \boldsymbol{O}\left( (\boldsymbol{\log N})^{\boldsymbol{3}} \left( \text{polylog}(\log N) \right) \right) \qquad \square \qquad (127)$$

---

[23]Then why do I include conditions 1, 3 and 4 at all? the answer is to be make the generalized conjecture as similar to the baseline conjecture as possible.
This potential improvement (as to whether conditions 1, 3 and 4 can be dropped) is being actively investigated.

## § Section 13.2 : `Cyclotomic and ϒ polynomials can be fooled`

Continuing with the notation from prior sections, suppose that given an $N$, we find the canonical parameter $m$ which is a prime or a power $> 1$ of a single base prime that does not divide $(N-1)$.
Let $\Phi_m(x)$ denote the $m$-th cyclotomic polynomial.
Suppose that $\Upsilon_m(x)$ and $\Psi_m(x)$ are determined as per **Primality Lemma 3** .
Let $d$ be the degree of the polynomial $\Upsilon_m(\cdot)$.

Experimental data show that there are many **composite** numbers $N$
that simultaneously satisfy the following 5 conditions:

$$\left\{ \left[ (1+x)^N - 1 - (x)^N \right] \mod \Upsilon_m(x) \right\} \mod N \equiv 0 \tag{128}$$

**AND**

$$\left\{ \left[ (1+x)^N - 1 - (x)^N \right] \mod \Phi_m(x) \right\} \mod N \equiv 0 \tag{129}$$

**AND**

$$\left[ (x)^{(N^d-1)} \mod \Upsilon_m(x) \right] \mod N \equiv 1 \tag{130}$$

**AND**

$$\left[ (x)^{(N^d-1)} \mod \Phi_m(x) \right] \mod N \equiv 1 \tag{131}$$

**AND**

$$\left[ (x)^{(N^d-1)} \mod \Psi_m(x) \right] \mod N \equiv \begin{cases} 1 & \text{if } P_m = 2 \\ \texttt{Jacobi\_Symbol}(N, P_m) & \text{otherwise} \end{cases} \tag{132}$$

What saves the Generalized Primality Conjecture is the (experimentally observed) fact that in all those cases

$$\left\{ \left[ (1+x)^N - 1 - (x)^N \right] \mod \Psi_m(x) \right\} \mod N \not\equiv 0 \tag{133}$$

We illustrate the preceding claim using the composite number

$$N_c = 6368689 = 1129 \times 5641 \tag{134}$$

This $N$ happens to be the smallest one in the list of pseudo-primes $< 10^{13}$
(available from G.E. Pinch's excellent web-site [1, 34]), that displays this peculiar behavior.

For this $N$, it can be verified that $m = 5$, so that

$\Phi_5(x) = x^4 + x^3 + x^2 + x + 1$

$\Upsilon_5(x) = x^2 + x - 1$

     and

$\Psi_5(x) = x^4 + 5x^2 + 5$

Then, it can be verified that

$\left\{ \left[ (1+x)^{6368689} - 1 - (x)^{6368689} \right] \mod (x^2 + x - 1) \right\} \mod 6368689 \equiv 0$

$\left\{ \left[ (1+x)^{6368689} - 1 - (x)^{6368689} \right] \mod (x^4 + x^3 + x^2 + x + 1) \right\} \mod 6368689 \equiv 0$

$\left[ (x)^{\left((6368689)^2 - 1\right)} \mod (x^2 + x - 1) \right] \mod 6368689 \equiv 1$

$\left[ (x)^{\left((6368689)^2 - 1\right)} \mod (x^4 + x^3 + x^2 + x + 1) \right] \mod 6368689 \equiv 1$

     and

$\left[ (x)^{\left((6368689)^2 - 1\right)} \mod (x^4 + 5x^2 + 5) \right] \mod 6368689 \equiv 1 = \textbf{Jacobi\_Symbol}(6368689, 5)$

What saves the day is the fact that

$$\left. \begin{array}{l} \left[ \left( (1+x)^{6368689} - 1 - (x)^{6368689} \right) \mod (x^4 + 5x^2 + 5) \right] \mod 6368689 \\ \equiv \ (3749099)\, x^3 + (2597822)\, x^2 + (1348224)\, x + 6150380 \quad \not\equiv \quad 0 \end{array} \right\} \tag{135}$$

In the experimental data gathered thus far, we have not encountered any composite number that
(i) satisfies (or in other words, fools) the FGPC-condition (126)
       but is identified as a composite because
(ii) it does not satisfy one or more of conditions (128) thru (132).

⋆⋆ **1** ⋆⋆ **:**     First, note that if a `QNR` $q \neq -1$ is encountered; the algorithm needs no more than
2 scalar modular exponentiations :
(1) To check whether the Euler Criterion is satisfied ;   and
      if the Euler Criterion is satisfied ;          then
(2) the second modular exponentiation with symbolic computation (using $\sqrt{q}$ as a "symbol")
to check the whether the modular Binomial Congruence is satisfied.
Or equivalently if implemented with polynomial remainder operations, then
the Degree of the divisor polynomial $\mathcal{D}(x) = x^2 - q$ is a small fixed constant = $2 = O(1)$ ,
independent of the size of the input $N$.

In other words,
`if a  QNR is found;  then`
`⟨i⟩ the number of modular exponentiations needed`
`is at most "2" = a fixed (and small) constant = ` $O(1)$ **,**
`independent of the size (or bit-length) of` $N$ **.**
`    and`
`⟨ii⟩ the degree of the divisor polynomial is also`
`a small fixed constant = 2 = ` $O(1)$ **.**

**Therefore the overall complexity (when a  QNR is found) becomes as low**
**as** $O\big((\log N)^{\mathbf{2}} \cdot \log\log N\big)$ **.**

In contrast, in the AKS [10] and its latest and fastest/most efficient
derivatives/variants [6, 25, 26, 27, 28] ;  the number of modular exponentiations that need to be
performed are on the order of
at least a 4-th degree polynomial of the bit-length of $N$; i.e.; $O\left( [\lg N]^{(4+\epsilon)} \right)$ [6].

Additionally, the degree $r$ of the divisor polynomial $(x^r - 1)$ in the AKS family of algorithms is
at least $r > (\lg N)^2$

The other state-of-the art methods used in practice are
the Miller-Rabin test [36, 37] ;   or
the Baillie–PSW primality test (see [44] and references therein for further details).
Both of these are probabilistic.  Therefore, when the number being tested is a prime; the
Miller-Rabin method has to execute a sufficient number of iterations to guarantee that the
probability of $N$ being a composite masquerading as a prime is driven to drastically low values.
The same is true of the Baillie–PSW test : even though it works in practice with a small average
number of iterations required to find a witness if the number is composite ; it is not known
whether the worst-case bound is as small as the average or at least within an order magnitude of
the average value or is or some low-degree polynomial of $(\lg N)$.
(if that was the case; then those tests would be deterministic).

All that effort that goes into executing a sufficient number of iterations (each iteration, in turn includes one modular exponentiation) to render the probability of wrong outcome to be drastically small is obviated by our method if a `QNR` is found !!!

⋆⋆ **2** ⋆⋆ :     Second, note that $\frac{11}{12} = 91.67\%$  of all odd integers do not require a search for a `QNR` .

⋆⋆ **3** ⋆⋆ :     It is possible to further restrict (in reality, substantially lower) the number of cases wherein an explicit search for a `QNR` is needed ; by adding checks for conditions under which other small primes are `QNR` s. For example, check whether 3, 5 and/or 7 are `QNR` s w.r.t. $N$ (the additional checks are needed only if $N \mod 24 = 1$).

⋆⋆ **4** ⋆⋆ :     Moreover, experimental data demonstrate that for all the numbers hitherto considered difficult cases for primality testing (including ALL Carmichael Numbers we could find in the literature [45, 46] and on the Internet) ; we never needed more than 22 iterations at most to find a `QNR` . (for further details, please refer to **Table 7 in Part/Article 2** .)
For example, the 397 decimal digits long Carmichael number `Arnault_N4`  , which is hand constructed in Arnault's landmark article [47] ;
`DOES NOT NEED a search for a` `QNR ;` `2 is a` `QNR`  modulo that huge 3-factor number.

⋆⋆ **5** ⋆⋆ :     Note that before initiating the big loop on exhaustive modular exponentiations (for every single integer in a non-trivially large range); the AKS method requires an explicit check to verify that $N$ is not an integer-power of another integer.  This entails an explicit check for all prime powers
from $2, 3, 5, 7, 11, 13, \cdots,$ to $E_p$  where ;    $E_p \leq \lceil \lg N \rceil$

In contrast; our method needs only one single check for a perfect square.
And that too is needed only for $\frac{1}{12} \approx 8.33\%$ ; or a small minority of all odd integers.

⋆⋆ **6** ⋆⋆ :     The baseline conjecture together with **Primality Lemma 1**  enable clean, synergistic fusion of Miller-Rabin iterations into our methods.

Even if the generalized versions of the conjecture had not been discovered, the hybrid algorithms presented in **Section 7**  are interesting in their own right, and substantially improve the Miller-Rabin method.

★★ **7** ★★ :   **The most exciting developments have been the discovery of the Generalized versions of the Primality Conjecture.**

**The** **Furthermost Generalized Primality Conjecture (FGPC)** **enables the**
`PPTA_INR` **algorithm to achieve deterministic worst case complexity =** $O\left((\log N)^{\mathbf{3}}\,(\mathrm{polylog}(\log N))\right)$

★★ **8** ★★ :   In principle, the deterministic low complexity achieved by the `PPTA_INR` algorithm obviates the need for Miller-Rabin iterations.

Yet the generalized conjectures leave room to integrate Miller-Rabin iterations (for example, if the canonical parameter $m$ happens to be a `QR` ), to seamlessly and synergistically harness the full power/capabilities of randomized methods;
**without affecting the worst case complexity.**

Such an integration of the best attributes of randomized algorithms will drive the
**average/expected complexity toward**
         **the lower bound =** $O\left((\log N)^{\mathbf{2}}\,(\mathrm{polylog}(\log N))\right)$

★★ **9** ★★ :   The generalized versions of the primality conjecture do not even need to screen out exact squares of other integers (although it is advantageous to perform this check as explained before).

Thus, a strength of the generalized versions of the conjecture is is that those handle all types of inputs $N$ uniformly and efficiently ; independent of whether or not it is
(i) a Carmichael Number
(ii) exact square of another integer
(iii) square free
(iv) an (odd) power of a single prime
(v) odd power of any arbitrary integer.

★★ **10** ★★ :   Potential relations between the results and methods unveiled in this set of articles and other state-of-the art topics (such as connections to the Baillie–PSW test) is a topic for future investigation(s) after the full analytic proof(s) of the conjectures are completed.

It is well known [10] that, an integer $N$ is a prime number **iff** the following Modular Binomial Expansion Congruence (**MBEC**) holds :

$$\textbf{MBEC} \qquad : \qquad (a+x)^N \bmod N = (a + x^N \bmod N) \bmod N = a + (x^N \mod N) \qquad (136)$$

where, $a$ is any integer relatively co-prime w.r.t. $N$ ; and $x$ is a variable, i.e., an indeterminate.

The proof is based on another well known fact [10, 6, 5]:

$$(^N\boldsymbol{C}_k) \bmod N \;=\; 0 \quad \text{for} \quad k=1,\cdots,(N-1) \;\textbf{ iff}\;\; N \text{ is a prime, } \text{ wherein} \qquad (137)$$
$$(^N\boldsymbol{C}_k) \;=\; \text{ the binomial coefficient representing } N\text{--choose--}k \qquad\qquad\qquad (138)$$

Purely symbolic direct verification of the **MBEC** (i.e., Eqn. (136) ) leaving $x$ as a true indeterminate symbol is not possible for all but small toy values of $N$.

The ingenuity of the now famous AKS deterministic primality test [10] lies in demonstrating that a verification of the above **MBEC** for all integer values of $x$ and $a$ can be guaranteed tho hold as long as the congruence holds for all integer values of $x$ within a range that is upper bounded by a polynomial of $(\lg N)$ ; when the value of $a$ is selected as specified in the AKS algorithm [10] and its derivatives [6, 25, 26, 27, 28].

§ **Section 15.1 : Intuition that led to the Baseline Conjecture**

Note that if the **MBEC** could be verified at some numerical value of $x$ such that none of the powers of that numerical value exist as integers; then that single verification would be sufficient to conclude that $N$ must be a prime number. For example, if the **MBEC** could be verified at a single transcendental real value of $x$ (for example, $x = \pi^{24}$ ; or $x = e =$ the base of natural logarithms ; or any real number that is not an algebraic integer); then that single verification should be sufficient to conclude that $N$ must be a prime number.

However, it is not immediately clear how to efficiently compute the floating-point "Remainders" that would arise in such a numerical verification. It is highly likely that the precision (and consequently the total amount of computations) required to verify the **MBEC** at transcendental values of $x$ is impossibly large for all but small toy values of $N$.

In an attempt to circumvent these difficulties and get the best of both worlds, i.e.

(i) the exact accuracy of integer computations

together with

(ii) a verification of the **MBEC** ideally needed only at a single value of the indeterminate $x$ ;

---

[24]Also known as "Archimedes's" constant = 3.1415⋯

I tried a different approach : I deployed two mechanisms:

⟨1⟩ Use "Matrices" together with scalar integers :

To better mimic the effect of a true indeterminate, I replaced the scalars $x$ and $a$ with arbitrary (scalar) multiples of Matrices. To keep matters simple I considered only $2 \times 2$ square matrices (turns out that they are sufficient for our purpose.)

The intuition behind the use of a matrix instead of a scalar variable $x$ in Eqn. (136) is that since a matrix is inherently more complex (it is a two-dimensional entity); therefore, if the **MBEC** could be tested with an arbitrary matrix, it could potentially lead to faster deterministic methods to test primality.

One complication with matrices is that `a matrix product is not commutative` .
Therefore a Binomial Expansion implicit in the **MBEC** is not valid for arbitrary matrices. However,
requiring that the matrix that replaces integer $a$ must be a scalar multiple of the $2 \times 2$ Identity matrix
ensures commutativity ; and then the Binomial Expansion in terms of the Binomial Coefficients is valid.

⟨2⟩ Use symbolic computation leveraging values that do not exist as integers :

The intuition is to leverage the full power of the binomial expansion identity; which is valid irrespective of whether or not $x$ is an integer.
Note that the modulo or remaindering operation with pure integers causes enormous complications. I wanted to stay as far away from all the nasty and complicated
"integer domain effects/Integer Traps"[25]
Therefore I started experimenting with symbols or values that do not exist as integers[26] Accordingly, I allowed the elements of the matrix (that is used in place of the indeterminate $x$), as well as the scalar coefficients (multiplying) the matrices ; to be algebraic-integers of small order; or in other words; roots of small degree polynomials with rational coefficients. (Note that this definition of algebraic integers includes plain/scalar integers as well as complex numbers).

After each computation, if there are any pure scalar integer coefficients or elements of the matrices that exceed $N$, they are reduced modulo $N$.

The matrix method evolved rapidly, mainly since 2017 spring (March 2017).
It evolved from 16 matrices needed for the algorithm : in other words, the **MBEC** needed to be verified at 16 specific matrix values of indeterminate $x$ to (experimentally) guarantee primality of $N$ ;
down to 8 ; to 4 $\cdots$ ;
finally down to the desired goal: one verification needed only at a single ($2 \times 2$ matrix) value of the indeterminate $x$.

I probed deeper and deeper and found out experimentally that the new method has ALWAYS WORKED;

---

[25]For example, when the variables are allowed to assume non-integer values; Linear Programming can be solved in Polynomial Time; but Integer programming turns out to be NP (to the best of my knowledge).

[26]For a striking example that corroborates this intuition, please see **Section 23 in Companion Part/Article 2** and **Remark 23.3 at the end of that section** .

for every single number I have tried it on to date ; The details of the extensive experiments can be found in the  **Companion manuscript Part/Article 2**.

Initially, in addition to square-roots of quadratic-non-residues ( `QNR` s); I also tried higher order algebraic integers as the entries of matrix at which the `MBEC` is tested. For example: cube-roots of cubic non-residues; and/or 4th-roots of quartic non-residues were tested as some key elements of the matrix (that replaces the variable $x$). As expected, a mixture of roots of different orders (ex: cube-roots and square-roots) as the algebraic-integer elements of the matrix always turned out to be more robust; albeit at the cost of more computations. Encouraged by the success of the experiments, I kept on thinking that square roots of Quadratic Non Residues should suffice. I am glad that that intuition has turned out to be correct.

Finally around Dec. of 2017, I realized that the matrices were really not needed at all. Likewise, there was no need for an algebraic integer that is the root of a polynomial equation (with rational coefficients) of degree greater than 2; or in other words; square-root of a single `QNR` was always turning out to be sufficient.

Accordingly, as of December 2017, the only ingredient that was missing from the conditions in the baseline conjecture was the Euler-Criterion Check.

I was fully aware that in the post AKS universe, unless the new method is both

1 Analytically provably deterministic

    AND

2 Substantially faster than the best known variant(s) of the AKS;

the whole exercise would be futile.
So it was an enormous risk; a huge gamble: what is the probability of latching onto something that meets all the stated criteria ??

But the success of numerical experiments is what kept me going. Numerically (in experiments) the method was always turning out to be fast, correct and ROBUST.
For example, initially; I thought that the smallest prime number $q$ that happens to be a `QNR` is needed. If that were true it would be a big setback.
But fortunately ; it turns out that **any** `QNR` **(except** $q = -1$**) is sufficient.**

The literature search and especially the Wikipedia Web Pages that I looked up in attempting to search for a direction for analytic proof (around Dec 2017 thru March 2018) reinforced the usefulness and simplicity of the Euler Criterion Check. My immediate thought was
"how and why does the Miller-Rabin method   NOT   include the Euler Criterion Check ?
especially since the Jacobi Symbol is so easy to compute ; an `ECC`   could be added before giving up on a non-witness and trying another number??"

**Even more important, the experiments clearly demonstrated that the combination of Euler Criterion Check and the Binomial Congruence Check with $\sqrt{q}$ was impenetrable.**

**Accordingly, on or about the 15-th of March 2018; I identified all the precise conditions,**
**C–1 thru C–5 that when satisfied together, are   SUFFICIENT   to guarantee primality as per the Baseline Primality Conjecture.**

Since then, up till about 24-th May 2019, I have spent a huge amount of immensely focused and intense effort to try to come up with an analytic proof of the Baseline Primality Conjecture. A summary of the outcomes is presented in the next sub-section.

**§   Section 15.2 :   Current Status of Baseline Primality Conjecture:**
                      **Analytic proof(s) done for some critical special cases**

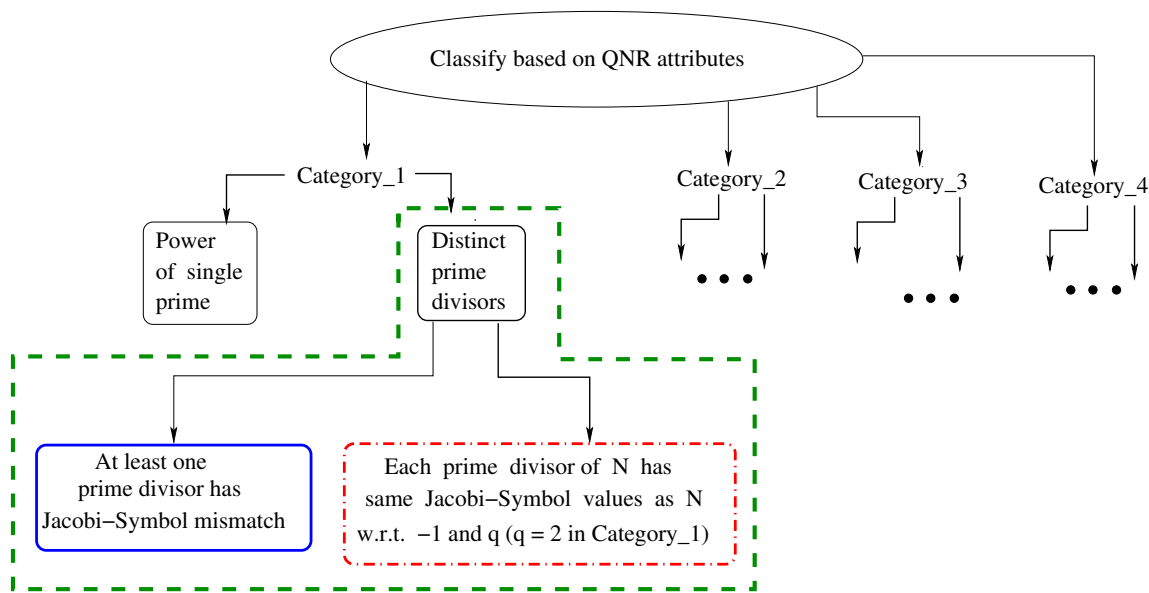

**Figure 1.** Illustration of one possible partitioning of all paths in the proof-space.

The solid line-styled blue colored box indicates the region wherein all cases that fall within that box have been analytically proved (see **Section 32**  in
**Companion Manuscript Part/Article 3**  for details).

The red colored box with dash-dots line-style indicates the region where we have partial proofs (i.e. not all cases within that box have been analytically proved yet ;  see **Section 33**  in  **Companion Manuscript Part/Article 3** ).
The same is true of the green colored box with dashes line-style (not all cases within that boundary are done since the red box has only been partially covered).

As seen in Figure  1 , it turns out that all odd integers fall into one of the four (disjoint) categories listed in the next Table :

| | | | | |
|---|---|---|---|---|
| **Category_1 :** | $(-1$ is a **QNR** $)$ | **AND** | $(2$ is also a **QNR** $)$ | $\Rightarrow$  $N = 8I + 3$ |
| **Category_2 :** | $(-1$ is a **QNR** $)$ | **AND** | $(2$ is not  a **QNR** $)$ | $\Rightarrow$  $N = 8I + 7$ |
| **Category_3 :** | $(-1$ is not a **QNR** $)$ | **AND** | $(2$ is a **QNR** $)$ | $\Rightarrow$  $N = 8I + 5$ |
| **Category_4 :** | $(-1$ is not a **QNR** $)$ | **AND** | $(2$ is not a **QNR** $)$ | $\Rightarrow$  $N = 8I + 1$ |

**Table 2.** All possible input types

In the above table, the number of interest ($q = -1$ or $q = 2$) is a  **QNR**  w.r.t.  $N$  if the `Jacobi-Symbol`  of that number w.r.t.  $N$  evaluates to  "$-1$".

Given that the algorithm heavily relies on $q = 2$ whenever it happens to be a `QNR` ; Alan Sherman suggested the divide-and-conquer strategy to try to develop a proof separately for each of the categories in the above table.

Also, given the strong dependence on `QNR` values; I decided to try and tackle the first case in the above Table first, since it has maximum number of `QNR` parameters (for this category, the most valuable parameter = integer 2 is a `QNR` w.r.t. $N$ $\Rightarrow$ $q = 2$ and additionally $-1$ is also a `QNR` w.r.t. $N$).

For each category, the proof-space can be further sub-divided into 2 distinct cases :

1. $N$ is an odd power of a single prime

2. At least 2 distinct primes divide $N$ : This case can be further sub-sub-divided into

**2**–⟨a⟩ : At least one prime divisor of $N$ has a different value of its Jacobi-Symbol w.r.t. $-1, 2$ or $q$. (obviously, the number of such divisors and/or their powers in the prime factorization of $N$ would have to be even; so as to mask Jacobi-symbol mismatch(es).
Further details can be found in **Companion Manuscript Part/Article 3** ).

   and

**2**–⟨b⟩ : All prime divisors $P_i$ of $N$ have the same values of `Jacobi-symbols` w.r.t. $-1, 2$ (and $q$ if the input falls under category_4 ; where $q \neq 2$ and $q \neq -1$) as those of $N$.

As of today : August 19, 2019 (this date automatically generated by the "`\today`" LATEX macro) the following steps (toward a full analytic proof) have been completed :

Circa 31st of May 2018 : The proof for 2 out of 3 total (sub)-cases that arise in the sub-space represented by the blue colored solid line-styled box

(It is shown in Figure 1 under Category_1 ; within the dashed green-colored boundary (which the right-side child box under Category_1 with distinct prime-divisors) and under that box ; the left-side sub-child of that box ; which corresponds to the case where at least one prime divisor of $N$ has a Jacobi-Symbol mismatch) was completed.

As of the 15-th April 2019 : The proofs of all the remaining cases inside the blue colored box have been completed.

Additionally proofs of a substantial number of sub-cases that fall within the red-colored box with dashed line-style have also been completed.

We hope that the above mentioned proofs (albeit only for special cases) that are presented in **Companion Manuscript Part/Article 3** in this document will give the readers and the reviewers, the same confidence that the authors share; viz. that things don't work out to this degree over an extended period of intense experimental and analytic scrutiny by chance or by fluke... ;
that the underlying **Baseline Primality Conjecture** must be correct ; its only a matter of time before the remaining parts of the proof are completed.

In parallel with the efforts to complete the remaining cases of the analytic proof; experiments with ever increasing number and types of input integers have (and are being) continued. Those are described in detail in **Companion Manuscript Part/Article 2** .

We close this subsection with a time-line of the most exciting developments: the discovery of the generalized versions of the Baseline Primality Conjecture, the algorithm(s) they enable and the current status of the overall work.

§   **Section 15.3 :   Time-line of Generalized Conjectures**

How and when did the generalized versions evolve?
The extremely succinct and truthful answer is:
"under immense pressure in the last few months since about the 24-th of May 2019".

Toward the end of May 2019 as I was finalizing the write-up for potential submissions and dissemination; I decided to re-visit the complexity issue and was stunned to discover that I was oblivious to the fact that the construction of, or a search for explicit `QNR` value(s) and more generally, construction of an irreducible polynomial for an arbitrary composite $N$ happen to be unsolved open problems even today (to the best of my knowledge). In other words, no deterministic polynomial effort/cost method is known to solve either of those problems, for arbitrary composite numbers $N$.
It suddenly became clear that all the good experimental data, improvements over Miller-Rabin, and everything else done up to that point did not meet the ideal goals of achieving "low complexity" and "Deterministic" polynomial cost guarantees.

In desperation, I tried to salvage as much as possible. In the process, I stumbled upon Sze's outstanding work. Fortunately, I was able to circumvent the requirement of having an explicit value of a `QNR` (or more generally, any higher order non-residue modulo-$N$). The rest is documented in Sections 8 and onward...

§   **Section 15.4 :   Cross-validation of Baseline Conjecture via Generalized versions**

In a large number of cases an explicit value of a `QNR` $q$ modulo-$N$ also happens to satisfy

$$N \mod q \neq 1 \quad \text{and } q \text{ is a prime} \tag{139}$$

In all such cases, $q$ itself can also be used as the "canonical parameter" $m$; which provides the opportunity to verify that the outcomes of the tests in the Generalized conjectures are consistent with the outcomes of tests (the ECC and BCC) in the Baseline Conjecture.

Even if the explicit value of a `QNR` cannot be used directly as the value of canonical parameter $m$ (for example when $N \mod 8 = 7$ and our algorithms set $q = (-2 \mod N) \equiv (N-2)$ ); it is possible to find the value of $m$ and then cross-verify that the PGPC results are consistent with the PBPC results.

Such comprehensive cross-validations experiments are almost finished (on the main data sets we have used in **Companion Manuscript Part/Article 2** ).
As expected, no inconsistency has been found.

In closing, I would like to point out that polynomials are not scalars but "vectors" or 1-dimensional arrays (when specified via the coefficients of each degree). In principle, therefore the methods that have worked; do end up harnessing the full power of the the Modular Binomial Expansion Congruence with 1-D arrays or vectors. In other words trying to invoke 2-dimensional entities (Matrices) turns out to be unnecessary, but invoking 1-Dimensional entities is essential and works well. I think that this fact is a strong corroboration of the original intuition.

# § Section 16 : Concluding Remarks

It appears that the way I approached the primality detection problem is in some sense a complement of the way the AKS method approaches that problem. They (AKS) tackled the difficulty head-on and proved it in the hard-direction. Hats off to their unparalleled originality.

My approach was incremental: note that for a composite number $N$; the total number of distinct non-witnesses of various types such as
(i) Fermat Witnesses, (ii) Euler-Witnesses, (iii) Binomial-Witnesses and (iv) Miller-Rabin witnesses is substantially higher than the number of integers in $Z_N^*$ that can defy/fool these tests (individually), which is why the Miller-Rabin method is so effective in practice. However, it appears that the prior methods used only one or two independent types of tests at most (for example, Miller-Rabin method combines Fermat's Little Theorem Test with whether a non-trivial square root of 1 can be found).
One intuitively straight-forward way to further strengthen these tests is to apply more of such tests (beyond just the two that Miller-Rabin method deploys). In principle, all known tests should be tried with the same base candidate/argument before giving up on it and going over to the next candidate. The hope was that there should be some combination of tests that could be devised, such that no composite could satisfy/fool all of those tests together.
Thus, my initial focus was on finding some type of witness,
or in other words it was a witness-centric approach
(because the total number of witnesses of some sort is substantially higher than totally inert non-witnesses. So, it seems logical to put more effort toward finding an element from the the majority, i.e., toward finding a witness of some sort).

The other main intuition was to try to create tests that could be performed (in essence) with NON-INTEGER values and potentially simultaneously also invoking higher dimensional entities. In some sense this was an attempt to "amplify" the number of witnesses by taking the problem to higher dimensional spaces and domains where non-integer values/symbols/entities could also serve as witnesses...

Fortunately it all seems to have worked out !!!

I hope that readers and reviewers will feel the same level of excitements as the authors !!!
and jump in and complete the analytic proofs and help us "retire the problem of primality detection" as a "solved-problem" once and for all.

**Basic Number Theory Background Fact 3 :**

**Let $N > 1$ be an odd positive integer that is not a square of some other integer. Then,**

$$\left.\begin{array}{l}\textbf{Number of ints } q \textbf{ in } Z_N^* \textbf{ with } \texttt{Jacobi\_Symbol(}q, N\texttt{)} = -1 \\ \qquad = \\ \textbf{Number of ints } r \textbf{ in } Z_N^* \textbf{ with } \texttt{Jacobi\_Symbol(}r, N\texttt{)} = +1 \end{array}\right\} = \left(\frac{\phi(N)}{2}\right) \qquad (140)$$

**where,**

$$\phi(N) = \texttt{Euler\_Totient\_Function}(N) \qquad (141)$$

**Proof :** Let

$$n_q = \text{Number of ints } q \text{ in } Z_N^* \text{ with } \texttt{Jacobi\_Symbol(}q, N\texttt{)} = -1 \qquad (142)$$

and,

$$n_r = \text{Number of ints } r \text{ in } Z_N^* \text{ with } \texttt{Jacobi\_Symbol(}r, N\texttt{)} = +1 \qquad (143)$$

and,

$$\mathcal{S}_\phi = \{\text{set of all ints in } Z_N^* \text{ that are relatively co-prime w.r.t. } N\} \qquad (144)$$

Then it is clear that

$$n_q + n_r = \|\mathcal{S}_\phi\| = \text{ cardinality of } \mathcal{S}_\phi = \phi(N) \qquad (145)$$

Assume that $n_q > 0$ and pick any one value say $q_1$ from the subset that contains all $\texttt{QNR}$ s. Consider the set

$$\mathcal{S}_\phi^{'} = \{x = q_1 \times s_k \mod N\} \quad \forall s_k \in \mathcal{S}_\phi \qquad (146)$$

In other words,

$$\mathcal{S}_\phi^{'} = \text{the set obtained by multiplying each element of set } \mathcal{S}_\phi \text{ by } q_1, \quad \text{modulo-}N \qquad (147)$$

Then use the facts that

$$\texttt{QNR} \times \texttt{QNR} = \texttt{QR} \qquad (148)$$

and,

$$\texttt{QNR} \times \texttt{QR} = \texttt{QNR} \qquad (149)$$

to obtain the fact that

$$n_q^{'} = \text{number of } \texttt{QNR} \text{ s in } \mathcal{S}_\phi^{'} = n_r \qquad (150)$$

and,

$$n_r^{'} = \text{number of } \texttt{QR} \text{ s in } \mathcal{S}_\phi^{'} = n_q \qquad (151)$$

But it is also a well known fact that

$$\mathcal{S}'_\phi = \text{merely a permutation of all the elements in } \mathcal{S}_\phi \quad \Rightarrow \quad \mathcal{S}'_\phi = \mathcal{S}_\phi \tag{152}$$

The preceding relation in-turn implies that

$$n'_q = n_q = n_r = n'_r = \left( \frac{\phi(N)}{2} \right) \qquad \square \tag{153}$$

Note that the proof assumed that $n_q > 0$, given that $N$ is not a perfect square of some other integer.

We leave that last proof to the reader. It is fairly straightforward and unnecessarily tedious, adding little extra value or insights. It is therefore omitted for the sake of brevity and clarity[27]

Then why not omit the appendix altogether ?
because we think that the proof is cute, using only symmetry arguments and covers primes as well as composites (except perfect squares) under a unified line of reasoning.

We hope that the readers will agree with the decision to include the good part of the proof but leave out the unnecessarily tedious stuff ...

---

[27]I am afraid that that appears to be a conundrum given that this set of manuscripts is already super-long, spanning $> 200$ pages in total.

We are aware that it will be a hard task to find reviewers willing to go thru such a long document.
Our only hope is that it will NOT turn out to be NP-hard = impossibly hard.

# PPT : New Low Complexity Deterministic Primality Tests Leveraging Explicit and Implicit Non-Residues

## A Set of Three Companion Manuscripts

### PART/Article 2 : Substantial Experimental Data and Evidence[28]

**Dhananjay Phatak**    (**phatak@umbc.edu**)

and

Alan T. Sherman   and   Steven D. Houston   and   Andrew Henry
(CSEE  Dept. UMBC, 1000 Hilltop Circle, Baltimore, MD 21250, U.S.A.)

First identification of the Baseline Primality Conjecture @ $\approx 15^{\text{th}}$ March 2018

First identification of the Generalized Primality Conjecture @ $\approx 10^{\text{th}}$ June 2019

Last document revision date  (time-stamp) = August 19, 2019

---

[28]  **No counter example has been found**

**§ Section 18 : Introduction: Software and Data Sets used**

I used the "Maple" software package [9] for all our experiments. The main reasons for selecting it (over "Mathematica" or "Mat-lab", ... etc.) were the following :

**1 :**   Maple has built in support for arbitrarily large bit-length operands and operations [29]

**2 :**   Strong robust tools available for number-theory experiments.

**3 :**   Maple has strong and robust built-in support for symbolic computations. For example; after an assignment such as "s := sqrt(2)"; the entity "s" is treated as an expression or a symbol in all computations. Maple will not substitute the numerical value unless specifically instructed to do so. This emphasis on "retaining symbols" as far as possible has been a conscious design choice right from the get-go; when Maple was created. Maple therefore is an ideal tool for symbolic modular computations (which is the main thrust of this work).

**4 :**   It was created at and is maintained extremely well by an academic institution (Univ. of Toronto, Canada)

**5 :**   As a result, our institution has been providing access to MAPLE free of charge to all faculty, staff and students; on campus servers; for a long time.

**6 :**   An excellent, detailed, and recent (October 2018) article [2] "hand-constructs" composite numbers that get wrongly classified as primes by some of the current software libraries (for example : GMP, Java, Python etc).

In that article, at the end of the section titled "(Appendix) J.2 Maple"; on page 298, in column 1, in the 4-th paragraph from the top ;  immediately preceding the next sub-section titled "J.3 SageMath"
the authors write :

" To fool Maple's primality testing for numbers larger than $5 \times 10^9$, we would need a composite  $n$  passing a Lucas test and 5 rounds of Miller-Rabin testing. We do not currently know any such  $n$."

In other words, for large numbers ($> 5 \times 10^9$), the authors of that (excellent) article could not find a single (counter) example that fools Maple. This was the case despite the fact that the team of authors expended a  HUGE  amount of computational effort (guided by equal amount of smart intellectual analytical effort).

The article was published just 10 months ($<$ one year) ago !
Therefore, in hind-sight it is a fair assessment to say that "Maple is not a bad choice after all".

---

[29]there is an upper limit on operand length; but it is so high, certainly above billion bits ; that even a copy operation starts requiring humanly noticeable delay, and the maple execution quits with the error message essentially saying "there is not enough memory available to complete the operation". We encountered this limit when we tried to directly test the 33-rd Fermat Number $F_{33} = \left( [2]^{(2^{33})} + 1 \right)$  using the PPTA_EQNR  Algorithm.

In a real sense, the tool used is secondary or even immaterial; it should be the
(1) quality and then
(2) the quantity of the data generated and the conclusions it supports that should matter;
not the specifics of the software used (as long as it is widely available to the readers for independently
cross-checking the experimental results presented herein. Maple certainly meets this criterion).

Additionally I shall be glad to provide all the maple code I wrote, to any interested reader upon request.

§  **Section 18.1 :  Data Sets**

(i) The main data sets I used are available via an excellent repository created by number theory
researcher G.E. Pinch  [1, 34, 48]. The repository contains 4 main data files (all available via
G. E. Pinch's data-site)

The first 3 files contain an exhaustive list of every single Carmichael Number $< 10^{(18)}$.
The fourth file contains a list of Pseudo-primes (including strong Pseudo-primes) $< 10^{(13)}$.

Further details about this data set and the outcomes can be found in  **Section 20** .

(ii) The second data item is list of about 30 big Carmichael Numbers of lengths from 20 decimal
digits all the way to 397 decimal digits; that I gathered from various sources. See  **Section 21**  and
**Appendix A.2**  for details.

(iii) The third data set contains gigantic "probable primes" that are between 30,000 and 60,000 decimal
digits long.  These were taken from another excellent web-site created and actively maintained by
Number Theory Researcher Henri Lifchitz at the url
www.primenumbers.net/prptop/prptop.php  (also accessible via reference [49]).

See  **Section 22**  for details about experimental outcomes for this data set.

(iv) The last data set contains all the big numbers that appear in a recent landmark article [2], titled
"Primes and Prejudice: Primality Testing Under Adversarial Conditions".
The composites in that article are carefully constructed to fool the probabilistic primality tests
implemented in one of the currently used software-libraries (such as GNU-MP, Java, Python ..., etc.)

See  **Section 24**  and  **Appendix A.3**  for details regarding the results for that data set.

Before presenting a summary of all the experimental data that corroborate the Conjectures ; we first show that the conditions  C–3  thru  C–5  stated on  " **article 1 – page 16** "  are necessary.

**Remark 19.1:**  We would like to point out that not every condition is necessary for every $N$. However, if one of those conditions is dropped then there are composites that get incorrectly classified as primes.

In other words, all those conditions when satisfied together are  SUFFICIENT  to conclude primality of any/every input value $N$; however, no smaller set of conditions has been found to be sufficient to guarantee the primality of ALL input values.

**(I) :  Demonstrate the necessity of condition   C–3  in  PBPC :**

Let        $N = 3215031751 = 151 \times 751 \times 28351$ (154)

(it is the 1018-th Carmichael Number ; see the extensive archive of Carmichael Numbers and Pseudo-primes compiled and made available on-line by Richard G. E. Pinch [1, 34, 48]) for further details.)
Then, it can be verified that  $N \bmod 4 = 3$  $\Rightarrow$  `Jacobi_Symbol(-1,N) = -1` .
Accordingly, set $q = -1$.    Then it is clear that

$$q^{\left(\frac{N-1}{2}\right)} = (-1)^{(\texttt{odd\_number})} = -1 \mod N = \texttt{Jacobi\_Symbol}(-1,N)$$ (155)

The preceding Equation shows that the Euler Criterion is (trivially) satisfied by $q = -1$ .
Finally, it can also be verified that

$$(1+\sqrt{-1})^N \mod N = 1+(\sqrt{-1})^N \mod N  \Rightarrow  \texttt{BCC}(-1,N) = 0$$ (156)
 $\Rightarrow$  `MBEC`   check is also satisfied

Thus all conditions except C–3 are satisfied ; but $N$ is a composite number.
This counter-example clearly demonstrates that condition C–3 cannot be ignored; it is necessary.

For this $N$ ; it can be verified that
$N \bmod 8 = 7$  $\Rightarrow$  $2$  is a  `QR`  w.r.t. $N$ ;
However; $-1$ is a  `QNR`   $\Rightarrow$   $q = -2 \mod N = (N-2) = 3215031750$   is a  `QNR` .

It turns out that `ECC(-2,N) = 0`  $\Rightarrow$   the Euler criterion is satisfied.
Then, it is the last step in the `PPTA_EQNR` algorithm; i.e. ;  the  `MBEC`   check ; `BCC(`$\sqrt{-2}$`,N)` that detects the composite-ness of $N$ : it can be verified that :

$$\texttt{BCC(}\sqrt{-2}\texttt{,N)} \equiv 569101174 - 2730810320\sqrt{-2}$$ (157)

---

[30] Equivalently; if instead of using $\sqrt{-2}$ as the symbol, if we use $\sqrt{N-2}$ as the symbol (in the modular exponentiations

A second counter example is

$$N = 2047 = 23 \times 89 \tag{160}$$

$N \mod 4 = 3 \quad \Rightarrow \quad -1 \text{ is a } \texttt{QNR}$

$\quad \Rightarrow \quad$ Euler Criterion is trivially satisfied

(It is possible that this trivial satisfaction of the `ECC` by the special value "$-1$" negates what that check/condition is intended to screen for.)

Finally, it can be verified that

$$\texttt{BCC}(-1, N) \equiv \left[ (1 + \sqrt{-1})^N - 1 - (\sqrt{-1})^N \right] \mod N = 0$$

For this $N$ it can be verified that

$$\texttt{Jacobi\_Symbol}(3, N) = -1 \quad \text{and} \quad \texttt{ECC}(3, N) = 1566 \neq 0 \quad \Rightarrow \quad 3 \text{ is an Euler Witness}$$

## (II) : Demonstrate the necessity of condition $\quad$ C–4 in PBPC :

Let $\qquad N = 561 = 3 \times 11 \times 17 \tag{161}$

(this is the very first, the smallest Carmichael Number that exists)
For this $N$, it can be verified that
$\texttt{Jacobi\_Symbol}(389, 561) = -1$ ; $\quad$ Accordingly, set $\; q = 389$
Then it can be verified that

$$(1 + \sqrt{389})^{561} \mod 561 = 1 + (\sqrt{389})^{561} \mod 561 \quad \Rightarrow \quad \texttt{BCC} = 0 \tag{162}$$

$\quad \Rightarrow \quad$ `MBEC` $\;$ is satisfied

In this counter example ; all conditions except C–4 are satisfied and $N$ is composite ; thereby demonstrating that condition C–4 (the all important Euler Criterion) cannot be dropped.

## (III) : Demonstrate the necessity of condition $\quad$ C–5 in PBPC :

Let $\qquad N = 2047 = 23 \times 89 \tag{163}$

For this $N$, it can be verified that
$N \mod 8 = 7 \quad \Rightarrow \quad -2 \text{ is a } \texttt{QNR}$ .
Accordingly, set $\; q = (-2 \mod N) = (N - 2) = 2045$

in the Binomial Congruence Check) ; then it can be verified that

$$\texttt{BCC}(\sqrt{3215031749}, N) = 569101174 + 484221431\sqrt{3215031749} \tag{158}$$

The preceding relation is the same as Eqn. (157) since $\; -2730810320 \mod N = 484221431 \tag{159}$

Then, it can be verified that

$$
\left.
\begin{aligned}
(2045)^{1023} \mod N = 2046 \equiv -1 \mod 2047 = \texttt{Jacobi\_Symbol(}q, N\texttt{)} \\
\Rightarrow \quad \text{Euler Criterion is satisfied}
\end{aligned}
\right\} \tag{164}
$$

Finally, it can be verified that

$$
(1+\sqrt{2045})^N \mod N = 1522 + 1068\sqrt{2045} \quad \Rightarrow \quad \texttt{BCC} \neq 0 \quad \Rightarrow \quad N \text{ is composite} \tag{165}
$$

In this counter example all conditions except C–5 are satisfied and $N$ is composite ; thereby demonstrating that condition C–5 (the main new condition we have identified) must be satisfied.

This value $N = 2047$ turns out to be the smallest odd integer that is correctly identified as a composite (only) by the $\texttt{MBEC}$ check; which is the last step within the $\texttt{PPTA\_EQNR}$ Algorithm. In other words, this is the smallest odd integer for which the new component in the $\texttt{PPTA\_EQNR}$ algorithm, viz, the $\texttt{MBEC}$ check is needed.

We also note the following coincidence : "2047" happens to be the smallest integer in the OEIS Sequence A014233 . That web-page (also accessible via ref. [50]) lists the sequence of odd numbers such that the $k$-th number in that sequence has the following property: It is the "Smallest odd number for which Miller-Rabin primality test on bases $\leq k$-th prime does not reveal composite-ness".

2047 corresponds to $k = 1$ ; so that all bases $\leq 1$-st prime number
i.e., all bases $\leq 2$ are NON–witnesses for the Miller-Rabin test for $N = 2047$ .
This fact can be verified based on the following :

$$
\frac{N-1}{2} = 1023 \qquad \text{and}
$$
$$
2^{(1023)} \mod 2047 = 1 \tag{166}
$$

Accordingly, it is not surprising that 2047 also appears in several related web-sites (such as those listed in [35, 51])[31]

---

[31] In addition to the fact that 2047 was used twice in this sub-section : as a counter example to demonstrate that condition C-3 is required AND condition C-5 is also necessary;

it should be noted that 2047 is the smallest Mersenne number (which is defined as a number of the form $(2^p - 1)$ where the exponent $p$ is a prime number) which is not a prime number itself. All smaller Mersenne numbers $2^2 - 1, 2^3 - 1, 2^5 - 1$ and $2^7 - 1$ are primes themselves. The next Mersenne number $= 2^{11} - 1 = 2047$ is the smallest such number which is not a prime itself.

So, 2047 could be a strong competitor for a pageant or a contest for being the

"smallest most-interesting two-factor composite with non-trivial factors; where non-trivial means > 10)"

Just joking as usual. I strongly believe that if a document does not make an effort to put a smile on the reader's face at least once, then it is not even worth the fonts it is typeset in

# § Section 20 : Experimental Outcomes from the Main Data Sets: All Carmichaels $< 10^{18}$ and Pseudoprimes $< 10^{13}$ (Details in Appendix A.1 )

G.E. Pinch has created an excellent repository [1, 34, 48] that contains 4 main data files (all available via  G. E. Pinch's data-site :)

File/Set/List 1 : Every single Carmichael Number $\mathcal{C}$  (in ascending order of magnitude); where
$$\mathcal{C} < 10^{16}$$

File/Set/List 2 : Every single Carmichael Number $\mathcal{C}$ in the range $10^{16} \leq \mathcal{C} < 10^{17}$ and

File/Set/List 3 : Every single Carmichael Number $\mathcal{C}$ in the range $10^{17} \leq \mathcal{C} < 10^{18}$

File/Set/List 4 : Additionally, there is a list of pseudo-primes $\mathcal{S} < 10^{13}$
The pseudo-primes list also includes Carmichael Numbers. We therefore removed the overlap and created a list of only pseudo-primes that are not Carmichaels.

The results obtained by testing primality of each of the numbers in the first set of Carmichaels with the `PPTA_EQNR`  Algorithm are summarized in Tables 3  and 4 .

See Appendix  A.1  for the source code and the screen capture of the execution.

In Table  3 , the first column indicates the total number of cases seen thus far (i.e., up to the point where that print statement was encountered) in the execution.  Roughly 10 such printouts were generated at equally spaced intervals in order to facilitate the rendering of the data in a graphical format as plots . (Further details can be found in the screen-capture included in Appendix A.1 : follow **link-to-first-print-instance** . Overall, there are 10 such prints. Accordingly the number of rows in the table is 10.

The middle column in this table shows the actual count (i.e., the number) of cases out of the total number of cases seen thus far (which is the number in the same row in the first column); as well as the equivalent fractional value. For example, from the first row in that Table, we see that out of first 24668 Carmichaels tested ;  21726 required a search for a  `QNR` . This means that $\frac{21726}{24668} = 0.8807 \equiv 88.07\,\%$ of the Carmichael numbers seen/tested (up to that point in the execution) were of the form $(8I + 1)$  and therefore needed a search to find a  `QNR` .

The last column indicates the mean or average number of iterations needed to find a  `QNR` when the simplest brute-force exhaustive march thru all prime numbers starting at 3 was used to find a  `QNR` .
(we comment on the trends seen at the end of this section ;  after the data for all sets of Carmichael Numbers and Pseudoprimes has been presented).

Table  4  shows the fractions resolved by some of the individual component mechanisms within the `PPTA_EQNR`  Algorithm. (again, for brevity, we defer comments to the end of this section.)

| Total no. of cases = index of last Carmichael number tested | No. of cases that need a search for a **QNR** | Average number of iterations needed to find a **QNR** |
|---|---|---|
| 24668 | 21726 $\longleftrightarrow$ 0.8807 | 4.0622 |
| 49336 | 44043 $\longleftrightarrow$ 0.8927 | 4.1615 |
| 74004 | 66479 $\longleftrightarrow$ 0.8983 | 4.2169 |
| 98672 | 89118 $\longleftrightarrow$ 0.9032 | 4.2554 |
| 123340 | 111844 $\longleftrightarrow$ 0.9068 | 4.2857 |
| 148008 | 134619 $\longleftrightarrow$ 0.9095 | 4.3101 |
| 172676 | 157444 $\longleftrightarrow$ 0.9118 | 4.3312 |
| 197344 | 180328 $\longleftrightarrow$ 0.9138 | 4.3491 |
| 222012 | 203277 $\longleftrightarrow$ 0.9156 | 4.3620 |
| 246683 | 226188 $\longleftrightarrow$ 0.9169 | 4.3753 |

**Table 3.** Carmichaels Set 1 : Raw count and the fraction of cases where a search for a **QNR** is needed ; and the average number of iterations needed to find a **QNR** $q$ by a brute-force-stepping thru all primes from 3 onward.

| Total no. of cases = Index of last Carmichael number tested | Number of cases resolved by mechanisms within the **PPTA_EQNR** Algorithm | | |
|---|---|---|---|
| | **JS** $= 0$ during **QNR** search | Euler Criterion | only by **BCC** |
| 24668 | 8905 $\longleftrightarrow$ 0.3610 | 15575 $\longleftrightarrow$ 0.6314 | 188 $\longleftrightarrow$ 0.0076 |
| 49336 | 17855 $\longleftrightarrow$ 0.3619 | 31162 $\longleftrightarrow$ 0.6316 | 319 $\longleftrightarrow$ 0.0065 |
| 74004 | 26969 $\longleftrightarrow$ 0.3644 | 46592 $\longleftrightarrow$ 0.6296 | 443 $\longleftrightarrow$ 0.0060 |
| 98672 | 36136 $\longleftrightarrow$ 0.3662 | 61987 $\longleftrightarrow$ 0.6282 | 549 $\longleftrightarrow$ 0.0056 |
| 123340 | 45305 $\longleftrightarrow$ 0.3673 | 77403 $\longleftrightarrow$ 0.6276 | 632 $\longleftrightarrow$ 0.0051 |
| 148008 | 54624 $\longleftrightarrow$ 0.3691 | 92666 $\longleftrightarrow$ 0.6261 | 718 $\longleftrightarrow$ 0.0049 |
| 172676 | 63945 $\longleftrightarrow$ 0.3703 | 107939 $\longleftrightarrow$ 0.6251 | 792 $\longleftrightarrow$ 0.0046 |
| 197344 | 73303 $\longleftrightarrow$ 0.3714 | 123184 $\longleftrightarrow$ 0.6242 | 857 $\longleftrightarrow$ 0.0043 |
| 222012 | 82668 $\longleftrightarrow$ 0.3724 | 138405 $\longleftrightarrow$ 0.6234 | 939 $\longleftrightarrow$ 0.0042 |
| 246683 | 91970 $\longleftrightarrow$ 0.3728 | 153696 $\longleftrightarrow$ 0.6231 | 1017 $\longleftrightarrow$ 0.0041 |

**Table 4.** Carmichaels in Set 1 : Counts and fractions resolved by mechanisms within the **PPTA_EQNR** Algorithm.

We believe that a picture is worth a 1000 words. Therefore, we also illustrate the data from Tables 3 and 4 in a graphical format; as plots.

To that end, Figure 2 illustrates the data from Table 3 with two distinct plots.

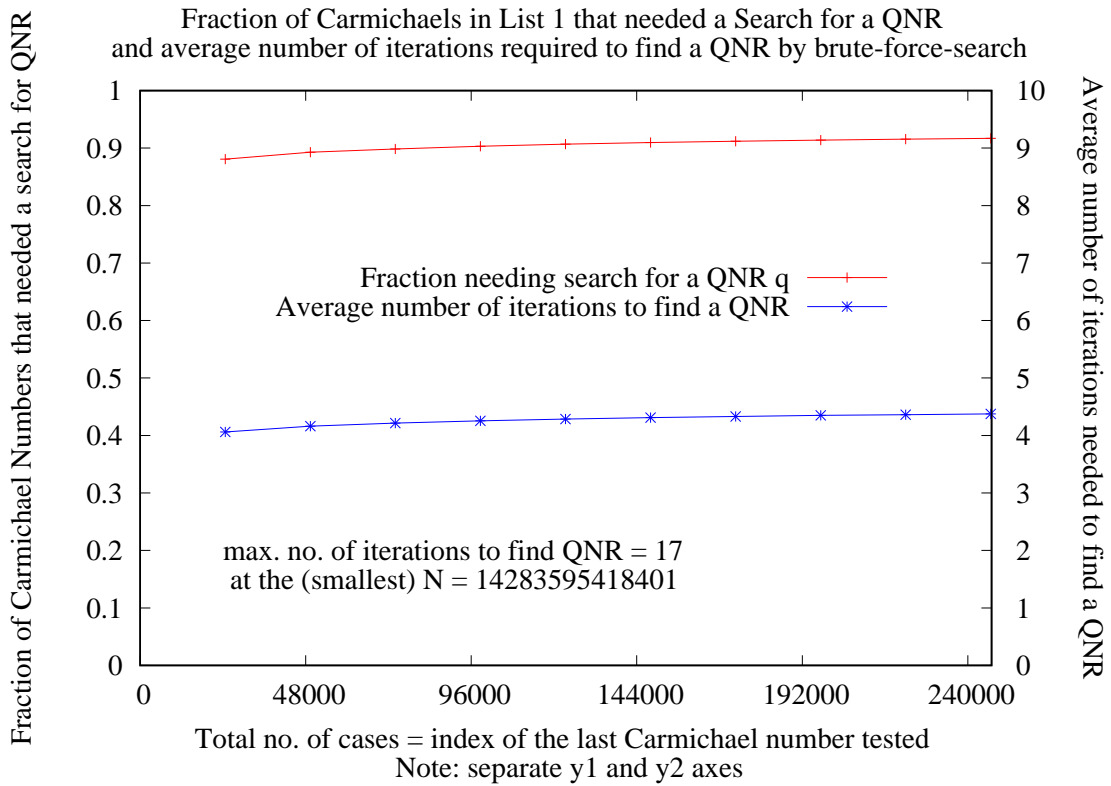

**Figure 2.** Fraction of cases where a search for a **QNR** is needed ($Y_1$ axis on the left) ;    and the average number of iterations needed to find a **QNR** $q$ by a brute-force-stepping thru primes from 3 onward ($Y_2$ axis on the right)

In this Figure, there are two distinct $Y$ axes corresponding to two distinct plots.

The red colored plot uses the Y1-axis on the left side to indicate the fraction of values that needed a search for a **QNR** .

The blue colored plot uses the Y-2 axis on the right to track the average number of iterations required to find a **QNR** only in cases where a search was needed (otherwise the average values would have been even lower).

The one extra informational item the Figure (which does not appear in the Table) is a label that indicates the maximum number of iterations that were required to find a **QNR** for any of the $N$ value(s) in the entire set; and the smallest value of $N$ which required that maximum number of iterations.

The index of $N = 14283595418401$ in the list is 21890. In other words, in the first set of Carmichaels;

the maximum number of iterations required to find a `QNR` is 17; and the smallest value of $N$ that needs 17 iterations to find a `QNR` happens to be the

**21890-th Carmichael number** $N = 14283595418401$ .

Next, we illustrates the Data in Table 4 with Figures 3 and 4 .

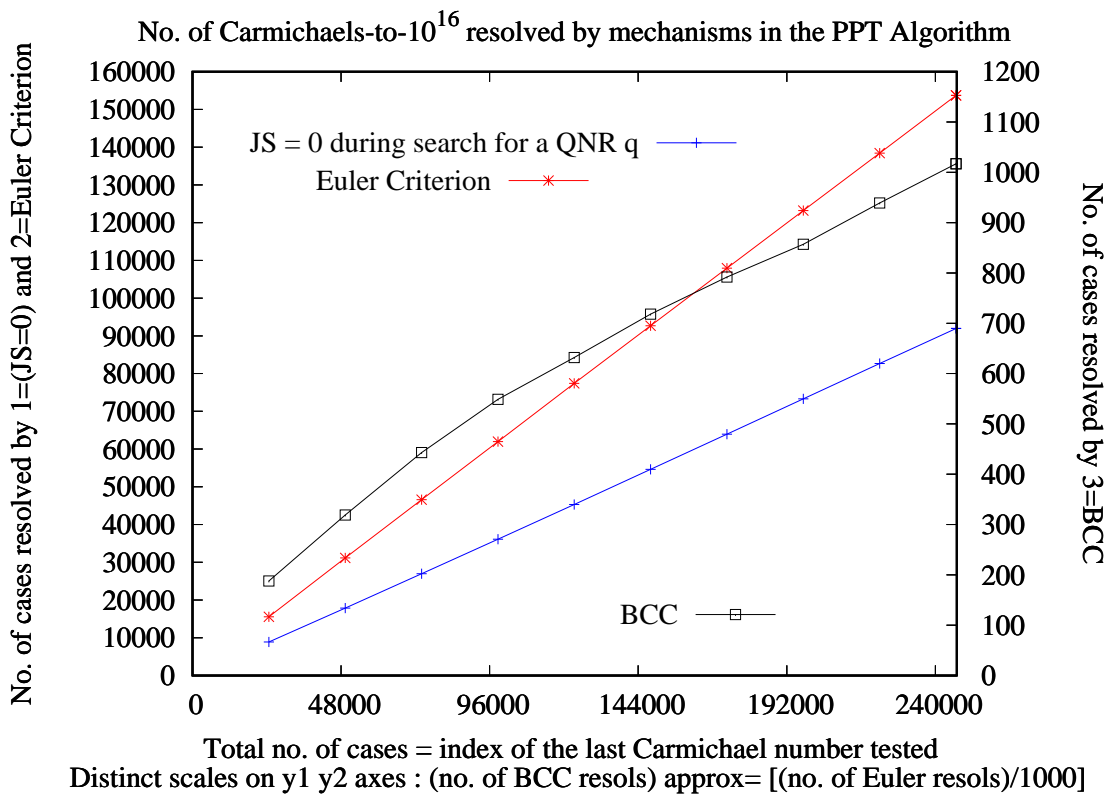

**Figure 3.** Set 1 : Number of cases resolved by individual mechanisms within the `PPTA_EQNR` Algorithm.

Figure 3 illustrates the raw counts (i.e., the integers) in each of the three columns in Table 4 with a distinct plot. Accordingly, there are three plots in this Figure.

The first column (corresponding to `JS = 0`) is plotted with a the blue colored graph ; and
The second column (Euler Criterion) is plotted with a the red colored graph.
These two graphs use the Y-1 axis on the left side.

The third column (in Table 4) is plotted with a the black colored graph ; and uses the Y-2 axis on the right.

Finally, the fractions appearing in the columns in Table 4 are illustrated in Figure 4 as a box-plot. The authors think that a box-plot is probably the best way to illustrate the contribution of each of the three relevant mechanisms within the `PPTA_EQNR` Algorithm.

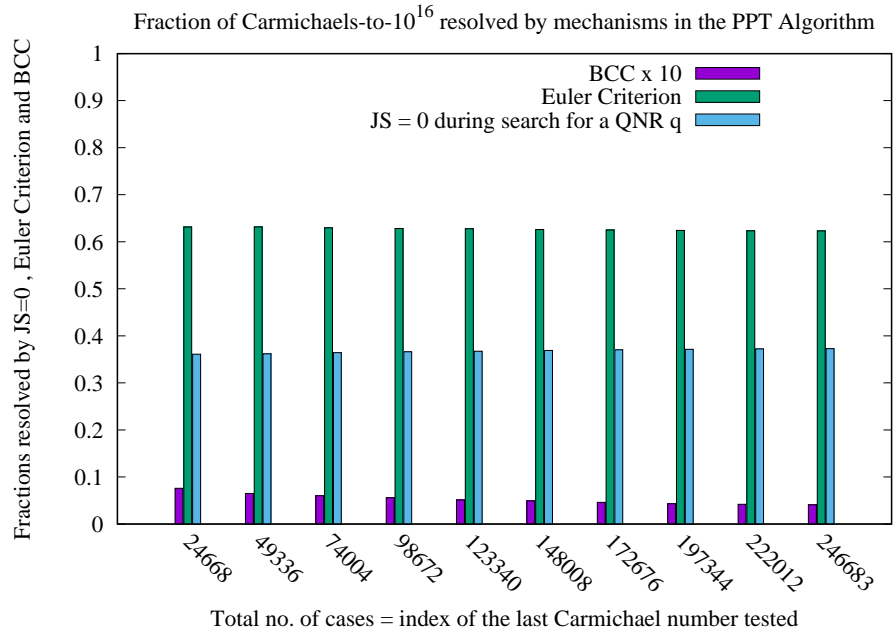

**Figure 4.** Set 1 : Fractions resolved by individual mechanisms within the `PPTA_EQNR` Algorithm illustrated with a box-plot.

In this Figure there is only a single Y axis on the left. Also note that the values (in the `BCC` column) are multiplied by a factor 10 in order to make them appear on the plot (the actual fractions are too small to appear as a boxes of non-zero height on a common Y-axis scale).

Next, similar results (i.e., results for the same metrics) for the second and third sets of Carmichael Numbers are summarized in

Tables 5 ; 6 (for **Set 2)** and
Tables 7 ; 8 (for **Set 3)** ; Respectively.

| No. of cases = index of last Carmichael number tested | No. of cases that need a search for a `QNR` | Number of iterations needed to find a `QNR` | |
|---|---|---|---|
| | | Average | Max seen up-to this point |
| 33867 | 31622 ⟷ 0.9337 | 4.4975 | 19 |
| 67734 | 63164 ⟷ 0.9325 | 4.5028 | 19 |
| 101601 | 94723 ⟷ 0.9323 | 4.52 | 19 |
| 135468 | 126354 ⟷ 0.9327 | 4.5258 | 19 |
| 169335 | 157990 ⟷ 0.9330 | 4.5314 | 19 |
| 203202 | 189669 ⟷ 0.9334 | 4.5367 | 19 |
| 237069 | 221418 ⟷ 0.9340 | 4.5436 | 19 |
| 270936 | 253167 ⟷ 0.9344 | 4.5485 | 19 |
| 304803 | 284822 ⟷ 0.9344 | 4.5532 | 19 |
| 338672 | 316707 ⟷ 0.9351 | 4.5577 | 22 |

**Table 5.** Carmichaels in Set 2 : Raw count and the fraction of cases where a search for a `QNR` is needed ; and the average and maximum number of iterations needed to find a `QNR` $q$ by a brute-force-stepping thru all primes from 3 onward.

| Total no. of cases = Index of last Carmichael number tested | Number of cases resolved by mechanisms within the `PPTA_EQNR` Algorithm | | |
|---|---|---|---|
| | `JS` $= 0$ during `QNR` search | Euler Criterion | only by `BCC` |
| 33867 | 12923 ⟷ 0.3816 | 20847 ⟷ 0.6156 | 97 ⟷ 0.0029 |
| 67734 | 25864 ⟷ 0.3818 | 41681 ⟷ 0.6154 | 189 ⟷ 0.0028 |
| 101601 | 38886 ⟷ 0.3827 | 62442 ⟷ 0.6146 | 273 ⟷ 0.0027 |
| 135468 | 51663 ⟷ 0.3814 | 83453 ⟷ 0.6160 | 352 ⟷ 0.0026 |
| 169335 | 64602 ⟷ 0.3815 | 104291 ⟷ 0.6159 | 442 ⟷ 0.0026 |
| 203202 | 77416 ⟷ 0.3810 | 125275 ⟷ 0.6165 | 511 ⟷ 0.0025 |
| 237069 | 90423 ⟷ 0.3814 | 146046 ⟷ 0.6160 | 600 ⟷ 0.0025 |
| 270936 | 103316 ⟷ 0.3813 | 166935 ⟷ 0.6161 | 685 ⟷ 0.0025 |
| 304803 | 116259 ⟷ 0.3814 | 187767 ⟷ 0.6160 | 777 ⟷ 0.0025 |
| 338672 | 129112 ⟷ 0.3812 | 208706 ⟷ 0.6162 | 854 ⟷ 0.0025 |

**Table 6.** Carmichaels in Set 2 : Counts and fractions resolved by mechanisms within The `PPTA_EQNR` Algorithm.

There is the following minor difference between the first table for Set 1 (i.e., Table 3) ; and the corresponding ones for Sets 2 and 3 : there is one more column in Tables 5 and 7 than in Table 3. The additional column is the last column. It tracks the maximum number of iterations required to find a `QNR` for any of the $N$ values tested up to that point (in the execution of the big loop testing all the values in that Set/List in order; one at a time).

| No. of cases = index of last Carmichael number tested | No. of cases that need a search for a **QNR** | Number of iterations needed to find a **QNR** | |
|---|---|---|---|
| | | Average | Max seen up-to this point |
| 81629 | 76679 ⟷ 0.9394 | 4.6023 | 22 |
| 163258 | 153443 ⟷ 0.9399 | 4.615 | 22 |
| 244887 | 230287 ⟷ 0.9404 | 4.6238 | 22 |
| 326516 | 307136 ⟷ 0.9406 | 4.6295 | 22 |
| 408145 | 384210 ⟷ 0.9414 | 4.6345 | 22 |
| 489774 | 461152 ⟷ 0.9416 | 4.6391 | 22 |
| 571403 | 538286 ⟷ 0.9420 | 4.6437 | 22 |
| 653032 | 615372 ⟷ 0.9423 | 4.6476 | 22 |
| 734661 | 692563 ⟷ 0.9427 | 4.6524 | 22 |
| 816289 | 769814 ⟷ 0.9431 | 4.6524 | 22 |

**Table 7.** Carmichaels in Set 3 : Raw count and the fraction of cases where a search for a **QNR** is needed ; and the average and maximum number of iterations needed to find a **QNR** $q$ by a brute-force-stepping thru all primes from 3 onward.

| Total no. of cases = Index of last Carmichael number tested | Number of cases resolved by mechanisms within the **PPTA_EQNR** Algorithm | | |
|---|---|---|---|
| | **JS** $= 0$ during **QNR** search | Euler Criterion | only by **BCC** |
| 81629 | 31231 ⟷ 0.3826 | 50230 ⟷ 0.6153 | 168 ⟷ 0.0021 |
| 163258 | 62464 ⟷ 0.3826 | 100472 ⟷ 0.6154 | 322 ⟷ 0.0020 |
| 244887 | 93501 ⟷ 0.3818 | 150913 ⟷ 0.6163 | 473 ⟷ 0.0019 |
| 326516 | 124619 ⟷ 0.3817 | 201258 ⟷ 0.6164 | 639 ⟷ 0.0020 |
| 408145 | 155991 ⟷ 0.3822 | 251358 ⟷ 0.6159 | 796 ⟷ 0.0020 |
| 489774 | 187146 ⟷ 0.3821 | 301679 ⟷ 0.6160 | 949 ⟷ 0.0019 |
| 571403 | 218532 ⟷ 0.3824 | 351755 ⟷ 0.6156 | 1116 ⟷ 0.0020 |
| 653032 | 249717 ⟷ 0.3824 | 402047 ⟷ 0.6157 | 1268 ⟷ 0.0019 |
| 734661 | 281015 ⟷ 0.3825 | 452244 ⟷ 0.6156 | 1402 ⟷ 0.0019 |
| 816289 | 312567 ⟷ 0.3829 | 502176 ⟷ 0.6152 | 1546 ⟷ 0.0019 |

**Table 8.** Carmichaels Set 3 : Counts and fractions resolved by mechanisms within The PPTA_EQNR Algorithm.

It turns out that when the data from Tables 5 thru 8 is graphed ; the plots show trends that are identical to those seen in Figures illustrating the results for the first set of Carmichaels. For the sake of brevity, we therefore omit the analogous plots for **Set 2** and **Set 3** of Carmichael Numbers.

Finally, the results for the same metrics for the set of Pseudo-primes ( **Set/List 4** mentioned above) are summarized in Tables  9  and  10

| Total no. of cases = index of last Pseudoprime tested | No. of cases that need a search for a QNR | Number of iterations needed to find a QNR | |
|---|---|---|---|
| | | Average | Max seen up-to this point |
| 24496 | 13598 ⟷ 0.5551 | 3.0131 | 13 |
| 48992 | 27713 ⟷ 0.5657 | 3.0828 | 14 |
| 73488 | 41773 ⟷ 0.5684 | 3.1143 | 17 |
| 97984 | 55980 ⟷ 0.5713 | 3.1379 | 17 |
| 122480 | 70261 ⟷ 0.5737 | 3.1595 | 22 |
| 146976 | 84533 ⟷ 0.5751 | 3.1778 | 22 |
| 171472 | 98824 ⟷ 0.5763 | 3.1905 | 22 |
| 195968 | 113206 ⟷ 0.5777 | 3.2013 | 22 |
| 220464 | 127512 ⟷ 0.5784 | 3.2086 | 22 |
| 244961 | 141834 ⟷ 0.5790 | 3.2161 | 22 |

**Table 9.** Pseudoprimes : Raw count and the fraction of cases where a search for a QNR is needed ;  and  the average and maximum number of iterations needed to find a QNR $q$ by a brute-force-stepping thru primes from 3 onward.

| Total no. of cases = Index of last Pseudoprime tested | Number of cases resolved by mechanisms within the PPTA_EQNR Algorithm | | |
|---|---|---|---|
| | JS $= 0$ during QNR search | Euler Criterion | only by BCC |
| 24496 | 3463 ⟷ 0.1414 | 17862 ⟷ 0.7292 | 3169 ⟷ 0.1294 |
| 48992 | 6914 ⟷ 0.1411 | 35815 ⟷ 0.7310 | 6261 ⟷ 0.1278 |
| 73488 | 10317 ⟷ 0.1404 | 53770 ⟷ 0.7317 | 9399 ⟷ 0.1279 |
| 97984 | 13770 ⟷ 0.1405 | 71686 ⟷ 0.7316 | 12526 ⟷ 0.1278 |
| 122480 | 17166 ⟷ 0.1402 | 89626 ⟷ 0.7318 | 15686 ⟷ 0.1281 |
| 146976 | 20503 ⟷ 0.1395 | 107684 ⟷ 0.7327 | 18787 ⟷ 0.1278 |
| 171472 | 23822 ⟷ 0.1389 | 125645 ⟷ 0.7327 | 22003 ⟷ 0.1283 |
| 195968 | 27229 ⟷ 0.1389 | 143610 ⟷ 0.7328 | 25127 ⟷ 0.1282 |
| 220464 | 30611 ⟷ 0.1388 | 161550 ⟷ 0.7328 | 28301 ⟷ 0.1284 |
| 244961 | 33818 ⟷ 0.1381 | 179633 ⟷ 0.7333 | 31508 ⟷ 0.1286 |

**Table 10.** Pseudoprimes : Counts and fractions resolved by mechanisms within The  PPTA_EQNR  Algorithm.

Figure 5 illustrates the data from Table 9.

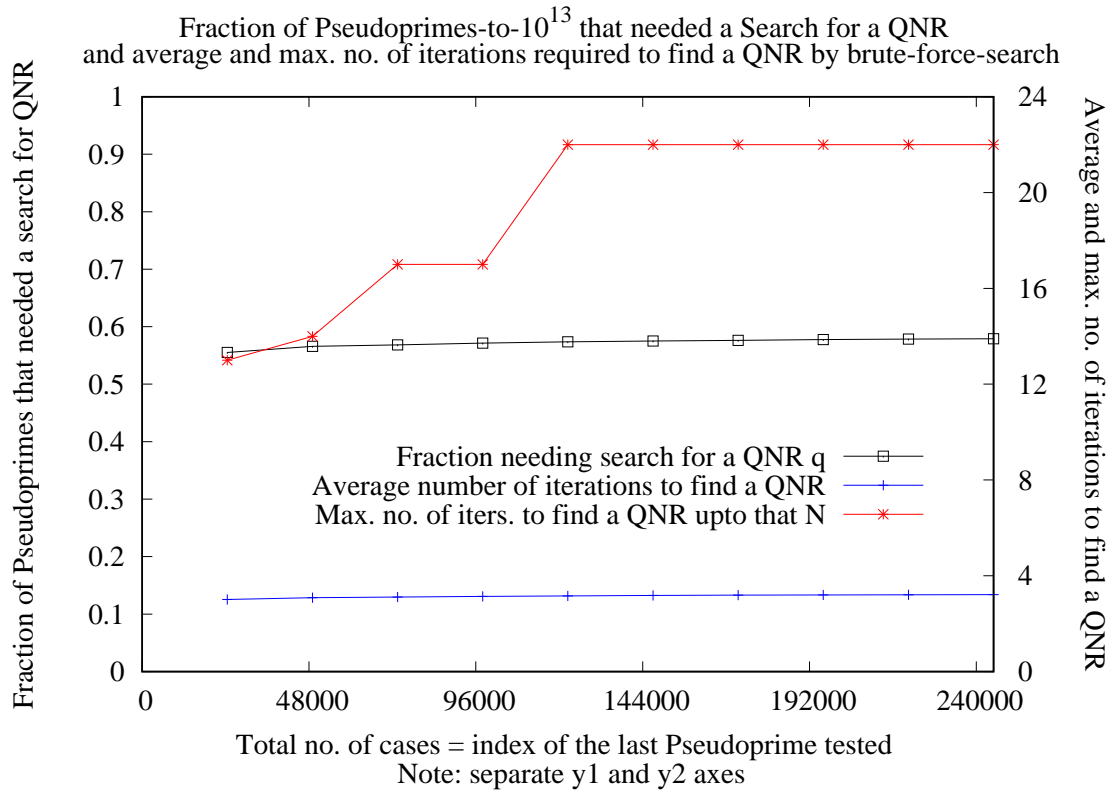

**Figure 5.** Fraction of cases where a search for a **QNR** is needed ($Y_1$ axis on the left) ;     and
the average and maximum number of iterations needed (among all $N$'s tested up to this point) to find a **QNR** $q$ by a
brute-force-stepping thru primes from 3 onward ($Y_2$ axis on the right)

This figure includes 3 distinct plots:

Plot–1 : the black colored plot that tracks the fraction of cases wherein a search for a **QNR** was needed.
            It uses the Y-1 axis on the left.                              and

Plot–2 : the blue colored plot tracking the average number of iterations, and

Plot–3 : the red colored plot tracking the maximum number of iterations.

Plots 2 and 3 use the Y-2 axis on the right.

Figure 6 illustrates the raw counts (i.e., the integers in Table 10) indicating the number of cases resolved by the three relevant mechanisms within the `PPTA_EQNR` Algorithm.

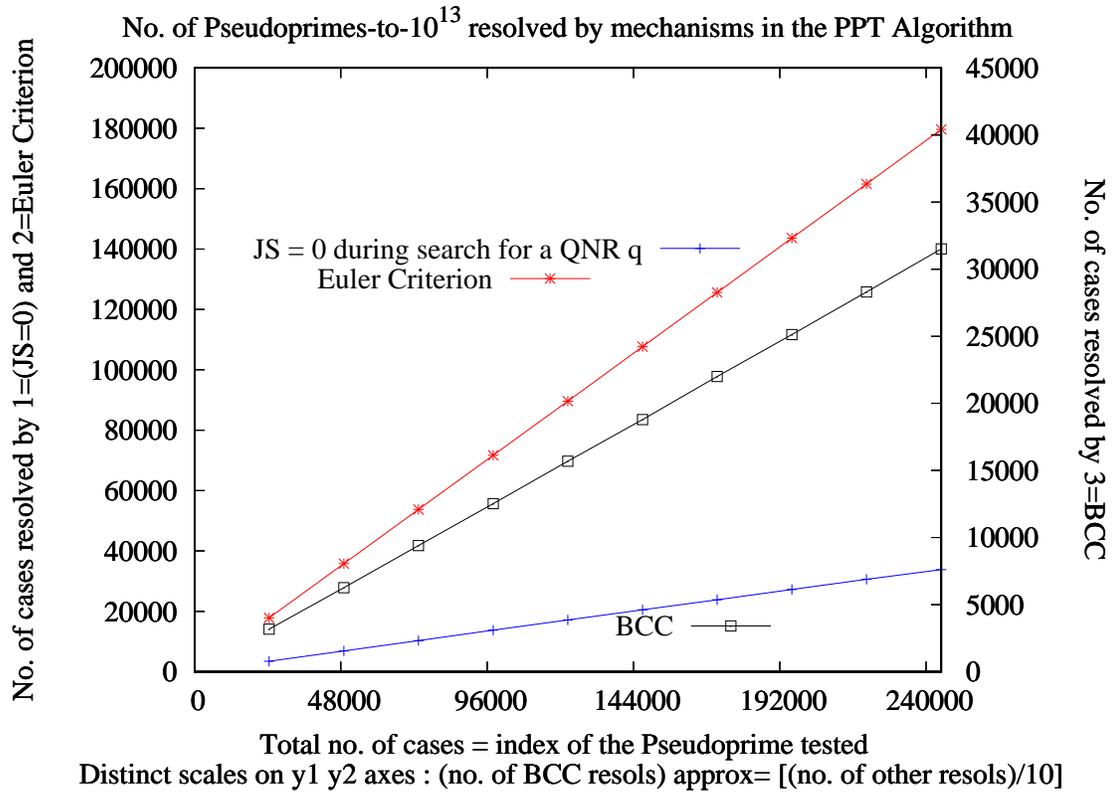

**Figure 6.** Set 4 : Number of cases resolved by individual mechanisms within the `PPTA_EQNR` Algorithm.

This figure is analogous to Figure 3, and illustrates the same metrics for Pseudo-primes in Set 4. Accordingly, it also has 3 plots that follow the same color coding as in Figure 3 :

The blue colored plot indicates the number of cases resolved by **JS**= 0
The red colored plot indicates the number of cases resolved by the Euler Criterion. These two plots use the Y-1 axis on the left

and

The black colored plot tracks the number of cases resolved by **MBEC** check.

Finally, the fractions appearing in the columns in Table 10 are illustrated in Figure 7 with a box-plot.

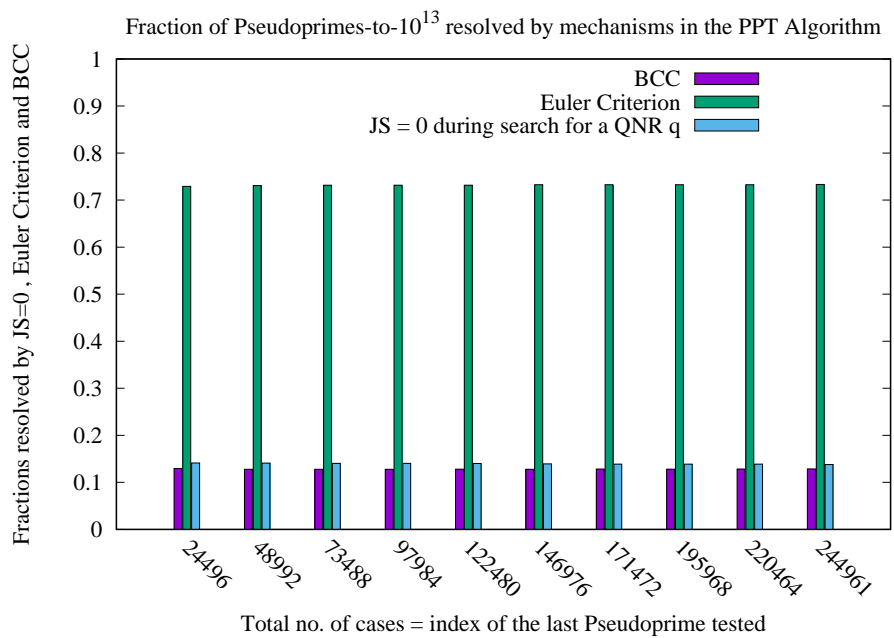

**Figure 7.** Set 4 : Fractions resolved by individual mechanisms within the `PPTA_EQNR` Algorithm illustrated with a box-plot.

This figure has only one common Y axis (shown on the left). In this case, there is no scaling of the `BCC` fractions.

We conclude this section by identifying the trends exhibited by the data.

**Trend 1 :** Overwhelming majority of Carmichael numbers do need a search for a `QNR` (which is a bad news of sorts...)

**Trend 2 :** However, the average number of iterations needed to find a `QNR` is very small $< 5$ ; this happens despite using the simplest, brute-force method that exhaustively steps thru successive primes starting at the first odd prime 3 in order to find a `QNR` (which is the good news....)

**Trend 3 :** An overwhelming majority of Carmichaels are resolved by the combination of `JS` $= 0$ and the Euler Criterion Check. Very few Carmichaels need to go to the last step to check if `BCC` $= 0$

**Trend 4 :** Similar behavior is exhibited by Pseudo-primes, wherein, the combination of `JS` $= 0$ and the Euler Criterion Check resolves a substantial majority of cases. However, the fraction of Pseudo-primes resolved by `BCC` is significantly greater than the corresponding fraction for Carmichaels.
This is somewhat counter intuitive because Carmichaels are by and large the hardest cases for primality testing.

**Trend 5:**    All the graphs are pretty much straight-lines; indicating that the fraction of cases that need a search for a `QNR` is is proportional to the size/magnitude of the input number $N$.

**Trend 6:**    For Carmichaels, the average number of iterations as well as the maximum number of iterations is almost constant; independent of the size of $N$ (especially for large values of $N$).

Pseudo-primes exhibit a similar trend; but appear to be a smidgen more sensitive to the size of $N$.

**§ Section 20.1: Experimental Outcomes from the Main Data Sets using Furthermost Generalized Primality Conjecture (FGPC) and the `PPT_INR` Algorithm**

The most interesting question is:

**what is the fraction of cases in which the generalized conjecture had to be invoked because the value of the canonical parameter $m$ is NOT a `QNR`?**

The answer:

(i) For Carmichaels, it turns out that a fixed fraction $\approx 0.28$ (or equivalently, approximately 28% ) of the cases end up using the generalized conjecture because the canonical parameter $m$ that was found did not happen to be a `QNR` modulo-$N$.

(ii) For the Pseudoprimes the corresponding fraction is $\approx 0.14$ (or equivalently, approximately 14% ) of the cases end up using the generalized conjecture.

The word "fixed" in the preceding answers means that if the fraction is plotted as a function of the number of cases seen up to that point, the value is pretty much constant (in other words the plot will be a horizontal line).

We close out this section by illustrating how the Furthermost Generalized Conjecture can substantially reduce the effort in cases wherein the number of iterations required to find a `QNR` is relatively large.

For that purpose, let $N_{22}$ be

$$N_{22} = 139 \times 353 \times 433 \times 461 \times 691 \times 19141 = 129545102216217601 \tag{167}$$

This $N$ is the smallest Carmichael number in File/Set 3, which requires the maximum number of iterations to find a `QNR` . That maximum number of iterations is "22" (therefore the subscript "22"). Thus, the smallest prime which is a `QNR` w.r.t. $N_{22}$ is $p = 83$.

It can be verified that

$$N_{22} - 1 = (2)^{10} \times (3)^3 \times (5)^2 \times (11) \times (23) \times (29) \times (25544579) \tag{168}$$

From the preceding equation, it is clear that the canonical parameter is $m = 7$.

The reader can verify that the procedure "Find_qnr_or_PGPC_parameter_m" returns this value $m = 7$ quickly, after only 4 iterations. Then

$$\Psi_m(u) = \Psi_7(u) = u^6 + 7u^4 + 14u^2 + 7 \tag{169}$$

And finally

$$\left. \begin{aligned}
&\left[ \left( (1+x)^{N_{22}} - 1 - (x)^{N_{22}} \right) \mod (x^6 + 7x^4 + 14x^2 + 7) \right] \mod N_{22} \\
&\quad \equiv \quad (7749118361454337)x^5 + (95250414045526348)x^4 + \\
&\quad (57509069882335769)x^3 + (33810842987621038)x^2 + \\
&\quad (106786570291725190)x + 119437295377094747 \\
&\quad \not\equiv \quad 0 \mod N_{22} \quad \Rightarrow \quad \text{we have a binomial witness proving that } N_{22} \text{ is a composite}
\end{aligned} \right\} \tag{170}$$

§ **Section 21 :**    `Other Carmichaels and Pseudo-primes` $> 10^{18}$ `that we could find on the Internet (Details in Appendix` **A.2** `)`

We did a search for big Carmichael Numbers and pseudo-primes on the Internet. In particular, we looked for links/sources wherein actual values were specified as decimal, hex or binary strings; or the list of factors of $N$ was specified.[32] We then created a list of Carmichaels $> 10^{(18)}$ that we could find in the literature or on-line.

That list has been coded in Maple; and is printed in the run-log screen-capture presented in **Appendix A.2** . The name of the list/array in the maple code is "**BIG_CARMS**" ;
and a printout of that entire list can be seen (when viewing a PDF version of this document on an electronic device) by following the preceding hyperlink, i.e., clicking on the green colored and underlined text BIG_CARMS within the quotation marks.

If the document is not being viewed electronically (can happen, for instance, one of the co-authors prefers a printed hard-copy); then the page number for Appendix A.2 and the (start of) that list is "Part/Article 2 - page 138" .

That list includes every single number mentioned as prime or composite in Arnault's Landmark article [47] ; including the carefully constructed composites :

**Arnault_N1**   (see Eqn. (6) on page 155 in [47] )   =   1253075960778449601058457{3923}

 = 1253075960778449601058457{3923}

**Arnault_N1**   (see Eqn. (6) on page 155 in [47] )   =   1253075960778449601058457{3923}

**Arnault_N2**   (Page 157 , in [47] ,          =   162930656995886348108319337637{8114-}\
the line immediately preceding Section            11498750450660078823067\
"4.4 Large Example")

**Arnault_N3**   (Page 156 , in [47],        =   1361818694691324890202933658522561-\
4-th line from the top)            82377286394691192846117390651100 30-\
              838492720163

**Arnault_N4**   (Page 157 , in [47] ,        =   2887148238050771212671429597130393-\
the 397 decimal digits long Carmichael number    9919776094592797227009265160241974-\
specified in Section "4.4 Large Example")     3230379915273311632898314463922594-\
                  1977803110929349655578418949441740-\
                  9338056151139799994215424169339729-\
                  0542371100275104208013496673175515-\
                  2859226962916775325475044445856101-\
                  9494042000399044321167766199496295-\
                  3925045269871932907037356403227370-\
                  1278453899126120309244841494728976-\
                  8854060249767681220770716879381217-\
                  0981132297802059565867

---

[32]We were less interested in algorithms in the literature that search for and find big Carmichaels; since our focus thus far has been on the identification and development of the new results and their analytic proof(s).

The super-long composite `Arnault_N4` is a product of 3 numbers :

$p_1 = 2967449566868551055015417464290533273077199179985304335099507553312\text{-}$
$76838753171770199594238596428121188033664754218345562493168782883$

$p_2 = 313(p_1 - 1) + 1$   and

$p_3 = 353(p_1 - 1) + 1$ 

<div align="right">(171)</div>

each of which is claimed to be a prime. Our new result ( **PBP Conjecture** ) allows a direct and independent verification of the fact that each integer specified in Eqn. (171) is indeed a prime number : it can be verified that

$p_i \mod 8 = 3$   for $i = 1, 2, 3$

so that $q = 2$ is a `QNR` w.r.t. $p_1, p_2, p_3$   $\Rightarrow$   there is no need to search for a `QNR` ; and

`ECC`$(2, p_i) = 0$   and   `BCC`$(2, p_i) = 0$

<div align="right">(172)</div>

The preceding relations imply that each of $p_1, p_2$ and $p_3$ is a prime as per the **Baseline Primality Conjecture** .

For the Arnault composites the following results can be verified numerically :

| Composite | Needs QNR search ? | QNR value | detection mechanism |
|---|---|---|---|
| `Arnault_N1` | No | 2 | non-zero `BCC(`$\sqrt{2}, N$`)` $= 10949201963878277750738064142 +$ <br> $50879926156062181307916934552\sqrt{2}$ |
| `Arnault_N2` | No | 2 | non-zero `BCC(`$\sqrt{2}, N$`)` $= 1096361218337115145969769966\text{-}$ <br> $41239725078504823956191975722 +$ <br> $78955348949647711714902277987537901323790653166456470522\sqrt{2}$ |
| `Arnault_N3` | No | 2 | non-zero `BCC(`$\sqrt{2}, N$`)` $= 9291005018667653687460739008812 5699\text{-}$ <br> $70574438800314276103299250949107847281779205 +$ <br> $43884472513057699993108838935719349291294632296723279\text{-}$ <br> $406005148048041842095343341\sqrt{2}$ |
| `Arnault_N4` | No | 2 | non-zero `BCC(`$\sqrt{2}, N$`)` $= 5837937961375539142975211348889524\text{-}$ <br> $8459875360786839303596599817183572210311892844364 4622\text{-}$ <br> $667482393527242942689626049259265148650707984748 38525\text{-}$ <br> $225994770147987026082539171140414651171956643284 90603\text{-}$ <br> $545190142901831100545190279970145189980401006473 01657\text{-}$ <br> $552096517783757740641730132638978249735989164584 14389\text{-}$ <br> $074863778417610058283442322728952655127229276133 46440\text{-}$ <br> $049057092166165065090134884890121965489076 75 +$ <br> $208477429132520623956768321905574238688834309200 67511\text{-}$ <br> $213043310124139034044607819034045165674902491518 25001\text{-}$ <br> $753646441335093600487139389010728687359918380651 615711\text{-}$ <br> $766351386400472425842916054036810022763121062810 63824\text{-}$ <br> $177876623201154599115467969024359442026753233268 33493\text{-}$ <br> $199754620219681017830943972353912759996570993769 05905\text{-}$ <br> $042687947010814837880067348849474908284904661267 19588\text{-}$ <br> $525712986312703965936610 66 $\sqrt{2}$ |

Next, we point out that the `PPTA_EQNR` algorithm seamlessly handles the "higher order Carmichaels" identified by Everett Howe [52, 53]

$$\texttt{howe\_carm1} \quad = \quad 17 \times 31 \times 41 \times 43 \times 89 \times 97 \times 167 \times 331 = 443372888629441 \qquad (173)$$

$$\texttt{howe\_carm2} \quad = \quad 59 \times 67 \times 71 \times 79 \times 89 \times 101 \times 113 \times 191 \times 233 \times 239 \times 307 \times 349 \times 379$$
$$\times\, 911 \times 2089 \times 5279 = 97723892848682923994567734100095132801 \qquad (174)$$

It turns out that `howe_carm1` $< 10^{16}$ and is therefore listed in Pinch's Set 1. It can be verified that when the `PPTA_EQNR` Algorithm is run with `howe_carm1` as the input; it gets detected as a composite during the search for a `QNR`. Specifically
in procedure `Find_QNR` ;
in the 6-th iteration of the **"for" loop on line 7** in the procedure specification ;
the algorithm evaluates the Jacobi-Symbol = **Jacobi_Symbol**$(17,$**howe_carm1**$) = 0$ ; thereby identifying 17 as a divisor of that input.

Finally, it can be verified that `howe_carm2` (which is included in the above mentioned list named **BIG_CARMS** in Appendix **A.2** ) ; also requires a search for a `QNR` ; and a brute-force steeping thru odd primes starting at 3 needs 9 iterations to discover that
**Jacobi_Symbol**$(31,$**howe_carm2**$) = -1 \quad \Rightarrow \quad q = 31$
and that $q = 31$ is an Euler-witness to the composite-nature of `howe_carm2` ; since
**ECC**$(31,$**howe_carm2**$) = 2 \neq 0 \quad \Rightarrow \quad$ The Euler Criterion is violated.

§ **Section 21.1 :** `Experimental Outcomes for the BIG_CARMS using`
`Generalized Primality Conjecture (PGPC)`
`and the` `PPTA_INR` `Algorithm`

It turns out that only 1 out of the 30 Carmichaels in that list of BIG_CARMS ends up using the Generalized Primality Conjecture.

The fraction is $\left(\frac{1}{30}\right) \approx 3.3\,\%$

which is a lot less than the corresponding (cumulative) fraction ($\approx 28\%$ of cases) that is observed when all existing carmichales $< 10^{(18)}$ were tested.

This is not unexpected, since the BIG_CARMS list is highly likely to be far from a uniform sampling of the space of all existing Carmichaels.

It turns out that `howe_carm1` also ends up invoking the generalized primality conjecture(s). However, as mentioned before, `howe_carm1` $< 10^{16}$ and therefore it is not included in the BIG_CARMS list, since it is already covered in Pinch's File/Set 1.

It can be verified that

$$(\texttt{howe\_carm1} - 1) = (2)^6 \times (3)^2 \times (5)^2 \times (7) \times (11) \times (47) \times (83) \times (211) \times (347)$$

$$\Rightarrow \quad m = 13 \quad \Rightarrow \quad \Psi_{13}(x) = x^{12} + 13x^{10} + 65x^8 + 156x^6 + 182x^4 + 91x^2 + 13 \tag{175}$$

And finally

$$\left[ \left( (1+x)^{(\texttt{howe\_carm1})} - 1 - (x)^{(\texttt{howe\_carm1})} \right) \mod \Psi_{13}(x) \right] \mod (\texttt{howe\_carm1}) \Bigg\}$$

$$\equiv \quad (178460653426619)x^{11} + (202985071765805)x^{10}$$
$$+(284129884605406)x^9 + (413820067085710)x^8$$
$$+(404383242454180)x^7 + (238837540418259)x^6$$
$$+(203557127787304)x^5 + (6061685505666)x^4$$
$$+(260655781077493)x^3 + (96773751187738)x^2$$
$$+(138141834928239)x + (37713192327089)$$

$$\not\equiv \quad 0 \mod (\texttt{howe\_carm1})$$
$$\Rightarrow \quad \text{we have a binomial witness proving that } (\texttt{howe\_carm1}) \text{ is a composite}$$

(176)

Finally, we would like to point out that `howe_carm2` is the only one (out of the 30) in the BIG_CARMS list that ends up invoking the generalized conjecture. It can be verified that

$$
\begin{aligned}
(\texttt{howe\_carm2} - 1) \;=\; & (2)^7 \times (3)^3 \times (5)^2 \times (7) \times (11) \times (13) \times (17) \times (19) \times (29) \times (907) \\
& \times (455683) \times (4031231) \times (72401009171)
\end{aligned}
\tag{177}
$$

$$
\Rightarrow \quad m = 23
\tag{178}
$$

$$
\Rightarrow \quad \Upsilon_m(t) = t^{11} + t^{10} - 10t^9 - 9t^8 + 36t^7 + 28t^6 - 56t^5 - 35t^4 + 35t^3 + 15t^2 - 6t - 1
\tag{179}
$$

and

$$
\begin{aligned}
\Psi_m(u) \;=\; & u^{22} + 23u^{20} + 230u^{18} + 1311u^{16} + 4692u^{14} + 10948u^{12} + 16744u^{10} \\
& + 16445u^8 + 9867u^6 + 3289u^4 + 506u^2 + 23
\end{aligned}
\tag{180}
$$

And

$$
\begin{aligned}
& \left[ \left( (1+x)^{(\texttt{howe\_carm2})} - 1 - (x)^{(\texttt{howe\_carm2})} \right) \mod \Upsilon_{23}(x) \right] \mod (\texttt{howe\_carm2}) \\[4pt]
& \equiv \quad (8719299886205396908954888784893978 6557)x^{10} \\
& + (4510081729187987401458272814613499 9166)x^9 \\
& + (5971404412880109720702630346314492 2987)x^8 \\
& + (2364156258517124699587029976738187 3515)x^7 \\
& + (6534432062545712155916663337158729 9314)x^6 \\
& + (8231827026209149177650677407006756 901)x^5 \\
& + (3500932525009892333226141598551100 9686)x^4 \\
& + (6808457395133463061223813938270606 01)x^3 \\
& + (7887218199426041816134092991206948 6270)x^2 \\
& + (3675168258391421289159481791485044 0257)x \\
& + (5671265871212424808514223760680306 8366) \\
& \not\equiv \quad 0 \mod (\texttt{howe\_carm2}) \\
& \Rightarrow \quad \text{we have seen a binomial witness proving that (\texttt{howe\_carm2}) is a composite}
\end{aligned}
\tag{181}
$$

We have deliberately shown the `BCC` check using the $\Upsilon_m$ polynomial as the divisor polynomial.

It can be verified that checking the single `BCC` congruence using the $\Psi_m$ polynomial (as the divisor polynomial) also yields a **non-zero finite field remainder**
**(as it must ; as per the Furthermost Generalized Primailty Conjecture (FGPC))**

$$[((1+x)^{(\text{howe\_carm2})} - 1 - (x)^{(\text{howe\_carm2})}) \mod \Psi_{23}(x)] \mod (\text{howe\_carm2})$$

$\equiv \quad (270726230958893405793612975647736974 7)x^{21}$
$+(2012037002463767876404473952478689997 5)x^{20}$
$+(6572174429544680154190655180188369429 1)x^{19}$
$+(8019216436929095858208935082561107921 3)x^{18}$
$+(6931406958848368417351029130767036574 3)x^{17}$
$+(8823114377293281636070223351044430984 7)x^{16}$
$+(7124361480989797266332967912363723965 0)x^{15}$
$+(9472780249361785167972502979442359393 4)x^{14}$
$+(4674135998523632307374580975774730606)x^{13}$
$+(6264433132915980270734141185143709604 9)x^{12}$
$+(8346598595291730660274466297272326071 0)x^{11}$
$+(2066015786010499071589883030682113601 0)x^{10}$
$+(7504760268081820423069346962220784817 8)x^{9}$
$+(6563546394791824693230088573759595996 4)x^{8}$
$+(1148806937452764628044615322048724384 1)x^{7}$
$+(7042346876685529152978510220621228068 7)x^{6}$
$+(4259174137124217710479209413824344822 2)x^{5}$
$+(2579573669759840321800790102748081283 3)x^{4}$
$+(1948129449126918699478968748376109450 8)x^{3}$
$+(2413124604549343406336985447364410880 5)x^{2}$
$+(66936809587641527577040507932524540850)x$
$+(26349438634724092046157710083990549904)$
$\not\equiv \quad 0 \mod (\text{howe\_carm2})$
$\Rightarrow \quad$ we have another binomial witness to the composite nature of $(\text{howe\_carm2})$

$\left.\begin{array}{c} \\ \\ \\ \\ \\ \\ \\ \\ \\ \\ \\ \\ \\ \\ \\ \\ \\ \\ \\ \\ \\ \\ \\ \\ \\ \end{array}\right\}$ (182)

Number Theory researcher Henri Lifchitz maintains an excellent repository of gigantic probable prime numbers at the url www.primenumbers.net/prptop/prptop.php (also accessible via reference [49]). To demonstrate the computational efficiency of our new method; we selected 10 numbers that were posted to that list recently (on or after 1-st November 2018).

The selection criteria were :

1. Decimal digit length of each number should not be more than about 60,000 digits (otherwise it takes a day or longer just to execute the 2 modular exponentiations required in the `PPTA_EQNR` algorithm).
2. The formula for the numbers should be as diverse as possible.

The numbers we picked are listed in Table 11. Each of these numbers was tested with the `PPTA_EQNR` algorithm. The relevant parameters of each run together with the final output (a decision indicating whether $N$ is a prime number or a composite) are illustrated in the next Table (i.e., Table 12).

Note that printing a sequence of 30,000+ digits of each number $N$ does not serve any purpose. Therefore, we have only reported the MD5-hash of the long string of decimal digits of each number $N$ in Table 12 (this would be critical when peers in the community replicate or re-verify or cross-check the results we have presented herein).

| Formulae for $N_i$ | locators at web-site | | | |
|---|---|---|---|---|
| | decimal length | Rank | added by | date |
| $N_1 = 39924 \times [10]^{(30000)} + 3 \times [2]^{(99654)} + 455$ | 30,005 | 84522 | Gilbert Mozzo | Nov. 2018 |
| $N_2 = [23291]^{(7901)} - [23290]^{(7901)}$ | 34,505 | 19681 | Lelio R Paula | Dec. 2018 |
| $N_3 = [2]^{(144269)} + 2 \times 1496526 - 1$ | 43,430 | 13414 | Matt Stath | Dec. 2018 |
| $N_4 = [156]^{(23552)} + 155$ | 51,653 | 10127 | Norbert Schneider | Dec. 2018 |
| $N_5 = 4019 \times [10]^{(30023)} + 3 \times [2]^{(99732)} - 25$ | 30,027 | 93431 | Gilbert Mozzo | Nov. 2018 |
| $N_6 = 58662 \times [10]^{30003} + [2]^{99667} - 1$ | 30,008 | 93558 | Gilbert Mozzo | Dec. 2018 |
| $N_7 = ([2]^{100699} + 168315) / 4139 / 16187 / \backslash$ $44273 / 8165960225173$ | 30,289 | 91716 | Toney Prest | Nov. 2018 |
| $N_8 = [2]^{(101336)} + 1920241$ | 30,506 | 25122 | Tony Prest | Nov. 2018 |
| $N_9 = [7]^{(36295)} + 216094$ | 30,673 | 22858 | Jorge Menendez | Nov. 2018 |
| $N_{10} = [8815]^{(14088)} - [14088]^{(8815)}$ | 55,581 | 9177 | Norbert Schneider | Nov. 2018 |

**Table 11.** The ten gigantic probable primes we tested with `PPTA_EQNR` algorithm

| Number | MD5 hash of $N$ | Needed search for `QNR` ? | no. of. iters to find $q$ | $q$ | output |
|--------|------------------|---------------------------|---------------------------|-----|--------|
| $N_1$ | `d1aea68b5345f8a4bc32595cb7c03d74` | Yes | 2 | 5 | prime |
| $N_2$ | `0dc6bbc85f353f874b1a4a341e567cad` | No | – | 2 | prime |
| $N_3$ | `3cfdf207dad1f2b765a95263985d5960` | No | – | 2 | prime |
| $N_4$ | `919fdea5b4e6bd40e862c55017575ea4` | No | – | 2 | prime |
| $N_5$ | `7fdfc79900dcb17462bdd89aee69da6f` | Yes | 1 | 3 | prime |
| $N_6$ | `acc4d9a4964f47a2b2816bc466055b19` | Yes | 1 | 3 | prime |
| $N_7$ | `4d5d4ab8ade6292842a98cf434bbe004` | Yes | 1 | 3 | prime |
| $N_8$ | `4507ef8409ae3bf20651077e6364f09f` | Yes | 1 | 3 | prime |
| $N_9$ | `5e7fd5150a84fb336ed03eacd23524f3` | No | – | 2 | prime |
| $N_{10}$ | `66d8319810ff4aaf7ee0eeb885611443` | Yes | 2 | 5 | prime |

**Table 12.** The parameters and output of `PPTA_EQNR` algorithm for each $N$ in the preceding Table

**Critical observations and questions**

**Observation 1:** A significant fraction of the cases (4 out of 10) do not need a search for a `QNR`.

**Observation 2:** In cases where a search is needed, the number of iterations required to find the `QNR` is extremely small.

**Observation 3:** I tested at least 20 more randomly selected $N$ values (each number had a length > 30,000 decimal-digits) reported/listed at that web-site. Each one of those additional 20 probable-primes actually turns out to be a prime (as per the `PPTA_EQNR` algorithm).

**Observation 3 in the preceding list made me curious as to why the probabilistic test (that must be passed before the $N$ value can be added to the list as a "probable prime") was so accurate. More fundamentally is there a counter example where that test FAILS ?**

The answer turns out to be YES: we have found a composite that passes the test presented at that web-site. We illustrate it in detail in the following section.

We first summarize the conditions $N$ must satisfy in order to be listed as a probable prime at Henri's web-site. To that end, We directly quote from  Henri's probable primes web-site :

``What is a PRP ?

A PRP is a probable prime number, a number that nobody knows how to prove
or disprove its primality. Here are some criterions that can be used to
recognize these numbers.

-- N does not have any prime factor up to 2^{32}
-- N cannot be trivially written into a product
-- Take 3 square-free consecutive bases such that
     jacobi_symbol(b1,N) = jacobi_symbol(b2,N) = jacobi_symbol(b3,N) = -1
-- N should be such that

$$[b1]^{(\frac{N-1}{2})} = [b2]^{(\frac{N-1}{2})} = [b3]^{(\frac{N-1}{2})} = -1 \bmod \; N$$

 -- Take a prime base $b$ such that
     jacobi_symbol(b,N) = -1    AND    $b \neq \{b1, b2, b3\}$     AND
     $N$ should pass the strong test in base $b$

-- or more (if someone has got an idea) !!!"

    We have the obvious idea:  replace the tests by the PPTA_EQNR Algorithm.

Note that besides being deterministic, the PPTA_EQNR  Algorithm should be significantly faster than Henri's tests because he needs 4  QNR values that satisfy rather stringent constraints.
In contrast the PPTA_EQNR  Algorithm needs only one single  QNR value (any value except $-1$)

It turns out that the composite number

$$N_{hc} = 3317044064679887385961981 = 1287836182261 \times 2575672364521 \tag{183}$$

    and bases

$$b_1 = 33 \qquad b_2 = 34 \qquad b_3 = 35 \qquad \text{and} \qquad b = 7 \tag{184}$$

Satisfy all of Henri's conditions listed above and is therefore a
counter-example which demonstrates that those conditions are
not deterministic guarantees of primality.

*** **How did I come up this** $N_{hc}$ **value ?**
**I did not ;**
**it happens to be the last number in the** OEIS Sequence A014233 **[50]** ***

Our new PPTA_EQNR algorithm correctly identifies this particular $N_{hc}$ to be a composite :
it can be verified that $N_{hc} \mod 8 = 5 \Rightarrow q = 2$ is a QNR modulo-$N_{hc}$; and therefore
**there is no need to search for a** QNR.
It can also be verified that

QNR $\quad q = 2 \quad$ satisfies the Euler Criterion $\hfill (185)$

$\qquad\qquad$ however ;

BCC($\sqrt{2}, N_{hc}$) $= 20605378916168 + 15454034187128\,\sqrt{2} \quad \neq \quad 0 \mod N_{hc}$ $\hfill (186)$

$\Rightarrow \quad N_{hc} \quad$ must be a composite as per the **Baseline Primality Conjecture.** $\qquad \square$

**Remark 23.1 :** Besides the 10 gigantic numbers listed in Table 12, and the 20 more randomly selected gigantic probable-primes that were tested as mentioned in **Observation 3 in the preceding section** ;
I have tested at least a 100 additional numbers (having smaller digit-lengths) listed at Henri's (excellent) web-site. Every single one turned out to be a prime (as per the new PP Testing Algorithm as well as Maple's "isprime()" function).

Accordingly ; despite the counter example, I am almost certain that every number listed as a **"PRP"** at Henri's site is actually a prime number.
Why is this happening ?
I guess that the reason is the following : The gigantic numbers listed at Henri's site are not arbitrary or uniformly distributed prime numbers. Rather, they are all generated by some underlying formula of very specific type. It is likely that for numbers that have such special forms, the conditions listed by Henri are sufficient for primality.

A analogy here is the fact that simplified special conditions often exist to deterministically prove or disprove the primality of Numbers that have special forms ; For instance :
– the renowned "Pepin's test" [54] is a fully deterministic primality test for Fermat Numbers [55].
– Likewise, simplified specialized tests exist to determine the primality of Mersenne numbers [56].

**Remark 23.2 :** It turns out that for this specific $N = N_{hc}$ ; the $b_1, b_2, b_3$ and $b$ values specified in Eqn. (184) are all `QNR` s modulo-$N$ ; and also satisfy the Euler Criterion :

$$
\left.
\begin{aligned}
\texttt{ECC}(b_1, N) &\equiv 0 \mod N \\
\texttt{ECC}(b_2, N) &\equiv 0 \mod N \\
\texttt{ECC}(b_3, N) &\equiv 0 \mod N \\
&\text{and} \\
\texttt{ECC}(b, N) &\equiv 0 \mod N
\end{aligned}
\right\} \quad \text{where ;} \quad N = N_{hc}
\tag{187}
$$

Therefore, 4 out of the 5 conditions that guarantee primality as per the **Baseline Primality Conjecture** are satisfied by each pair $(b_1, N_{hc})$, $(b_2, N_{hc})$, $(b_3, N_{hc})$ and $(b, N_{hc})$

However, $N_{hc}$ is a composite number. Therefore as per the **Baseline Primality Conjecture** none of those four pairs can satisfy the `BCC` .

Accordingly, it can be verified that

$$
\left.
\begin{aligned}
\texttt{BCC}(\sqrt{b_1}, N_{hc}) &\equiv 12438915242604310735103 48 + 12438915242552797287813 08\sqrt{33} \not\equiv 0 \\
\texttt{BCC}(\sqrt{b_2}, N_{hc}) &\equiv 20103297361750922662047 28 + 20103297361699409214756 88\sqrt{34} \not\equiv 0 \\
\texttt{BCC}(\sqrt{b_3}, N_{hc}) &\equiv 17560821518999578552455 47 + 17560821518948065105165 07\sqrt{35} \not\equiv 0 \\
&\text{and} \\
\texttt{BCC}(\sqrt{b}, N_{hc}) &\equiv 11056813549001642549593 81 + 11056813548950129102303 41\sqrt{7} \not\equiv 0
\end{aligned}
\right\}
\tag{188}
$$

**Remark 23.3 :**

I think that this example brings out the advantages of
"`staying as far away from pure integers as long as possible`" ;
because integer domain is a mine-field: extremely unique and bizarre phenomena always occur with integers (which is why number theory is an endless science).

The PP Tests therefore deliberately invoke symbols that do not exist as integers modulo-$N$ (for example, the square-root of a `QNR` ). As a result, even a single symbolic test of the Modular `BCC` is sufficient to distinguish a composite from a prime ; while lot more stringent conditions on not just one but 4 distinct `QNR` s (listed at Henri's web-site) still remain only probabilistic; and can be fooled as demonstrated.

**In essence, the best a `QNR` has to offer; viz., the property that its square-root does not exist as an integer modulo-$N$ which makes it an ideal vehicle to test the powerful Modular Binomial Expansion Congruence <u>symbolically</u> ; (thereby staying clear of the hazardous integer domain traps) has been missed by the primality tests proposed thus far. The PP Tests identify and fully leverages this most critical strength of a `QNR` (or Non Residues in general); which in my view is the most important and unique new contribution/strength of the PP Tests.**[33]

**Remark 23.4 :**

Note that none of the Gigantic PRP's listed in  Table 12  required more than 2 iterations to find a `QNR` . It is therefore clear that even if the `PPTA_INR`  algorithm was used, it would always find the `QNR` very quickly and take the least-effort path (i.e. use the Baseline Primality Conjecture). Therefore there is nothing special to report about the use of generalized methods except that the cross-validation has been consistent (as it must be, **ASSUMING THAT ALL THE CONJECTURES UNVEILED HEREIN, ARE TRUE**).

---

[33]This remark should not be misinterpreted. The symbolic computation with  $x = \sqrt{q}$  can also be implemented as a remainder of modular exponentiations in the **BCC** w.r.t. the divisor polynomial $(x^2 - q)$ ;
wherein; the coefficients of all polynomials are integers modulo-$N$. When implemented in this manner, there is no explicit reference to  "$\sqrt{q}$ ".
As a matter of fact, the initial matrix version of the PP Test was found to be unnecessary after I realized that the whole framework also worked with the characteristic polynomials of the matrices; wherein ; the coefficients of the polynomials were integers modulo-$N$.

§ **Section 24 :** **Every single** $N$ **that appears in the recent landmark article "Primes and Prejudice ..." [2] wherein the** $N$ **values are constructed to fool some current library implementation (Details in Appendix**  **)**

In a recent and excellent article in reference [2] ; the authors present a thorough and in depth analysis of the implementations of the state-of-the-art (probabilistic) primality testing algorithms in the current versions of many widely used extremely long-length integer computation libraries. These library functions are in-turn invoked by current cryptographic protocols underlying secure Internet protocols (such as https, ssh, $\cdots$ )

It turns out that even the latest and greatest variants of the fully deterministic AKS method are way too slow for currently deemed secure cryptographic key lengths of 1024-bits or higher. Therefore, all primarily testing functions actually deployed in practice today have to depend on faster but probabilistic methods such as the Miller Rabin test or the Baillie-PSW test and/or a combination of such tests/conditions and/or their building blocks (such as the Euler Criterion [3], the Lucas Lehmer test [57] ...)

As long as there are no deterministic (analytically provable) guarantees, the interesting question of generating special composite numbers that fool the primality tests used in popular (widely used) libraries will always exist. The authors of [2] have spent substantial analytic as well as computational effort to identify such special cases that fool many implementations (of primality tests) in current versions of widely used libraries.

We have tested each of these special composite values identified in [2], using the `PPTA_EQNR` algorithm and are glad to report that every single one of these is correctly identified as a composite (extremely quickly) by our Algorithm. A summary of those tests appears in Table 13.

The salient features (SFs) observed in Table 13 are :

**SF 1 :** None of the carefully constructed composites requires a search for a `QNR` .
This in turn implies that even if the more generalized version, i.e., `PPTA_INR` Algorithm is used, it takes the same (least effort) path as the simpler `PPTA_EQNR` Algorithm.

**SF 2 :** Unlike Carmichael numbers, a strong majority of these specially designed numbers (7 out of 10) get resolved by the new `BCC` mechanism we have introduced.

In closing this section we point out that each of the composites in Table 13 is a product of numbers that are claimed to be Primes. We have verified the primality of every factor of every composite using the `PPTA_EQNR` Algorithm.

| $N_i$ | where it appears in [2] | $\langle$ length = decimal-digits , MD–5 hash $\rangle$ | **QNR** $\langle$ search? , Iters, $q \bmod N_i$ $\rangle$ | det. mechanism |
|---|---|---|---|---|
| $N_1$ | Section "4.2 GNU MP", page 288, column 2 | $\langle$ 308 , 93d38ec043468720f299de331715fd81 $\rangle$ | $\langle$ No - $-2$ $\rangle$ | **ECC** $\neq 0$ |
| $N_2$ | Appendix section "A", page 295, column 2 | $\langle$ 155 , 5929270ff8d466c10098ffb3150171a9 $\rangle$ | $\langle$ No - 2 $\rangle$ | **BCC** $\neq 0$ |
| $N_3$ | Appendix section "B", page 295, column 2 | $\langle$ 617 , c4369e4f912d66b3943d4035ac0f9c7b $\rangle$ | $\langle$ No - $-2$ $\rangle$ | **ECC** $\neq 0$ |
| $N_4$ | Appendix section "D", page 296, column 2 | $\langle$ 891 , f493b1a9e1c756b38708ba307577aef5 $\rangle$ | $\langle$ No - $-2$ $\rangle$ | **BCC** $\neq 0$ |
| $N_5$ | Appendix section "E", page 296, column 2 | $\langle$ 1288 , 3307bd313d6d25e8547baa1b5d753bdd $\rangle$ | $\langle$ No - 2 $\rangle$ | **BCC** $\neq 0$ |
| $N_6$ | Appendix section "F", page 296, column 2 | $\langle$ 697 , 57e387afbcf7f4a0c7f0393cdbc3516b $\rangle$ | $\langle$ No - 2 $\rangle$ | **BCC** $\neq 0$ |
| $N_7$ | Appendix section "G", page 297, column 1 | $\langle$ 2115 , f18238fc918ac964e2833d3fed9417c7 $\rangle$ | $\langle$ No - 2 $\rangle$ | **BCC** $\neq 0$ |
| $N_8$ | right below $N_7$, page 297, column 1 | $\langle$ 309 , fa7b1a5060bd20a3f5b01889951fed3a $\rangle$ | $\langle$ No - 2 $\rangle$ | **BCC** $\neq 0$ |
| $N_9$ | Appendix section "H", page 297, column 1 | $\langle$ 309 , ede4ec1a7839cc4d061b4199d1f0d6ac $\rangle$ | $\langle$ No - $-2$ $\rangle$ | **ECC** $\neq 0$ |
| $N_{10}$ | Appendix section "J.4", page 298, column 2 | $\langle$ 309 , c3e0c36f3bde7353521c5ddd567ec385 $\rangle$ | $\langle$ No - 2 $\rangle$ | **BCC** $\neq 0$ |

**Table 13.** Summary of parameter values and outputs of runs of the PPTA_EQNR Algorithm on special composites identified in [2]

**§   Section 25 :   Concluding   Remarks**

**Conc. 1 :**      In this Part/Article 2 ;  we described our computational/numerical experiments in detail and summarized their outcomes. The data clearly corroborate all the conjectures introduced in Part/Article 1.

As we have repeatedly demonstrated throughout this set of manuscripts;
**No counter-example has been found.**

**Conc. 2 :**      The `PPTA_EQNR` Algorithm works correctly and extremely efficiently for a substantial majority of inputs.

**Conc. 3 :**      Moreover, the generalized version, i.e. the `PPTA_INR` Algorithm (which yields the low cost deterministic guarantees even in the worst case) does end up taking the least-effort path (i.e. the same path taken by the `PPTA_EQNR` algorithm) in almost all cases.

**§   Section 26 :   Current and Future work**

We have the following tasks on the "to-do" independent duel stacks  to be executed in parallel

**Stack 1 :**

**Task 1 :**      Search for or construct composite numbers that fool Maple's extremely robust
         `"isprime()"`  function.
It was a (relatively) long time ago (in the late 1990s) when the numbers constructed in Arnault's Landmark Article fooled Maple's primality tests. It is no longer the case.
We have not yet encountered any integer for which; Maple's output is different from
(i.e., contradicts) the output of our algorithms.

Since Maple's tests are probabilistic; it appears that composites that fool those tests should exist.
It is an extremely interesting (and potentially challenging) task to find at least one such example.

**Task 2 :**      More generally it seems as though the sets of numbers that are (at least relatively) hard to detect as composites (such as the Carmichael Numbers) do not contain a sizable number of cases where an explicit `QNR` search would take a very long time. It is our intent to create a repository of such numbers.

**Task 3 :**      We think that just like the connection to Miller-Rabin, there probably are strong connections between our methods and the Ballie-PSW test, or its components. Investigating such potential connections is the next item on the stack.

**Task 4 :**      Further explore integration of randomized strategies into the generalized version (i.e., `PPTA_INR` Algorithm).

**Stack 2 :**

Task 1 :    `Complete the analytic proofs of the Conjectures` .

Task 2 :    `Complete the analytic proofs of the Conjectures` .

Task 3 :    `Complete the analytic proofs of the Conjectures` .

We hope that the readers, reviewers and peers will find these ideas interesting and help us complete the tasks on the (never ending ; rather ; ever expanding) stack of "things to be done".

**§ Section 27: Part/Article 2 Appendices: Source Code and Screen Captures**

```
>>>>>>>>>  source code  <<<<<<<<<<<<
 ------ file = analyze-c16.mpl -----------

# print_BCC_cases := true ;
 print_BCC_cases := false ;

output_gnuplot_data := true ;

SIGLE_STEP_PRINT_THRESHOLD := 20;

print("testing carms the list in the file CARMS-TO-10-TO-16.mpl"):

NCASES := nops(CARMS_TO_10_TO_16):
print("number of values in the list = ", NCASES):

MYFOUND := false :
EW_and_not_MRW_list := [] ;
n_EW_and_not_MRW := 0 ;

#----  modify below this line for actual runs ---------
#
#START_INDEX := NCASES - 10 :  REVERSE_ORDER := 1 :
#START_INDEX := (198288 + 1);

START_INDEX source code
:= 1 :
REVERSE_ORDER := 0:

# START_INDEX := 118958 :
# REVERSE_ORDER := 1:

#tmp_iter_limit := 10: print_freq := 2:
#tmp_iter_limit := 100: print_freq := 10:

 tmp_iter_limit := 0: print_freq := 10:
```

```
#------- should not need to modify below this line ----------

gnuplot_data_rows := [];

n_composites :=  0 :
no_of_perfect_squares := 0 :
ncases_js_found_factor := 0 :
ncases_resolved_by_ECC := 0 :
ncases_resolved_by_BCC := 0 :

n_cases_needing_search_for_q := 0 :
avg_val_of_q := 0 :
sum_of_q_vals := 0 :

sum_iters_to_find_q := 0 :
max_no_of_iters_to_find_q := 0 :
value_of_n_that_needs_max_iters := 0 :
index_of_n_that_needs_max_iters := 0 :

n_qcases := 0 :
max_qsearch_iters_for_primes := 0 :
max_qsearch_iters_for_composites := 0 :

prime_that_needs_max_qsearch_iters := 0 :
index_of_prime_that_needs_max_qsearch_iters := 0 :

composite_that_needs_max_qsearch_iters := 0 :
index_of_composite_that_needs_max_qsearch_iters := 0 :

n_trivial_cases := 0 :

Fails := 0:

CARMS_to_16_resolved_by_BCC := [];

if tmp_iter_limit <> 0 then
   LOOP_LIMIT := START_INDEX + tmp_iter_limit - 1:
   print("**** temp_iter_limit set to ", tmp_iter_limit):
else
   LOOP_LIMIT := NCASES:
fi:

print_delta := round(evalf((LOOP_LIMIT - START_INDEX)/print_freq)):
print("print_delta = \n\n", print_delta):
```

```
print_stmt := 0:

print(" =========================================================="):

# for mycase from 1 to NCASES do
# tmp_iter_lim := 3;
# for mycase from 1 to tmp_iter_lim do

mycount := 0:
print_stmt := 0:
n_primes_found := 0:

for mycase_ind from START_INDEX to LOOP_LIMIT do

  mycount := mycount + 1:
  if (REVERSE_ORDER = 1) then
     mycase := LOOP_LIMIT - mycount + 1:
  else
     mycase := mycase_ind:
  fi:

  n := CARMS_TO_10_TO_16[mycase];
  mypow := n;
  mymod := n;

  if mycount = 1 then
     START_n := n:
  fi:

  n_is_prime := true :

#  printf("dbg 1 n = %d \n", n) ;

# initialize the boolean flags to false
#
  miller_rabin_witness := false :
  euler_witness := false :

  n_is_prime, trivial , triv_path_id ,
              perf_square , val_of_sqrt ,
              js_found_factor , val_of_factor ,
        resolved_by_ECC , ECC,
        resolved_by_BCC , BCC,
        needed_a_search_for_q , q , n_iter_to_find_q  :=
            DSP_PPT_NO_MILLER_RABIN(n) ;
```

```
#  printf("dbg 2 : n = %d, q = %d \n", n, q) ;

  maple_says_prime := isprime(n);

  if maple_says_prime <> n_is_prime then
     Fails := Fails + 1;
     print(" XXX at n = ", n, " maple_says_prime =  <> my result",
           maple_says_prime, n_is_prime);
     print("ERROR XXXXXXX breaking the loop");
     ERROR(" XXX at n = ", n, " maple_says_prime =  <> my result",
           maple_says_prime, n_is_prime);
     ERROR("ERROR XXXXXXX breaking the loop");
     break;
  fi :

###
  if n_is_prime = false then
    n_composites := n_composites + 1 ;

    if trivial = true then
       n_trivial_cases := n_trivial_cases + 1 ;
    fi ;

    if perf_square = true then
       no_of_perfect_squares := no_of_perfect_squares + 1 :
    fi :

    if js_found_factor = true then
       ncases_js_found_factor := ncases_js_found_factor + 1 :
    fi :

#####
    if resolved_by_ECC = true then

       ncases_resolved_by_ECC := ncases_resolved_by_ECC + 1 :
       euler_witness := true :

#        if MYFOUND = false then

          lof, tpow := lof_tpow(n-1):
  b := modexp(q, lof, n) :
  if b = (n-1) then b := -1 ; fi :

  for tmpii from 1 to tpow do
     s := b^2 mod n :
     if s = (n-1) then s := -1 ; fi :
```

```
        if s = 1 and b <> 1 and b <> -1 then
           miller_rabin_witness := true :
                    break :
        fi :

              b := s :

    od :

            if (tmpii - 1) = tpow and s <> 1 then # FLT is violated
#           if s <> 1 then
        miller_rabin_witness := true :
            fi :

            if miller_rabin_witness = false and euler_witness = true then
        MYFOUND := true :
        print(" -_-_-_ FOUND the case <<<< ") :
        printf("    ind = %d , n = %d , q = %d \n", mycase, n, q) :
        print(" MR witness , euler witness = ",  miller_rabin_witness,
                euler_witness) :
        n_EW_and_not_MRW := n_EW_and_not_MRW + 1 ;
        EW_and_not_MRW_list := [op(EW_and_not_MRW_list) ,
         [n_EW_and_not_MRW, mycase, n, q] ] ;

    fi :
########
#       fi :

#####
    fi:  # end of if resolved_by_ECC = true

    if resolved_by_BCC = true then
       ncases_resolved_by_BCC := ncases_resolved_by_BCC + 1 :
       CARMS_to_16_resolved_by_BCC :=
        [op(CARMS_to_16_resolved_by_BCC),
     [ncases_resolved_by_BCC, mycase,  n , q, BCC] ] ;

      if print_BCC_cases = true then
        printf(" ***--->>> BCC resol case_no= %d , index in c16 = %d , ",
          ncases_resolved_by_BCC, mycase) ;
         printf("n = %d , q = %d \n", n , q) ;
 print("BCC = \n\n", BCC) ;
      fi :

    fi :
```

```
   if max_qsearch_iters_for_composites < n_iter_to_find_q then
       max_qsearch_iters_for_composites := n_iter_to_find_q :
       composite_that_needs_max_qsearch_iters := n :
index_of_composite_that_needs_max_qsearch_iters := mycase :
   fi :
### end of if part of  if n_is_prime = false
 else
       n_primes_found := n_primes_found + 1;
       if max_qsearch_iters_for_primes < n_iter_to_find_q then
         max_qsearch_iters_for_primes := n_iter_to_find_q :
 prime_that_needs_max_qsearch_iters := n :
 index_of_prime_that_needs_max_qsearch_iters := mycase :
       fi:
 fi :

 if needed_a_search_for_q = true then

    n_cases_needing_search_for_q := n_cases_needing_search_for_q + 1 :
    sum_iters_to_find_q := sum_iters_to_find_q + n_iter_to_find_q ;

    if max_no_of_iters_to_find_q < n_iter_to_find_q then
     max_no_of_iters_to_find_q := n_iter_to_find_q :
       value_of_n_that_needs_max_iters := n :
       index_of_n_that_needs_max_iters := mycase :
    fi:

 fi :

 if q <> 0 then
    #
    # n_qcases = number of cases where the value of q returned is non zero
    # so it excludes perfect squares and possibly some of js_found_factor
    # cases
    #
    n_qcases := n_qcases + 1 :
    sum_of_q_vals := sum_of_q_vals + q ;
 fi :

if n_qcases <> 0 then
   avg_q_val := evalf(sum_of_q_vals/n_qcases) ;
else
   avg_q_val := 0 ;
fi ;

alt_n_composites := n_trivial_cases + no_of_perfect_squares +
      ncases_js_found_factor +
```

```
        ncases_resolved_by_ECC +
        ncases_resolved_by_BCC ;

   if alt_n_composites <> n_composites then
       printf("alt_n_composites = %d != n_composites = %d \n",
              alt_n_composites, n_composites) ;
       print(" XXX ERROR counting n_composites, quit -1");
       ERROR(" XXX ERROR counting n_composites, quit -1");
   fi;

   if (mycount mod print_delta = 0) then
         print_stmt := print_stmt + 1:

         printf("\n\n %s\n",FormatTime("at date = %Y-%m-%d :: time = %H:%M:%S")):
         printf("%d/%d ... count, mycase, n , n_primes_found = %d %d %d %d \n",
                print_stmt, print_freq, mycount, mycase, n, n_primes_found) :
         printf("number of perfect squares encountered = %d \n",
                no_of_perfect_squares);
         printf("N_cases resolved by jacobisymbol = 0 during qnr search = %d \n",
               ncases_js_found_factor);
         printf("no. of cases resolved by euler test = %d \n",
                ncases_resolved_by_ECC);
         printf("no. of cases resolved by BCC = %d \n",
             ncases_resolved_by_BCC) :
         printf("n_qcases = %d , sum_of_q_vals = %d , avg_q_val = %g \n",
                n_qcases, sum_of_q_vals, avg_q_val);
         printf("no of cases wherein search for q was needed = %d \n",
                n_cases_needing_search_for_q);
         printf("sum of q search iters = %d \n", sum_iters_to_find_q);

         if n_cases_needing_search_for_q <> 0 then
           avg_no_of_q_search_iters :=
               evalf(sum_iters_to_find_q/n_cases_needing_search_for_q);
         else
           avg_no_of_q_search_iters := 0;
         fi :

         printf("avg_no_of_q_search_iters = %g \n",
                avg_no_of_q_search_iters);
         printf("max q iters = %d | at n = %d , index of n = %d \n",
                max_no_of_iters_to_find_q, value_of_n_that_needs_max_iters,
         index_of_n_that_needs_max_iters) ;
         printf("composite that needs max qfind iters= %d,
                n_comp_max_iters= %d , index = %d \n",
          composite_that_needs_max_qsearch_iters,
max_qsearch_iters_for_composites,
```

```
               index_of_composite_that_needs_max_qsearch_iters) ;

########
        if print_freq < 26 and output_gnuplot_data = true then
          if n_composites <> 0 then
            frac_js_0 := evalf(ncases_js_found_factor/n_composites);
             frac_ECC_resol := evalf(ncases_resolved_by_ECC/n_composites);
      frac_BCC_resol := evalf(ncases_resolved_by_BCC/n_composites);
      frac_cases_needing_search_for_q :=
        evalf(n_cases_needing_search_for_q/mycount) ;

  else
           frac_js_0 := 0 ;
             frac_ECC_resol :=  0 ;
      frac_BCC_resol :=  0 ;
           fi ;

          gnuplot_data_rows := [op(gnuplot_data_rows),
    [mycase , [ncases_js_found_factor , frac_js_0] ,
                    [ncases_resolved_by_ECC ,frac_ECC_resol] ,
            [ncases_resolved_by_BCC , frac_BCC_resol] ,
                    [n_cases_needing_search_for_q ,
                        frac_cases_needing_search_for_q]      ,
      avg_no_of_q_search_iters , max_no_of_iters_to_find_q] ] :

        fi ;
########

  fi:

od:

if print_freq < 26 and output_gnuplot_data = true then

          if n_composites <> 0 then
            frac_js_0 := evalf(ncases_js_found_factor/n_composites);
             frac_ECC_resol := evalf(ncases_resolved_by_ECC/n_composites);
      frac_BCC_resol := evalf(ncases_resolved_by_BCC/n_composites);
      frac_cases_needing_search_for_q :=
        evalf(n_cases_needing_search_for_q/mycount) ;
  else
           frac_js_0 := 0 ;
             frac_ECC_resol :=  0 ;
      frac_BCC_resol :=  0 ;
           fi ;
```

```
        gnuplot_data_rows := [op(gnuplot_data_rows),
    [mycase , [ncases_js_found_factor , frac_js_0] ,
                    [ncases_resolved_by_ECC ,frac_ECC_resol] ,
            [ncases_resolved_by_BCC , frac_BCC_resol] ,
                    [n_cases_needing_search_for_q ,
                        frac_cases_needing_search_for_q]     ,
        avg_no_of_q_search_iters , max_no_of_iters_to_find_q] ] :
fi :

printf(" \n\n\n ------- at the exit of the big loop ");
printf("finished thru count = %d values \n", mycount):
printf(" current value of n = %d \n", n):
printf("start index = %d, loop limit = %d \n", START_INDEX, LOOP_LIMIT):
printf("reverse order = %d \n", REVERSE_ORDER):
printf("starting and ending n values = %d, %d \n", START_n, n):
printf(" number of primes found = %d \n", n_primes_found) :

#-----------
        printf("number of perfect squares encountered = %d \n",
                no_of_perfect_squares);
        printf("N_cases resolved by jacobisymbol = 0 during qnr search = %d \n",
            ncases_js_found_factor);
        printf("no. of cases resolved by euler test = %d \n",
                ncases_resolved_by_ECC);
        printf("no. of cases resolved by BCC = %d \n",
            ncases_resolved_by_BCC) :

        tdiff := n_composites - alt_n_composites ;
        printf("n_composites = %d , alt_n_composites = %d , diff = %d \n",
                n_composites , alt_n_composites , tdiff);

        printf("n_qcases = %d , sum_of_q_vals = %d , avg_q_val = %g \n",
                n_qcases, sum_of_q_vals, avg_q_val);

        printf("no of cases wherein search for q was needed = %d \n",
                n_cases_needing_search_for_q);
        printf("sum of q search iters = %d \n", sum_iters_to_find_q);

        avg_no_of_q_search_iters :=
            evalf(sum_iters_to_find_q/n_cases_needing_search_for_q);
        printf("avg_no_of_q_search_iters = %g \n",
                avg_no_of_q_search_iters);
        printf("max q iters = %d :: at n = %d , index = %d \n",
                max_no_of_iters_to_find_q , value_of_n_that_needs_max_iters ,
        index_of_n_that_needs_max_iters) ;
```

```
      printf("prime that needs max q finditers = %d ,
             n_prime_max_iters = %d , index = %d \n",
          prime_that_needs_max_qsearch_iters ,
   max_qsearch_iters_for_primes ,
   index_of_prime_that_needs_max_qsearch_iters);

      printf("composite that needs max qfind iters= %d,
             n_comp_max_iters = %d, index = %d \n",
        composite_that_needs_max_qsearch_iters ,
max_qsearch_iters_for_composites ,
index_of_composite_that_needs_max_qsearch_iters) :

if print_BCC_cases = true then
   print("\n\n ***  list of comps resolved only by BCC follows") :
   print(CARMS_to_16_resolved_by_BCC) :
fi:

if print_freq < 26 and output_gnuplot_data = true then
   printf("\n\n -----  Rows for gnuplot data table --------------\n") :
   n_rows := nops(gnuplot_data_rows) :
   for myrow from 1 to n_rows do
       unassign('v');
       v := gnuplot_data_rows[myrow] ;
  printf("%d | %d (%1.4f) , %d (%1.4f) , %d (%1.4f) | %d (%1.4f) , %.5g , %d\n",
             v[1] , v[2][1] , v[2][2] , v[3][1] , v[3][2] ,
                 v[4][1] , v[4][2] , v[5][1] , v[5][2] ,
             v[6] , v[7] ) ;
   od:
fi:

if MYFOUND = false then
  printf("\n\n -_-_-_ did not find any N that is EW but not MRW \n\n"):
else
  printf("\n\n -_-_-_ No. of Ns that are Euler Witnesses but not MRWs = %d \n",
         n_EW_and_not_MRW) ;
  print(" list of Ns that are EWs but not MRWs is printed from the next line ");
  print(EW_and_not_MRW_list) ;
fi:

if (Fails > 0) then
   print(" \/\/\/ another one bites the dust at case, n = ", mycase, n):
else
   printf("\n\n ***** nO overall failures, method worked in this run !!! \n\n"):
fi:

 >>>>>>>>> end of source code  <<<<<<<<<<<<
```

```
set MYMAPLE = /home/phatak/maple2017/bin/maple

and now the execution snapshot
start time = Sun Dec 16 14:47:21 EST 2018

the nohup maple run status = 0

 set NUMBER_OF_CARMICHAELS_THRU_10_TO_16 = 246683

                        print_BCC_cases := false

                      output_gnuplot_data := true

                    SIGLE_STEP_PRINT_THRESHOLD := 20

          "testing carms the list in the file CARMS-TO-10-TO-16.mpl"

                "number of values in the list = ", 246683

                       EW_and_not_MRW_list := []

                        n_EW_and_not_MRW := 0

                       gnuplot_data_rows := []

                    CARMS_to_16_resolved_by_BCC := []

"print_delta =

    ", 24668

        " ==========================================================="

 at date = 2018-12-16 :: time = 14:47:36

1/10 ...  count, mycase, n , n_primes_found = 24668 24668 19796660551201 0

number of perfect squares encountered = 0
N_cases resolved by jacobisymbol = 0 during qnr search = 8905
```

```
no. of cases resolved by euler test = 15575
no. of cases resolved by BCC = 188
n_qcases = 15763 , sum_of_q_vals = 851545113638570 , avg_q_val = 5.40218e+10
no of cases wherein search for q was needed = 21726
sum of q search iters = 88256
avg_no_of_q_search_iters = 4.06223
max q iters = 17 | at n = 14283595418401 , index of n = 21890
composite that needs max qfind iters= 14283595418401, n_comp_max_iters= 17 ,
index = 21890

 at date = 2018-12-16 :: time = 14:47:46
2/10 ... count, mycase, n , n_primes_found = 49336 49336 131018290365601 0
number of perfect squares encountered = 0
N_cases resolved by jacobisymbol = 0 during qnr search = 17855
no. of cases resolved by euler test = 31162
no. of cases resolved by BCC = 319
n_qcases = 31481 , sum_of_q_vals = 9002807580426840 , avg_q_val = 2.85976e+11
no of cases wherein search for q was needed = 44043
sum of q search iters = 183283
avg_no_of_q_search_iters = 4.16146
max q iters = 17 | at n = 14283595418401 , index of n = 21890
composite that needs max qfind iters= 14283595418401, n_comp_max_iters= 17 ,
index = 21890

 at date = 2018-12-16 :: time = 14:47:55
3/10 ... count, mycase, n , n_primes_found = 74004 74004 388513194641881 0
number of perfect squares encountered = 0
N_cases resolved by jacobisymbol = 0 during qnr search = 26969
no. of cases resolved by euler test = 46592
no. of cases resolved by BCC = 443
n_qcases = 47035 , sum_of_q_vals = 35733126522835083 , avg_q_val = 7.59714e+11
no of cases wherein search for q was needed = 66479
sum of q search iters = 280333
avg_no_of_q_search_iters = 4.21687
max q iters = 17 | at n = 14283595418401 , index of n = 21890
composite that needs max qfind iters= 14283595418401, n_comp_max_iters= 17 ,
index = 21890

 at date = 2018-12-16 :: time = 14:48:05
4/10 ... count, mycase, n , n_primes_found = 98672 98672 844192728662401 0
number of perfect squares encountered = 0
N_cases resolved by jacobisymbol = 0 during qnr search = 36136
no. of cases resolved by euler test = 61987
```

```
no. of cases resolved by BCC = 549
n_qcases = 62536 , sum_of_q_vals = 84990282006597468 , avg_q_val = 1.35906e+12
no of cases wherein search for q was needed = 89118
sum of q search iters = 379231
avg_no_of_q_search_iters = 4.25538
max q iters = 17 | at n = 14283595418401 , index of n = 21890
composite that needs max qfind iters= 14283595418401, n_comp_max_iters= 17 ,
index = 21890

 at date = 2018-12-16 :: time = 14:48:16
5/10 ... count, mycase, n , n_primes_found = 123340 123340 1540625721742801 0
number of perfect squares encountered = 0
N_cases resolved by jacobisymbol = 0 during qnr search = 45305
no. of cases resolved by euler test = 77403
no. of cases resolved by BCC = 632
n_qcases = 78035 , sum_of_q_vals = 193845622998950570 , avg_q_val = 2.48409e+12
no of cases wherein search for q was needed = 111844
sum of q search iters = 479332
avg_no_of_q_search_iters = 4.28572
max q iters = 17 | at n = 14283595418401 , index of n = 21890
composite that needs max qfind iters= 14283595418401, n_comp_max_iters= 17,
index = 21890

 at date = 2018-12-16 :: time = 14:48:26
6/10 ... count, mycase, n , n_primes_found = 148008 148008 2516872212769201 0
number of perfect squares encountered = 0
N_cases resolved by jacobisymbol = 0 during qnr search = 54624
no. of cases resolved by euler test = 92666
no. of cases resolved by BCC = 718
n_qcases = 93384 , sum_of_q_vals = 326573080075700655 , avg_q_val = 3.49710e+12
no of cases wherein search for q was needed = 134619
sum of q search iters = 580219
avg_no_of_q_search_iters = 4.31008
max q iters = 17 | at n = 14283595418401 , index of n = 21890
composite that needs max qfind iters= 14283595418401, n_comp_max_iters= 17,
 index = 21890

 at date = 2018-12-16 :: time = 14:48:37
7/10 ... count, mycase, n , n_primes_found = 172676 172676 3822912038756161 0
number of perfect squares encountered = 0
N_cases resolved by jacobisymbol = 0 during qnr search = 63945
no. of cases resolved by euler test = 107939
no. of cases resolved by BCC = 792
```

```
n_qcases = 108731 , sum_of_q_vals = 593583980562458048 ,
avg_q_val = 5.45920e+12
no of cases wherein search for q was needed = 157444
sum of q search iters = 681917
avg_no_of_q_search_iters = 4.33117
max q iters = 17 | at n = 14283595418401 , index of n = 21890
composite that needs max qfind iters= 14283595418401, n_comp_max_iters= 17,
index = 21890

 at date = 2018-12-16 :: time = 14:48:47
8/10 ... count, mycase, n , n_primes_found = 197344 197344 5495699006127001 0
number of perfect squares encountered = 0
N_cases resolved by jacobisymbol = 0 during qnr search = 73303
no. of cases resolved by euler test = 123184
no. of cases resolved by BCC = 857
n_qcases = 124041 , sum_of_q_vals = 1001435780159122607 ,
avg_q_val = 8.07343e+12
no of cases wherein search for q was needed = 180328
sum of q search iters = 784263
avg_no_of_q_search_iters = 4.34909
max q iters = 17 | at n = 14283595418401 , index of n = 21890
composite that needs max qfind iters= 14283595418401, n_comp_max_iters= 17,
index = 21890

 at date = 2018-12-16 :: time = 14:48:59
9/10 ... count, mycase, n , n_primes_found = 222012 222012 7538828395667401 0
number of perfect squares encountered = 0
N_cases resolved by jacobisymbol = 0 during qnr search = 82668
no. of cases resolved by euler test = 138405
no. of cases resolved by BCC = 939
n_qcases = 139344 , sum_of_q_vals = 1441536256078147071 ,
avg_q_val = 1.03452e+13
no of cases wherein search for q was needed = 203277
sum of q search iters = 886703
avg_no_of_q_search_iters = 4.36204
max q iters = 17 | at n = 14283595418401 , index of n = 21890
composite that needs max qfind iters= 14283595418401, n_comp_max_iters= 17,
index = 21890

 at date = 2018-12-16 :: time = 14:49:09
10/10 ... count, mycase, n , n_primes_found = 246680 246680 9999447614343265 0
number of perfect squares encountered = 0
N_cases resolved by jacobisymbol = 0 during qnr search = 91968
```

```
no. of cases resolved by euler test = 153695
no. of cases resolved by BCC = 1017
n_qcases = 154712 , sum_of_q_vals = 1986250261097238013 ,
avg_q_val = 1.28384e+13
no of cases wherein search for q was needed = 226185
sum of q search iters = 989638
avg_no_of_q_search_iters = 4.37535
max q iters = 17 | at n = 14283595418401 , index of n = 21890
composite that needs max qfind iters= 14283595418401, n_comp_max_iters= 17,
index = 21890

 ------- at the exit of the big loop finished thru count = 246683 values
 current value of n = 9999924433632001
start index = 1, loop limit = 246683
reverse order = 0
starting and ending n values = 561, 9999924433632001
 number of primes found = 0
number of perfect squares encountered = 0
N_cases resolved by jacobisymbol = 0 during qnr search = 91970
no. of cases resolved by euler test = 153696
no. of cases resolved by BCC = 1017
                              tdiff := 0

n_composites = 246683 , alt_n_composites = 246683 , diff = 0
n_qcases = 154713 , sum_of_q_vals = 1986250261097238030 ,
avg_q_val = 1.28383e+13
no of cases wherein search for q was needed = 226188
sum of q search iters = 989655
                   avg_no_of_q_search_iters := 4.375364741

avg_no_of_q_search_iters = 4.37536
max q iters = 17 :: at n = 14283595418401 , index = 21890
prime that needs max q finditers = 0 , n_prime_max_iters = 0 , index = 0

composite that needs max qfind iters= 14283595418401, n_comp_max_iters= 17, index
= 21890

 -----  Rows for gnuplot data table --------------
24668 | 8905 (0.3610), 15575 (0.6314), 188 (0.0076) | 21726 (0.8807), 4.0622, 17
49336 | 17855 (0.3619), 31162 (0.6316), 319 (0.0065) | 44043 (0.8927), 4.1615, 17
74004 | 26969 (0.3644), 46592 (0.6296), 443 (0.0060) | 66479 (0.8983), 4.2169, 17
98672 | 36136 (0.3662), 61987 (0.6282), 549 (0.0056) | 89118 (0.9032), 4.2554, 17
123340 | 45305 (0.3673), 77403 (0.6276), 632 (0.0051) | 111844 (0.9068), 4.2857, 17
148008 | 54624 (0.3691), 92666 (0.6261), 718 (0.0049) | 134619 (0.9095), 4.3101, 17
172676 | 63945 (0.3703), 107939 (0.6251), 792 (0.0046) | 157444 (0.9118), 4.3312, 17
```

```
197344 | 73303 (0.3714), 123184 (0.6242), 857 (0.0043) | 180328 (0.9138), 4.3491, 17
222012 | 82668 (0.3724), 138405 (0.6234), 939 (0.0042) | 203277 (0.9156), 4.362, 17
246680 | 91968 (0.3728), 153695 (0.6231), 1017 (0.0041) | 226185 (0.9169), 4.3753, 17
246683 | 91970 (0.3728), 153696 (0.6231), 1017 (0.0041) | 226188 (0.9169), 4.3753, 17

 -_-_-_ did not find any N that is EW but not MRW

 ***** nO overall failures, method worked in this run !!!

       "maple timestamp at exit || date = 2018-12-16 :: time = 14:49:09"
```

**§ Section 27.2 : A.2 : Run-log from testing "other big Carmichaels"**

```
set MYMAPLE = /home/phatak/maple2017/bin/maple

and now the execution snapshot
start time = Sat Jan 12 15:55:12 EST 2019

the nohup maple run status = 0

 BIG_CARMS := [3778118040573702001, 3825123056546413051, 7156857700403137441,

    1791562810662585767521, 8767496993623482137601, 6553130926752006031481761,

    1590231231043178376951698401, 1253075960778449601058457923,

    3523786921171888954731064241, 3280942684035956499117717272754241,

    281086456263536842600526814261601, 3494075153422874350506032047195872201,

    9772389284868292399456773410009513280,

    12586188784963996984763868103868078736,

    12758106140074522771498516740500829830401,

    23333793365462164081311115337105403499032,

    29457179106737538988590723908950340861856001,

    13091296197431676772386520145418795505617841560,

    135130930814893808401886512466750320670111400792,

    748289593771326239288330694917291704892806812920640,

    1320340354477450170682291329830138947225695029536281601,

    1629306569958863481083193376378114149875045066007882306,

    379382381447399527322618466130154668512652910714224209601,

    704168871425331764173904119314839931241207857013952964240,
```

2884167509593581480205474627684686008624483147814647841436801,

4754868377601046732119933839981363081972014948522510826417784001,

1334733877147062382486934807105197899496002201113849920496510541601,

260849323075371835669784094383812120359260783810157225730623388382401,

```
13618186946913248902029336585225618237728639469119284611739065110030838492\
    720163,
```

```
28871482380507712126714295971303939919776094592797227009265160241\
974323037991527331163289831446392259419778031109293496555784189494417409\
338056151139799994215424169339729054237110027510420801349667317551528592\
269629167753254750444458561019494042000399044321167766199496295392504526\
987193290703735640322737012784538991261203092448414947289768854060249767\
68122077071687938121709811322297802059565867]
```

                        N_Big_Carms := 30

                       print_BCC_cases := true

                      output_gnuplot_data := true

                    SIGLE_STEP_PRINT_THRESHOLD := 20

           "testing carms the list in the file Big-Carms.mpl"

               "number of values in the list = ", 30

                      EW_and_not_MRW_list := []

                        n_EW_and_not_MRW := 0

                       gnuplot_data_rows := []

                  SPECIAL_CARMS_resolved_by_BCC := []

                           print_delta := 1

        " =========================================================="

 at date = 2019-01-12 :: time = 15:55:12
1/30 ... count, mycase, n , n_primes_found = 1 1 3778118040573702001 0

```
number of perfect squares encountered = 0
 THIS CASE RESOLVED BY : JS-found-factor
N_cases resolved by jacobisymbol = 0 during qnr search = 1
no. of cases resolved by euler test = 0
no. of cases resolved by BCC = 0
n_qcases = 0 , sum_of_q_vals = 0 , avg_q_val = 0
no of cases wherein search for q was needed = 1
sum of q search iters = 4
avg_no_of_q_search_iters = 4
max q iters = 4 | at n = 3778118040573702001 , index of n = 1

 ***--->>> BCC resol case_no= 1 , index in LIST = 2 , n = 3825123056546413051 ,
 q = 2

"BCC =
                              1/2
     ", 670906496244617370 2    + 500131578192727760

 at date = 2019-01-12 :: time = 15:55:12
2/30 ... count, mycase, n , n_primes_found = 2 2 3825123056546413051 0
number of perfect squares encountered = 0
 THIS CASE RESOLVED BY : BCC
N_cases resolved by jacobisymbol = 0 during qnr search = 1
no. of cases resolved by euler test = 0
no. of cases resolved by BCC = 1
n_qcases = 1 , sum_of_q_vals = 2 , avg_q_val = 2
no of cases wherein search for q was needed = 1
sum of q search iters = 4
avg_no_of_q_search_iters = 4
max q iters = 4 | at n = 3778118040573702001 , index of n = 1

 at date = 2019-01-12 :: time = 15:55:12
3/30 ... count, mycase, n , n_primes_found = 3 3 7156857700403137441 0
number of perfect squares encountered = 0
 THIS CASE RESOLVED BY : JS-found-factor
N_cases resolved by jacobisymbol = 0 during qnr search = 2
no. of cases resolved by euler test = 0
no. of cases resolved by BCC = 1
n_qcases = 1 , sum_of_q_vals = 2 , avg_q_val = 2
no of cases wherein search for q was needed = 2
sum of q search iters = 8
avg_no_of_q_search_iters = 4
```

```
max q iters = 4 | at n = 3778118040573702001 , index of n = 1

 at date = 2019-01-12 :: time = 15:55:12
4/30 ... count, mycase, n , n_primes_found = 4 4 1791562810662585767521 0
number of perfect squares encountered = 0
 THIS CASE RESOLVED BY : JS-found-factor
N_cases resolved by jacobisymbol = 0 during qnr search = 3
no. of cases resolved by euler test = 0
no. of cases resolved by BCC = 1
n_qcases = 1 , sum_of_q_vals = 2 , avg_q_val = 2
no of cases wherein search for q was needed = 3
sum of q search iters = 12
avg_no_of_q_search_iters = 4
max q iters = 4 | at n = 3778118040573702001 , index of n = 1
composite that needs max qfind iters= 3778118040573702001,
n_comp_max_iters= 4 , index = 1

 at date = 2019-01-12 :: time = 15:55:12
5/30 ... count, mycase, n , n_primes_found = 5 5 8767496993623482137601 0
number of perfect squares encountered = 0
 THIS CASE RESOLVED BY : JS-found-factor
N_cases resolved by jacobisymbol = 0 during qnr search = 4
no. of cases resolved by euler test = 0
no. of cases resolved by BCC = 1
n_qcases = 1 , sum_of_q_vals = 2 , avg_q_val = 2
no of cases wherein search for q was needed = 4
sum of q search iters = 15
avg_no_of_q_search_iters = 3.75
max q iters = 4 | at n = 3778118040573702001 , index of n = 1
composite that needs max qfind iters= 3778118040573702001,
n_comp_max_iters= 4 , index = 1

 at date = 2019-01-12 :: time = 15:55:12
6/30 ... count, mycase, n , n_primes_found = 6 6 6553130926752006031481761 0
number of perfect squares encountered = 0
 THIS CASE RESOLVED BY : JS-found-factor
N_cases resolved by jacobisymbol = 0 during qnr search = 5
no. of cases resolved by euler test = 0
no. of cases resolved by BCC = 1
n_qcases = 1 , sum_of_q_vals = 2 , avg_q_val = 2
no of cases wherein search for q was needed = 5
sum of q search iters = 19
avg_no_of_q_search_iters = 3.8
```

```
max q iters = 4 | at n = 3778118040573702001 , index of n = 1

 at date = 2019-01-12 :: time = 15:55:12
7/30 ... count, mycase, n ,
n_primes_found = 7 7 15902312310431783769516984401 0
number of perfect squares encountered = 0
 THIS CASE RESOLVED BY : JS-found-factor
N_cases resolved by jacobisymbol = 0 during qnr search = 6
no. of cases resolved by euler test = 0
no. of cases resolved by BCC = 1
n_qcases = 1 , sum_of_q_vals = 2 , avg_q_val = 2
no of cases wherein search for q was needed = 6
sum of q search iters = 25
avg_no_of_q_search_iters = 4.16667
max q iters = 6 | at n = 15902312310431783769516984401 , index of n = 7

  ***--->>> BCC resol case_no= 2 , index in LIST = 8 ,
n = 1253075960778449601058457392301 , q = 2
                                          1/2
"BCC = ", 5087992615606218130791693452 2    + 10949201963878277507380641428

 at date = 2019-01-12 :: time = 15:55:12
8/30 ... count, mycase, n , n_primes_found = 8 8
1253075960778449601058457392301 0
number of perfect squares encountered = 0
 THIS CASE RESOLVED BY : BCC
N_cases resolved by jacobisymbol = 0 during qnr search = 6
no. of cases resolved by euler test = 0
no. of cases resolved by BCC = 2
n_qcases = 2 , sum_of_q_vals = 4 , avg_q_val = 2
no of cases wherein search for q was needed = 6
sum of q search iters = 25
avg_no_of_q_search_iters = 4.16667
max q iters = 6 | at n = 15902312310431783769516984401 , index of n = 7

 at date = 2019-01-12 :: time = 15:55:12
9/30 ... count, mycase, n , n_primes_found = 9 9
35237869211718889547310642241 0
number of perfect squares encountered = 0
 THIS CASE RESOLVED BY : JS-found-factor
N_cases resolved by jacobisymbol = 0 during qnr search = 7
no. of cases resolved by euler test = 0
no. of cases resolved by BCC = 2
```

```
n_qcases = 2 , sum_of_q_vals = 4 , avg_q_val = 2
no of cases wherein search for q was needed = 7
sum of q search iters = 30
avg_no_of_q_search_iters = 4.28571
max q iters = 6 | at n = 15902312310431783769516984O1 , index of n = 7

 at date = 2019-01-12 :: time = 15:55:12
10/30 ... count, mycase, n , n_primes_found = 10 10
32809426840359564991177172754241 0
number of perfect squares encountered = 0
 THIS CASE RESOLVED BY : JS-found-factor
N_cases resolved by jacobisymbol = 0 during qnr search = 8
no. of cases resolved by euler test = 0
no. of cases resolved by BCC = 2
n_qcases = 2 , sum_of_q_vals = 4 , avg_q_val = 2
no of cases wherein search for q was needed = 8
sum of q search iters = 35
avg_no_of_q_search_iters = 4.375
max q iters = 6 | at n = 15902312310431783769516984O1 , index of n = 7

 at date = 2019-01-12 :: time = 15:55:12
11/30 ... count, mycase, n , n_primes_found = 11 11
28108645626353684260052681426160O1 0
number of perfect squares encountered = 0
 THIS CASE RESOLVED BY : JS-found-factor
N_cases resolved by jacobisymbol = 0 during qnr search = 9
no. of cases resolved by euler test = 0
no. of cases resolved by BCC = 2
n_qcases = 2 , sum_of_q_vals = 4 , avg_q_val = 2
no of cases wherein search for q was needed = 9
sum of q search iters = 40
avg_no_of_q_search_iters = 4.44444
max q iters = 6 | at n = 15902312310431783769516984O1 , index of n = 7

 at date = 2019-01-12 :: time = 15:55:12
12/30 ... count, mycase, n , n_primes_found = 12 12
3494075153422874350506O3204719587201 0
number of perfect squares encountered = 0
 THIS CASE RESOLVED BY : JS-found-factor
N_cases resolved by jacobisymbol = 0 during qnr search = 10
no. of cases resolved by euler test = 0
no. of cases resolved by BCC = 2
n_qcases = 2 , sum_of_q_vals = 4 , avg_q_val = 2
```

```
no of cases wherein search for q was needed = 10
sum of q search iters = 44
avg_no_of_q_search_iters = 4.4
max q iters = 6 | at n = 15902312310431783769516984001 , index of n = 7

 at date = 2019-01-12 :: time = 15:55:12
13/30 ... count, mycase, n , n_primes_found = 13 13
97723892848668292399456773410009513280l 0
number of perfect squares encountered = 0
 THIS CASE RESOLVED BY : Euler Criterion
N_cases resolved by jacobisymbol = 0 during qnr search = 10
no. of cases resolved by euler test = 1
no. of cases resolved by BCC = 2
n_qcases = 3 , sum_of_q_vals = 35 , avg_q_val = 11.6667
no of cases wherein search for q was needed = 11
sum of q search iters = 54
avg_no_of_q_search_iters = 4.90909
max q iters = 10 | at n = 97723892848668292399456773410009513280l ,
index of n = 13

 at date = 2019-01-12 :: time = 15:55:12
14/30 ... count, mycase, n , n_primes_found = 14 14
125861887849639969847638681038680787361 0
number of perfect squares encountered = 0
 THIS CASE RESOLVED BY : JS-found-factor
N_cases resolved by jacobisymbol = 0 during qnr search = 11
no. of cases resolved by euler test = 1
no. of cases resolved by BCC = 2
n_qcases = 3 , sum_of_q_vals = 35 , avg_q_val = 11.6667
no of cases wherein search for q was needed = 12
sum of q search iters = 59
avg_no_of_q_search_iters = 4.91667
max q iters = 10 | at n = 97723892848668292399456773410009513280l ,
index of n = 13

 at date = 2019-01-12 :: time = 15:55:12
15/30 ... count, mycase, n , n_primes_found = 15 15
127581061400745227714985167405008298304.01 0
number of perfect squares encountered = 0
 THIS CASE RESOLVED BY : JS-found-factor
N_cases resolved by jacobisymbol = 0 during qnr search = 12
no. of cases resolved by euler test = 1
no. of cases resolved by BCC = 2
```

```
n_qcases = 3 , sum_of_q_vals = 35 , avg_q_val = 11.6667
no of cases wherein search for q was needed = 13
sum of q search iters = 64
avg_no_of_q_search_iters = 4.92308
max q iters = 10 | at n = 9772389284868292399456773410009513280 1 ,
index of n = 13

 at date = 2019-01-12 :: time = 15:55:12
16/30 ... count, mycase, n , n_primes_found = 16 16
23333793365462164081311115337105403499032 01 0
number of perfect squares encountered = 0
 THIS CASE RESOLVED BY : JS-found-factor
N_cases resolved by jacobisymbol = 0 during qnr search = 13
no. of cases resolved by euler test = 1
no. of cases resolved by BCC = 2
n_qcases = 3 , sum_of_q_vals = 35 , avg_q_val = 11.6667
no of cases wherein search for q was needed = 14
sum of q search iters = 68
avg_no_of_q_search_iters = 4.85714
max q iters = 10 | at n = 9772389284868292399456773410009513280 1 ,
index of n = 13

 at date = 2019-01-12 :: time = 15:55:12
17/30 ... count, mycase, n , n_primes_found = 17 17
29457179106737538988590723908950340861856 0001 0
number of perfect squares encountered = 0
 THIS CASE RESOLVED BY : JS-found-factor
N_cases resolved by jacobisymbol = 0 during qnr search = 14
no. of cases resolved by euler test = 1
no. of cases resolved by BCC = 2
n_qcases = 3 , sum_of_q_vals = 35 , avg_q_val = 11.6667
no of cases wherein search for q was needed = 15
sum of q search iters = 72
avg_no_of_q_search_iters = 4.8
max q iters = 10 | at n = 9772389284868292399456773410009513280 1 ,
index of n = 13

 at date = 2019-01-12 :: time = 15:55:12
18/30 ... count, mycase, n , n_primes_found = 18 18
13091296197431676772386520145418795505617 8415601 0
number of perfect squares encountered = 0
 THIS CASE RESOLVED BY : JS-found-factor
N_cases resolved by jacobisymbol = 0 during qnr search = 15
```

```
no. of cases resolved by euler test = 1
no. of cases resolved by BCC = 2
n_qcases = 3 , sum_of_q_vals = 35 , avg_q_val = 11.6667
no of cases wherein search for q was needed = 16
sum of q search iters = 76
avg_no_of_q_search_iters = 4.75
max q iters = 10 | at n = 9772389284868292399456773410009513280l ,
index of n = 13

 at date = 2019-01-12 :: time = 15:55:12
19/30 ... count, mycase, n , n_primes_found = 19 19
1351309308148938084018865124667503206701114007920l 0
number of perfect squares encountered = 0
 THIS CASE RESOLVED BY : JS-found-factor
N_cases resolved by jacobisymbol = 0 during qnr search = 16
no. of cases resolved by euler test = 1
no. of cases resolved by BCC = 2
n_qcases = 3 , sum_of_q_vals = 35 , avg_q_val = 11.6667
no of cases wherein search for q was needed = 17
sum of q search iters = 80
avg_no_of_q_search_iters = 4.70588
max q iters = 10 | at n = 9772389284868292399456773410009513280l ,
index of n = 13

 at date = 2019-01-12 :: time = 15:55:12
20/30 ... count, mycase, n , n_primes_found = 20 20
7482895937713262392883306949172917048928068129206401 0
number of perfect squares encountered = 0
 THIS CASE RESOLVED BY : JS-found-factor
N_cases resolved by jacobisymbol = 0 during qnr search = 17
no. of cases resolved by euler test = 1
no. of cases resolved by BCC = 2
n_qcases = 3 , sum_of_q_vals = 35 , avg_q_val = 11.6667
no of cases wherein search for q was needed = 18
sum of q search iters = 85
avg_no_of_q_search_iters = 4.72222
max q iters = 10 | at n = 9772389284868292399456773410009513280l ,
index of n = 13

 at date = 2019-01-12 :: time = 15:55:12
21/30 ... count, mycase, n , n_primes_found = 21 21
13203403544774501706822913298301389472256950295362816 01 0
number of perfect squares encountered = 0
```

```
 THIS CASE RESOLVED BY : JS-found-factor
N_cases resolved by jacobisymbol = 0 during qnr search = 18
no. of cases resolved by euler test = 1
no. of cases resolved by BCC = 2
n_qcases = 3 , sum_of_q_vals = 35 , avg_q_val = 11.6667
no of cases wherein search for q was needed = 19
sum of q search iters = 90
avg_no_of_q_search_iters = 4.73684
max q iters = 10 | at n = 9772389284868292399456773410095132801 ,
index of n = 13

 ***--->>> BCC resol case_no= 3 , index in LIST = 22 ,
n = 16293065699588634810831933763781141498750450660078823067 , q = 2

"BCC = "
                                                          1/2
    7895534894964771171490227798753790132379065316645647052 2

    +  10963612183371151459697699664123972507850482395619197572

 at date = 2019-01-12 :: time = 15:55:12
22/30 ... count, mycase, n , n_primes_found = 22 22
16293065699588634810831933763781141498750450660078823067 0
number of perfect squares encountered = 0
 THIS CASE RESOLVED BY : BCC
N_cases resolved by jacobisymbol = 0 during qnr search = 18
no. of cases resolved by euler test = 1
no. of cases resolved by BCC = 3
n_qcases = 4 , sum_of_q_vals = 37 , avg_q_val = 9.25
no of cases wherein search for q was needed = 19
sum of q search iters = 90
avg_no_of_q_search_iters = 4.73684
max q iters = 10 | at n = 9772389284868292399456773410095132801 ,
index of n = 13

 at date = 2019-01-12 :: time = 15:55:12
23/30 ... count, mycase, n , n_primes_found = 23 23
37938238144739952732261846613015466851265291071422420 9601 0
number of perfect squares encountered = 0
 THIS CASE RESOLVED BY : JS-found-factor
N_cases resolved by jacobisymbol = 0 during qnr search = 19
no. of cases resolved by euler test = 1
no. of cases resolved by BCC = 3
n_qcases = 4 , sum_of_q_vals = 37 , avg_q_val = 9.25
```

```
no of cases wherein search for q was needed = 20
sum of q search iters = 96
avg_no_of_q_search_iters = 4.8
max q iters = 10 | at n = 9772389284868292399456773410009513 2801 ,
index of n = 13

 at date = 2019-01-12 :: time = 15:55:12
24/30 ... count, mycase, n , n_primes_found = 24 24
7041688714253317641739041193148399312412078570139 5296424001 0
number of perfect squares encountered = 0
 THIS CASE RESOLVED BY : JS-found-factor
N_cases resolved by jacobisymbol = 0 during qnr search = 20
no. of cases resolved by euler test = 1
no. of cases resolved by BCC = 3
n_qcases = 4 , sum_of_q_vals = 37 , avg_q_val = 9.25
no of cases wherein search for q was needed = 21
sum of q search iters = 100
avg_no_of_q_search_iters = 4.7619
max q iters = 10 | at n = 9772389284868292399456773410009513 2801 ,
index of n = 13

 at date = 2019-01-12 :: time = 15:55:12
25/30 ... count, mycase, n , n_primes_found = 25 25
2884167509535814802054746276846860086244831478146 47841436801 0
number of perfect squares encountered = 0
 THIS CASE RESOLVED BY : JS-found-factor
N_cases resolved by jacobisymbol = 0 during qnr search = 21
no. of cases resolved by euler test = 1
no. of cases resolved by BCC = 3
n_qcases = 4 , sum_of_q_vals = 37 , avg_q_val = 9.25
no of cases wherein search for q was needed = 22
sum of q search iters = 106
avg_no_of_q_search_iters = 4.81818
max q iters = 10 | at n = 9772389284868292399456773410009513 2801 ,
index of n = 13

 at date = 2019-01-12 :: time = 15:55:12
26/30 ... count, mycase, n , n_primes_found = 26 26
4754868377601046732119933839981363081972014948522 510826417784001 0
number of perfect squares encountered = 0
 THIS CASE RESOLVED BY : JS-found-factor
N_cases resolved by jacobisymbol = 0 during qnr search = 22
no. of cases resolved by euler test = 1
```

```
no. of cases resolved by BCC = 3
n_qcases = 4 , sum_of_q_vals = 37 , avg_q_val = 9.25
no of cases wherein search for q was needed = 23
sum of q search iters = 112
avg_no_of_q_search_iters = 4.86957
max q iters = 10 | at n = 9772389284868292399456773410095132801 ,
index of n = 13

 at date = 2019-01-12 :: time = 15:55:12
27/30 ... count, mycase, n , n_primes_found = 27 27
133473387714706238248693480710519789949600220111384992049651054601 0
number of perfect squares encountered = 0
 THIS CASE RESOLVED BY : JS-found-factor
N_cases resolved by jacobisymbol = 0 during qnr search = 23
no. of cases resolved by euler test = 1
no. of cases resolved by BCC = 3
n_qcases = 4 , sum_of_q_vals = 37 , avg_q_val = 9.25
no of cases wherein search for q was needed = 24
sum of q search iters = 118
avg_no_of_q_search_iters = 4.91667
max q iters = 10 | at n = 9772389284868292399456773410095132801 ,
index of n = 13

 at date = 2019-01-12 :: time = 15:55:12
28/30 ... count, mycase, n , n_primes_found = 28 28
260849323075371835669784094383812120359260783810157225730623388382401 0
number of perfect squares encountered = 0
 THIS CASE RESOLVED BY : JS-found-factor
N_cases resolved by jacobisymbol = 0 during qnr search = 24
no. of cases resolved by euler test = 1
no. of cases resolved by BCC = 3
n_qcases = 4 , sum_of_q_vals = 37 , avg_q_val = 9.25
no of cases wherein search for q was needed = 25
sum of q search iters = 124
avg_no_of_q_search_iters = 4.96
max q iters = 10 | at n = 9772389284868292399456773410095132801 ,
index of n = 13

 ***--->>> BCC resol case_no= 4 , index in LIST = 29 ,
n = 1361818694691324890202933658522561823772863946911928461173906511003081-\
    38492720163 , q = 2

"BCC = "
     43884472513057699931088389357193492912946322967232794060051480480418\
```

```
209534341 2    + 9291005018667653687460739008812569970574438800314276103\
29925094910784728177 9205
```

```
 at date = 2019-01-12 :: time = 15:55:12
29/30 ... count, mycase, n , n_primes_found = 29 29
1361818694691324890202933658522561823772863946911928461173906511003 08-\
38492720163 0
number of perfect squares encountered = 0
 THIS CASE RESOLVED BY : BCC
N_cases resolved by jacobisymbol = 0 during qnr search = 24
no. of cases resolved by euler test = 1
no. of cases resolved by BCC = 4
n_qcases = 5 , sum_of_q_vals = 39 , avg_q_val = 7.8
no of cases wherein search for q was needed = 25
sum of q search iters = 124
avg_no_of_q_search_iters = 4.96
max q iters = 10 | at n = 9772389284868292399456773410009513280 1 ,
index of n = 13
```

```
 ***--->>> BCC resol case_no= 5 , index in LIST = 30 ,
```

```
n = 2887148238050771212671429597130393991977609459279722700926516024-\
1974323037991527331163289831446392259419778031109293496555784189494 4-\
1740933805615113979999421542416933972905423711002751042080134966731 7-\
5515285922696291677532547504444585610194940420003990443211677661994 9-\
6295392504526987193290703735640322737012784538991261203092448414947 2-\
8976885406024976768122077071687938121709811322297802059565867 ,
```

```
q = 2 ,
```

```
"BCC =
    ", 2084774291325206239567683219055742386888343092006751121304331012413903\
    4044607819034045165674902491518250017536464133509360048713938901072868 73\
    5991838065161571176635138640047242584291605403681002276121062810638241 7\
    7876623201154599115467969024359442026753233268334931997546202196810178 30\
    9439723539127599965709937690590504268794701081483788006734884947490828 49\
```

```
    0466126719588525712986312703965936610 66 2    + 5837937961375539142975211\
    3488895248459875360786839303596599817183572210311892844364462266748239 35\
    2724294268962604925926514865070798748738525225994770147987026082539171 14\
    0414651171956643284906035451901429018311005451902799701451899804010064 73\
    0165755209651778375774064173013263897824973598916458414389074863778417 61\
    0058283442322728952655127229276133464400490570921661650650901348848901 21\
```

```
    96548907675

 at date = 2019-01-12 :: time = 15:55:12
30/30 ... count, mycase, n , n_primes_found = 30 30

28871482380507712126714295971303939917760945927972270092651024-\
19743230379915273311632898314463922594197780311092934965557841894944-\
17409338056151139799942154241693397290542371100275104208013496677317-\
55152859226962916775325475044445856101949404200039904432116776619949-\
62953925045269871932907037356403227370127845389912612030924484149472-\
89768854060249767681220770716879381217098113222978020595565867 , 0
number of perfect squares encountered = 0
 THIS CASE RESOLVED BY : BCC
N_cases resolved by jacobisymbol = 0 during qnr search = 24
no. of cases resolved by euler test = 1
no. of cases resolved by BCC = 5
n_qcases = 6 , sum_of_q_vals = 41 , avg_q_val = 6.83333
no of cases wherein search for q was needed = 25
sum of q search iters = 124
avg_no_of_q_search_iters = 4.96
max q iters = 10 | at n = 97723892848682923994567734100095132801 ,
index of n = 13

 ------- at the exit of the big loop finished thru count = 30 values
 current value of n =
28871482380507712126714295971303939917760945927972270092651024-\
19743230379915273311632898314463922594197780311092934965557841894944-\
17409338056151139799942154241693397290542371100275104208013496677317-\
55152859226962916775325475044445856101949404200039904432116776619949-\
62953925045269871932907037356403227370127845389912612030924484149472-\
89768854060249767681220770716879381217098113222978020595565867
start index = 1, loop limit = 30
reverse order = 0
starting and ending n values = 3778118040573702001,
28871482380507712126714295971303939917760945927972270092651024-\
19743230379915273311632898314463922594197780311092934965557841894944-\
17409338056151139799942154241693397290542371100275104208013496677317-\
55152859226962916775325475044445856101949404200039904432116776619949-\
62953925045269871932907037356403227370127845389912612030924484149472-\
89768854060249767681220770716879381217098113222978020595565867
 number of primes found = 0
number of perfect squares encountered = 0
N_cases resolved by jacobisymbol = 0 during qnr search = 24
```

```
no. of cases resolved by euler test = 1
no. of cases resolved by BCC = 5
                              tdiff := 0

n_composites = 30 , alt_n_composites = 30 , diff = 0
n_qcases = 6 , sum_of_q_vals = 41 , avg_q_val = 6.83333
no of cases wherein search for q was needed = 25
sum of q search iters = 124
               avg_no_of_q_search_iters := 4.960000000

avg_no_of_q_search_iters = 4.96
max q iters = 10 :: at n = 9772389284868292399456773410095132801 ,
index = 13

" ***  list of comps resolved only by BCC follows"

                                                  1/2
[[1, 2, 3825123056546413051, 2, 670906496244617370 2    + 500131578192727760],

    [2, 8, 1253075960778449601058457 3923, 2,

                         1/2
    5087992615606218130791693452 2    + 10949201963878277507380641428], [3, 22,

    16293065699588634810831933763781141498750450660078823067, 2,

                                                  1/2
    7895534894964771171490227798753790132379065316645647052 2

    + 1096361218337115145969769966412397250785048239561 9197572], [4, 29, 136\
    18186946913248902029336585225618237728639469119284611739065110030838492 7\
    20163, 2, 4388447251305769999310883893571934929129463229672327940600514 80\

                 1/2
    4804184209534341 2    + 929100501866765368746073900881256997057443880031\
    4276103299250949107847281779205], [5, 30, 2887148238050771212671429597130\
    3939919776094592797227009265160241974323037991527331163298931446392259 41\
    9778031109293496557841894944174093380561511397999942154241693397290542 3\
    7110027510420801349667317551528592269629167753254750444458561019494042 00\
    0399044321167766199496295392504526987193290703735640322737012784538991 26\
    1203092448414947289768854060249767681220770716879381217098113222978020 59\

    565867, 2, 2084774291325206239567683219055742386888343092006751121304331 0\
    1241390340446078190340451656749024915182500175364641335093600487139389 01\
```

072868735991838065161571176635138640047242584291605403681002276312106281\
063824177876623201154599115467969024359442026753233268334931997546202196\
810178309439723539127599965709937690590504268794701081483788006734884947\

$$\frac{1/2}{}$$

490828490466126719588525712986312703965936610662\ \ \ + 58379379613755391\
429752113488895248459875360786839303596599817183572210311892844364462266\
748239352724294268962604925926514865070798748738525225994770147987026082\
539171140414651171956643284906035451901429018311005451902799701451899804\
010064730165755209651778375774064173013263897824973598916458414389074863\
778417610058283442322728952655127229276133464400490570921661650650901348\
8489012196548907675]]

  ----- Rows for gnuplot data table ---------------
1 | 1 (1.0000), 0 (0.0000), 0 (0.0000) | 1 (1.0000), 4, 4
2 | 1 (0.5000), 0 (0.0000), 1 (0.5000) | 1 (0.5000), 4, 4
3 | 2 (0.6667), 0 (0.0000), 1 (0.3333) | 2 (0.6667), 4, 4
4 | 3 (0.7500), 0 (0.0000), 1 (0.2500) | 3 (0.7500), 4, 4
5 | 4 (0.8000), 0 (0.0000), 1 (0.2000) | 4 (0.8000), 3.75, 4
6 | 5 (0.8333), 0 (0.0000), 1 (0.1667) | 5 (0.8333), 3.8, 4
7 | 6 (0.8571), 0 (0.0000), 1 (0.1429) | 6 (0.8571), 4.1667, 6
8 | 6 (0.7500), 0 (0.0000), 2 (0.2500) | 6 (0.7500), 4.1667, 6
9 | 7 (0.7778), 0 (0.0000), 2 (0.2222) | 7 (0.7778), 4.2857, 6
10 | 8 (0.8000), 0 (0.0000), 2 (0.2000) | 8 (0.8000), 4.375, 6
11 | 9 (0.8182), 0 (0.0000), 2 (0.1818) | 9 (0.8182), 4.4444, 6
12 | 10 (0.8333), 0 (0.0000), 2 (0.1667) | 10 (0.8333), 4.4, 6
13 | 10 (0.7692), 1 (0.0769), 2 (0.1538) | 11 (0.8462), 4.9091, 10
14 | 11 (0.7857), 1 (0.0714), 2 (0.1429) | 12 (0.8571), 4.9167, 10
15 | 12 (0.8000), 1 (0.0667), 2 (0.1333) | 13 (0.8667), 4.9231, 10
16 | 13 (0.8125), 1 (0.0625), 2 (0.1250) | 14 (0.8750), 4.8571, 10
17 | 14 (0.8235), 1 (0.0588), 2 (0.1176) | 15 (0.8824), 4.8, 10
18 | 15 (0.8333), 1 (0.0556), 2 (0.1111) | 16 (0.8889), 4.75, 10
19 | 16 (0.8421), 1 (0.0526), 2 (0.1053) | 17 (0.8947), 4.7059, 10
20 | 17 (0.8500), 1 (0.0500), 2 (0.1000) | 18 (0.9000), 4.7222, 10
21 | 18 (0.8571), 1 (0.0476), 2 (0.0952) | 19 (0.9048), 4.7368, 10
22 | 18 (0.8182), 1 (0.0455), 3 (0.1364) | 19 (0.8636), 4.7368, 10
23 | 19 (0.8261), 1 (0.0435), 3 (0.1304) | 20 (0.8696), 4.8, 10
24 | 20 (0.8333), 1 (0.0417), 3 (0.1250) | 21 (0.8750), 4.7619, 10
25 | 21 (0.8400), 1 (0.0400), 3 (0.1200) | 22 (0.8800), 4.8182, 10
26 | 22 (0.8462), 1 (0.0385), 3 (0.1154) | 23 (0.8846), 4.8696, 10
27 | 23 (0.8519), 1 (0.0370), 3 (0.1111) | 24 (0.8889), 4.9167, 10
28 | 24 (0.8571), 1 (0.0357), 3 (0.1071) | 25 (0.8929), 4.96, 10
29 | 24 (0.8276), 1 (0.0345), 4 (0.1379) | 25 (0.8621), 4.96, 10
30 | 24 (0.8000), 1 (0.0333), 5 (0.1667) | 25 (0.8333), 4.96, 10
30 | 24 (0.8000), 1 (0.0333), 5 (0.1667) | 25 (0.8333), 4.96, 10

```
-_-_-_ did not find any N that is EW but not MRW

***** nO overall failures, method worked in this run !!!

        "maple timestamp at exit || date = 2019-01-12 :: time = 15:55:12"
```

## § Section 27.3 : A.3 : Run-log from testing $N$s in "Primes and Prejudice ..."[2]

```
set MYMAPLE = /home/phatak/maple2017/bin/maple

and now the execution snapshot
start time = Thu Jan 24 17:13:30 EST 2019

the nohup maple run status = 0

 *** Testing all hand crafted numbers in primes-and-prejudice ***

 --- N1 : Section 4.2 GNU MP ;  page 288 column 2

      string1_N1 := "0x00000000000000000000000000000081d564fbdd20b406"

      string2_N1 := "0x750af7bd334dcf547b131a1d8f8235fd603dba44e22e0775"

      string3_N1 := "0x0ecf755051d33cb8895413f5d69f5a3df701889e3a69f92e"

      string4_N1 := "0xdd3f5f36662521877231ba4753a3e7185a89ddb0b2d73a35"

      string5_N1 := "0x9e976a9bcfeae1a7c026d74bc7515a5010f3cd62c69fa9ad"

      string6_N1 := "0x7b699f40e7a85192e1a4aa95537363fcb93d789aee32bbbf"

       rawstr1_N1 := "00000000000000000000000000000081d564fbdd20b406"

       rawstr2_N1 := "750af7bd334dcf547b131a1d8f8235fd603dba44e22e0775"

       rawstr3_N1 := "0ecf755051d33cb8895413f5d69f5a3df701889e3a69f92e"

       rawstr4_N1 := "dd3f5f36662521877231ba4753a3e7185a89ddb0b2d73a35"

       rawstr5_N1 := "9e976a9bcfeae1a7c026d74bc7515a5010f3cd62c69fa9ad"
```

```
           rawstr6_N1 := "7b699f40e7a85192e1a4aa95537363fcb93d789aee32bbbf"

                    my_coef_tpow1 := 9355494833841550342

   my_coef_tpow2 := 2869882152402220034260078593867335798174380081616857728885

    my_coef_tpow3 := 3631495544982228622589935170233024059056134248912069573 58

   my_coef_tpow4 := 5424974056917995850954352115224679481452390044646424132149

   my_coef_tpow5 := 3888651541004072378986093199844272359891555548033814866349

      my_const := 302606781023228287446235034302023220643944327557720 5898175

                              tpow1_N1 := 960

                              tpow2_N1 := 768

                              tpow3_N1 := 576

                              tpow4_N1 := 384

                              tpow5_N1 := 192

N1 := 91172234887816366836097425833085947590883244148569730116910168859458 82\
      72593762845182023711893193846438389419570189953953879727667482485082645 9\
      06798847756809083320529285386430903407807541069345091761181412272133887 9\
      81990337812539644611817999766010592289473141340883221339294271083689346 1\
      192337881585852682950 3

                              len_N1 := 308

  decimal digits length of N1 = 308

  MD5--hash of N1 = 93d38ec043468720f299de331715fd81

                              N1_maple := false

N1_ppt, trivial, triv_path_id, perf_square, val_of_sqrt, js_found_factor,

     val_of_factor, resolved_by_ECC, ECC, resolved_by_BCC, BCC,

     needed_a_search_for_q, q, n_iter_to_find_q := false, false, -1, false, 0,

     false, 0, true, 35343712878021351962777081987004796507706918379115381428 3\
```

000296580026238786638339163799674522363945932146797792108306959833660578\
881160740349486921641098623450503834292101039638088068732896127666785526\
217492545714802924320071891272771243873429001359775986848952358640710558\

(1/3)

584164922191382632287706502044445191, false, 7     , false, 91172234887816\
366836097425833085947590883244148569730116910168859458827259376284518202\
371189319384643838941957018995395387972766748248508264590679884775680908\
332052928538643090340780754106934509176118141227213388798199033781253964\
461181799976601059228947314134088322133929427108368934611923378815858526\

829501, 0

"both maple and ppt match for all components , done with this case"

--- N2 : Appendix A AN OVERVIEW OF ARNAULT's METHOD; page 295 col 2

P1 := 14244538716141548240482636541817596226668913330066163

P2 := 58402608736180347785978809821452144529342544553252643

P3 := 14386984103302963722887462907235772188935602433622363

N2 := 1196879422460471829354990810475951820434393065275928852982987578098131192\
705057270518153987329384847623539323031465491272992065786463031797156272\
7057595285667

len_N2 := 155

 decimal digits length of N2 = 155

 MD5--hash of N2 = 5929270ff8d466c10098ffb3150171a9

P1_N2_maple := true

P2_N2_maple := true

P3_N2_maple := true

N2_maple := false

P1_N2_ppt, trivial, triv_path_id, perf_square, val_of_sqrt, js_found_factor,

```
        val_of_factor, resolved_by_ECC, ECC, resolved_by_BCC, BCC,

        needed_a_search_for_q, q, n_iter_to_find_q :=

        true, false, -1, false, 0, false, 0, false, 0, true, 0, false, 2, 0

P2_N2_ppt, trivial, triv_path_id, perf_square, val_of_sqrt, js_found_factor,

        val_of_factor, resolved_by_ECC, ECC, resolved_by_BCC, BCC,

        needed_a_search_for_q, q, n_iter_to_find_q :=

        true, false, -1, false, 0, false, 0, false, 0, true, 0, false, 2, 0

P3_N2_ppt, trivial, triv_path_id, perf_square, val_of_sqrt, js_found_factor,

        val_of_factor, resolved_by_ECC, ECC, resolved_by_BCC, BCC,

        needed_a_search_for_q, q, n_iter_to_find_q :=

        true, false, -1, false, 0, false, 0, false, 0, true, 0, false, 2, 0

N2_ppt, trivial, triv_path_id, perf_square, val_of_sqrt, js_found_factor,

        val_of_factor, resolved_by_ECC, ECC, resolved_by_BCC, BCC,

        needed_a_search_for_q, q, n_iter_to_find_q := false, false, -1, false, 0,

        false, 0, false, 0, true, 95071353938334627422713004625126246621917461111\
        371816939266168759791402431582969357328835116136788557608798967030195044\
                                    1/2
        7065870215001447869562327724160498 8 2   + 89560570158156424421667682662\
        733061510314239376158825902438911416728786536127376615961819503383496457\

        26238659498880649074845956812600899791420763406497312, false, 2, 0

        "both maple and ppt match for all components , done with this case"

-- N3 : Appendix B LARGE STRONG LUCAS PSEUDOPRIME; page 295 col 2

        rawstr1_N3_p1 := "000000000000000000000000bc508ae6dacc43b138c0e9f22d"
```

```
        rawstr2_N3_p1 := "fb99b146bedd0ac93f84e8cfe2780a881fdbad85918a6b75"

        rawstr3_N3_p1 := "bd3af841123bad7438fe08c5433ec8b5fa7b0a1b149876bf"

        rawstr4_N3_p1 := "5af73cd9a608317066029e0cff4171ce336ff0b666344757"

                N3_coef_576 := 14919821275410909962861925757485

    N3_coef_384 := 616922290768388271916085858904664191416728734648623983477

    N3_coef_192 := 4639914694831736733475869981465266661365701879542501111487

      N3_const := 2230474245517876999429259763345976716843784064002268874583

P1_N3 := 369012538595434689365878622051913500627130245213169388019826598097\
        107079718295481926241398412699320815932808015860263240282855670239765686\
        869973444864115322609857375876438922226372746215468824202413623127

P2_N3 := 1143938869645847537034223728836093185194410376016082510286146245410\
        103194712671599397134833507936789452939170484916681604487685257774327362\
        92969176790787575000905578652169606589017555132679533550274822316967

P3_N3 := 1586753915960369164273278075482322805269666005441662836848525437181\
        756044278867057228283801317460707950851107446819913193321627938203099245\
        35408858129156958872223867162686873655734028087265159440703785794503

N3 := 6698129179250022303680418276550844853471546552467132588517485097081200\
        90047758152011512279001301539029474811303447198490991280789655006979985\
        617043973491020680240984777302624055937148011571160086698984525170773780\
        646150387925023280436219006757821606926619787915180974323526158281333102\
        221358792942524316309648612582551007693655624280569040000189913850390091\
        949941495106930906440830519675652462869368493804414578514532782117418093\
        303329308939479432896367346791865204279430029135550046807910943237629686\
        817425767454872759214278220289803110224677554440281119960826668392507282\
        5828225074019194302318324623049819212337927

                        len_N3 := 617

 decimal digits length of N3 = 617

  MD5--hash of N3 = c4369e4f912d66b3943d4035ac0f9c7b

                              P1_N3_maple := true

                              P2_N3_maple := true
```

```
                    P3_N3_maple := true

                    N3_maple := false

P1_N3_ppt, trivial, triv_path_id, perf_square, val_of_sqrt, js_found_factor,

    val_of_factor, resolved_by_ECC, ECC, resolved_by_BCC, BCC,

    needed_a_search_for_q, q, n_iter_to_find_q := true, false, -1, false, 0,

    false, 0, false, 0, true, 0, false, 3690125385954346893658786222051913500\
    627130245213169388019826598097107079718295481926241398412699320815932808\
    015860263240282855670239765686869973444864115322609857375876438922226372\

    746215468824202413623125, 0

P2_N3_ppt, trivial, triv_path_id, perf_square, val_of_sqrt, js_found_factor,

    val_of_factor, resolved_by_ECC, ECC, resolved_by_BCC, BCC,

    needed_a_search_for_q, q, n_iter_to_find_q := true, false, -1, false, 0,

    false, 0, false, 0, true, 0, false, 1143938869645847537034223728836093185\
    194410376016082510286146245410103194712671599397134833507936789452939170\
    484916681604487685257774327362929691767907875750009055786521696065890175\

    551326795335502748223169 65, 0

P3_N3_ppt, trivial, triv_path_id, perf_square, val_of_sqrt, js_found_factor,

    val_of_factor, resolved_by_ECC, ECC, resolved_by_BCC, BCC,

    needed_a_search_for_q, q, n_iter_to_find_q := true, false, -1, false, 0,

    false, 0, false, 0, true, 0, false, 1586753915960369164273278075482322805\
    269666005441662836848525437181756044278867057228283801317460707950851107\
    446819913193321627938203099245354088581291569588722238671626868736557340\

    280872651594407037857945 01, 0

N3_ppt, trivial, triv_path_id, perf_square, val_of_sqrt, js_found_factor,

    val_of_factor, resolved_by_ECC, ECC, resolved_by_BCC, BCC,

    needed_a_search_for_q, q, n_iter_to_find_q := false, false, -1, false, 0,
```

false, 0, true, 6224977537991991418823356036668353673406251698148446598951\
18410102910239473563859881027852920854452657763859857737489051983566506501\
04436904755836521767735373048430644257879630520258730620873831505972014501\
01062383676851340781811554034364002563101094196533416872545257903471650301\
49668164504046227199690438999496401278711299633075577905999417976373946801\
52291113590884975343901785247453091264661150967503315494190934367955125501\
18002586311818677782089857383347037067888598410577562188792753172914875901\
71733300185614505870922633377008214228863630156680565432317591300987931501\

                                                                   (1/3)
59281816330039551665926561414388314790731622382094282858, false, 7        ,

false, 66981291792500223036804182765508448534715465524671325885174850970801\
12009004775815201151227900130153990294748113034471984909912807896550069701\
99856170439734910206802409847773026240559371480115711600866988452517077\1
37806461503879250232804362190067578216069266197879151809743235261582813301\
31022213587929425243163096486125825510076936556242805690400001899138503901\
00919499414951069309064408305196756524628693684938044145785145327821174101\
80933033293089394794328963673467918652042794300291355500468079109432376201\
96868174257674548727592142782202898031102246775544402811199608266683925001\

72825828225074019194302318324623049819212337925, 0

    "both maple and ppt match for all components , done with this case"

-- N4: Appendix  D A PSEUDOPRIME FOR MINI-GMP; page 296 col 2

        str1_N4 := "000000000000000000000000000000000000000000002e394"

        str2_N4 := "1a2fe4aa9e66358347f63732494d08635ccc9ae0a3c17764"

        str3_N4 := "a8e266f4d26758ab804a702c235f63b1e109a81fc007f94b"

        str4_N4 := "ec5158f231a30b1cbf96a7fc444c09be62f5a809f049cc5d"

        str5_N4 := "e94b84275c38885c9b61a6bdc44111501527722a8ac87ea2"

        str6_N4 := "a5d4498caa2d9d07b34001a508fa53063991206268c547d7"

     C0 := 406612131192679292350254679233316564668498311732292044759

    C192 := 572037639378422293119041101906974763567524937807146152702

    C384 := 579449469966088845342144995452937815361464590575543057101

```
       C576 := 41410330339005861875372900182655636842840228375501225 31147

       C768 := 6421054049386492694177406455593550138571123395462567 99588

                          C960 := 189332

P1_N4 := 18451007892866429572074399803913400598378263776577783815 74059909391\
         65788567210674400738559414678928974295448928973552165437508 2912307460745\
         30306832236498094289559289864907209205632134037557265349860 3028629642809\
         74895976723223042953579681030092115433265733241996206865291 5629542244550\
         523065026519

P2_N4 := 20179867332428014022977771065540086234446307092443122159 27549322901\
         65622955958314592087762431834344619186932493618374003339002 8181190669817\
         13796582417057965724490995325249014708199864996876381113142 2132412240341\
         02243729742189042078330097142611746649362732446771251448569 3824030352864\
         9070762195027367

P3_N4 := 20770299584999739769284151859265315053594411533293611241 37919240002\
         18928190109056172911396333104070346364386859345527672633003 0834384508560\
         98766401048625904741756892600926045402780093286078213604337 7429328388910\
         93440400997332179452844646935574694343227235910515130068258 7124175704690\
         5238143003513127

N4 := 77335903220984981782945972728419722785273792891382592170000 19543260045\
       25878170332715339538607434577045467520428981748227664906871 4623093811730\
       44357881783690719189625122948558420253842911801106709301661 7044278092622\
       11074645768868382831894592597507905056898817609580247804023 2113674255454\
       79524927714767976618196813780000893460737873950446791539213 1457995108940\
       64248263115284732365286696826328149564873312892129145130728 8750982768550\
       93294464042040009799455448075839714304610324904748412008967 8934672930207\
       38038122436622318741039414885239209158504741711406519455870 4705533669046\
       93548727523722322742199207312727211561591799622751920104747 8011260771656\
       07779061973720971561630454757753766411274749598012117555341 8973301929593\
       18082822391331280666951023492477281082002698374662346118687 6875869499802\
       09135536149968648813589546499066420951934804075954574465911 3341809203948\
       6086031794198579492563 6324071

                          len_N4 := 891

 decimal digits length of N4 = 891

 MD5--hash of N4 = f493b1a9e1c756b38708ba307577aef5

                          P1_N4_maple := true
```

```
                    P2_N4_maple := true

                    P3_N4_maple := true

                     N4_maple := false

P1_N4_ppt, trivial, triv_path_id, perf_square, val_of_sqrt, js_found_factor,

    val_of_factor, resolved_by_ECC, ECC, resolved_by_BCC, BCC,

    needed_a_search_for_q, q, n_iter_to_find_q := true, false, -1, false, 0,

    false, 0, false, 0, true, 0, false, 18451007892866429572074399803913400059\
    83782637765777838157405990939165788567210674400738559414678928974295448 9\
    28973552165437508291230746074530306832236498094289559289864907209205632 1\
    34037557265349860302862964280974895976723223042953579681030092115433265 7\

    332419962068652915629542244550523065026517, 0

P2_N4_ppt, trivial, triv_path_id, perf_square, val_of_sqrt, js_found_factor,

    val_of_factor, resolved_by_ECC, ECC, resolved_by_BCC, BCC,

    needed_a_search_for_q, q, n_iter_to_find_q := true, false, -1, false, 0,

    false, 0, false, 0, true, 0, false, 20179867332428014022977771065540086 23\
    44463070924431221592754932290165622955958314592087762431834344619186932 4\
    93618374003339002818119066981713796582417057965724490995325249014708199 8\
    64996876381113142213241224034102243729742189042078330097142611746649362 7\

    3244677125144856938240303528649070762195027365, 0

P3_N4_ppt, trivial, triv_path_id, perf_square, val_of_sqrt, js_found_factor,

    val_of_factor, resolved_by_ECC, ECC, resolved_by_BCC, BCC,

    needed_a_search_for_q, q, n_iter_to_find_q := true, false, -1, false, 0,

    false, 0, false, 0, true, 0, false, 20770299584999739769284151859265315 05\
    35944115332936112413791924000218928190109056172911396333104070346364386 8\
    59345527672633003083438450856098766401048625904741756892600926045402780 0\
    93286078213604337742932838891093440400997332179452844646935574694343227 2\

    3591051513006825871241757046905238143003513125, 0

N4_ppt, trivial, triv_path_id, perf_square, val_of_sqrt, js_found_factor,
```

val_of_factor, resolved_by_ECC, ECC, resolved_by_BCC, BCC,

needed_a_search_for_q, q, n_iter_to_find_q := false, false, -1, false, 0,

false, 0, false, 0, true, 6132282778905121066824121401411706087240111505 8\
8176232235422465048246068128144271144657008237799933320241218873256758 42\
4378644277228371824765948421344572711469603176987458557611233898133022 62\
8952513827065938257923079207914989145636639887340489944560683437564371 1\
6699236716970071083573051789105720408527457788075860255998767765664806 2\
2450275165418658135892720059445587104324724817496368188007405816548664 87\
3833132535664032996309616819640590896086711792500417267924203206226018 18\
4073714478647000960909367044325928297311841845944218267190745337799848 23\
8347049657679327234189064602202142197469264791561864737466835103933809 32\
1201279082202381049302145310210616788141809552022397436028928739769607 98\
5562589246427306458655709001453747031949228045800675701480869005045943\
0071703687242817219188346171525286986335855129035858497189043940220607 5\

8814337975088568735480861758738921910841102852763299 7733590322098481 782\
9459727284197227852737928913825921700001954326004525878170332715339538 60\
7434577045467520428981748227664906871462309381173044357881783690719189 62\
5122948558420253842911801106709301661704427809262211074645768868382831 89\
4592597507905056898817609580247804023211367425545479524927714767976618 19\
6813780000893460737395044679153921135479951089406424826311528473236526\
8696826328149564873312892129145130728875098276855093294464042040009799 45\
5448075839714304610324904748412008967893467293020738038122436622318741 03\
9414885239209158504741711406519455870470553366904693548727523722322742 19\
9207312727211561591799622751920104747801126077165607779061973720971561 63\
0454757753766411274749598012117555341897330192959318082822391331280666 95\
1023492477281082002698374662346118687687586949980209135536149968648813 58\
9546499066420951934804075954574465911334180920394860860317941985794925 63\

    1/2\
6324069    + 31048869581849270065481499656550323044590813439413708561 862\
3233668802375217362442266166902111318271778336981425479999554969501521 17\
3958554990173008625423888748065399350304614608226483774667161802148425 98\
2561595760110114720497112278554830769874486624110644302342018101511951 2\
4169749503848051298696396666523709771078132679507482245431628686227439 77\
2028919132603928719465671512141762573133642668102035315665791875411433 61\
3378321814105122733023686491905370854934187945808567280990121130713010 1\
3724665079811243024698319649716068469664653942360219460436133621369229 37\
1823764187890076871428413420985651688825190419492487309860688382740739 15\
4751182379439813962949681827071922407184218574035980176339246452433911 2\
0900568752876648035027655292220303588611760603448734000633091681774926 90\
6939807626835671163171265770735581375435669760892192842503458316691460 46\

0930117241189024630509645363720142090646, false, 7733590322098498178294 59\

7272841972278527379289138259217000019543260045258781703327153395386074434\
5770454675204289817482276649068714623093811730443578817836907191896251224\
9485584202538429118011067093016617044278092622110746457688638283189459244\
5975079050568988176095802478040232113674255454795249277147679766181968133\
7800008934607378739504467915392131457995108940642482631152847323652686964\
8263281495648733128921291451307288750982768550932944640420400097994554484\
0758397143046103249047484120089678934672930207380381224366223187410394144\
8852392091585047417114065194558704705533669046935487275237223227421992074\
3127272115615917996227519201047478011260771656077790619737209715616304544\
7577537664112747495980121175553418973301929593180828223913312806669510234\
4924772810820026983746623461186876858694998020913553614996864881358954644\
4990664209519348040759545744659113341809203948608603179419857949256363244

069, 0

"both maple and ppt match for all components , done with this case"

-- N5: App. E AN EXAMPLE PSEUDOPRIME FOR JSBN; page 296 col 2

        str1344 := "00000000000000000000000000000083dda18eb04a7597ca3"

        str1152 := "c6bc877df8a08eec6725fa0832cba270c42adc358bc0cf50"

         str960 := "c82cb10f2733c3fb8875231fc1498a7b14cb675fac1bf3c5"

         str768 := "127a76fc11e5d20e27940c95ceba671fe1c4232250b74cbd"

         str576 := "f8448c90321513324c0681afb4ba003353b1afb0f1e8b91c"

         str384 := "60af672a5a6f4d06dd0070a4bc74e425f3eae90379e57754"

         str192 := "82d26e80e247464a4bb817dfcf7572f89f8b9cacd059b584"

           str0 := "0e4389c8af84f6a6ea15a3ea5d62cb994b082731ba4cde73"

      C0 := 349747877319125519155117987369980154709454871524061011571

    C192 := 320774607322505182348990701566033550822920466432596598515 6

    C384 := 237071341952275911704544028879316459819868604019500315221 2

    C576 := 608750800316536425926351990059457930963947080307917145526 0

    C768 := 453088511710529505206151022326189686496423403702720089277

C960 := 490826633921674588949903817181645243436429359503824938 2853

C1152 := 487300338967201120051709752920014375155494049638544 4728656

C1344 := 389189888207583737601 63

P1_N5 := 149446543543083645457036059621052921393541243836750548706481925 2343\
7530482472401251725664068684814949588779644083289776272581318830344 62411\
0152295134965367218621910337465901747035296579855521540699486195373 54089\
9785897644915651031264255203023455677284343404880972742346387613171 33356\
0275622453586156561794632202184300886820826624562851321402814660588 72641\
8255372307360994387346859948919643309365459081627814846050589608870 99352\
3

P2_N5 := 151389348609143732847977528396126609371657280006628305839666190 2624\
2218378744542467998097701577717543933433779456372543364124875975139 10422\
3584274971719916992463995171852958469746755435339364332072857951591 339693\
1483114314299554494670690520662760601089039869144425387996890652142 56089\
6559205545482776590979624208126967983494973706821638885810512511763 7986\
1692692147356687314382369128255598672387210049688976439049242727378 631643\
7787

P3_N5 := 306813753893950724123295030402021647620940173596848876494407392 5061\
7250080515839769792788333009925091505764609302993910687609447558697 51329\
8142661912083898899830781922817496286663463878443385723056045159101 87946\
7260447865011831567185515931807154505464757010220637040037133769840 74779\
9245852897212379421364379911084369720643157060227533762839978498188 65533\
6678279347012121477223103475132027714127287494581903878941860467012 14970\
0667

N5 := 694154302139271373043166838706899903298750581362862622367895126500 9922\
9791714948896196209196292181729380500227126503902311089953072371692 69667\
6440983333126847764693992411011978952753216125693365611281810962094 5798\
4578041435866298282960712662700332884952795233638965081310836973104 9080\
7471838369135928929519335607653452044687904196919591053055172673786 8378\
3128727158433123200640736745710012951061193730973314456890957226558 47864\
1910508997802906399221809833295978576247571877019433173473143183426 30851\
8887827243403093654898284478765039702245960961631901755866648386557 52084\
8114326438654543475409306234610422754846201955355349982116992680801 631254\
1697331792145166796984714867320957576327029857490042676686184075961 94879\
0102358112050697733453701515402945607070679129119726171178725689175 74563\
7975104247236362528993727600408985277723349354239223273512995662819 32266\
2566751325027023227604354605365713093025865394693108614066225103681 33332\
9718040292864685422602353962042710457949298274746364675506552110365 7096\
1196187597469821934390308094064989478697705887172422900312995739714 11197\
9821667173517579542660010663841323729942640584293345632914966574695 58139\

```
    808375030168417007373133166391119553220773091860698215980720042621281779\
    381114805103592526045494482900527480395872621990534353634675251867

                                len_N5 := 1288

decimal digits length of N5 = 1288

MD5--hash of N5 = 3307bd313d6d25e8547baa1b5d753bdd

                           P1_N5_maple := true

                           P2_N5_maple := true

                           P3_N5_maple := true

                            N5_maple := false

P1_N5_ppt, trivial, triv_path_id, perf_square, val_of_sqrt, js_found_factor,

    val_of_factor, resolved_by_ECC, ECC, resolved_by_BCC, BCC,

    needed_a_search_for_q, q, n_iter_to_find_q :=

    true, false, -1, false, 0, false, 0, false, 0, true, 0, false, 2, 0

P2_N5_ppt, trivial, triv_path_id, perf_square, val_of_sqrt, js_found_factor,

    val_of_factor, resolved_by_ECC, ECC, resolved_by_BCC, BCC,

    needed_a_search_for_q, q, n_iter_to_find_q :=

    true, false, -1, false, 0, false, 0, false, 0, true, 0, false, 2, 0

P3_N5_ppt, trivial, triv_path_id, perf_square, val_of_sqrt, js_found_factor,

    val_of_factor, resolved_by_ECC, ECC, resolved_by_BCC, BCC,

    needed_a_search_for_q, q, n_iter_to_find_q :=

    true, false, -1, false, 0, false, 0, false, 0, true, 0, false, 2, 0

N5_ppt, trivial, triv_path_id, perf_square, val_of_sqrt, js_found_factor,

    val_of_factor, resolved_by_ECC, ECC, resolved_by_BCC, BCC,
```

needed_a_search_for_q, q, n_iter_to_find_q := false, false, -1, false, 0,

false, 0, false, 0, true, 50708147828199248844848999063749862277393182121\
287875609695143937818149871646894566084125329070280843078280772760838857\
3007567624651281620847062240388537085217069570583029563451679934665682\
250880343233285761539112585876515968300430134560042335748800521032995506\
374218900403402038858892864764999702707317867710473631838524516476541951\
372183528990597634372860815525072170414005013656277524396993197746308324\
141886237094228588380593548364161963531165209024325058234779380380174784\
924903408760923424813955422414904058961236581266326488832397356345288210\
809162888013042120616364590642807815791351423089259757981196342001225737\
984808160678587271186954806344805318849877819379673459307104883053678203\
119835992755652700025963495926734883554002330815670651455870302753669258\
37955531467593491641603781289354740902656676615543258692437477927523793\
860982776448973052706729466029973465394621774323331324530085472346068713\
93495171257609565345172551662377162703333422542686245615687106765530867\
910658407368107259447234966757662648799894222090672846440994645403697541\
839551601190072519107506278728665642455177308443006733104720791428695154\
583115782651659366473900239079843976186475802032401913052708504663883721\
171915118631936883396801618879438994749874732289965319238049252442111757\

                  1/2
44183214401068441 2    + 67767454198558474355676014530743668380327115480\
114473810358770426375734084048479571912806302321137904054415556019405730\
103269053822286694164654805498313371577898128734939098031724209603952260\
685663623364101658455307622486324133222898863279493385968397844246728178\
697666903849802103916876498674019573307515933562204431443482629633799638\
802152330624833062025643267285596823672714022328696955558409912682677534\
484662253167463976563662323918987849227474588077487836577780675382677064\
690408898549878712907770597067329174317407054887816288009562573222361172\
778897632814369040742394356267547060039693895808316693014641914478230463\
514595563021450144467795821536547835514793540050035881622612789709783653\
153403170693475046901564623248237974315733856415621166197054390801722422\
270649732149204939945232236109827085832275179488655260873663908932094841\
099278773799045312434877824787422086686198757701675399950273388254223801\
926215141654024267140695143644668327515518817436821026624200351312662293\
913578352002171422947782246447532764431932347652479062344741044083504685\
267014300789892937518003274336902733098823426008266048153380523047072285\
555185217564330011630331576761380046862064597636166428713710363705284818\
707478116509466674560134597970558034631174995587767240496245028595662066\

62334631596462235, false, 2, 0

  "both maple and ppt match for all components , done with this case"

-- N6: App. F AN EXAMPLE PSEUDOPRIME FOR CRYPTLIB; page 296 col 2

```
          str576 := "24a027808260908b96d740bef8355ded63f6edb7f70de9a9"

          str384 := "b99c408f131cef3855b4b0aea6b17a4469ed5a7ec8b2be62"

          str192 := "66c3a9eae83a6769e175cb2598256da977b9e191b9b847a7"

           str0 := "e2cf4750d9bc2d64ccd3406f5db662c22c3fc65e3c56eff3"

        C0 := 55613572192479451536119286391985445387001652605290 70698483

       C192 := 25197735858214425675646336884455273611098672133772 17513383

       C384 := 45511527868331496894595148594763899381490571474556 23315042

       C576 := 89805716639628982450338334342567395632912039964704 2300329

P1_N6 := 22211683951059300447425330752354393066860502377626791487125168133999\
         39729863340244743625878412836415901800725569032731901731279834484776044 7\
         61998664069644323493578946903234375700665946172079826882759862218316218 8\
         17048571800819527667

P2_N6 := 14237689412629011586799637012259165955857582024058773343247232773899\
         01366842401096880664188062628142593054265089749981149009750373904741444 6\
         92441143668642011359384104964973234824126871496301690318490716819406962 \
         61728134524325317233907

P3_N6 := 15037310034867146402906948919343924106264560109653337836783738826711\
         39197117481345691434719685490253565519091210235159497472076447946193382 3\
         03873095575149207005152947053489672349350845558498042799628426721800080 1\
         39141883109154820229883

N6 := 47554449008989626405708050796224079455131964320119119856484100222400888\
      80898588920037609578752790123924575606815854816408064641949819393836322 3\
      75187804097517865165295593068331909051558956511495004634031513757833278 4\
      42565239485719299919124413007990606077189704231818226156523193015783627 6\
      91153498978453275291294761019331074402549345029398192480499301518250255 5\
      79024958799675231548536117989526030089139543982247618576952343421528613 5\
      93802903966090452696167081552767096127450386003813203809365323894442795 \
      44356582095228370593938065643830918025255480869101953568190898362321466 4\
      51603411147069570302064248587595965158321361946659493291847667065163381 6\
      99459977254847876852676401257595441432298417328862 7

                       len_N6 := 697
```

```
decimal digits length of N6 = 697

 MD5--hash of N6 = 57e387afbcf7f4a0c7f0393cdbc3516b

                        P1_N6_maple := true

                        P2_N6_maple := true

                        P3_N6_maple := true

                         N6_maple := false

P1_N6_ppt, trivial, triv_path_id, perf_square, val_of_sqrt, js_found_factor,

    val_of_factor, resolved_by_ECC, ECC, resolved_by_BCC, BCC,

    needed_a_search_for_q, q, n_iter_to_find_q :=

    true, false, -1, false, 0, false, 0, false, 0, true, 0, false, 2, 0

P2_N6_ppt, trivial, triv_path_id, perf_square, val_of_sqrt, js_found_factor,

    val_of_factor, resolved_by_ECC, ECC, resolved_by_BCC, BCC,

    needed_a_search_for_q, q, n_iter_to_find_q :=

    true, false, -1, false, 0, false, 0, false, 0, true, 0, false, 2, 0

P3_N6_ppt, trivial, triv_path_id, perf_square, val_of_sqrt, js_found_factor,

    val_of_factor, resolved_by_ECC, ECC, resolved_by_BCC, BCC,

    needed_a_search_for_q, q, n_iter_to_find_q :=

    true, false, -1, false, 0, false, 0, false, 0, true, 0, false, 2, 0

N6_ppt, trivial, triv_path_id, perf_square, val_of_sqrt, js_found_factor,

    val_of_factor, resolved_by_ECC, ECC, resolved_by_BCC, BCC,

    needed_a_search_for_q, q, n_iter_to_find_q := false, false, -1, false, 0,

    false, 0, false, 0, true, 3580614380385190753029787870489756486169000240\2\
    6827454383793500613229987443197266492239816171965316461251634985206510000\
    0844361129446255168217747124358491285896268340440690926492336307986199720\
    4860924625227909209190667658601672671217955176844045800723863599930963900\
```

```
81206792402453943185812600754941650366391667256455892117366184425141935 0\
417532122144973140009318109500905344215150665507507504345458395327698234\
297783321942125204117681588464518920251170049333529037249330984714088556\
969586321401780389203066395184822910810403268244464156227174281665677235\
347588906960908694643515880836799859051562596931960151404915272250357626\
690743049645530202282149340817363045868927768373403932451510456167421290\

        1/2
27 2     + 16070583013336108359130601338469442998867366746970249465316769\
623743746833845614186219238604798761146547045605939966888730155539292761\
463197210818490605538491085882682865948932503316961203227189607215468852\
813340513469066165485224654908962977150475836026615350768382143931273097\
907631715044005303663124770895898128047446742136535896008759592073288672\
935162475825128986338657498772595117862588182095787866136157701497520756\
471145571340873057466828828421006814278608614793409071695571790784452863\
718432663962890557664842095051392500601781878626158274004336750498428676\
627499810302885443188967369448769200419529088962044415695733651445299202\

338231258244418327549981377022161281777651806155778431406 73, false, 2, 0

    "both maple and ppt match for all components , done with this case"

-- N7: App. G PSPs for LIBTOMMATH, LIBTOMCRYPT, WOLFSSL; page 297 col 1

        str_2304 := "00000000000000000000000000000000000000001e46d6a8"

        str_2112 := "4d42d684ddb3415e871b661303b1c60f0388dfb9e525f8bc"

        str_1920 := "51c9de3c9f45627608de2f75dee580d9d4d97cab6fa86dad"

        str_1728 := "9e6bbfd721f297472480a9bed9508aa884bda9dc56833752"

        str_1536 := "fac8e89f413a9517d14731277148789987806654a8723593"

        str_1344 := "a452f960facc9b65f6962cb26131b42650c29c8735083c7e"

        str_1152 := "6c3a220d77d1cbe7f9628885a7b79465287d4b02ad546007"

        str_960 := "1d43306a8813836de5ccd162fbeca4f117552dba01975451"

        str_768 := "2f7684e32b0377e76f87b96906f8fa276381db612f76c2c7"

        str_576 := "dd97ab4380042c991a4719884377c70065a3614237a41289"
```

```
          str_384 := "24a1017fbb529443b0ad43c5424753db5b518cee5a1fcd87"

          str_192 := "ea038ffcad33380db1d89cd4e0b15b480cf0c62e8999924d"

           str_0 := "0284af806081ea106f35f85a664456166b864650ef034cf3"

      C_0 := 6174860841637432935218200680472929128571150202276382641 9

     C_192 := 5738004519854949881241888767857027945176653248678453678669

     C_384 := 8981387287432649376898741821237290118577937933215234656 07

     C_576 := 5433431236515834234717472608998705293049472629430701789833

     C_768 := 1163788520414798662298766502689095521885596489029975982791

     C_960 := 7175133706672993271799228138135727899858689579295351327 53

    C_1152 := 2653720331545715928040800747136558446517368790351195955207

    C_1344 := 4029215642575367831400755699079036683513726996859926887550

    C_1536 := 6149225391980774567922890311910818842447217677533646566803

    C_1728 := 3884469067239225277582363758523667716322245838985324869458

    C_1920 := 2005449344854747580093184108758428824002119851081485282733

    C_2112 := 1894436311542447017110189405632958787308305069751551785148

                      C_2304 := 507958952

P1_N7 := 190080965196429793284539657366893779873436830111672366248881646565 3\
         850457169635536238329413823488311896159501537950387704882660206924124702\
         319767792112295081202063239825692483258274851577725555804284352026976128\
         284526935861630031791154642395256538340471624900082495610151339985167 92\
         689132376286139252916898534542631569361562095241253775712906006443172754\
         937344955794702245521682959506315285400771587080099118585628535260790657\
         982364170344198148111227942260024011356635573591350816373257142371334822\
         961251413464688009750724419940054051790800253090303365833973258551947851\
         236063024914180757185378753321451274470828789328966010596890419576693866\
         086998173674081322567916936236887943395514342216042478259443

P2_N7 := 500483181362199645718192917847031322406759173684033340333305375406 6\
         588253727650366915521346597244725222587967549423370826956044324831220341\
         207948596631672948805032510461048308419037684204151388432680698887028145\
```

7731594221236718737061101730166710465450461788361917210941528478180947 15\
15048554676140465293019384145074892212899299677022119145208151496487386 3\
75002926860745101245859123238012814646023158878190097923595993334166180 2\
46756486051627372397686317197064322190202146526602669951078605586372458 8\
85697497165252352967365739770216231836517706638676876224085158976727869 2\
30455394459903793366910225749538120568169220230316750590161247474543494 9\
40706619128385612232132529311172595496038926305483984525711078 7

P3_N7 := 1117866156320203614306377724974702319435681997886745185909672963 451\
02945386146265886176152826959347622613140285446862300924149246769207773 7\
43425543854124073725493339134148974940419144021286039936849962742706466 1\
04413029098022462169637804510106503701980313626037385156683300030452772 5\
78047875049387849464042802816452162594153466821138134549676002238922989 7\
17865256850286439059130174848566401934419377036180629164020814158687098 5\
95942836857942293090421315284312012107883738082907341510911252542858200 9\
38351195625858301853440103136674578785816962884240740944695967335440053 1\
31192866495202703300721244828345494516294411004364910832031255753053662 1\
64576362593772722580219195020091379951090198465725458146437784 03

N7 := 1063452078062964842316680439100667853997956255631409884703724996175 6\
11913155370425059004666755351126370765757325069220154090475525113111331 5\
34155822504132427954121926028977897326016452032632643031621775313813263 \
10998970713694521304856783427108211035837719173191888424497687716945903 8\
47840832601158220338315116311569535166611157453976183644379418363862188 \
08101529448483687495168944238726821330736995943973596941803455012007421 \
09710969089802656966120128959989924056907838239415947883597253487288911 2\
65130861076995785143322559232949628580082040985968381890859811544811963 2\
50166807225739883175758540592938038640525057077635260377283817570121188 5\
06720763322577723854603213904763949026451035886610117960695498777558294 9\
14095190286423434103595325322044355050505888390419892966130432853384420 0\
25991603809604285550077998959033403957562518661694503445321270371924756 2\
78334786001733438835926798599715925477942244937628383988529536076933049 3\
98733022131632892984494328799535905089231306547424707999668050846457897 2\
92105617153237859066140231207977706483096753560267776977308513770417672 7\
19215980272226316038096369457350381579229798547106736019854318870385754 1\
41371733858517502044218801634760265587498814280073328060552232720804872 9\
21969797735332629722017664747892121030708748712733180842486503682755515 4\
81434761547891191431690407378061063299887695616928512486921716155822944 3\
78708863947668462062195571745339003058779915940858554595239390172360173 7\
69656161537253021381577803596097455028941144286297184282046724577533468 3\
48343936096263977229393852616124982192888432604948830175933041671162863 3\
93593616389115022094081675889914163162852300506873558662214056728175728 8\
39204797446701664449699367321558065141345504373533192399722226922311408 6\
12622529245371708935701643178960931071329253928802738471610262504489056 8\
99252740549408250528391001108587552456472577625704595893048563596869581 9\
67781113019047808212452621728173827862405874506295664794063113108306419 8\

```
528915906368753667317327626645537256342437825075237709498791366843288703\
723646982443682876999808864021794699452314576420850628038065790946742573\
18340352731256782194724089323
```

                          len_N7 := 2115

 decimal digits length of N7 = 2115

 MD5--hash of N7 = f18238fc918ac964e2833d3fed9417c7

                          P1_N7_maple := true

                          P2_N7_maple := true

                          P3_N7_maple := true

                           N7_maple := false

P1_N7_ppt, trivial, triv_path_id, perf_square, val_of_sqrt, js_found_factor,

    val_of_factor, resolved_by_ECC, ECC, resolved_by_BCC, BCC,

    needed_a_search_for_q, q, n_iter_to_find_q :=

    true, false, -1, false, 0, false, 0, false, 0, true, 0, false, 2, 0

P2_N7_ppt, trivial, triv_path_id, perf_square, val_of_sqrt, js_found_factor,

    val_of_factor, resolved_by_ECC, ECC, resolved_by_BCC, BCC,

    needed_a_search_for_q, q, n_iter_to_find_q :=

    true, false, -1, false, 0, false, 0, false, 0, true, 0, false, 2, 0

P3_N7_ppt, trivial, triv_path_id, perf_square, val_of_sqrt, js_found_factor,

    val_of_factor, resolved_by_ECC, ECC, resolved_by_BCC, BCC,

    needed_a_search_for_q, q, n_iter_to_find_q :=

    true, false, -1, false, 0, false, 0, false, 0, true, 0, false, 2, 0

N7_ppt, trivial, triv_path_id, perf_square, val_of_sqrt, js_found_factor,

    val_of_factor, resolved_by_ECC, ECC, resolved_by_BCC, BCC,

needed_a_search_for_q, q, n_iter_to_find_q := false, false, -1, false, 0,

false, 0, false, 0, true, 3522945064982865085783910756853806304725026149 0\
6878861480557894424364242286338661223335230860951957519292182654102945866\
1074260167780684408803438367745809682730765556418352621661953500308375 03\
7346094070307206273787052515020013570695048100400763089887269766412184 78\
0804535175741949382700501902556728282105409111290247037188175826042024 16\
3895327942654764450958917891184214762798834825734468825946074283077641 05\
1673664938222512434517257037374071061839038739088546025915844154130781 81\
7189438343334915403042673360291758681920944575395498599629231356535010 00\
2953550256473124797370028134622503943509840698645931881617219355611544 98\
1255143852450448069560717426199395959254859314770426313182547816473 43\
7489133732396227864019010386590555120283802647972396986253463874436946 0\
6351284760539258115947641484051964937592785103577292694168621380341979 75\
4807286753217829083694444076929676303712829892581643088717083072545247 36\
2026350830963471848485618445716305573473361986988723861974528321776493 07\
8172167143563110236681106568471499875354821528035113299097401682795733 3\
1730995501110820525719881016915408586915452136976727079441625808437826 94\
1607006243571916589342023195919400983485941278410214757223186915365608 64\
9783853644446197549456872477348677267962323902240058108464317871740554 72\
9318401200147003692703605393617102175972189309154809343830453592094147 31\
6142932726815563786686364897160953135169996290645202110184798499933974 6\
2432666773395222631619847272544583550773219894504813299879063737194389 7\
8064890234228358545632585912870318085738815589350211967978769541522844 27\
2998809118768738187223336470688316833202067758727293088090716798451498 86\
0567786769192593169322763573269453570203247130985076368983621995967089 57\
6878653538477859610850334417684009950861744343744886287461915764055615 94\
3232104210436381827644305071736777464864150369712762076511498290165077 3\
2809019588778606914395614903952039922343519054302584637961133639242946 0\
1977678832020515232328258161753221386488411272503101092526775831676171 03\
3194622943810502548135365984709827007221137208226219300278906528934095 71\

                                                    1/2
3693859398022707551985261894194891273081551505764 96   2     + 3794025502562\
5829105531567145882682096521485651953998665725198442157293929704692581 17\
3562836686173989773049842844116880065792894704070044979533029483770678 23\
3004141034462765219583888480911441540519251917237871782330021267916594 34\
2123387702605114416363169045924917210842063795487303408060828486596317 88\
8479637228163136116783416876547998413937905542944158369621958343429594 22\
6114966278339798391531378005046665069779322720202904436528597729064156 1\
4461065378049839536241437649253733480309085268039850595915669135970934 03\
7320421870969137203407131578914822186690958360627698142280763139954547 98\
0574193134229836919463166544710616949433551295183897717713008223422817 24\
7298837580037081133278602656618782052468983404304276183894857598196074 41\
4607583468916622594091811720993243198147656541998121364690813505598585 19\
2212456149434370499465666005712233009680483915499687578087711265088805 58\

6438748333307854795739605381158924342368572427486271286361567701642535 94\
0884923109692976554836650672539085033692487414638397115582092698212078 4\
2291903095025698936973438075419309559319939366273771582040495862743972 95\
7049279615199199681178300505887685097922645608418480061438723970397445 56\
2160205127825622108800158658827804044724336534022056055177589217310989 28\
4662183892815717754875110120867122940008400613253132022282535463188570 91\
0447539109763784054383522862306965168598577699524011735568410634309300 60\
9029918112867768214014252667257512029305452435494894335384706978916497 1\
2095243513342225255800077598839378910420814196598536995417987768069267 68\
8629528881056435515211426320633741811827962751354715980677189786481608 78\
5608680584010414222506032377136092558821683594304834371061392015391190 63\
4298810857097584886803783879422998384902273738979270836258710373417122 66\
2431153225883676304839566057887449291255752793073441002689626121506207 49\
2855862158945471553956143579627148248714568735474958387146234395217691 27\
9478714063256733163559443102176094477676600610042668618050385717261117 0\
1268769954340966018819483476288413647838318418020354657459771283862723 68\
0421067633924082514917184762775021345673492455605973283912073308066126 84\

4261611469705, false, 2, 0

"both maple and ppt match for all components , done with this case"

-- N8: right below N7 on the same page = 297 same col = 1

        str_192 := "000000000000e17516504450e648b6aedb0c0784e17dda33"

         str_0 := "63e1956a843076a9e5b6d15a819cf0907a96154d47662d0b"

      C_0 := 24490795584672899753279859367347698585833006940667955233 39

            C_192 := 19640107031361074622158075258139564465576499

P1_N8 := 12328294992973675251091628263800256032862628743324146786474137 88593\
        43760091491850110380631340085554443

P2_N8 := 28724927333628663335043493854645459655656992497194526201248474 127422\
        7096101317601075718687102239934184987

P3_N8 := 29711190933066557355130824115758617039198935271411193755402672 30510\
        1846182049535876601732152960618620523

N8 := 10521605559439088484043832497276931939972259404665136039207007179 49734\
        2353018847108786785541918881316495456114022714597785551433698574 62509893\
        6631894049079858371059715172007542738743794053576739529627253214 93970655\

```
902673038736203513210730585029200327705228367266690052620882639642154558\
6903174091231320127043
```

                              len_N8 := 309

 decimal digits length of N8 = 309

 MD5--hash of N8 = fa7b1a5060bd20a3f5b01889951fed3a

                              P1_N8_maple := true

                              P2_N8_maple := true

                              P3_N8_maple := true

                               N8_maple := false

P1_N8_ppt, trivial, triv_path_id, perf_square, val_of_sqrt, js_found_factor,

    val_of_factor, resolved_by_ECC, ECC, resolved_by_BCC, BCC,

    needed_a_search_for_q, q, n_iter_to_find_q :=

    true, false, -1, false, 0, false, 0, false, 0, true, 0, false, 2, 0

P2_N8_ppt, trivial, triv_path_id, perf_square, val_of_sqrt, js_found_factor,

    val_of_factor, resolved_by_ECC, ECC, resolved_by_BCC, BCC,

    needed_a_search_for_q, q, n_iter_to_find_q :=

    true, false, -1, false, 0, false, 0, false, 0, true, 0, false, 2, 0

P3_N8_ppt, trivial, triv_path_id, perf_square, val_of_sqrt, js_found_factor,

    val_of_factor, resolved_by_ECC, ECC, resolved_by_BCC, BCC,

    needed_a_search_for_q, q, n_iter_to_find_q :=

    true, false, -1, false, 0, false, 0, false, 0, true, 0, false, 2, 0

N8_ppt, trivial, triv_path_id, perf_square, val_of_sqrt, js_found_factor,

    val_of_factor, resolved_by_ECC, ECC, resolved_by_BCC, BCC,

```
      needed_a_search_for_q, q, n_iter_to_find_q := false, false, -1, false, 0,

      false, 0, false, 0, true, 617009628208409003298717267283754967203246836\
      2238588585260300677302676188352707827873260032306992389720778747204439\
      218556944601092815962637851519788853075505461957882310428029310204059032\
      0831237838530390422171176369873146341607529849534655709171924193739228\
```

                                           1/2
```
      795215272952817935780472631289140285606336556 2    + 6059913486562955526\
      210796954823663916179256922382103857901454501100777746075504593443712952\
      9673751839810918554913349707128593289066649907374569865837321055905093\
      702153641752704461134681295366496534776809770314817759478634723360324157\
      801321316448592488662495121229238733606998654065515681479241465544720196\
```

      4, false, 2, 0

         "both maple and ppt match for all components , done with this case"

-- N9: H AN EXAMPLE PSEUDOPRIME FOR GOLANG PRE-1.8
   The number right below N8 on the same page = 297 ;
   same column = 1

```
          str_960 := "00000000000000000000000000000000ff7d428a8a9f9ffc"

          str_768 := "2ea178501115ec855f1154c054f5f67e15967a139a92fe15"

          str_576 := "ddf2c49b044820ea8c58551b74f81b45b116da4e1f11b926"

          str_384 := "93e0cdc58006bc2052eb9b2fc32c71dd041d1907225e2814"

          str_192 := "ebe18736f626fea57c965b67b296a6461455226b39aba263"

          str_0 := "3faeb483847a715c6a01d8d0e401a4aaf8f3d22121fd142f"

      C_0 := 1561488932408212556712777294024384356114169102621699019823

      C_192 := 578378454202212458214469064512064477936378968472761425161\
```
9

```
      C_384 := 362596143794134373009452486995439084847725567421355276085\
```
2

```
      C_576 := 544215678642073255795592500295468962138701037614988065002\
```
2

```
      C_768 := 114338246880522876083543071094380524784937142284602120552\
```
5
```



```
                    C_960 := 18409944014559092732

N9 := 17941068535417227506556388369618767790767456525140396263114670609143523
      2423534967377144418465326071185099924376638206704525347388987104534565878
      4961599929629773751338809103769895525886635199389623976739941293398562566
      3195187481628011954040150091462071420597614071649073054718632348584811148
      2325012250364387662750357

                         len_N9 := 309

  decimal digits length of N9 = 309

  MD5--hash of N9 = ede4ec1a7839cc4d061b4199d1f0d6ac

                         N9_maple := false

N9_ppt, trivial, triv_path_id, perf_square, val_of_sqrt, js_found_factor,

    val_of_factor, resolved_by_ECC, ECC, resolved_by_BCC, BCC,

    needed_a_search_for_q, q, n_iter_to_find_q := false, false, -1, false, 0,

    false, 0, true, 48220641829233937406798055624322714048979931044235206028693
    89589655004851147963555966296104713070344001561137775307152762979545789895
    13381150077068791800849422831117979961080704053726894736144458735813205298
    06402765206838423181993632063515923957498413507328407353203082378250925

                                          (1/3)
    68013368045160126060855667556827363, false, 7     , false, 17941068535417937
    22750655638836961876779076745652514039626311467060914352242353496737714445
    41846532607118509992437663820670452534738898710453456587496159992962977375
    75133880910376989552588663519938962397673994129339856256319518748162801195
    95404015009146207142059761407164907305471863234858481114232501225036438756
    6627501, 0

        "both maple and ppt match for all components , done with this case"

-- N10: App. J.4 SymPy; page 298 col 2

            str_192 := "000000000000f8ae31e07964373e4997647e75fa186dd5e7"

             str_0 := "e42ada869da0b3a333813f8102b1fb5f20623d6543e78a3b"
```



```
        C_0 := 5594648294098026397800267786824731705334810462505461582395

            C_192 := 21663122340629185708954260333800701968635367

P1_N10 := 135981622838257435299056640532900774348225733374487447290922279391\
      475152577516907500556991301144775227

P2_N10 := 327715711040200419070726503684290866179224017432514747971122693333\
      45511771181574707634234903575890829467

P3_N10 := 34947277069432160871857556616955490074940134772432739537670258036\
      09114212421845227643146764394207233083

N10 := 155736648906806895927301993984793243377807755233473437818352553026191\
      565941961295330074319845684137503282213067287251154605239452087365454765\
      249451467057011833184935343597293108348882323965716163212396526196339616\
      849737969679440721383292277856598404280416828081993981335648754907549763\
      536484837691309893859747

                      len_N10 := 309

 decimal digits length of N10 = 309

 MD5--hash of N10 = c3e0c36f3bde7353521c5ddd567ec385

                          P1_N10_maple := true

                          P2_N10_maple := true

                          P3_N10_maple := true

                          N10_maple := false

P1_N10_ppt, trivial, triv_path_id, perf_square, val_of_sqrt, js_found_factor,

      val_of_factor, resolved_by_ECC, ECC, resolved_by_BCC, BCC,

      needed_a_search_for_q, q, n_iter_to_find_q :=

      true, false, -1, false, 0, false, 0, false, 0, true, 0, false, 2, 0

P2_N10_ppt, trivial, triv_path_id, perf_square, val_of_sqrt, js_found_factor,

      val_of_factor, resolved_by_ECC, ECC, resolved_by_BCC, BCC,
```



```
      needed_a_search_for_q, q, n_iter_to_find_q :=

      true, false, -1, false, 0, false, 0, false, 0, true, 0, false, 2, 0

P3_N10_ppt, trivial, triv_path_id, perf_square, val_of_sqrt, js_found_factor,

      val_of_factor, resolved_by_ECC, ECC, resolved_by_BCC, BCC,

      needed_a_search_for_q, q, n_iter_to_find_q :=

      true, false, -1, false, 0, false, 0, false, 0, true, 0, false, 2, 0

N10_ppt, trivial, triv_path_id, perf_square, val_of_sqrt, js_found_factor,

      val_of_factor, resolved_by_ECC, ECC, resolved_by_BCC, BCC,

      needed_a_search_for_q, q, n_iter_to_find_q := false, false, -1, false, 0,

      false, 0, false, 0, true, 115518000965508185306404574408533737743714485601\
      202746062603174304549915963346599547638639758340371478038782131683172843\
      690063745704658053468808397920793869219096012573483075988178567927744995\
      597730169555555159661760393483505033390055770840186321905568044852507156\
```

$$\frac{1}{2}$$

```
      406905416854328878990334533305613780144993739 5 2    + 1009047101388798021\
      036181475116871881130358842595656812686336319594610605144070096754595122\
      302473429589828383952864954142565055023894144736022666034749214436472804\
      936731291983799779583058478294838037251899703071754945720377217822963472\
      168379296639984273770619608103296620211409848031824456844454830926208005\

      904, false, 2, 0

      "both maple and ppt match for all components , done with this case"

      "maple timestamp at exit || date = 2019-01-24 :: time = 17:13:36"
```



# PPT : New Low Complexity Deterministic Primality Tests Leveraging Explicit and Implicit Non-Residues

## A Set of Three Companion Manuscripts

**PART/Article 3 :** **Analytic Proofs of** Baseline Primality Conjecture **for Special Cases**


**Dhananjay Phatak**    (phatak@umbc.edu)

and

Alan T. Sherman   and   Steven D. Houston   and   Andrew Henry
(CSEE Dept. UMBC, 1000 Hilltop Circle, Baltimore, MD 21250, U.S.A.)


First identification of the Baseline Primality Conjecture @ $\approx 15^{th}$ March 2018

First identification of the Generalized Primality Conjecture @ $\approx 10^{th}$ June 2019

Last document revision date (time-stamp) = August 19, 2019



# § **Section 28 :** **A Road Map of the Proofs**

Before diving into the details, we re-state the conditions required by the conjecture :

**C–1 :**     The NUT  $N$  is a positive odd number and is not a perfect square of any integer.

**C–2 :**     `Jacobi_Symbol`$(q, N) = -1$  $\Rightarrow$  $q$  is a  `QNR`  modulo-$N$.

**C–3 :**     There is only one single exclusion : $q \not\equiv -1 \mod N$. Any other integer in the interval $[2, N-2]$ which is a  `QNR`  w.r.t. $N$ works.

**C–4 :**     $q$ satisfies the Euler Criterion, i.e.,

$$q^{\left(\frac{N-1}{2}\right)} \mod N = \text{Jacobi\_Symbol}(q, N) = -1 \mod N$$

**C–5 :**     $q$ also satisfies the Modular Binomial Congruence :

$$(1 + \sqrt{q})^N \mod N = [1 + (\sqrt{q})^N] \mod N$$

**The claim is that if all of the above conditions are satisfied then $N$ must be a prime number, it cannot be a composite.**

**As repeatedly mentioned (in prior sections) we have been able to prove the claim in some critical special cases. The method of demonstration is "proof by contradiction"**

**In other words, in addition to all the above conditions**

**we assume the following : $N$ is composite**

**and show that that assumption/hypothesis together with conditions  C–1  thru  C–5  leads to contradictions.**



## § Section 29 : **Basic Results heavily used in the Proofs**

**Lemma 1 (Conjugation Lemma) :** For any positive integers $L$ and $N$

**If**  $(a + b\sqrt{q})^L \mod N \;=\; (x + y\sqrt{q}) \mod N$  **where**

$\qquad\qquad\qquad\qquad = \; a, b, x, y$  **are integers modulo-**$N$  **so that**  $0 \le \{a, b, x, y\} \le N-1$

$\qquad\qquad\qquad\qquad$ **and**  $q$  **is an integer which is Quadratic Non Residue ( QNR ) w.r.t.**  $N$

$\qquad$ **Then**  $(a - b\sqrt{q})^L \;=\; (x - y\sqrt{q}) \mod N$

**Proof :** Straightforward : simply expand the left-hand sides using the Binomial Theorem and collect even and odd terms: Let

$$(a + b\sqrt{q})^L \mod N \;=\; \big[ a^L + (^L C_1)\, a^{(L-1)} \cdot \sqrt{q} + (^L C_2)\, a^{(L-2)} \cdot q + (^L C_3)\, a^{(L-3)} \cdot q\sqrt{q} + \cdots$$
$$+ (^L C_{L-1})\, a \cdot [\sqrt{q}]^{(L-1)} + [\sqrt{q}]^L \big] \mod N$$
$$= \; (x + y\sqrt{q}) \mod N \qquad \text{where}$$
$$x \;=\; [a^L + (^L C_2)\, a^{(L-2)} \cdot q + (^L C_4)\, a^{(L-4)} \cdot q^2 + \cdots] \mod N \qquad \text{and}$$
$$y \;=\; [(^L C_1)\, a^{(L-1)} + (^L C_3)\, a^{(L-3)} \cdot q + (^L C_5)\, a^{(L-5)} \cdot q^2 + \cdots] \mod N$$

$\qquad\qquad\qquad$ then it follows that

$$(a - b\sqrt{q})^L \mod N \;=\; a^L - (^L C_1)\, a^{(L-1)}\sqrt{q} + (^L C_2)\, a^{(L-2)} q - (^L C_3)\, a^{(L-3)} q\sqrt{q} + \cdots$$
$$+ (^L C_{L-1}) a \cdot [-\sqrt{q}]^{(L-1)} + [-\sqrt{q}]^L$$
$$= \; \big[ a^L + (^L C_2) a^{(L-2)} q + \cdots \big] - \big[ (^L C_1) a^{(L-1)} + (^L C_3)\, a^{(L-3)} q + \cdots \big]\sqrt{q}$$
$$= \; (x - y\sqrt{q}) \mod N \qquad \square$$

This result is robust and holds independent of whether $L$ is even or odd; and also irrespective of whether $N$ is even or odd. Likewise, the result holds independent of whether $L = N$ or $L$ is unrelated to $N$.

**Lemma 2 (Conjugation Corollary) :** For integers $a, b, q, L$ and $N$ (satisfying the same conditions as in Lemma 1)

**If**  $(a + b\sqrt{q})^L \mod N = [(a^L + (b\sqrt{q})^L] \mod N$

$\qquad\qquad\qquad$ **i.e.,** $(a + b\sqrt{q})$ **satisfies the Modular Binomial Congruence**

**Then** $\qquad\qquad\qquad (a - b\sqrt{q})$ **also satisfies the Binomial Congruence, so that**

$\qquad (a - b\sqrt{q})^L \mod N = \Big[ \big( a^L + \{ (-b) \cdot \sqrt{q} \} \big)^L \Big] \mod N$



**Proof :**  Immediately follows from **Lemma 1**      □

**Lemma 3 :**

    **If**  $q$  **is a**  `QNR`  **w.r.t.**  $N$      **Then**  $q^{(2i+1)} \neq 1 \mod N$

    **in other words, no "odd" power of** $q$ **equals 1 modulo-**$N$

**Proof :**

        If  $q^{(2i+1)} = 1 \mod N$  then

$q \;=\; \dfrac{1}{(q^i)^2} \mod N$

$\;=\; x^2 \mod N$   where ;   $x = \left[ \left( \dfrac{1}{q^i} \right) \mod N \right]$

    which contradicts the hypothesis that   $q$   is a  `QNR`

    $\because$ the definition of a  `QNR`   requires that  $\nexists$  an integer  $x$  that satisfies  $x^2 = q \mod N$      □

Next, we state a few more basic facts/results (that are well known).

**Lemma 4 :**  Let $\mathcal{D} \neq 1$  be any integer divisor of a composite integer  $N$.
Then;  for any expression $\mathcal{E}$ (including integer coefficients and algebraic integers treated as symbols);

$\mathcal{E} \mod \mathcal{D} = (\mathcal{E} \mod N) \mod \mathcal{D}$         (189)

**Proof :**

Let  $N = k \times \mathcal{D}$   for some integer  $k$
Then  $\mathcal{E} = \mathcal{Q} \times N + \mathcal{R}$    $=$    $\mathcal{Q} \times (k \times \mathcal{D}) + \mathcal{R}$         (190)
taking the remainder of the preceding Equation  w.r.t.  $N$ yields
$\mathcal{E} \mod N = \mathcal{R}$         (191)
taking the remainder (of that same Equation/Relation) w.r.t. $\mathcal{D}$ yields
$\mathcal{E} \mod \mathcal{D} = \mathcal{R} \mod \mathcal{D} = [\mathcal{E} \mod N] \mod \mathcal{D}$   □         (192)



Even though it is obvious, we list following special cases of the previous lemma as separate basic lemmas in their own right without proofs (the proofs are trivial).

**Lemma 5 :** For any expression $\mathcal{E}$ (including integer coefficients and algebraic integers treated as symbols)

**If** $\mathcal{E} \mod N = r$ ; where $r$ is a modulo-N integer ; **Then**
$$\mathcal{E} \mod \mathcal{D} = (r \mod \mathcal{D})$$
$$= r \quad \text{if } |r| < \mathcal{D} \ ; \ \text{where} \ ; \ |r| \text{ denotes the absolute value of } r$$

**Lemma 6 :** In particular,

**If** $\mathcal{E} \mod N = \pm 1 \mod N$                  (193)

**Then** $\mathcal{E} \mod \mathcal{D} = \pm 1 \mod \mathcal{D}$            (194)

**Lemma 7 : If** $N$ **is a prime number then the Modular Binomial Congruence is satisfied for arbitrary algebraic integers, integers or an arbitrary mix from the two categories (regular integers and algebraic integers)**

**Proof :** Let $\mathcal{X}$ and $\mathcal{Y}$ be algebraic integers. Then the Binomial Theorem gives

$$(\mathcal{X} + \mathcal{Y})^N \ = \ \left[ \mathcal{X}^N + (^N\boldsymbol{C}_1)\, \mathcal{X}^{(N-1)} \cdot \mathcal{Y} + (^N\boldsymbol{C}_2)\, \mathcal{X}^{(N-2)} \cdot \mathcal{Y}^2 + \cdots + \right.$$
$$\left. + (^N\boldsymbol{C}_{N-1})\, \mathcal{X} \cdot \mathcal{Y}^{N-1} + \mathcal{Y}^N \right] \tag{195}$$

As long as the product of the entities $\mathcal{X}$ and $\mathcal{Y}$ is well defined and commutative, the above expansion for positive integer exponents $N$ holds.

Since $(^N\boldsymbol{C}_r) = \dfrac{N \cdot (N-1) \cdots (N-r+1)}{1 \cdot 2 \cdots r}$                 (196)

Therefore, if $N$ is a prime number, then all the non-trivial Binomial coefficients viz.

$(^N\boldsymbol{C}_1), \cdots, (^N\boldsymbol{C}_{N-1})$       evaluate to 0 modulo-N $\Rightarrow$

**if** $N$ **is a prime number, then** $(\mathcal{X} + \mathcal{Y})^N \mod N = [\mathcal{X}^N + \mathcal{Y}^N] \mod N$    $\square$     (197)



To conclude this section on background results, we state a couple of well known results from elementary number theory.

**Lemma 8-a (Even–8 Lemma) :**  The product of an **even** number of distinct or identical factors; each of which is of the form $(8k_i + r)$ for some even number of integers $k_i = k_{i_1}, k_{i_2}, k_{i_3}, \cdots, k_{i_{(2t)}}$ ; and $r = 1$ or $r = 3$ or $r = 5$ or $r = 7$ ; is an integer of the form $(8m + 1)$ for some integer $m$.

**Proof** : Note that for arbitrary (positive) integers $i$ and $j$ ;

$$
\begin{aligned}
(8i + r) \times (8j + r) \mod 8 \quad &\equiv \quad [64 \cdot (i \cdot j) + 8r(i + j) + r^2] \mod 8 \\
&\equiv \quad r^2 \mod 8 \\
&\equiv \quad 1 \quad \text{for } r \in \{1, 3, 5, 7\} \qquad \square
\end{aligned}
\tag{198}
$$

**Lemma 8-b (Odd–8 Lemma) :**  The product of an **odd** number of distinct or identical factors; each of which is of the form $(8k_i + r)$ for some odd number of integers $k_i = k_{i_1}, k_{i_2}, k_{i_3}, \cdots, k_{i_{(2t+1)}}$ ; and $r = 1$ or $r = 3$ or $r = 5$ or $r = 7$ ; is integer of the same form $(8m + r)$ for some integer $m$.

**Proof** : Split the product into two terms : select any one of the factors by itself as the first term; and let the product of all the remaining factors be the second term. Then, by Lemma 8-a the second factor must be of the form $(8m + 1)$. Then, note that for arbitrary (positive) integers $k$ and $m$ ;

$$
\begin{aligned}
(8k + r) \times (8m + 1) \mod 8 \quad &\equiv \quad [64 \times (k \cdot m) + 8 \times (m \cdot r + k) + r] \mod 8 \\
&\equiv \quad r \mod 8 \qquad \square
\end{aligned}
\tag{199}
$$



**§ Section 30 :**     **Step 1 : Show the existence of at least one canonical**
                              **prime divisor** $\mathcal{P}$ **with specific attributes**

Let the prime factorization of the NUT $N$ be[34]

$$N = F_1 \times F_2 \times \cdots \times F_t \tag{200}$$

$$= (P_1)^{e_1} \times (P_2)^{e_2} \times \cdots \times (P_t)^{e_t} \tag{201}$$

$$\text{where} ; \quad P_1, P_2, \cdots, P_t \quad \text{are the distinct prime divisors of } N ; \text{ and}$$

$$F_i = (P_i)^{e_i} \tag{202}$$

$$\text{and therefore}$$

$$\gcd(F_i, F_j) = 1 \quad \text{for} \quad i \neq j \tag{203}$$

Then, the Jacobi-Symbol of any integer $q$ w.r.t. $N$; (denoted as `Jacobi_Symbol(`$q, N$`)`) satisfies

$$\texttt{Jacobi\_Symbol(}q, N\texttt{)} = \big[\texttt{Jacobi\_Symbol(}q, P_1\texttt{)}\big]^{e_1} \times \big[\texttt{Jacobi\_Symbol(}q, P_2\texttt{)}\big]^{e_2} \times \cdots$$
$$\times \big[\texttt{Jacobi\_Symbol(}q, P_t\texttt{)}\big]^{e_t} \tag{204}$$

$$\text{wherein} ;$$

$$\texttt{Jacobi\_Symbol(}q, P_i\texttt{)} = \text{the Jacobi–Symbol of } q \text{ w.r.t. the prime divisor } P_i \text{ for } i = 1, 2, \cdots t$$

$$\text{and}$$

$$\texttt{Jacobi\_Symbol(}q, P_i\texttt{)} = \pm 1$$

As a result,

**if** `Jacobi_Symbol(`$q, N$`)` $= -1$ **then**

**at least one of the component Jacobi-Symbols in Eqn. (204); say** $\big[\texttt{Jacobi\_Symbol(}q, P_j\texttt{)}\big] = -1$
$$\tag{205}$$

**AND the corresponding exponent** $e_j$ **must be an odd number** $\tag{206}$

To simplify the notation; we denote that "at least one prime $= P_j$" (whose jacobi-symbol w.r.t. $q = -1$) with the letter "$\mathcal{P}$"

$$P_j \;\stackrel{\triangle}{=}\; \mathcal{P} \tag{207}$$

; and ; its exponent in the prime-factorization of $N$; viz.; $e_j$ ; by the letter "$\theta$"

$$e_j \;\stackrel{\triangle}{=}\; \theta \tag{208}$$

Note that the subscript small-case "$j$" in Eqn. (205) is just a dummy index;
it should not be (mistakenly) construed to tag or indicate the $j$-th prime number.

---

[34] subsumes the case where $N$ is an odd power of a single prime base, where $P_k = F_k = 1$ for all $k \neq j$; and; $F_j = (P_j)^{e_j}$ for the single prime divisor $P_j$



Likewise, even though the list of prime divisors $\left\{ P_{j_1}, P_{j_2}, \cdots, P_{j_{(2m+1)}} \right\}$ of the NUT $N$ ;
whose Jacobi-Symbols w.r.t. $q$ evaluate to $-1$ ;
can assumed to be sorted in the ascending order of magnitude; the sub-script $j$ does not refer to $j$-th element of such a sorted list.

Rather, $P_j$ is just any one of the prime factors whose Jacobi-Symbols w.r.t. $q = -1$

Likewise, the choice to denote $P_j$ with $\mathcal{P}$ completely arbitrary; it is merely for notational convenience; an attempt to minimize super and sub-scripts if not necessary.

In essence, we arbitrarily pick any one of those factors **whose Jacobi-Symbols w.r.t. $q$ evaluate to $-1$;** to be the "at least one" canonical factor that must satisfy a substantial number of relations we derive throughout the rest of this document.

Thus far, we have demonstrated that

If $q$ has Jacobi_Symbol $= -1$ w.r.t. a composite number $N$ then

$\qquad \exists$ at least one prime $\mathcal{P}$ that stratifies the conditions

$\texttt{Jacobi\_Symbol}(q, \mathcal{P}) = -1 \Rightarrow q$ must be a $\texttt{QNR}$ w.r.t. $\mathcal{P}$ $\qquad\qquad$ (209)

$\qquad$ AND

$\theta \overset{\triangle}{=}$ the highest power of $\mathcal{P}$ that divides $N$ satisfies the following constraint :

$\theta = (2l + 1)$ for some integer $l$ ; **i.e., $\theta$ is an odd number** $\qquad\qquad$ (210)

Since $N$ is odd; and the prime number $\mathcal{P}$ divides $N$; it follows that

$\mathcal{P} > 2$ $\qquad\qquad$ (211)

Also, the Jacobi-Symbol of any integer w.r.t. a prime number equals its Legendre-Symbol w.r.t. that prime.

Therefore Eqn. (209) $\qquad \Rightarrow \qquad q^{\left(\frac{\mathcal{P}-1}{2}\right)} = -1 \mod \mathcal{P}$ $\qquad\qquad$ (212)

Let $\mathcal{M}$ denote the product of all factors of $N$ that are relatively co-prime w.r.t. $\mathcal{P}$ ; so that

$N = \mathcal{M} \times (\mathcal{P})^\theta$ $\qquad\qquad$ (213)

$\qquad\qquad\qquad$ AND

$\gcd(\mathcal{M}, \mathcal{P}) = 1$ $\qquad\qquad$ (214)

Eqn. (213) yields the following important relation :

for any integer $a$

$\left. \begin{array}{ll} \texttt{Jacobi\_Symbol}(a, N) & = \texttt{Jacobi\_Symbol}(a, \mathcal{M}) \times \left[\texttt{Jacobi\_Symbol}(a, \mathcal{P})\right]^\theta \\ & = \texttt{Jacobi\_Symbol}(a, \mathcal{M}) \times \texttt{Jacobi\_Symbol}(a, \mathcal{P}) \end{array} \right\}$ $\quad$ (215)

$\because$ $\quad$ Jacobi-Symbol $\in \{-1, 0, 1\}$ and $\theta$ is odd



More generally, it is well-known that

**if** `Jacobi_Symbol(`$q, N$`) = -1` **then**

$\langle$ bkg–1 $\rangle$ Multiple distinct factors can have their individual Jacobi-Symbols = $-1$

 subject to the constraints

$\langle$ bkg–2 $\rangle$ the total number of factors that have their Jacobi-Symbols = $-1$   must be an odd number

and

$\langle$ bkg–3 $\rangle$ the exponent of the prime-base of each of those factors must also be an odd number

## § Section 31 :  **Step 2 :  Building a Repertoir of tools**

### § Section 31.1 :  **Tool Set 1 :  Canonical Relations modulo** $\mathcal{P}$

In  Eqn. (15)  note that

$$(\sqrt{q})^N = q^{\left(\frac{N}{2}\right)} = q^{\left(\frac{N-1}{2}\right)+\frac{1}{2}} = q^{\left(\frac{N-1}{2}\right)} \cdot \sqrt{q} = (-1) \cdot \sqrt{q} \tag{216}$$

Where the "$-1$" in the very last (right-most) expression in the preceding equality-chain results from the fact that  $q$ is a QNR w.r.t.  $N$  (see Condition **C–3**  and  Eqn. (14) ).

Substitute from the above equation into  Eqn. (15)  to obtain

**Fundamental Condition 1 :** $(1+\sqrt{q})^N \mod N = [1+(\sqrt{q})^N] \mod N$

$$= (1-\sqrt{q}) \mod N \tag{217}$$

It therefore follows (from the  **conjugation lemma**   and the prior relations shown above) that

$$(1-\sqrt{q})^N \mod N = (1+\sqrt{q}) \mod N \tag{218}$$

Multiply (the left-hand-sides as well as the right-hand-sides of the) preceding two equations to obtain

$(1+\sqrt{q})^N \cdot (1-\sqrt{q})^N = (1-\sqrt{q}) \cdot (1+\sqrt{q}) \mod N \Rightarrow$

$(1-q)^N = (1-q) \Rightarrow$

$$(1-q)^{N-1} = 1 \mod N \tag{219}$$

Next, multiply (both sides of)  Eqn. (217)  by  $(1+\sqrt{q})$  to obtain

$$(1+\sqrt{q})^{N+1} = (1-\sqrt{q}) \cdot (1+\sqrt{q}) = (1-q) \mod N \tag{220}$$

Likewise, multiply (both sides of)  Eqn. (218) by  $(1-\sqrt{q})$  to obtain

$$(1-\sqrt{q})^{N+1} = (1+\sqrt{q}) \cdot (1-\sqrt{q}) = (1-q) \mod N \tag{221}$$



Note that multiplying the preceding two equations yields

$$(1-q)^{N+1} = (1-q)^2 \quad \Rightarrow \quad (1-q)^{N-1} = 1 \mod N$$

which is the same as Eqn. (219) before.

Next, using Lemma 4 ; take the remainder of both sides of Eqn. (217) w.r.t. $\mathcal{P}$ to obtain

$$[(1+\sqrt{q})^N \mod N] \mod \mathcal{P} = [(1-\sqrt{q}) \mod N] \mod \mathcal{P} \Rightarrow$$

$$(1+\sqrt{q})^N \mod \mathcal{P} = (1-\sqrt{q}) \mod \mathcal{P} \tag{222}$$

In an analogous manner, take the remainder of both sides of Eqn. (218) w.r.t. $\mathcal{P}$ to obtain

$$[(1-\sqrt{q})^N \mod N] \mod \mathcal{P} = [(1+\sqrt{q}) \mod N] \mod \mathcal{P} \Rightarrow$$

$$(1-\sqrt{q})^N \mod \mathcal{P} = (1+\sqrt{q}) \mod \mathcal{P} \tag{223}$$

Now, since $\mathcal{P}$ is a prime, and `Jacobi_Symbol(`$q,\mathcal{P}$`)` $= -1$; then; analogous to the derivations of Eqn. (217) and Eqn. (218) , we obtain

$$(1+\sqrt{q})^{\mathcal{P}} \mod \mathcal{P} = (1-\sqrt{q}) \mod \mathcal{P} \tag{224}$$

and

$$(1-\sqrt{q})^{\mathcal{P}} \mod \mathcal{P} = (1+\sqrt{q}) \mod \mathcal{P} \tag{225}$$

From the previous two Equations, it is clear that

$$
\begin{aligned}
(1+\sqrt{q})^{\mathcal{P}^2} &= \left[(1+\sqrt{q})^{\mathcal{P}}\right]^{\mathcal{P}} = [1-\sqrt{q}]^{\mathcal{P}} \\
&= (1+\sqrt{q}) \mod \mathcal{P}
\end{aligned} \tag{226}
$$

so that

$$(1+\sqrt{q})^{\mathcal{P}^3} = (1-\sqrt{q}) \mod \mathcal{P} \cdots \text{ and so on ;} \tag{227}$$

Or, in other words

$$(1+\sqrt{q})^{(\text{even\_ power\_ of\_}\mathcal{P})} = (1+\sqrt{q}) \mod \mathcal{P} \tag{228}$$

and

$$(1+\sqrt{q})^{(\text{odd\_ power\_ of\_}\mathcal{P})} = (1-\sqrt{q}) \mod \mathcal{P} \tag{229}$$

For conciseness; let

$d$ denote any positive odd integer ; and $\tag{230}$

$e$ denote any positive even integer $\tag{231}$



With these symbols, Eqn. (228) and Eqn. (229) can be re-stated; respectively; as

$$\left[1+\sqrt{q}\right]^{\left([\mathcal{P}]^{\,e}\right)} \quad = \quad \left(1+\sqrt{q}\right) \mod \mathcal{P} \tag{232}$$

and

$$\left[1+\sqrt{q}\right]^{\left([\mathcal{P}]^{\,d}\right)} \quad = \quad \left(1-\sqrt{q}\right) \mod \mathcal{P} \tag{233}$$

In an exactly analogous manner, the corresponding "conjugate" relations are

$$\left[1-\sqrt{q}\right]^{\left([\mathcal{P}]^{\,e}\right)} \quad = \quad \left(1-\sqrt{q}\right) \mod \mathcal{P} \tag{234}$$

and

$$\left[1-\sqrt{q}\right]^{\left([\mathcal{P}]^{\,d}\right)} \quad = \quad \left(1+\sqrt{q}\right) \mod \mathcal{P} \tag{235}$$

Eqn. (233) and Eqn. (235) can be more concisely stated as the following single equation

$$\left[1\pm\sqrt{q}\right]^{\left([\mathcal{P}]^{\,d}\right)} = \left(1\mp\sqrt{q}\right) \mod \mathcal{P} \tag{236}$$

Finally, multiplying the previous Eqn. by $\left(1\pm\sqrt{q}\right)$ yields;

$$\left[1+\sqrt{q}\right]^{\left(1+[\mathcal{P}]^{\,d}\right)} \quad = \quad \left(1-q\right) \mod \mathcal{P} \tag{237}$$

and

$$\left[1-\sqrt{q}\right]^{\left(1+[\mathcal{P}]^{\,d}\right)} \quad = \quad \left(1-q\right) \mod \mathcal{P} \tag{238}$$

$$\Rightarrow$$

$$\left(1-q\right)^{\left(1+[\mathcal{P}]^{\,d}\right)} \quad = \quad \left(1-q\right)^2 \mod \mathcal{P} \tag{239}$$

$$\Rightarrow$$

$$\left(1-q\right)^{\left(\left\{[\mathcal{P}]^{\,d}\right\}-1\right)} \quad = \quad 1 \mod \mathcal{P} \tag{240}$$

Substitute $N=\mathcal{M}\cdot(\mathcal{P})^{\theta}$ in Eqn. (217) to obtain

$$\left(1+\sqrt{q}\right)^{(\mathcal{M}\cdot\mathcal{P}^{\theta})} = \left(1-\sqrt{q}\right) \mod \mathcal{P} \quad \Rightarrow$$

$$\left(\left[\left[\left(1+\sqrt{q}\right)^{\mathcal{P}}\right]^{\mathcal{P}}\right]^{\cdot^{\cdot^{\cdot}}}\right)^{\mathcal{M}} = \left(1-\sqrt{q}\right) \mod \mathcal{P} \tag{241}$$

Note that as per Eqn. (210), $\theta$ is an odd number. Therefore

$$\left(1+\sqrt{q}\right)^{(\mathcal{P}^{\theta})} = \left(\left[\left[\left(1+\sqrt{q}\right)^{\mathcal{P}}\right]^{\mathcal{P}}\right]^{\cdot^{\cdot^{\cdot}}}\right) = \left(1-\sqrt{q}\right) \mod \mathcal{P} \tag{242}$$



Substitute the immediate previous relation into the left-hand-side of Eqn. (241) to obtain

$$(1 - \sqrt{q})^{\mathcal{M}} = (1 - \sqrt{q}) \quad \Rightarrow$$
$$[1 - \sqrt{q}]^{(\mathcal{M}-1)} = 1 \mod \mathcal{P} \tag{243}$$

The analogous conjugate relation is the following :

$$[1 + \sqrt{q}]^{(\mathcal{M}-1)} = 1 \mod \mathcal{P} \tag{244}$$

Multiplying the preceding two relations yields

$$[(1 - \sqrt{q}) \cdot (1 + \sqrt{q})]^{(\mathcal{M}-1)} = 1 \mod \mathcal{P} \qquad \text{or after simplifying}$$
$$[1 - q]^{(\mathcal{M}-1)} = 1 \mod \mathcal{P} \tag{245}$$

**However, since $\mathcal{M}$ is an odd divisor of $N$; then $(\mathcal{M}-1)$ is an even number.**

**It turns out that an overwhelming majority of the contradictions (that we are looking for) arise when chain of successive square-roots of $[(1 + \sqrt{q})]^{(\mathcal{M}-1)}$ and chain of successive square-roots of $[(1 + \sqrt{q})]^{(N+1)}$ are analyzed.**
**Therefore, we call/label/designate each of these chains of square-roots to be the "Meta-Path" rooted at it's starting value.**





**Square roots in Meta–path I : the $\mathcal{M}$-path**

Figure 8 illustrates the square roots that can occur in meta–path I; which starts at the exponent = $(\mathcal{M} - 1)$

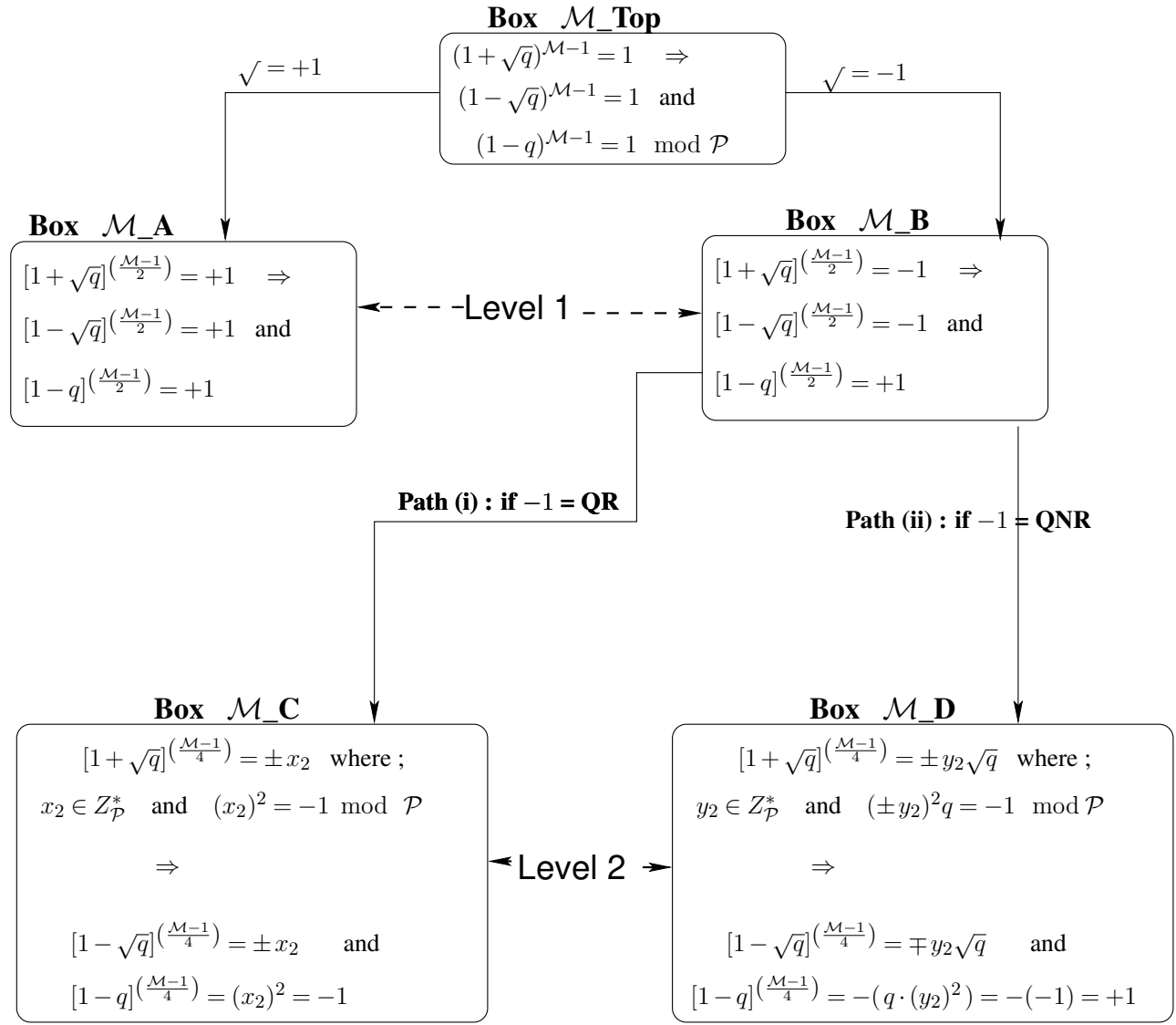

**Figure 8.** Illustration of the paths feasible in the first two levels of square roots of $[1 + \sqrt{q}]^{(\mathcal{M}-1)}$



Next we explain the derivations of the relations in each of the boxes (Box $\mathcal{M}$_A, Box $\mathcal{M}$_B, Box $\mathcal{M}$_C and Box $\mathcal{M}$_D);  in Figure  8 .

Before proceeding; we would like to point out that not all levels can occur in all paths. The number of successive square roots that need to be considered can be as high as the largest power of 2 that divides the canonical exponents $(\mathcal{M}-1)$,    $(N+1)$,  and  $(\mathcal{P}+1)$.

**Level 1 square roots in the $\mathcal{M}$_meta_path**

Since $(\mathcal{M}-1)$ is even ;  $\left(\dfrac{\mathcal{M}-1}{2}\right)$   is an integer.     so let  $I_1 = \dfrac{(\mathcal{M}-1)}{2}$ (246)

   and

Let  $(1+\sqrt{q})^{I_1} = (x_1 + y_1\sqrt{q})  \mod  \mathcal{P}$ (247)

Then from Lemma 1 it follows that

$(1-\sqrt{q})^{I_1} = (x_1 - y_1\sqrt{q})  \mod  \mathcal{P}$ (248)

Square both sides of  Eqn. (247)  and  Eqn. (248)  to obtain, respectively;

$\left[(1+\sqrt{q})^{I_1}\right]^2 = (x_1 + y_1\sqrt{q})^2 = 1  \mod  \mathcal{P}$ (249)

   and

$\left[(1-\sqrt{q})^{I_1}\right]^2 = (x_1 - y_1\sqrt{q})^2 = 1  \mod  \mathcal{P}$ (250)

Then, subtract  Eqn. (250)  from  Eqn. (249)  to get

$(x_1 + y_1\sqrt{q})^2 - (x_1 - y_1\sqrt{q})^2 = 0  \mod  \mathcal{P}$ (251)

Using the difference of squares forumlae

$a^2 - b^2 = (a+b)\cdot(a-b)$ (252)

   in  Eqn. (251)  yields

$(2x_1)\cdot(2y_1\sqrt{q}) = 0  \mod  \mathcal{P}$ (253)

Since $\mathcal{P}$ is a prime, the preceding equation leads to 3 cases to consider

**Case 0 :** $x_1 = 0  \mod  \mathcal{P}$  AND  $y_1 = 0  \mod  \mathcal{P}$ (254)

   OR

**Case 1 :** $x_1 = 0  \mod  \mathcal{P}$  AND  $y_1 \neq 0  \mod  \mathcal{P}$ (255)

   OR

**Case 2 :** $x_1 \neq 0  \mod  \mathcal{P}$  AND  $y_1 = 0  \mod  \mathcal{P}$ (256)

Next we show that 2 out of the 3 cases above lead to contradictions; leaving only one feasible path.



**Case 0 :  Leads to immediate contradiction**

If we plug in $x_1 = 0$  AND  $y_1 = 0$ in  Eqn. (249) , we obtain

$$0^2 = 0 = 1 \mod \mathcal{P}; \quad \textbf{which is absurd} \quad \square \tag{257}$$

Note that substitution in  Eqn. (250)  also yields the same contradiction.

**Case 1 :  Also leads to a contradiction**

If we plug in $x_1 = 0$ in  Eqn. (249)  or  Eqn. (250) , we obtain

$$(\pm y_1 \sqrt{q})^2 = 1 \mod \mathcal{P} \quad \Rightarrow (y_1)^2 \cdot q = 1 \quad \Rightarrow$$
$$q = \left(\frac{1}{y_1}\right)^2 \mod \mathcal{P}; \tag{258}$$

**the preceding Equation contradicts  Eqn. (209)  showing that  $q$  must be a QNR modulo-$\mathcal{P}$**  $\square$

**Case 2 :   This is the only path that survives (without yielding any contradictions) up to this point**

If we plug in $y_1 = 0$ in  Eqn. (249)  or  Eqn. (250) , we obtain

$$(x_1)^2 = 1 \mod \mathcal{P} \quad \Rightarrow$$
$$x_1 = \pm 1 \mod \mathcal{P} \tag{259}$$

Selecting the "+" sign in  Eqn. (259)  leads to box $\mathcal{M}$_A; (the left child of the top box in Figure  8 .); wherein ;

$$(1 + \sqrt{q})^{\left(\frac{\mathcal{M}-1}{2}\right)} = +1 \quad \text{and} \quad (1 - \sqrt{q})^{\left(\frac{\mathcal{M}-1}{2}\right)} = +1 \quad \Rightarrow \quad (1 - q)^{\left(\frac{\mathcal{M}-1}{2}\right)} = +1 \tag{260}$$

Likewise, selecting the "−" sign in  Eqn. (259)  leads to box $\mathcal{M}$_B; (the right child of the top box in Figure  8 .); wherein ;

$$(1 + \sqrt{q})^{\left(\frac{\mathcal{M}-1}{2}\right)} = -1 \quad \text{and} \quad (1 - \sqrt{q})^{\left(\frac{\mathcal{M}-1}{2}\right)} = -1 \quad \Rightarrow \quad (1 - q)^{\left(\frac{\mathcal{M}-1}{2}\right)} = +1 \tag{261}$$

**Level 2 square roots if $\left(\frac{\mathcal{M}-1}{2}\right)$ is even**

Note that the level 2 square-roots under Box $\mathcal{M}$_A can be derived simply by repeating the arguments in the derivation of all the Eqns in Box $\mathcal{M}$_A itself.  Therefore we do not show those paths in the Figure; and accordingly; the following analysis only considers the children of box $\mathcal{M}$_B (i.e. Boxes $\mathcal{M}$_C and $\mathcal{M}$_D in Figure  8 ).

If  $\left(\frac{\mathcal{M}-1}{2}\right)$  is an even number; then    let  $I_2 = \frac{(\mathcal{M}-1)}{4}$ \tag{262}

and

Let  $(1 + \sqrt{q})^{I_2} = (x_2 + y_2 \sqrt{q}) \mod \mathcal{P}$ \tag{263}

$\Rightarrow (x_2 + y_2 \sqrt{q})^2 = (1 + \sqrt{q})^{\left(\frac{\mathcal{M}-1}{2}\right)} = -1$ \tag{264}



Then from Lemma 1 it follows that

$$(1 - \sqrt{q})^{I_2} = (x_2 - y_2\sqrt{q}) \mod \mathcal{P} \quad \text{and} \tag{265}$$

$$(x_2 - y_2\sqrt{q})^2 = (1 - \sqrt{q})^{\left(\frac{\mathcal{M}-1}{2}\right)} = -1 \tag{266}$$

and then; analogous to the derivation of Level 1 square roots above; we obtain

$$(x_2 + y_2\sqrt{q})^2 - (x_2 - y_2\sqrt{q})^2 = [(-1) - (-1)] = 0 \mod \mathcal{P} \tag{267}$$

$$\Rightarrow \quad (2x_2) \cdot (2y_2\sqrt{q}) = 0 \mod \mathcal{P} \tag{268}$$

Since $\mathcal{P}$ is a prime, the preceding equation leads to 3 cases to consider (just as before);
However, if we assume that both $x_2 = 0 \mod \mathcal{P}$ AND $y_2 = 0 \mod \mathcal{P}$; then plugging these values into Eqn. (264) yields

      $0 = -1 \mod \mathcal{P}$ which is absurd; thereby terminating that path    □.

Therefore we are left with the other two cases to consider

**Case_1 :**     $x_2 \neq 0$    AND    $y_2 = 0$      which leads to Box $\mathcal{M}\_C$ in Figure 8 ;     (269)

     OR

**Case_2 :**     $x_2 = 0$    AND    $y_2 \neq 0$      which leads to Box $\mathcal{M}\_D$ in Figure 8 .     (270)

**Relations in Box $\mathcal{M}\_C$**

In this case, substitute $y_2 = 0$ in Eqn. (263) to obtain

$$(1 + \sqrt{q})^{\left(\frac{\mathcal{M}-1}{4}\right)} = x_2 \quad \Rightarrow \quad (x_2)^2 = -1 \tag{271}$$

The corresponding conjugate relations are

$$(1 - \sqrt{q})^{\left(\frac{\mathcal{M}-1}{4}\right)} = x_2 \quad \Rightarrow \quad (x_2)^2 = -1 \quad \Rightarrow \tag{272}$$

$$(1 - q)^{\left(\frac{\mathcal{M}-1}{4}\right)} = (x_2)^2 = -1 \mod \mathcal{P} \tag{273}$$

which completes the derivation of all Eqns. in Box $\mathcal{M}\_C$ in Figure 8 .    □

⋆⋆ **Note that this step/path that leads to Box $\mathcal{M}\_C$ is feasible only if "−1" is a Quadratic Residue (QR) w.r.t. $\mathcal{P}$ ⋆⋆**



**Analysis of Box $\mathcal{M}$_D**

In this case, substitute $x_2 = 0$ in Eqn. (263) to obtain

$$\left(1 + \sqrt{q}\right)^{\left(\frac{\mathcal{M}-1}{4}\right)} = y_2\sqrt{q} \quad \Rightarrow \quad \left(y_2\sqrt{q}\right)^2 = (y_2)^2 q = \left(1 + \sqrt{q}\right)^{\left(\frac{\mathcal{M}-1}{2}\right)} = -1 \tag{274}$$

The analogous conjugate is

$$\left(1 - \sqrt{q}\right)^{\left(\frac{\mathcal{M}-1}{4}\right)} = -y_2\sqrt{q} \quad \Rightarrow \quad \left(-y_2\sqrt{q}\right)^2 = (y_2)^2 q = \left(1 - \sqrt{q}\right)^{\left(\frac{\mathcal{M}-1}{2}\right)} = -1 \tag{275}$$

⋆⋆ **Note that Eqn. (274) ; as well as its conjugate Eqn. (275) ; both imply that**

$$(y_2)^2 q = -1 \quad \Rightarrow \tag{276}$$

$$\left.\begin{array}{rl} \texttt{Jacobi\_Symbol}(-1, \mathcal{P}) &= \texttt{Jacobi\_Symbol}((y_2)^2 q, \mathcal{P}) \\ &= \left(\texttt{Jacobi\_Symbol}(y_2, \mathcal{P})\right)^2 \cdot \texttt{Jacobi\_Symbol}(q, \mathcal{P}) \end{array}\right\} \tag{277}$$

**Since $q$ is QNR the preceding Eqn implies that**

$$\texttt{Jacobi\_Symbol}(-1, \mathcal{P}) = (\pm 1)^2 \cdot (-1) = -1 \tag{278}$$

**Or that "$-1$" must be a Quadratic Non-Residue (QNR) w.r.t. $\mathcal{P}$**

**It is therefore clear that the two possible paths (the children of Box $\mathcal{M}$_B) are mutually exclusive :**
**one is possible when $-1$ is a QR, the other is possible when $-1$ is a QNR.**

**This fact substantially mitigates the explosion of paths if they get longer/deeper (which happens if $(\mathcal{M}-1)$ is divisible by a large power of 2).**

Another interesting point to note before we move onto Level 3 is that multiplying Eqn. (274) and its conjugate Eqn. (275) yields

$$\left(1 - q\right)^{\left(\frac{\mathcal{M}-1}{4}\right)} = \left(y_2\sqrt{q}\right) \cdot \left(-y_2\sqrt{q}\right) = -((y_2)^2 q) = -(-1) = +1 \tag{279}$$

which completes the derivation of all Eqns. in Box $\mathcal{M}$_D in Figure 8 .  □

**Analysis of $\mathcal{M}$_meta_paths : Level 3 and beyond**

Next, we outline the analysis of only one more level; viz.; Level 3. (It turns out that the essential trends/features of the paths are seen by the 3-rd level; explicit derivations at deeper levels are not needed).

We would like to point out that Figure 9 does not include all possible paths. For example, the paths corresponding to $\left(1 + \sqrt{q}\right)^{\left(\frac{\mathcal{M}-1}{2}\right)} = x_2$ are shown; but those corresponding to $\left(1 + \sqrt{q}\right)^{\left(\frac{\mathcal{M}-1}{2}\right)} = -x_2$ (which is the other value of the square-root at the previous level) are not explicitly shown. The reason is that the derivations for those case are identical and therefore do not lead to new or useful insights. Accordingly, the ensuing analysis also does not re-do those derivations.



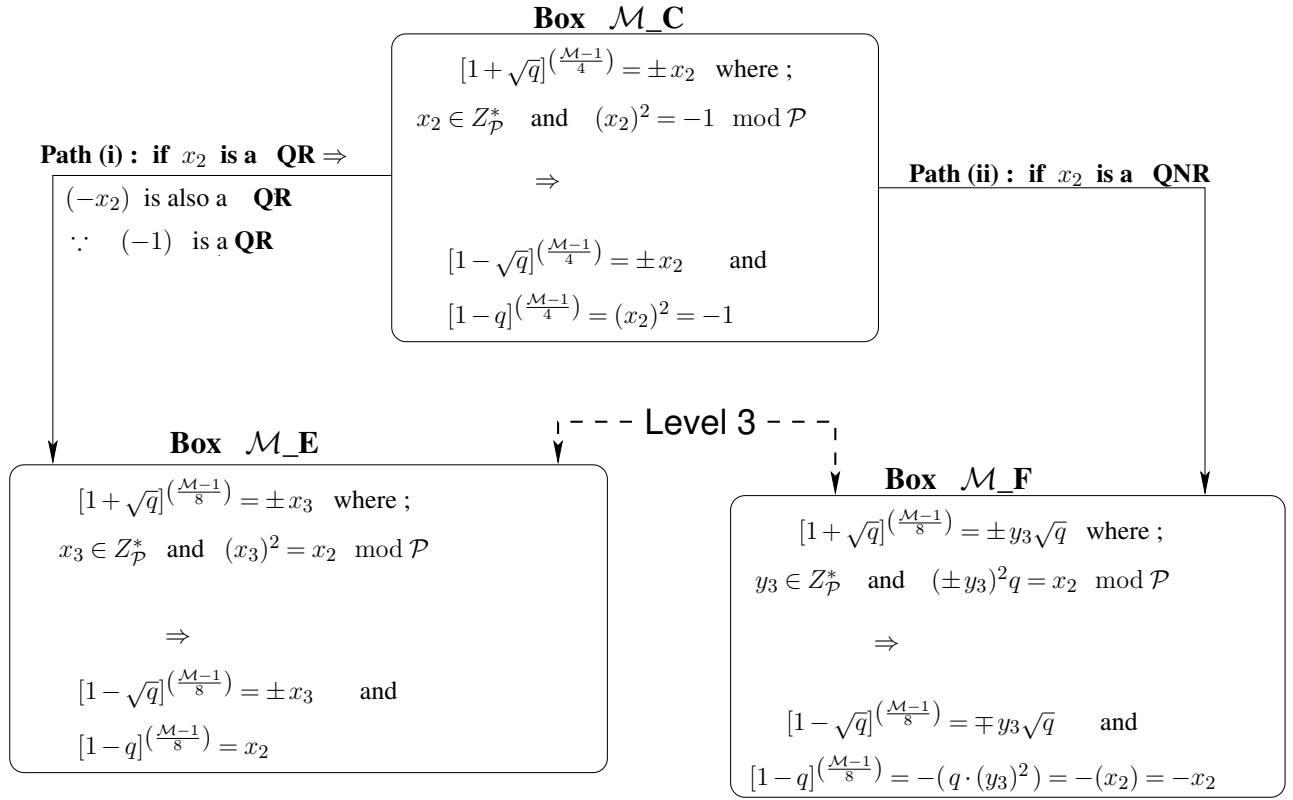

**Figure 9.** Illustration of the two types of paths feasible in the third level: under Box $\mathcal{M}\_C$

If $\left(\frac{\mathcal{M}-1}{4}\right)$ is an even number; then

$$\text{Let } I_3 = \frac{(\mathcal{M}-1)}{8} \tag{280}$$

and

$$\text{Let } (1+\sqrt{q})^{I_3} = (x_3 + y_3\sqrt{q}) \mod \mathcal{P} \tag{281}$$

$$\Rightarrow (x_3 + y_3\sqrt{q})^2 = (1+\sqrt{q})^{\left(\frac{\mathcal{M}-1}{4}\right)}$$

$$= x_2 \quad \text{Which leads to Box } \mathcal{M}\_E \text{ in Figure } \underline{9} . \tag{282}$$

$$\text{OR} \tag{283}$$

$$= y_2\sqrt{q} \quad \text{Which leads to Box } \mathcal{M}\_F \text{ in Figure } \underline{9} . \tag{284}$$

The corresponding conjugate relations are

$$(1-\sqrt{q})^{I_3} = (x_3 - y_3\sqrt{q}) \mod \mathcal{P} \tag{285}$$

$$\Rightarrow (x_3 - y_3\sqrt{q})^2 = (1+\sqrt{q})^{\left(\frac{\mathcal{M}-1}{4}\right)} = \pm 1 \quad \text{OR} \quad -y_2\sqrt{q} \tag{286}$$

**Relations in Box $\mathcal{M}\_E$**

Note that if $x_2$ in Box $\mathcal{M}\_C$ is a $\texttt{QR}$ w.r.t. $\mathcal{P}$; then $-x_2$ is also a $\texttt{QR}$ ; since $-1$ must be a $\texttt{QR}$ to reach box $\mathcal{M}\_C$ in the prior step. The Figure therefore indicates that $-x_2$ path will also fall under **case (i)**



(although that path not sketched in that Figure)

Likewise; since "$-1$" is a `QR` ;  (in this path) then

$x_2$ is a `QNR` $\Rightarrow$    $-x_2$  is also a `QNR` .

Having indicated how the paths for $-x_2$ evolve; further explicit relations for those paths are omitted for the sake of clarity and brevity.

Moreover, it is easy to verify that the derivations of all the relations in Box $\mathcal{M}\_E$ in Figure  9   are exactly analogous the derivations of relations in Box $\mathcal{M}\_A$.

**Relations in Box $\mathcal{M}\_F$**

These mirror the derivations of all the relations in Box $\mathcal{M}\_D$ in Figure  8 .



**Level 3 paths under Box D**

**Box  $\mathcal{M}\_D$**

$$[1 + \sqrt{q}]^{\left(\frac{\mathcal{M}-1}{4}\right)} = y_2\sqrt{q} \quad \text{where ;}$$

$$y_2 \in Z_{\mathcal{P}}^* \quad \text{and} \quad (y_2)^2 q = -1 \mod \mathcal{P}$$

$$\Rightarrow$$

$$[1 - \sqrt{q}]^{\left(\frac{\mathcal{M}-1}{4}\right)} = -y_2\sqrt{q} \quad \text{and}$$

$$[1 - q]^{\left(\frac{\mathcal{M}-1}{4}\right)} = -((y_2)^2 q) = -(-1) = +1$$

**Path type (iii) :  more general case; where;**
**the square also contains symbol(s)**

**Level 3**          **Box  $\mathcal{M}\_G$**

$$[1 + \sqrt{q}]^{\left(\frac{\mathcal{M}-1}{8}\right)} = (x_3 + y_3\sqrt{q}) \quad \text{where ;}$$

$$x_3 \text{ and } y_3 \in Z_{\mathcal{P}}^* \quad \text{and} \quad (x_3 + y_3\sqrt{q})^2 = y_2\sqrt{q}$$

$$\Rightarrow$$

$$[1 - \sqrt{q}]^{\left(\frac{\mathcal{M}-1}{8}\right)} = (x_3 - y_3\sqrt{q}) \quad \text{and}$$

$$[1 - q]^{\left(\frac{\mathcal{M}-1}{8}\right)} = (x_3)^2 - (y_3)^2 q$$

**Figure 10.** Illustration of the third type of path that occurs at the third level under Box $\mathcal{M}\_D$

Finally, we illustrate a more general type of path (which we call "**path type (iii)**" in Figure  10  . Note that this is the more general case; wherein the square (as well as the square-root) both contain the "symbol" (i.e., the "algebraic-integer" used in the symbolic computation).



Next, we derive some relations including the variables in Box $\mathcal{M}\_G$.

In this case the relations are

$$\left(1+\sqrt{q}\right)^{\left(\frac{\mathcal{M}-1}{8}\right)} = \left(x_3 + y_3\sqrt{q}\right) \Rightarrow \tag{287}$$

$$\left(x_3 + y_3\sqrt{q}\right)^2 = \left(1+\sqrt{q}\right)^{\left(\frac{\mathcal{M}-1}{4}\right)} = y_2\sqrt{q} \tag{288}$$

and the corresponding conjugates

$$\left(1-\sqrt{q}\right)^{\left(\frac{\mathcal{M}-1}{8}\right)} = \left(x_3 - y_3\sqrt{q}\right) \Rightarrow \tag{289}$$

$$\left(x_3 - y_3\sqrt{q}\right)^2 = \left(1-\sqrt{q}\right)^{\left(\frac{\mathcal{M}-1}{4}\right)} = -y_2\sqrt{q} \tag{290}$$

These in turn yield:

$$\left((x_3)^2 + (y_3)^2 q\right) = 0 \tag{291}$$

$$2(x_3 \cdot y_3) = y_2 \tag{292}$$

and

$$(1-q)^{\left(\frac{\mathcal{M}-1}{8}\right)} = (x_3)^2 - (y_3)^2 q \tag{293}$$

Multiplying Eqns ([288]) and ([290]) yields

$$\left[(x_3)^2 - (y_3)^2 q\right]^2 = -((y_2)^2 q) = +1 \mod \mathcal{P} \Rightarrow \tag{294}$$

$$\left[(x_3)^2 - (y_3)^2 q\right] = \pm 1 \mod \mathcal{P} \tag{295}$$

Subtracting Eqn.([295]) from ([291]) yields

$$2 \cdot (y_3)^2 q = \mp 1 \Rightarrow \tag{296}$$

$$y_3 = \sqrt{\frac{\mp 1}{(2 \cdot q)}} \tag{297}$$

Adding Eqn. ([295]) to ([291]) yields

$$2 \cdot (x_3)^2 = \pm 1 \Rightarrow \tag{298}$$

$$x_3 = \sqrt{\frac{\pm 1}{2}} \tag{299}$$



***Square roots in Meta–path II : the*** $(N+1)$***-path***

Figure 11 illustrates the square roots that can occur in meta–path II; which starts at the exponent = $(N+1)$

## Level 1 paths under Box N_Top

Next, we indicate how to establish the relations shown in each box in Figure 11 .

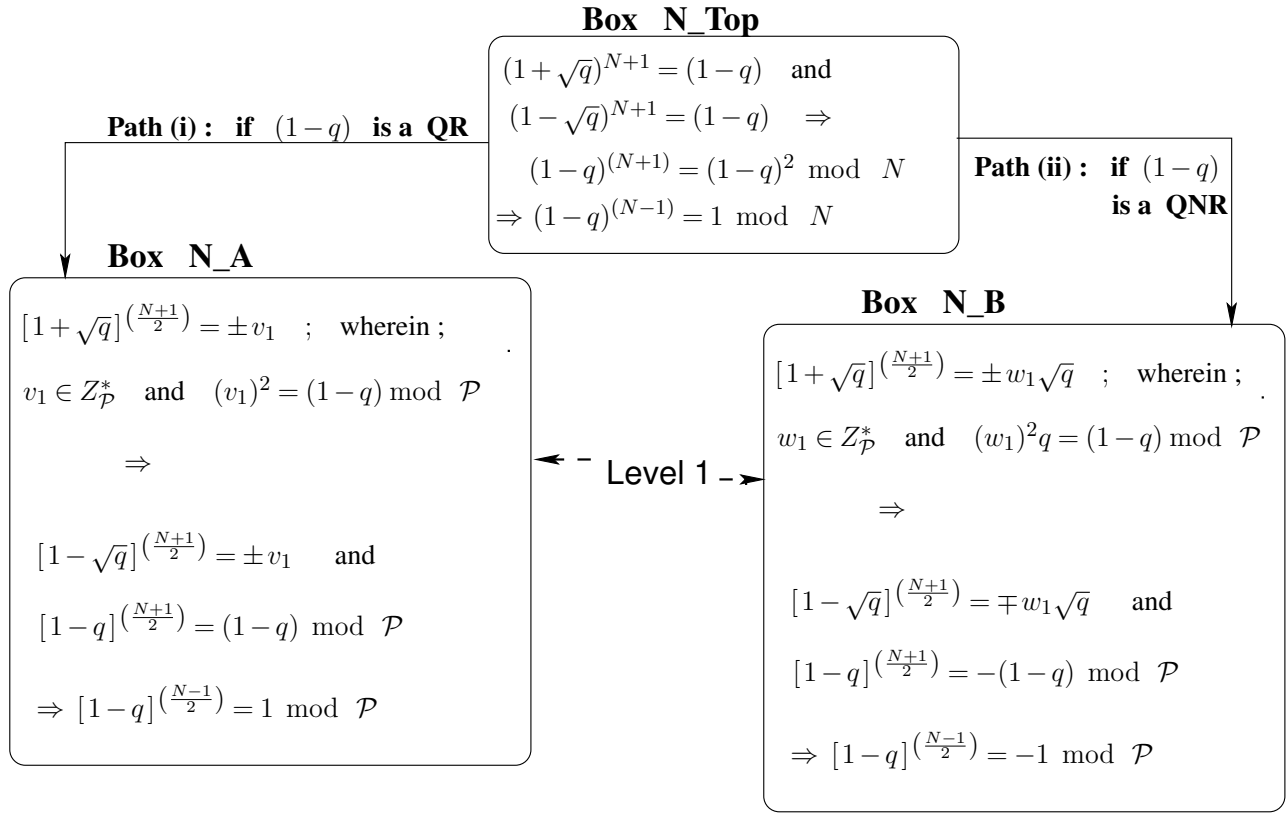

**Figure 11.** Illustration of the paths feasible at the first level of square roots of $[1+\sqrt{q}]^{(N+1)}$

## Box N_A

Since $(N+1)$ is even ; $\left(\dfrac{N+1}{2}\right)$ is an integer. so let $J_1 = \dfrac{(N+1)}{2}$     (300)

and

Let $(1+\sqrt{q})^{J_1} = (v_1 + w_1\sqrt{q}) \mod N$     (301)

Then from Lemma 1 it follows that

$(1-\sqrt{q})^{J_1} = (v_1 - w_1\sqrt{q}) \mod N$     (302)



Then, it is clear that

$$\left[(1+\sqrt{q})^{J_1}\right]^2 = (v_1 + w_1\sqrt{q})^2 = (1-q) \mod N \tag{303}$$

and

$$\left[(1-\sqrt{q})^{J_1}\right]^2 = (v_1 - w_1\sqrt{q})^2 = (1-q) \mod N \tag{304}$$

Subtract Eqn. (304) from Eqn. (303) to get

$$(v_1 + w_1\sqrt{q})^2 - (v_1 - w_1\sqrt{q})^2 = (1-q) - (1-q) = 0 \mod N \tag{305}$$

Then from Lemma 4 it follows that

$$(v_1 + w_1\sqrt{q})^2 - (v_1 - w_1\sqrt{q})^2 = (2v_1)\cdot(2w_1\sqrt{q}) = 0 \mod \mathcal{P} \tag{306}$$

This transition from the modulus $N$ to $\mathcal{P}$ is critical and required. (otherwise there will be an unnecessary increase in the number of paths depending upon which (hypothetical) factor of $N$ divides $v_1$ and/or $w_1$).

Since $\mathcal{P}$ is a prime, the preceding equation leads to 3 cases to consider (as seen before). Substitution of $v_1 = 0$ and $w_1 = 0$ in Eqn. (303) yields

$$0^2 = 0 = (1-q) \mod \mathcal{P} \quad \Rightarrow q = 1 \mod \mathcal{P} \tag{307}$$

$$\Rightarrow \quad q^3 = 1 \mod \mathcal{P} \text{ which contradicts Lemma 3} \quad \therefore \quad q \text{ is a } \mathbf{QNR} \qquad \square \tag{308}$$

Hence, there are only 2 cases to consider as shown in the Figure.

**Case 1 :** $v_1 \neq 0 \mod \mathcal{P}$ AND $w_1 = 0 \mod \mathcal{P}$ \hfill (309)

OR

**Case 2 :** $(v_1 = 0 \mod \mathcal{P}$ AND $w_1 \neq 0 \mod \mathcal{P})$ \hfill (310)

**Case 1 :** Plug in $w_1 = 0$ in Eqn. (303) yields

$$(v_1)^2 = (1-q) \mod \mathcal{P} \tag{311}$$

$$\Rightarrow (1-q) \text{ is a } \mathbf{QR} \text{ modulo-}\mathcal{P} \tag{312}$$

Thus, it is demonstrated that

If $(1-q)$ is a $\mathbf{QR}$ w.r.t. $\mathcal{P}$ ; \hfill (313)

then the square root takes **path (i)** ; wherein; \hfill (314)

$$(1+\sqrt{q})^{\left(\frac{N+1}{2}\right)} = \pm v_1 \tag{315}$$

$$(1-\sqrt{q})^{\left(\frac{N+1}{2}\right)} = \pm v_1 \qquad \text{where ;} \tag{316}$$

$$v_1 \in Z_{\mathcal{P}}^* \quad \text{and} \quad (v_1)^2 = (1-q) \mod \mathcal{P} \tag{317}$$



The three preceding relations yield

$$(1-q)^{\left(\frac{N+1}{2}\right)} = (v_1)^2 = (1-q) \mod \mathcal{P} \quad \Rightarrow \tag{318}$$

$$(1-q)^{\left(\frac{N-1}{2}\right)} = 1 \mod \mathcal{P} \tag{319}$$

which completes the demonstration of all Relations in Box N_A $\quad \square$.

**Box N_B**

Note that this box corresponds to **case (ii)** wherein; $v_1 = 0$.
Accordingly; plug in $v_1 = 0$ in Eqns ([301](#)) and ([302](#)); respectively; to obtain

$$(1+\sqrt{q})^{\left(\frac{N+1}{2}\right)} = w_1\sqrt{q} \tag{320}$$

$$(1-\sqrt{q})^{\left(\frac{N+1}{2}\right)} = -w_1\sqrt{q} \qquad \text{where ;} \tag{321}$$

$$w_1 \in Z_{\mathcal{P}}^* \quad \text{and} \tag{322}$$

$$(\pm w_1\sqrt{q})^2 = (w_1)^2 q = (1-q) \mod \mathcal{P} \tag{323}$$

The four preceding Eqns. lead to

$$(1-q)^{\left(\frac{N+1}{2}\right)} = -((w_1)^2 q) = -(1-q) \mod \mathcal{P} \quad \Rightarrow \tag{324}$$

$$(1-q)^{\left(\frac{N-1}{2}\right)} = -1 \mod \mathcal{P} \tag{325}$$

which completes the demonstration of all Relations in Box $N\_B$ $\quad \square$





We state and prove a series of claims that culminate into the proofs.

**<u>Claim 1</u> : The "at least one canonical prime divisor"** $\mathcal{P}$ **of** $N$ **must be**
**an integer of the same form** $(8k+3)$ **as** $N$

**Proof :** Note that for Category_1 , both "$-1$" as well as "2" are **QNR** s w.r.t. $N$

$-1$ is a **QNR** $\Rightarrow$ **Jacobi_Symbol**$(-1,N) = -1$ $\Rightarrow$ $N \mod 4 = 3$ $\qquad$ (326)

$\therefore$ **Jacobi_Symbol**$(-1,\mathcal{P}) = -1$ $\Rightarrow$ $\mathcal{P} \mod 4 = 3$ and $\qquad$ (327)

**Jacobi_Symbol**$(-1,\mathcal{M}) = +1$ $\Rightarrow$ $\mathcal{M} \mod 4 = 1$ $\qquad$ (328)

Further,

$\because$ **Jacobi_Symbol**$(2,N) = -1$ $\therefore$ 2 is a **QNR** Modulo-$N$ $\therefore$ $q = 2$ and $\qquad$ (329)

$(N \mod 8) = 3$ or $5$ $\qquad$ (330)

However ; $\because$ $(N \mod 4) = 3$ $\therefore$ $(N \mod 8) \neq 5$ $\qquad$ (331)

$\therefore (N \mod 8) = 3$ $\Rightarrow$ $N = (8I + 3)$ for some integer $I$ $\qquad$ (332)

Note that for the category we are considering;

$q = 2$ $\Rightarrow$ $(1 - q) = -1$ which is a **QNR** $\qquad$ (333)

Therefore Level 1 square roots in the $(N + 1)$ meta-path go thru box N_B in Figure **11** ; wherein; Eqn. (**323**) implies that

$$(\textbf{Jacobi\_Symbol}(w_1,\mathcal{P}))^2 \times \textbf{Jacobi\_Symbol}(q,\mathcal{P}) = \textbf{Jacobi\_Symbol}(1-q,\mathcal{P}) \Rightarrow (334)$$
$$(\pm 1)^2 \cdot \textbf{Jacobi\_Symbol}(q,\mathcal{P}) = \textbf{Jacobi\_Symbol}(1-q,\mathcal{P})$$
plug in $q = 2$ in the preceding Eqn. to obtain
$$\textbf{Jacobi\_Symbol}(2,\mathcal{P}) = \textbf{Jacobi\_Symbol}(-1,\mathcal{P}) = -1 \quad (335)$$

This last relation (i.e., Eqn. (**335**) ); together with Eqn. (**327**) ; implies that

$\mathcal{P} \mod 8 = 3$ $\Rightarrow$ $\mathcal{P} = (8L + 3)$ for some integer $L$ $\quad \square$ $\qquad$ (336)

Next, note that Eqn. (**332**) implies that

$\left( \dfrac{N-1}{2} \right)$ is odd $\qquad$ (337)

and

$\left( \dfrac{N+1}{2} \right)$ is even $\Rightarrow$ $\qquad$ (338)

$\left( \dfrac{N+1}{4} \right)$ is an integer $\qquad$ (339)



Moreover,

$$\because \quad N = 8I + 3 \qquad \text{see Eqn. (332)}$$

$$\therefore \left( \frac{N+1}{4} \right) = \left( \frac{8I+4}{4} \right) = (2I+1) \quad = \quad \textbf{an odd number} \tag{340}$$

Likewise, Eqn. (336) implies that

$$\left( \frac{\mathcal{P}-1}{2} \right) \text{ is odd} \tag{341}$$

and

$$\left( \frac{\mathcal{P}+1}{2} \right) \text{ is even} \quad \text{and} \tag{342}$$

$$\left( \frac{\mathcal{P}+1}{4} \right) \textbf{ is an odd integer} \tag{343}$$

**<u>Claim 2</u> : Relations analogous to Eqn. (341) thru Eqn. (343) hold for all odd powers of $\mathcal{P}$**

**proof :** Note that

$$[\mathcal{P}]^d = (8L+3)^d \tag{344}$$

Expand the r.h.s. using binomial theorem to obtain

$$[\mathcal{P}]^d = 8(JJ) + [3]^d \quad \text{wherein ;} \quad JJ = \text{ some integer} \tag{345}$$

$$\Rightarrow$$

$$\left( [\mathcal{P}]^d - 1 \right) \quad \Rightarrow \quad 8(JJ) + (3^d - 1) \tag{346}$$

and, note that

$$(3^d - 1) = (3-1) \times \left( \underbrace{[3]^{(d-1)} + [3]^{(d-2)} + \cdots [3]^2 + [3]}_{\text{sum of an even number of odd values = even num}} + 1 \right) \tag{347}$$

$$= 2 \times \text{odd\_number} \tag{348}$$

$$\Rightarrow \quad \frac{\left( [\mathcal{P}]^d - 1 \right)}{2} = [4(JJ) + \text{odd\_number}] = \text{an odd\_number} \qquad \square$$

It then follows that

$$\frac{\left( [\mathcal{P}]^d - 1 \right)}{2} + 1 = \frac{\left( [\mathcal{P}]^d + 1 \right)}{2} \quad \text{must be an even number} \tag{349}$$

Finally, **<u>Lemma 8-a</u>** and **<u>Lemma 8-b</u>** immediately imply that

$$\frac{\left( [\mathcal{P}]^d + 1 \right)}{4} \quad \text{must be an odd number} \qquad \blacksquare$$



**<u>Claim 3</u> :** $\mathcal{M}_\mathcal{P}$ **is an integer of the form** $(8J+1)$

**proof :** Relation (215) ; together with Eqns. (329) and (335) yields

$$\texttt{Jacobi\_Symbol}(2, \mathcal{M}_\mathcal{P}) = 1 \tag{350}$$

which, together with Eqn. (328) implies

$$\mathcal{M}_\mathcal{P} = (8J+1) \qquad \text{for some integer } J \qquad \square \tag{351}$$

---

The preceding claim imposes no further restrictions on $J$; thereby raising the possibility that an arbitrary power of 2 could divide the integer $J$[35] in the preceding Equation.

Or equivalently, in other words, the preceding claim does not impose any restriction on the length of a $\mathcal{M}$ meta-path (which must at rooted at some $\mathcal{M}-1$ value) in the Category under consideration.

The next claim demonstrate that this is not true; that there are very specific restrictions on any $\mathcal{M}$ meta_path in this category.

Before proceeding any further; An extremely important point must be clarified: In all the figures as well as all the derivations;
the word/term "Level" is the depth in the square-root-chain in any meta-path.

For the $\mathcal{M}$-meta paths "`Level_0`" ($\mathcal{M}$_Top) starts at the exponent $E_0 = (\mathcal{M}-1)$, and the value of the expression; $\mathcal{X} = (1 + \sqrt{q})^{\mathcal{M}-1} = 1 \mod \mathcal{P}$ (from Eqn. (245) )

Accordingly, the next level (first level of square root-chain) =
`Level_1` has the exponent $\frac{\mathcal{M}-1}{2}$;
`Level_2` = the second level has the exponent $\frac{\mathcal{M}-1}{4}$;
`Level_3` = the third level has the exponent $\frac{\mathcal{M}-1}{8}$;

$\vdots$ and so on.

Let $L_\mathcal{M}$ denote the length of the meta_path corresponding to $\mathcal{M}$.
Then, we derive the constraints that bootstrap the $\mathcal{M}$_meta_path for the special case under consideration;

**<u>Claim 4</u> : For** $N$ **in Category_1,**

   **If** $L_\mathcal{M} < 3$ **then all** $\mathcal{M}$**_paths lead to contradictions, and we are done.**

**Proof :**

Since $\mathcal{M}-1$ is an even number; then $L_\mathcal{M} \geq 1$.

Case 1 : Suppose $L_\mathcal{M} = 1$

---

[35]of course the power must be $< \lg N$



Then $\frac{\mathcal{M}-1}{2}$ is an odd number and

$$\left(1+\sqrt{2}\right)^{\frac{\mathcal{M}-1}{2}} = \pm 1 \quad \text{and} \quad \left(1-\sqrt{2}\right)^{\frac{\mathcal{M}-1}{2}} = \pm 1$$

so that multiplying the 2 preceding relations yields

$$(1-2)^{(\text{odd\_number})} = -1 = (\pm 1)^2 = +1 \qquad \text{Which is absurd} \qquad \square$$

Case 2 : Suppose $L_\mathcal{M} = 2$

Then $\frac{\mathcal{M}-1}{4}$ is an odd number. Note that at Level 1 if the square-root selected is +1; then we obtain the same contradiction as in the preceding case and we are done.

Therefore, the square-root at Level-1 must be a $-1$.
Then

$$\left(1+\sqrt{2}\right)^{\frac{\mathcal{M}-1}{4}} = \pm y_2\sqrt{2}$$

and

$$\left(1-\sqrt{2}\right)^{\frac{\mathcal{M}-1}{4}} = \mp y_2\sqrt{2}$$

wherein

$$(\pm y_2\sqrt{2})^2 = \left(1+\sqrt{2}\right)^{\frac{\mathcal{M}-1}{2}} = -1 \quad \text{multiplying the 2 relations preceding the previous line yields}$$

$$(1-2)^{(\text{odd\_number})} = -1 = (\pm 1) \times (\mp 1)(y_2\sqrt{2})^2 = (-1) \times (-1) = +1 \qquad \text{Which is absurd} \qquad \square$$

Therefore $L_\mathcal{M} \geq 3$.



**<u>Claim 5</u>** : **There may be a choice $\mathcal{P}$ of canonical prime such that considering relations modulo only that one $\mathcal{P}$ is NOT sufficient to derive the contradictions necessary to complete the proof.**

**Proof** : **We show an actual counter example (from G.E. Pinch's lists of all Carmichaels $< 10^{19}$) :**

$$N_1 = 288507171198331 = 59 \times 211 \times 307 \times 1531 \times 49307 = P_1 \times P_2 \times P_3 \times P_4 \times P_5 \tag{352}$$

wherein; both $-1$ and $2$ are $\mathtt{QNR}$ s w.r.t. each prime factor $P_i$ ; $i = 1..5$;
and $q = 2$ satisfies the Euler Criterion modulo-$N$.

**Therefore, as per the <span style="color:green">Baseline Primality Conjecture</span> ;**

$$
\begin{aligned}
\mathtt{BCC}(2, N_1) \quad &\overset{\Delta}{\equiv} \quad \left[ (1 + \sqrt{2})^{N_1} - 1 - (\sqrt{2})^{N_1} \right] \mod N_1 \\
&= \quad 105145291046221 + 258732005897941\,\sqrt{2} \quad \neq \quad 0 \mod N_1
\end{aligned}
\tag{353}
$$

**thereby demonstrating/proving that $N_1$ is a composite and that the conjecture works even in all such super specific and critical cases we have found and tested to date.**

However, let the "<span style="color:green">**at least one canonical prime divisor of $N_1$**</span>" be
$\mathcal{P} = P_2 = 211 =$ the second prime factor.

Then, it can be verified that

$$\left[ (1 + \sqrt{2})^{N_1} - 1 - (\sqrt{2})^{N_1} \right] \mod P_2 \quad = \quad 0 \mod P_2 \qquad \square \tag{354}$$

$\star\star$ This was what I had always suspected: analysis modulo a single prime divisor
is unlikely to be sufficient to complete the proof.$\star\star$



However, including analysis of paths modulo-$\mathcal{M}$ and its prime divisors
leads to contradictions in a substantial number of cases that fall under the
green colored box drawn in dashed-linestyle in **Figure 1** wherein;
$N$ is assumed to have at least 2 distinct prime divisors.

To that end, note that as per relation (328) `Jacobi_Symbol(`$-1,\mathcal{M}$`)` is $+1$ ;
AND from Eqn. (350) `Jacobi_Symbol(`$2,\mathcal{M}$`)` is $+1$

**Therefore, there are two cases to consider** [36]

⟨**a**⟩ **At least one prime divisor of $\mathcal{M}$ has at least one Jacobi-Symbol value mismatch.**
**This case is represented by the blue colored box drawn in solid line-style in Figure 1**

**or**

⟨**b**⟩ **All prime divisors of $\mathcal{M}$ have the same values of Jacobi-Symbols w.r.t. $-1$ AND 2; as the**
**corresponding values of Jacobi-Symbols w.r.t. $N$.**
**This case is corresponds to the red colored box drawn with dash-dots line-style in Figure 1**

---

[36] the third case where is $N$ is a power of a single prime has not been analytically proved yet. Accordingly, that Box is outside the dashed Green colored boundary in Figure 1 .



We handle **case ⟨a⟩** first

## § Section 32.1 : Proof that no prime with at least one Jacobi_Symbol mismatch can divide $N$

Since $N$ is in Category_1; then $q = 2$ and therefore the Jacobi_Symbol mismatch is either for $-1$ or 2 or both. Next couple of claims rule out any prime divisors with any Jacobi_Symbol Mismatches.

First we show that a mismatch of the Jacobi_Symbol w.r.t. $-1$ is not possible.

**Claim 6 :**    **Let $N$ be in Category_1.**

         **Then, no prime divisor $P$ of $\mathcal{M}$ can have `Jacobi_Symbol(`$-1, P$`)` $= +1$**

**Proof by contradiction :**    Assume that

`Jacobi_Symbol(`$-1, P$`)` $= +1 \Rightarrow -1$ is a `QR` w.r.t. $P$

$$\Rightarrow \ \exists \ \text{an integer } s \ \text{such that} \ s^2 = -1 \mod P \tag{355}$$

Plug-in $q = 2$ in Eqn. (220) ; and replace the modulus $N$ with $P$ (this is possible due to **Lemma 4** , **Lemma 5** , and **Lemma 6** , and ); to obtain

$$[1 + \sqrt{2}]^{(N+1)} = (1 - \sqrt{2}) \cdot (1 + \sqrt{2}) = (1 - 2) = -1 \mod P \tag{356}$$

and in an analogous manner, the corresponding "conjugate" relation

$$[1 - \sqrt{2}]^{(N+1)} = -1 \mod P \tag{357}$$

Since $(N+1)$ is even; and $-1$ is a `QR` (Quadratic-Residue) the square root of Relation (356) yields

$$\left(1 + \sqrt{2}\right)^{\left(\frac{N+1}{2}\right)} = s \mod P \tag{358}$$

where ; $s$ is an integer that satisfies $s^2 = -1 \mod P$ $\tag{359}$

The corresponding conjugate of the above relation is

$$\left(1 - \sqrt{2}\right)^{\left(\frac{N+1}{2}\right)} = s \mod P \tag{360}$$

Multiply Eqn. (358) and Eqn. (360) to obtain

$$(1 - 2)^{\left(\frac{N+1}{2}\right)} = s^2 = -1 \tag{361}$$

**The preceding relation is absurd since $\frac{N+1}{2}$ is even; as per Relation (338)**    □



**In other words, we have demonstrated that even though the Jacobi-Symbol**
`Jacobi_Symbol(-1,M) = +1;`
$\mathcal{M}$ **cannot have any prime divisor of the form** $(8i+1)$
   **AND**
$\mathcal{M}$ **cannot have any prime divisor of the form** $(8i+5)$
**(because for primes of any of those two forms; their remainder w.r.t. 4 is +1 which in turn implies that** $-1$ **is a** `QR` **w.r.t. that prime; which is ruled out by the preceding claim)**[37].

**That leaves only one single sub-case yet to be proved : sub-case** $\underline{\langle \mathbf{b-2} \rangle}$ **; which is equivalent to proving that a prime of the last possible remaining form** $(8i+7)$ **cannot divide** $\mathcal{M}$ **either. The following claim takes care of that case.**

**<u>Claim 7</u> :     Let** $N$ **be in Category_1.**
**                    Then, no prime** $P$ **of the form** $(8i+7)$ **can divide** $N$**.**

**Proof**[38] **:**

Let $N$ be of Category_1, so that some prime $P = 8I + 3$ divides $N$ by **<u>Claim 1</u>** . Suppose that some prime $Q = 8J + 7$ also divides $N$. Since $N$ is odd, it has only odd prime divisors, which must be of the form $8K + r$ for remainders $r \in \{1, 3, 5, 7\}$. Further, as per the preceding claim ; since $N$ has no prime divisors of the forms $8K + 1$ or $8K + 5$, all prime divisors of $N$ must take one of the forms $8K + 3$ or $8K + 7$. We therefore write

$$N = T \times S,$$

where

$$T = P_1^{t_1} \times P_2^{t_2} \times \cdots \times P_m^{t_m}, \quad S = Q_1^{s_1} \times Q_2^{s_2} \times \cdots \times Q_k^{s_k},$$

each $P_i = 8I_i + 3$, each $Q_j = 8J_j + 7$, and the existences of $P$ and $Q$ imply that $m$ and $k$ are both at least 1. We set

$$t = \sum_{i=1}^{m} t_i \quad \text{and} \quad s = \sum_{j=1}^{k} s_j,$$

so that $T$ is the product of $t$ factors of the form $8K + 3$ and $S$ is the product of $s$ factors of the form $8K + 7$. Accordingly, the Lemmas 8 imply that the parities of $t$ and $s$ determine the forms of $T$ and $S$, respectively ; which in turn determine the form of $N$, as shown in Table 14.

Since $N$ is of Category_1, it is of the form $8K + 3$, and therefore as per row 3 in Table 14 ;
 $t$ must be odd, $s$ must be even  so that
$T$ is of the form $8K + 3$, and
$S$ is of the form $8K + 1$.

---

[37]This part of the proof was completed as of 31st May 2018
[38]this proof was completed on the 14-th of April, 2019



| Parity of $t$ | Parity of $s$ | Form of $T$ | Form of $S$ | Form of $N$ |
|---|---|---|---|---|
| even | even | $8K+1$ | $8K+1$ | $8K+1$ |
| even | odd | $8K+1$ | $8K+7$ | $8K+7$ |
| odd | even | $8K+3$ | $8K+1$ | $8K+3$ |
| odd | odd | $8K+3$ | $8K+7$ | $8K+5$ |

**Table 14.** Parities of $t$ and $s$ determine the forms of $T$ and $S$ and therefore the form of $N$ as illustrated in this Table.

Now, we consider

$$\Psi_{Q_k} = \frac{N}{Q_k^{s_k}} = T\frac{S}{Q_k^{s_k}} = TS^*,$$

where $S^* = S/Q_k^{s_k}$ is the product of $(s - s_k)$ factors of the form $8K+7$, and therefore itself takes one of the forms $8K+1$ or $8K+7$. Since $T$ is of the form $8K+3$,
$\Psi_{Q_k}$ therefore takes one of the two forms $8K+3$ or $8K+5$.
In either case $L_{\mathcal{M}_{Q_k}} < 3$ which in turn results in contradictions as per **Claim 4** . $\qquad\square$

**The preceding 2 claims demonstrate that no prime with a Jacobi_Symbol mismatch can divide $N$.**

That brings us to the remaining part of the proof space under Category_1 and under "distinct primes with matching Jacobi_Symbols. Note that this part of the proof space corresponds to **Case ⟨b⟩** mentioned above.





**Claim 8 :**  **Let P be a prime with matching Jacobi_Symbols that divides** $N$**. Then the power of**
**P in the prime factorization of** $N$ **cannot be an even number.**

**Proof :**

Split the prime factorization of $N$ into two groups :
Group_1 : the powers of all of the prime-bases in the factorization of $N$ are odd
Group_2 : the powers of all of the prime-bases are even

In other words; let

$$N = \overbrace{\left\{ [P_{o_1}]^{(2i_1+1)} \times [P_{o_2}]^{(2i_2+1)} \times \cdots \times [P_{o_m}]^{(2i_m+1)} \right\}}^{\text{Group\_1}} \times$$

$$\underbrace{\left\{ [P_{e_1}]^{(2j_1)} \times [P_{e_2}]^{(2j_2)} \times \cdots \times [P_{e_k}]^{(2j_k)} \right\}}_{\text{Group\_2}}$$

$$(362)$$

The key point to note is that since the Jacobi_Symbol values of 2 as well as $-1$  w.r.t  $N$ are both $-1$ ;
**then the number of distinct primes** $m$ **in Group_1 must be** **odd**.

Consider the case where
Group_2 is non-empty    $\Rightarrow$     at least one base prime exists in Group_2.

Arbitrarily pick any one base-prime in Group_2 and let it be denoted by the symbol $P_e$.
Then

$$\mathcal{M}_{P_e} = \text{(product of all factors of } N \text{ that are co-prime w.r.t. } P_e)$$
$$= \{\text{product of all terms in Group\_1}\} \times \{\text{A\_square}\} \qquad (363)$$

Then using **Lemmas 8**,   obtain

$$\text{product of all terms in Group\_1} = (8I+3) \quad \text{for some integer } I \qquad (364)$$
$$\text{and}$$
$$\{\text{A\_square}\} = (8J+1) \quad \text{for some integer } J \qquad (365)$$

Plug-in the expressions from Eqns. (364) and (365) into  Eqn. (363)  to obtain

$$\mathcal{M}_{P_e} = (8\mathbf{K}+3) \quad \text{for some integer } \mathbf{K} \quad \Rightarrow \qquad (366)$$
$$L_{\mathcal{M}_{P_e}} < 3 \qquad (367)$$



**The preceding relation implies that we can obtain contradictions as per [Claim 4](#) .** ☐



# ACKNOWLEDGMENTS


The authors would like to thank Professors Ronald Rivest and Andrew Sutherland from the Massachusetts Institute of Technology for their encouraging feedback on an early/old version in the first week of February 2018. That version still had "Matrices" in it and was definitely incomplete, vague and potentially actually wrong. If the early feedback were to be negative, I might have simply given up.

We would also like to thank our colleague and Adjunct Professor, Dr. David DeLatte for his comments also an a very early version that still had matrices in it.

We acknowledge the sanity-checks (on the complete draft) by my former colleague and friend, Prof. Mircea Stan from the ECE Dept. at the Univ. of Virginia at Charlottesville (UVA).

Finally, the authors would like to thank our colleagues Professors Kostas Kalpakis and Ryan Robucci for stylistic comments.

Special thanks are due to Prof. Ryan Robucci for letting us use the machines in his labs for our numerical experiments. He also helped with software packages (installation/maintenance/upgrades etc.). Most important: he helped us wade through some very involved, obscure problems with LaTeX(even more than the number theory conjectures; those last minute document formatting/compiling issues turned out to be the hardest problems; I think that those were NP-hard !!!)